\documentclass[11pt,a4paper,twoside,openright]{report}

\usepackage{amsfonts,amsmath,amssymb,caption,enumerate,float,graphicx,setspace}
\usepackage{epsfig}
\usepackage{siunitx}
\usepackage{acronym}
\usepackage{fancyhdr}
\graphicspath{{figures/}}

\usepackage{url,array,csquotes,multirow,tabularx}
\usepackage{booktabs}

\usepackage[usenames,dvipsnames]{xcolor}   
\usepackage{float}      
\usepackage{slashed} 
\usepackage{empheq}
\newcommand{\ditto}{\raisebox{-0.5ex}{''}}
\usepackage{pdfpages}
\usepackage{wrapfig} 
\usepackage{enumitem}
\usepackage{comment}
\usepackage{tikz-feynman}
\tikzfeynmanset{compat=1.0.0}
\usepackage[theorems,skins]{tcolorbox}
\newtcolorbox{mymathbox}[1][]{colback=white, sharp corners, #1}

\newtcbox{\othermathbox}[1][]{nobeforeafter, math upper, tcbox raise base, enhanced, sharp corners, colback=black!10, colframe=red!30!black, drop fuzzy shadow, left=1em, top=2em, right=3em, bottom=4em}

\usepackage{longtable}
\usepackage{lscape}
\usepackage{geometry}
\usepackage{dsfont}
\usepackage[times]{quotchap}

\usepackage[nottoc]{tocbibind}
\usepackage[titles]{tocloft}
\makeatletter
\newlength{\@chapterlength} 
\newcommand\@chapterheadsmark{1}
\ifnum \@chapterheadsmark = 1
\fi

\def\l@section{\@dottedtocline{1}{\@chapterlength}{2.3em}}

\newlength{\@subsectionlength}
\def\l@subsection{\@dottedtocline{2}{\@subsectionlength}{3.2em}}

\newlength{\@subsubsectionlength}
\def\l@subsubsection{\@dottedtocline{3}{\@subsubsectionlength}{0em}}

\makeatother

\renewcommand\cftchappresnum{\chaptername~}
\newlength\mylength
\usepackage{afterpage}
\usepackage[acronym,nomain,toc]{glossaries}
\makeglossaries
\usepackage[intoc]{nomencl}
\makenomenclature

\usepackage[toc,page]{appendix}

\usepackage{datetime}
\newdateformat{thisdate}{%
	\twodigit{\THEDAY}}
\newdateformat{thismonthdigit}{%
	\twodigit{\THEMONTH}}
\newdateformat{thismonth}{%
	\monthname[\THEMONTH]}
\newdateformat{thisyear}{%
	\THEYEAR}
 
\providecommand{\keywords}[1]
{
    \large
    \textbf{\textit{Keywords---}} #1
}

\usepackage[backend=biber, sorting=none,
            defernumbers=false, style=numeric-comp, doi=true]{biblatex}
\usepackage{hyperref}

\hypersetup{linktoc=all, colorlinks=true, linkcolor=blue, citecolor=teal, urlcolor=cyan}

\newcommand{\beqa}{\begin{eqnarray}}
\newcommand{\eeqa}{\end{eqnarray}}
\newcommand{\be}{\begin{equation}}
\newcommand{\ee}{\end{equation}}
\newcommand{\ba}{\begin{array}} 
\newcommand{\ea}{\end{array}}
\usepackage{color}
\usepackage{graphicx}
\usepackage{textcomp}
\usepackage{amsmath,amssymb,array}
\usepackage{enumerate}
\usepackage{hhline}
\newcommand{\bds}{$B^0_{d,s}$-$\overline{B}^0_{d,s}$\,}

\newcommand{\dd}{$D^0$-$\overline{D}^0$\,}
\newcommand{\nn}{$n$-$\overline{n}$\,}
\usepackage{type1cm}
\usepackage{lettrine}
\allowdisplaybreaks
\newcommand{\nl}{\nonumber \\}
\newcommand{\cc}{\, {\mathcal{C}}^{-1}\,}
\newcommand{\hc}{\;+\; \rm{H.c.}}
\newcommand{\so}{$SO(10)$}
\newcommand{\lag}{\mathcal{L}}
\newcommand{\fs}{\mathbf{16}}
\newcommand{\ft}{\mathbf{10}}
\newcommand{\ff}{\mathbf{\overline{5}}}
\newcommand{\fn}{\mathbf{1}}
\usepackage{pifont}

\newcommand{\xmark}{\ding{55}}
\newcommand{\smg}{$SU(3)\,\times\,SU(2)\,\times U(1)$}
\newcommand{\sm}{$SU(3)_{\mathrm{C}}\,\times\, SU(2)_{\mathrm{L}}\,\times\,U(1)_{\mathrm{Y}}$}
\usepackage{pdfpages}
\newcommand{\eq}{&=&}
\newcommand{\ad}{&+&}
\newcommand{\mi}{&-&}
\newcommand{\hcn}{&+& \rm{H.c.}}

\newcommand{\hh}{{\mathrm{H}}}
\newcommand{\su}{$SU(5)$}
\newcommand{\vv}{{\mathrm{V}}}
\usepackage{xfrac}
\usepackage[sc,noBBpl]{mathpazo}

\newcommand{\mh}{M^2_{\hat{H}}}

\newcommand{\md}{M^2_{\hat{\Delta}}}

\usepackage{Zallman}

\begin{document}

\pagenumbering{gobble}
\begin{titlepage}
\begin{center}

{\LARGE \bfseries Unravelling the Scalar Sector of Grand Unification: \\[0.5ex] Phenomenology \& Implications}\\
\vspace*{1cm}

{\large \bfseries \itshape A thesis submitted 
in partial fulfilment of the requirements\\ for the award of the degree of}

{\fontencoding{T1}\fontfamily{pzc}\fontseries{m}\fontshape{n}\selectfont
\huge \bfseries Doctor of Philosophy}\\
\vspace*{0.5cm}

{\large \itshape {by}}\\
\vspace*{0.5cm}

{\Large \bfseries Saurabh Kumar Shukla  \\}
\vspace*{0.1cm}
(Roll No. 19330018)\\
\vspace*{0.5cm}

{\large Under the guidance of\\}
\vspace*{0.5cm}

{\Large \bfseries Dr. Ketan M. Patel\\}
\vspace*{0.1cm}
{\large Theoretical Physics Division\\
\vspace*{0.1cm}
Physical Research Laboratory, Ahmedabad, India\\}


\vspace*{1.0cm}

\includegraphics[scale=0.07]{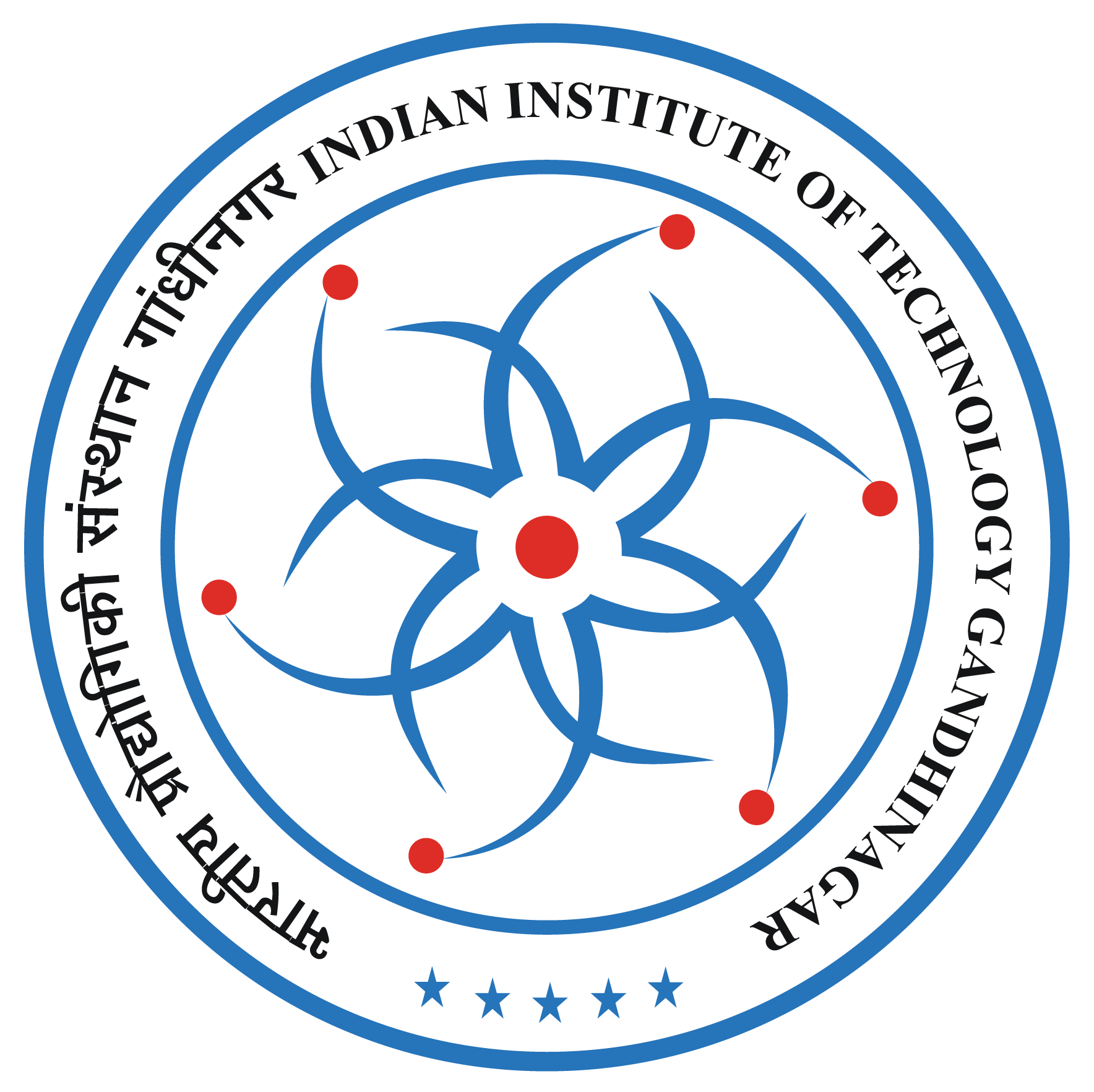}
\vspace*{0.5cm}

\textsc{\Large \bfseries{Department of Physics\\
Indian Institute of Technology Gandhinagar\\
}}
\vspace*{1cm}
{\Large \bfseries September 2024\\}
\vspace*{1cm}

\end{center}
\end{titlepage}

\newpage\null\newpage

\newpage



\graphicspath{{02_Approval/}}
\vspace*{0.08in}
\begin{center}
    \begin{huge}    
    \textbf{THESIS APPROVAL}\\[-1.6ex]
    \rule{7.5cm}{1pt}\\[-1.2ex]\rule{7.5cm}{1pt}
    \end{huge}
\end{center}
\vspace{0.5in}

\noindent Certified that the thesis entitled\\
\begin{center}
\textit{\textbf{Unravelling the Scalar Sector of Grand Unification:\\ Phenomenology \& Implications}},\\
\end{center}
submitted by \textbf{Saurabh Kumar Shukla} (Roll no.: \textbf{19330018}), to the \textbf{Indian Institute of Technology Gandhinagar}, is approved for the degree of \textit{Doctor of Philosophy}. \\

\noindent Date: September 10, 2024\hfill Place: Ahmedabad, Gujarat 

\vspace*{25pt}
\begin{center}
    \begin{minipage}[t]{0.35\textwidth}
    \begin{flushleft}
        \hrule\vspace{2ex}
        {\small {\bfseries Prof. Sudhir K. Vempati}\\
        External Examiner\\
        Centre for High Energy Physics\\
        Indian Institute of Science\\
        Bengaluru, India}
    \end{flushleft}
    \end{minipage}
    \hfill
    \begin{minipage}[t]{0.34\textwidth}
    \begin{flushright}
        \hrule\vspace{2ex}
        {\small {\bfseries Dr. Baradhwaj Coleppa}\\
        Member, FDC\\
        Physics Department\\
        Indian Institute of Technology\\
        Gandhinagar, India }
    \end{flushright}
    \end{minipage}
    \hfill
\end{center}
\vspace{1cm}
\begin{center}
    \begin{minipage}[t]{0.32\textwidth}
    \begin{flushleft}
        \hrule\vspace{2ex}
        {\small {\bfseries Prof. Srubabati Goswami}\\
        Member, DSC\\
        Theoretical Physics Division\\
        Physical Research Laboratory\\ Ahmedabad, India}
    \end{flushleft}
    \end{minipage}
    \hfill
    \begin{minipage}[t]{0.32\textwidth}
    \begin{center}
        \hrule\vspace{2ex}
        {\small {\bfseries Prof. Namit Mahajan}\\
        Member, DSC\\
        Theoretical Physics Division\\
        Physical Research Laboratory\\ Ahmedabad, India}
    \end{center}
    \end{minipage}
    \hfill
    \begin{minipage}[t]{0.32\textwidth}
    \begin{flushright}
        \hrule\vspace{2ex}
        {\small {\bfseries Dr. Vishal Joshi}\\
        Member, DSC\\
       Astronomy \& Astrophysics Division\\
        Physical Research Laboratory\\ Ahmedabad, India}
    \end{flushright}
    \end{minipage}
\end{center}

\vspace*{25pt}
\begin{center}
    \begin{minipage}[t]{0.4\textwidth}
    \begin{center}
        \rule{4.8cm}{0.4pt}\\
        {\small {\bfseries Dr. Ketan M. Patel}\\
        PhD Supervisor\\
        Theoretical Physics Division\\
        Physical Research Laboratory\\ Ahmedabad, India}
    \end{center}
    \end{minipage}

\end{center}

\addcontentsline{toc}{chapter}{Thesis Approval Certificate}


\newpage\null
\thispagestyle{empty}
\cleardoublepage

\pagenumbering{roman}


\vspace*{0.1in}
\begin{center}
    \begin{huge}
           
    \textbf{DECLARATION}\\[-1.6ex]
    \rule{5.8cm}{1pt}\\[-1.2ex]\rule{5.8cm}{1pt}
    \end{huge}
\end{center}
\vspace{0.8in}
I declare that this written submission represents my ideas in my own words. Where others' ideas or words have been included, I have adequately cited and referenced the original sources. I also declare that I have adhered to all principles of academic honesty and integrity and have not misrepresented, fabricated, or falsified any idea/data/fact/source in my submission. I understand that any violation of the above will cause disciplinary action by the Institute and can also evoke penal action from the sources which have thus not been properly cited or from whom proper permission has not been taken when needed.

\vspace{1in}


\noindent
\begin{minipage}[t]{0.35\textwidth}
    Date: September 10, 2024\\
    Place: Ahmedabad, Gujarat
\end{minipage}%
\hfill
\noindent
\begin{minipage}[t]{0.4\textwidth}
\begin{flushright}
    \bfseries
    \rule{4.7cm}{1pt}\\
	Saurabh Kumar Shukla \\
	IIT Gn Roll No.: 19330018
\end{flushright}
\end{minipage}
\addcontentsline{toc}{chapter}{Declaration}
\newpage
\thispagestyle{empty}
\null\newpage





\vspace*{0.1in}
\begin{center}
    \begin{huge}
           
    \textbf{CERTIFICATE}\\[-1.6ex]
    \rule{5cm}{1pt}\\[-1.2ex]\rule{5cm}{1pt}
    \end{huge}
\end{center}
\vspace{0.8in}


\noindent It is certified that the work contained in the thesis titled\\

\begin{center}
\textit{\textbf{Unravelling the Scalar Sector of Grand Unification:\\ Phenomenology \& Implications,}}\\
\end{center}

\noindent by \textbf{Saurabh Kumar Shukla} (Roll no: \textbf{19330018}), has been carried out under my supervision, and this work has not been submitted elsewhere for a degree.

\noindent I have read this dissertation, and in my opinion, it is fully adequate in scope and quality as a dissertation for the degree of Doctor of Philosophy.

\vspace{1in}
\noindent
\begin{minipage}[t]{0.35\textwidth}
    Date: September 10, 2024 \\
    Place: Ahmedabad, Gujarat
\end{minipage}%
\hfill
\noindent
\vspace{1cm}
\begin{minipage}[t]{0.55\textwidth}
    \raggedleft
    \rule{5cm}{1pt}\\
    \textbf{Dr. Ketan M. Patel} \\
    PhD Supervisor\\
	Theoretical Physics Division\\
	Physical Research Laboratory\\ 
 Ahmedabad, Gujarat, India \\
	
\end{minipage}

\addcontentsline{toc}{chapter}{Certificate}

\newpage
\thispagestyle{empty}
\null\newpage

\addcontentsline{toc}{chapter}{Acknowledgements}
\begin{center}
    \begin{huge}
           
    \textbf{Acknowledgements}\\[-1.6ex]
    \rule{6.5cm}{1pt}\\[-1.2ex]\rule{6.5cm}{1pt}
    \end{huge}
\end{center}
\vspace{0.5in}

\sloppy
\textit{Nothing in this universe can be an outcome of a solo act (except perhaps its creation!). This work is a result of the collective support, guidance, and encouragement I have received over the years. The entire lustrum of the PhD has been an adomanic journey for me, and I am deeply grateful to all those who have accompanied me on this path.}

\textit{I want to start by expressing my profound gratitude to Dr. Ketan M. Patel for accepting me as his PhD student. His steadfast commitment, insightful advice, and vast knowledge have been invaluable and enriched my academic pursuits. He has always provided me with academic freedom and constantly motivated me to be an independent researcher. His approachability and readiness to assist have been immensely reassuring. If I can embody even a fraction of his dedication to work and respect for time, it would greatly enhance my future endeavours.}

\textit{My sincere thanks goes to the esteemed members of my DSC committee. Prof. Namit Mahajan has been a constant source of encouragement and motivation throughout these years. His insightful advice, helpful nature, and thought-provoking questions have been invaluable. Prof. Srubabati Goswami provided unwavering support, empathy, and invaluable feedback, always guiding me with her expertise. Additionally, I am grateful to Dr. Vishal Joshi for his continuous support and kindness. All three have greatly enriched my understanding of the subject.}

\textit{I would like to acknowledge the various courses taught by Prof. Dilip S. Angom, Prof. Jitesh R. Bhatt, Prof. Partha Konar, Prof. Varun Sheel, Prof. Navinder Singh, Dr. Arvind Singh, and Prof. Umashankar Singh at PRL and IIT Gandhinagar. Their teachings have positively contributed to my knowledge and understanding. Additionally, I appreciate the engaging interactions with Prof. Hiranmaya Mishra, whose sense of humour and insights have been a delightful addition to my academic experience. Furthermore, I would like to express my gratitude to Dr. Vineet Goswami, Dr. Amitava Guharay, and Dr. Satyajit Seth for supervising the various projects integral to my studies.  I am also thankful for the interactions with Dr. Paramita Dutta, which have been particularly pleasant.}

\textit{I extend heartfelt thanks to my school mathematics teacher, Mrs. Afsar Jahan. Her lifelong lessons helped inculcate an appreciation for mathematics and, thus, science, shaping my academic journey. I am also deeply grateful to Prof. Vivek K. Tiwari, who has always been kind and supportive in cultivating my keen interest in high-energy physics. I sincerely thank Prof. Sudhir K. Vempati for hosting me at IISc Bangalore and for his invaluable support and insightful academic discussions. I am grateful that he extended me the opportunity to interact and work with him and his group members}.

\textit{During my time at PRL, I spent countless hours with Supriya and Bharathiganesh, whose conversations greatly shaped my learning and experience. Their constant cheer and support uplifted me whenever I felt low. Discussions with Debashis and Gurucharan, whether academic or otherwise, were always engaging and valuable. Dayanand's help with academic matters was incredibly helpful, and his readiness to assist is always appreciated. The times we spent playing volleyball, cricket, and football together created unforgettable memories. Arup's humourous comments never failed to bring a smile to my face. Engaging discussions with Deepanshu at night while writing the thesis was always stimulating and expanded my horizons.}

\textit{I am also thankful to my colleagues Sourav, Anupam and Monal for their support. Furthermore, I wish to acknowledge all the other stalwarts of the theory division who have left PRL for their guidance and support throughout the years. My batchmates Gourav and Bijoy have been excellent lobby mates, providing friendship and support throughout. Enjoying several movie nights with Rithvik offered a delightful escape and helped maintain a balanced life.}

\textit{Whatever I am today or aspire to be in the future, it is solely because of my family's unwavering encouragement, enduring support, and blessings. Their contributions are beyond words. A heartfelt thanks to my adorable nieces, Anvi and Avisha, whose constant cheer and delightful laughter have brought immense joy and light into my life.}

\textit{Finally, I would like to express my sincere thanks to all the people who have directly or indirectly helped me in any form during my academic journey.} 

\vspace{35pt}
\noindent
\begin{minipage}[t]{0.35\textwidth}
    Date: \textit{September 10, 2024} \\
    Place: \textit{Ahmedabad, Gujarat}
\end{minipage}%
\hfill
\noindent
\begin{minipage}[t]{0.4\textwidth}
\begin{flushright}
    \bfseries
	\textit{Saurabh Kumar Shukla}\\
\end{flushright}
\end{minipage}


\setstretch{1.5}  
\newpage
\chapter*{\centering Abstract\\\vspace*{-1.1cm}
    \rule{3.25cm}{1pt}\\\vspace*{-1.25cm}\rule{3.25cm}{1pt}}

\addcontentsline{toc}{chapter}{Abstract}
\pagestyle{fancy}
\fancyhf{}
\fancyhead[LO,RE]{Abstract}
\fancyhead[LE,RO]{\thepage}

\sloppy

The syncretism of Standard Model (SM) gauge symmetries into a larger unified gauge symmetry often necessitates the inclusion of irreducible representations (irreps) containing numerous scalar fields. These scalar irreps are typically larger than the irreps that unify quarks and leptons within the same multiplet. Unified models based on orthogonal groups, such as \(SO(10)\), and unitary groups, such as \(SU(5)\), are suitable candidates for grand unification. \(SO(10)\) unifies all the SM fermions of one generation into a single sixteen-dimensional irrep $({\mathbf{16}})$ and encompasses \(SU(5)\) as a subgroup. Grand Unified Theories (GUTs) are well-known setups that can realistically reproduce low-energy observables like fermion masses and mixing angles. This capability makes GUTs, in both their renormalisable and non-renormalisable versions, the most appealing candidates for exploring the implications of scalars. These implications can be broadly categorised as \textit{direct} and \textit{indirect}, which we have explored in this thesis.

Irreps contributing to the renormalisable \(SO(10)\) Yukawa sector include the \(10_{\mathrm{H}}\), \(120_{\mathrm{H}}\), and \(\overline{126}_{\mathrm{H}}\) dimensional tensors, while numerous irreps can contribute to non-renormalisable Yukawa sector, the smallest being the \(16_{\mathrm{H}}\) dimensional irrep. These tensors, which contribute to both the renormalisable and non-renormalisable Yukawa sectors, consist of scalar fields in addition to the SM Higgs. These scalars are charged under \(B-L\) symmetry and act as primary sources of Baryon (B) and Lepton (L) number violations, which are forbidden in the SM. Further, these scalars are also the source of flavour $(F)$ violation.  The primary aim of this thesis is to systematically and comprehensively classify and analyse the \textit{direct phenomenological implications} of scalars, contributing to the renormalisable and non-renormalisable Yukawa interactions, in their ability to induce $B$, $L$, and $F$ violating novel processes. $B$ and $L$ violating processes affect the nucleon's stability by inducing phenomena like nucleon decays and neutral baryon-antibaryon oscillation. GUTs provide a constrained setup in which these yet-to-be-observed phenomena are related to the observed sector of fermion masses, as both emanate from the same Yukawa interactions.

We compute the SM invariant couplings of various scalars with SM fermions stemming from an \(SO(10)\) invariant renormalisable Yukawa Lagrangian. Among the $sixty$ \(B-L\) charged scalars, we identified only three, along with their conjugates, that possess diquark and leptoquark couplings necessary to induce mass-dimension six (\(D=6\)), \(B-L\) conserving effective operators at leading order. These scalars are: \(T\left(3,1,\sfrac{1}{3}\right)\), \({\cal T}\left(3,1,\sfrac{4}{3}\right)\), and \(\mathbb{T}\left(3,3,\sfrac{1}{3}\right)\), where parenthesis denotes their repsective SM charges. These scalars mediate nucleon decaying into an antilepton accompanied by mesons. Additionally, we identified scalars mediating tree level \(D=7\), \(B-L\) violating effective operators, which induce nucleon decay into a lepton together with mesons. These scalars are: $\Theta\left(3,1,\sfrac{2}{3}\right)$, $\Delta\left(3,2,\sfrac{1}{6}\right)$, and $\Omega\left(3,2,\sfrac{7}{6}\right)$. We also computed the coupling of scalars residing in $16_{\mathrm{H}}$ with SM fermions, which couples to $\mathbf{16}$-plet at the non-renormalisable level. The pair of scalars leading to effective $D=6$, $B-L$ conserving operators at leading order are: $\hat{\sigma}\left(1,1,0\right)-\hat{T}\left(3,1,\sfrac{1}{3}\right)$, $\hat{H}\left(1,2,-\sfrac{1}{2}\right)-\hat{T}$, and $\hat{H}-\hat{\Delta}\left(3,2,\sfrac{1}{6}\right)$.

In a class of realistic renormalisable \(SO(10)\) models based on \(10_{\mathrm{H}}\) and \(\overline{126}_{\mathrm{H}}\), the branching pattern of the leading two-body modes of the proton was computed. It was further argued that \(16_{\mathrm{H}}\) can mimic the role played by \(\overline{126}_{\mathrm{H}}\) in renormalisable GUTs, and the branching pattern of proton decays in a realistic non-renormalisable $SO(10)$ models. In both renormalisable and non-renormalisable scenarios, the proton favours decay into second-generation mesons accompanied by charged or neutral leptons, and the dominant mode is $p\to K^+\,\overline{\nu}$. This is in contrast to the known proton decay behaviour into first-generation mesons when mediated by heavy gauge bosons. This distinction between scalar-mediated and gauge boson-mediated decay arises from the flavour-sensitive hierarchical Yukawa couplings in the former case. The masses of the scalars were constrained from the lower bounds on the proton decay modes, revealing that scalars can lie below the GUT scale while adhering to proton decay constraints. Additionally, pairs of scalars inducing proton decays via non-renormalisable interactions can remain closer to the Electroweak scale than the GUT scale.

The other direct phenomenological implications of scalars contributing to the renormalisable \(SO(10)\) Yukawa sector were explored, focusing on those \(B-L\) charged scalars that do not induce tree-level proton decays, specifically sextet scalars. These sextet fields include \(\Sigma \sim (6,1,-\frac{2}{3})\), \(S \sim (6,1,\frac{1}{3})\), \(\overline{S}\sim(\overline{6},1,-\frac{1}{3})\), \({\cal S}\sim(6,1,\frac{4}{3})\), and \(\mathbb{S}\sim(\overline{6},3,-\frac{1}{3})\). These sextets only exhibit diquark couplings and induce neutral hadron-antihadron oscillation, such as meson-antimeson oscillation and baryon-antibaryon oscillation. They can also operationally explain the observed cosmological baryon asymmetry of the universe. The spectrum of these sextet scalars was constrained in a realistic \(SO(10)\) model, considering their capacity to induce various meson-antimeson oscillation, neutron-antineutron oscillation, and matter-antimatter asymmetry, demonstrating the potential of new physics to falsify a UV complete model. 

Conventional GUT models rely on multiple scalar irreps, and consequently more scalars, to yield realistic fermion masses. These scalar tensors provide a source of arbitrary and tunable parameters in the Higgs sector and bring the hierarchy problem into the framework. And, a GUT model with a single scalar irrep in the Yukawa sector is known to yield inconsistent mass relations. For instance, a single \(5_{\mathrm{H}}\) dimensional Higgs in the \(SU(5)\) Yukawa sector is known to yield generational mass degeneracy in the down-quark and charged lepton sector, resulting in \(Y_{d}=Y_e^T\). It has been shown that a realistic \(SU(5)\) GUT model can be achieved without expanding the scalar sector. By incorporating quantum corrections induced by heavy scalars and other heavy degrees of freedom in a minimally extended \(SU(5)\) model with singlet(s), the generational mass degeneracy between down quarks and charged leptons can be alleviated, provided the masses of the triplet and singlet differ by at least two orders of magnitude. This approach highlights the importance of \textit{indirect implications} of scalar fields in constructing a minimal GUT model and ameliorating the hierarchy problem. Furthermore, it necessitates revisiting the Yukawa sector of other conventional GUT models, such as \(SO(10)\). 

The thesis comprehensively analyses the scalar sector of grand unification, focusing on its ability to induce novel phenomena and its potential to construct minimal GUT models. This work advances our understanding of the scalars within the framework of grand unification and strengthens our ambitions to unify fundamental interactions.

\keywords{Beyond standard model, grand unification, hierarchy problem, matter-antimatter asymmetry, neutron-antineutron oscillation, proton decay, quantum correction, quark flavour violation, $SO(10)$, $SU(5)$, scalar fields}

\newpage
\thispagestyle{empty}
\null\newpage
\newpage

\pagestyle{fancy}
\fancyhf{}
\fancyhead[RE]{\bfseries References}
\fancyhead[LO]{\bfseries References}
\fancyhead[LE,RO]{\thepage}
\addcontentsline{toc}{chapter}{Publications}
\chapter*{Publications}

    
    
    
\section*{}
Published Articles\footnote{ The authors are listed alphabetically following the convention in High Energy Physics.}:
\begin{enumerate}[label=(\arabic*)]
 \item\noindent \textbf{Saurabh K. Shukla}, \textit{"Constraining scalars of $16_{\rm{H}}$ through proton decays in non renormalisable $SO(10)$ models``}\\ \href{https://arxiv.org/abs/2403.14331}{arXiv:2403.14331 [hep-ph]};\;\;\; \href{https://doi.org/10.1016/j.nuclphysb.2025.117034}{Nucl. Phys. B 1018 (2025) 117034}
\item \noindent Ketan M. Patel, \textbf{Saurabh K. Shukla},"\textit{Quantum corrections and the minimal Yukawa sector of $SU(5)$}",\\\href{https://arxiv.org/abs/2310.16563}{arXiv:2310.16563 [hep-ph]};\;\;\; \href{https://doi.org/10.1103/PhysRevD.109.015007}{ Phys.Rev.D 109 (2024), 1, 015007 } 
\item \noindent Ketan M. Patel, \textbf{Saurabh K. Shukla},"\textit{Spectrum of colored sextet scalars in realistic $SO(10)$ GUT}",\\\href{https://arxiv.org/abs/2211.11283}{arXiv:2211.11283 [hep-ph]};\;\;\; \href{https://doi.org/10.1103/PhysRevD.107.055008}{Phy.Rev.D 107 (2023), 5, 055008} 
\item \noindent Ketan M. Patel, \textbf{Saurabh K. Shukla}, "\textit{Anatomy of scalar mediated proton decays in $SO(10)$ models}", \\\href{https://arxiv.org/abs/2203.07748}{arXiv:2203.07748 [hep-ph]};\;\;\; \href{https://doi.org/10.1007/jhep08(2022)042}{JHEP 08 (2022) 042}

 \end{enumerate}
Conference Proceeding(s):
\begin{enumerate}[label=(\arabic*)]
\item \noindent 
       \textbf{Saurabh K. Shukla}, "\textit{Analysis of scalar mediated proton decays in non-SUSY SO(10) Models}",\\ \href{https://doi.org/10.1007/978-981-97-0289-3_222}{Springer Proc.Phys. 304 (2024) 863-865} .
\end{enumerate}

\clearpage


\nomenclature{$1\leq\alpha, \beta, \gamma...\leq 3$}{\makebox[3.6cm]{$\cdots\cdots$}\hspace{2.35cm} $SU(3)_{C}$ \, indices}
\nomenclature{$4 \leq a,b,...,m\leq 5$}{\makebox[3.5cm]{$\cdots\cdots$}\hspace{2.35cm} $SU(2)_{L}$\; indices}
\nomenclature{$1\leq n,o,...,z\leq 5$}{\makebox[3.5cm]{$\cdots\cdots$}\hspace{2.35cm} $SU(5)$\; indices}
\nomenclature{$1\leq \mu,\nu,...\leq 10$}{\makebox[3.5cm]{$\cdots\cdots$}\hspace{2.45cm} $SO(10)$\; indices}
\nomenclature{$0\leq \Dot{\mu},\Dot{\nu}...\leq 3$}{\makebox[3.5cm]{$\cdots\cdots$}\hspace{2.55cm} Lorentz indices}
\nomenclature{$1\leq A, \Dot{A}, M,\Dot{M}...\leq 2$}{\makebox[2.1cm]{$\cdots\cdots$}\hspace{2.35cm} Weyl Spinor indices}
\nomenclature{SM}{\makebox[8cm]{$\cdots\cdots$} Standard Model}
\nomenclature{GUT}{\makebox[8cm]{$\cdots\cdots$} Grand Unified Theory}
\nomenclature{$vev$}{\makebox[8cm]{$\cdots\cdots$} Vacuum Expectation Value}
\nomenclature{$B$}{\makebox[8cm]{$\cdots\cdots$} Baryon Number}
\nomenclature{$L$}{\makebox[8cm]{$\cdots\cdots$} Lepton Number}
\nomenclature{$B-L$}{\makebox[8cm]{$\cdots\cdots$} Baryon-Lepton Number}
\nomenclature{$EW$}{\makebox[8cm]{$\cdots\cdots$} Electroweak}
\nomenclature{$Y$}{\makebox[8cm]{$\cdots\cdots$} Hypercharge}
\nomenclature{$Q$}{\makebox[8cm]{$\cdots\cdots$} Electric Charge}
\nomenclature{irrep}{\makebox[8cm]{$\cdots\cdots$} Irreducible Representation}
\nomenclature{$\phi_{\hh}$}{\makebox[8cm]{$\cdots\cdots$} Scalar field $\phi$}
\nomenclature{${\boldsymbol{\phi}}$}{\makebox[8cm]{$\cdots\cdots$} Fermion Field $\phi$}
\nomenclature{$\phi_{\vv}$}{\makebox[8cm]{$\cdots\cdots$} Gauge Boson}
\nomenclature{LTR}{\makebox[8cm]{$\cdots\cdots$} Large Tensor Representation}

\setstretch{1.5}

\thispagestyle{empty}
\addcontentsline{toc}{chapter}{\contentsname}
\pagestyle{fancy}
\fancyhf{}
\fancyhead[LO,RE]{Table of Contents}
\fancyhead[LE,RO]{\thepage}
\tableofcontents

\pagestyle{fancy}
\fancyhf{}
\fancyhead[LO,RE]{List of Figures}
\fancyhead[LE,RO]{\thepage}
\listoffigures


\listoftables
\pagestyle{fancy}
\fancyhf{}
\fancyhead[LO,RE]{List of Tables}
\fancyhead[LE,RO]{\thepage}

\clearpage
\let\cleardoublepage


\printnomenclature
\thispagestyle{empty}
\pagestyle{fancy}
\fancyhf{}
\fancyhead[LO,RE]{Lexicons}
\fancyhead[LE,RO]{\thepage}

\newpage
\thispagestyle{empty}
\clearpage

\pagenumbering{arabic}
\setstretch{1.5}

\pagestyle{fancy}
\renewcommand{\sectionmark}[1]{\markright{\thesection~#1}{}}
\renewcommand{\chaptermark}[1]{\markboth{\thechapter~-~#1}{}}
\fancyhf{}
\fancyhead[RE]{\leftmark}
\fancyhead[LO]{\rightmark}
\fancyhead[LE,RO]{\thepage}

\clearpage


\pagestyle{fancy}
\fancyhf{} 
\fancyfoot[LE,RO]{\thepage} 

\fancyhead[RO]{%
  \begin{tikzpicture}[remember picture, overlay]
    \node[fill=gray!30, text width=2.5cm, align=center, font=\bfseries\Large, rotate=90, anchor=north east, minimum height=1.5cm, text=black] 
      at ([xshift=-1.5cm]current page.east) {\strut Chapter-\thechapter};  
  \end{tikzpicture}
}
\fancyhead[RE]{\nouppercase{\leftmark}} 
\fancyhead[LO]{\nouppercase{\rightmark}} 

\renewcommand{\chaptermark}[1]{ \markboth{Chapter \thechapter: #1}{}}
\renewcommand{\sectionmark}[1]{\markright{Section  \thesection: #1}}

\chapter{Introduction}
\label{ch:1}
\graphicspath{{Chapter_1/Vector/}{Chapter_1/}}

\lettrine[lines=2, lhang=0.33, loversize=0.15, findent=0.15em]{H}{UMANS ARE ALWAYS} fascinated by nature and wish to understand it at the shortest possible  distance scales. This intrinsic curiosity has led to the revelation of the fundamental forces that govern the universe. We know there are, \textit{at least}, four kinds of interactions $viz$ 1) gravity, 2) weak, 3) electromagnetism, 4) strong. Gravity, the most familiar of these, acts between all objects and is responsible for the structures of the universe, from planets orbiting stars to galaxies. Electromagnetism, however, deals with electric and magnetic fields that affect how charged particles interact and are necessary for the stability of the atom. The weak nuclear force is crucial in radioactive decay and nuclear fusion. Lastly, the strong nuclear force binds protons and neutrons together in the nuclei of atoms.

While vastly different in their scales and strengths, these forces share a common feature: they are all expressions of underlying symmetries in nature (may be even gravity!). Symmetry refers to a property where certain aspects of systems do not change under transformations. In the realm of particle physics, this concept becomes particularly powerful. The symmetries have implications for the conservation laws observed in nature—for example, energy conservation and momentum conservation. These principles are not merely mathematical abstractions but are observable realities that have profound implications for the behaviour of all physical systems.

Particle physics is the study of the most fundamental constituents of matter, and its beauty lies in revealing how different symmetries dictate the possible forms of interaction, predicting the existence of particles and their behaviours. Thus, understanding symmetry unlocks deeper insights into the universe's fundamental forces and workings, bridging the quest for knowledge with the empirical pursuit of contenting our curiosity. The theory that lies at the centre of particle physics is its standard model.

\section{Standard Model in a Nutshell}
{\label{sec:c1:totheSM}}

Standard Model (SM)~\cite{Glashow:1961tr,Salam:1968rm,Weinberg:1967tq} is the culmination of various successes and failures experienced throughout the 20th century and serves as the fundamental cornerstone of modern particle physics. It is a quantum field theory based on the principles of gauge invariance~\cite{maxwell1865,weyl1929}, locality~\cite{einstein1905,einstein1915,schwinger1951}, Lorentz invariance~\cite{Poincare1906}, renormalisability~\cite{dyson1949,feynman1948,tomonaga1946,schwinger1951}, spontaneous symmetry-breaking~\cite{nambu1960,nambu1961,goldstone1961,Higgs:1964ia}, and unitarity~\cite{heisenberg1925,dirac1930,vonneumann1955}.

Local gauge invariance implies that all physical laws remain invariant under coordinate-dependent gauge transformations. Lorentz symmetry ensures that the laws of physics are unchanged when reference frames change. Locality upholds the principle of causality in interactions, asserting that effects cannot precede their causes. Renormalisability guarantees that physical observables are calculable and finite, addressing infinities that may arise in theoretical predictions. Spontaneous symmetry breaking occurs when the symmetry of the action is preserved but broken by the ground state of a field. Unitarity ensures the conservation of probability, meaning the total probability of all possible outcomes of a scattering process sums to one.

SM provides an accurate description of the observed strong, weak, and electromagnetic interactions. It is based on three mutually commuting gauge groups~(cf. Appendix~\ref{app:2}) for group definition), which results from prolonged insightful theoretical developments and experimental findings. The SM gauge group is \sm.

The field content of the SM includes different irreducible representations $(irreps)$ of the Lorentz group: spin-one, half, and zero, which are gauge bosons, fermions, and scalar fields, respectively. In four-dimensional Minkowski space, fermions $(\Psi)$ are represented as a Dirac spinor with four degrees of freedom with the generators $(\gamma's)$  satisfying the Clifford algebra\footnote{Clifford Algebra is defined as: $\{\gamma_{\Dot{\mu}},\gamma_{\Dot{\nu}}\}\,\equiv\,\gamma_{\Dot{\mu}}\,\gamma_{\Dot{\nu}}\,+\,\gamma_{\Dot{\nu}}\gamma_{\Dot{\mu}}\,=\,2\,g_{\Dot{\mu}\Dot{\nu}}$}~\cite{clifford1878,Georgi:2000vve} (cf. Appendix~(\ref{app:2})). In a particular basis, called as a chiral basis\footnote{A basis in which the $\gamma_{5}\,\equiv\,i\,\gamma_0\,\gamma_1\,\gamma_2\,\gamma_3$, is diagonal}, $\Psi$ decomposes as follows;
\beqa{\label{eq:c1:chiral}}
\Psi\eq \begin{pmatrix}
    \varphi \\ \xi
\end{pmatrix}\eeqa
$\varphi$ and $\xi$ transform alike under rotations but transform oppositely under boosts and are two-dimensional \textit{Weyl}-spinor fields. One can define the following combination with $\gamma_{5}$;
\beqa{\label{eq:c1:PLPR}}
\Psi_{L}&\equiv&P_{L}\,\Psi\equiv \frac{1+\gamma_{5}}{2}\,\Psi\,=\, \begin{pmatrix}
    \varphi \\ 0
\end{pmatrix},\nl
\Psi_{R}&\equiv& P_{R}\,\Psi\equiv \frac{1-\gamma_{5}}{2}\,\Psi\,=\, \begin{pmatrix}
    0 \\ \xi
\end{pmatrix},
\eeqa
$\Psi_{L,R}$ are conventionally called as left $(\Psi_{L})$ and right $(\Psi_{R})$ four component representations of Weyl spinors.\footnote{The left and right-handedness of a Dirac spinor should not be confused with left and right chiral irreps of $SU(2)_L$.} 
\begin{table}[t]
    \centering
    \begin{tabular}{c|ccccc}
    \hline\hline
    \multirow{9}{*}{\rotatebox[origin=c]{90}{Spin - $\sfrac{1}{2}$~~~~~~~~~~~~~~~~~}}  & Fields & Symbol & SM Charges & B & L  \\
    \cline{1-6}
         &  &  &  &  &    \\

     &  & $q^{\alpha} = \begin{pmatrix} u^{\alpha} \\ d^{\alpha} \end{pmatrix},\,\, \begin{pmatrix} c^{\alpha} \\ s^{\alpha} \end{pmatrix},\,\, \begin{pmatrix} t^{\alpha} \\ b^{\alpha} \end{pmatrix}\,\,$ & $\left( 3,2,\sfrac{1}{6}\right)$ & $~~\sfrac{1}{3}$ & $0$ \\
     & Quarks & $u^C_{\alpha}$, $c^C_{\alpha}$, $t^C_{\alpha}$ & $\left(\overline{3},1,-\sfrac{2}{3}\right)$ & $-\sfrac{1}{3}$ & $0$  \\
     &  & $d^C_{\alpha}$, $s^C_{\alpha}$, $b^C_{\alpha}$ & $\left(\overline{3},1,\sfrac{1}{3}\right)$ & $-\sfrac{1}{3}$ & 0  \\
          &  &  &  &  &    \\

    \cline{2-6}
    &  &  &  &  &    \\
     & Leptons & $l^a = \begin{pmatrix} \nu_{e} \\ e \end{pmatrix},$\,\,$\begin{pmatrix} \nu_{\mu} \\ \mu \end{pmatrix},$\,\,$\begin{pmatrix} \nu_{\tau} \\ \tau \end{pmatrix}$ & $\left(1,2,-\sfrac{1}{2}\right)$ & $0$ & $~~1$  \\
     
     &  & $e^C$, $\mu^C$, $\tau^C$ & $\left(1,1,1\right)$ & $0$ & $-1$  \\
     &  &  &  &  &  \\
    \hline
    \multirow{4}{*}{\rotatebox[origin=c]{90}{Spin - $1$~~~ }} & \multirow{2}{*}{}   &  &  &  &    \\
    & Gluons & $G^{\alpha}$ & $\left(8,1,0\right)$ & $0$ & $0$  \\
         & $W$ Bosons & $W^a$ & $\left(1,3,0\right)$ & $0$ & $0$  \\
      & $B$ Bosons & $B$ & $\left(1,1,0\right)$ & $0$ & $0$  \\
     &  &  &  &  &    \\

    \hline
    \multirow{2}{*}{\rotatebox[origin=c]{90}{~Spin - $0$~}} &  &  &  &  &    \\
    & Higgs Boson & $H$ & $\left(1,2,-\sfrac{1}{2}\right)$ & $0$ & $0$  \\
    &  &  &  &  &   \\ [0.75ex]
    \hline\hline
    \end{tabular}
    \caption{The content of the SM together with their quantum number under \sm, Baryon Number $(B)$, Lepton Number $(L)$. Here, $\alpha\,\epsilon\,\{1,2,3\}$ represents the colour quantum numbers and $a\,\epsilon\,\{1,2\}$ stands for the weak isospin quantum numbers.}
    \label{tab:c1:SMcontent}
\end{table}


Theories that place different fermions within the same gauge multiplet require these fermions to possess the same chirality; otherwise, the gauge symmetry would not commute with Lorentz symmetry. Consequently, we define a conjugated spinor $\psi^C$, which transforms a right-handed spinor (left-handed spinor) into a left-handed (right-handed) one. The mathematical formulation of this transformation is expressed as follows:
\begin{equation}
\label{eq:c1:conjugation}
\psi^C \,\equiv \, {\cal C}\,\xi^{*}\,=\,i\,\sigma_{2}\,\xi^*,
\end{equation}
where, ${\cal C}$ is called the Charge conjugation matrix and follows the relation ${\cal C}\,=\,-{\cal C}^T\,=\,-{\cal C}^{\dagger}\,=\,{\cal C}^*$. $\psi^C$ transform as a left-chiral spinor. The reader is referred to~\cite{Dreiner:2008tw} for a comprehensive review of the two-component formalism

Armed with the necessary rudiments of spinors, we now present a concise and relevant review of the SM, which is essential for our purposes. In the purview of this thesis, we have opted to represent the SM fermion fields using two-component Weyl spinors, replacing right-handed fields with their left-chiral conjugates. For a comprehensive review of the SM using the conventional left-right handed notation, refer to~\cite{Cottingham:2007zz,Peskin:1995ev}. The particle content of the SM includes quarks and leptons, both left and left-chiral conjugated, as well as gauge bosons and the Higgs boson. Each category of particles carries distinct quantum numbers under the SM gauge symmetry. The particle content of the SM is shown in Tab.~(\ref{tab:c1:SMcontent}).

The Lagrangian (pedantically, Lagrangian density!) of the SM receives contributions from strong and Electroweak (EW) Lagrangian, as depicted below:
\beqa{\label{eq:c1:SMlag}}
{\cal  L}_{\mathrm{ SM}} \eq {\cal L}_{\mathrm {QCD }} \,+\, {\cal L}_{\mathrm {EW}}.\eeqa 
The first contribution to the SM Lagrangian originates from the strong interaction and results in Quantum Chromo Dynamics (QCD). This theory is based on local gauge invariance under the non-abelian $SU(3)_C$ group of colour space. It acts only on the quarks,  irrespective of their chirality. Lie algebra of $SU(3)_C$ group is spanned by eight linearly independent generators, corresponding to the number of gluons mediating the strong interaction. QCD Lagrangian can be expanded as follows;
\beqa{\label{eq:c1:QCD}}
{\cal L}_{\mathrm{QCD}} \eq -\frac{1}{4}\,G_{\Dot{\mu}\Dot{\nu}}\boldsymbol{\cdot}G^{\Dot{\mu}\Dot{\nu}} + \sum_{f}\,\zeta^{\dagger}_{\alpha\,f}\,i\overline{\sigma}^{\Dot{\mu}}\,{\cal D}_{\Dot{\mu}\,\beta}^{\alpha}\,\zeta^{\beta}_f.
\eeqa
The first term in Eq.~\eqref{eq:c1:QCD} is the kinetic energy term and represents self-interaction among gluons, and the second term signifies the interactions of gluons with the quarks. 

\begin{figure}[t]
    \centering
    \includegraphics[width=0.75\textwidth]{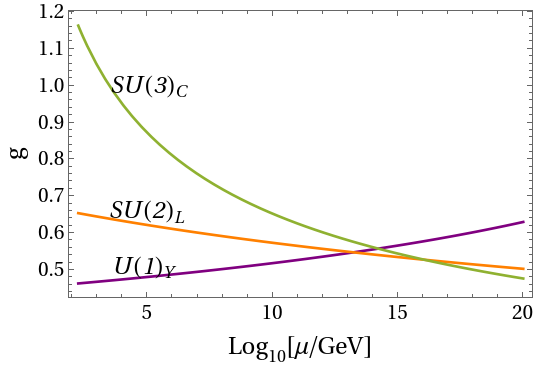}
    \caption{Running of SM gauge couplings at one loop.}
    \label{fig:c1:SMgaugerunning}
\end{figure}

Throughout the thesis, we denote Lorentz indices by dotted Greek alphabets $0\,\leq\Dot{\mu},\Dot{\nu}...\leq\,3$. 
$\zeta$ denotes various quarks (left and left chiral conjugates) fields, i.e. $\zeta \supset\,\{u,\,c,\,t,\, u^C,\, c^C,\,t^C,$ $d,\,s,\,b,\,d^C,\,s^C,\,b^C\}$ and $\alpha, \beta$ denotes $SU(3)$ indices. The equation also includes the sum over $f$, denoting the sum of all the kinetic terms of all the quarks, including all generations. Further, $G\boldsymbol{\cdot}G\equiv \sum _{A=1}^8\,G_A\,G_A$, $A$ being the number of $SU(3)$ generators. Also, $\overline{\sigma}^{\Dot{\mu}}\,=\,\left(\sigma^{0},-\vec{\sigma}\right)$ with $\sigma^0\,=\,\mathds{1}_{2\times 2}$ and $\vec{\sigma}\,\supset\,\{\sigma^1,\sigma^2,\sigma^3\}$ are the different Pauli matrices. The definition of the covariant field strength tensor is as follows;
\beqa{\label{eq:c1:FieldStrengthSU3}}
G^{A}_{\Dot{\mu}\Dot{\nu}}\eq \partial_{\Dot{\mu}}\,G^{A}_{\Dot{\nu}} - \partial_{\Dot{\nu}}\,G^{A}_{\Dot{\mu}} - g_{s}\,F^{ABC}\,G_{B\Dot{\mu}}\,G_{C\Dot{\nu}},\eeqa
provided, $F's$ are structure constants of $SU(3)$. The definition of covariant derivative used in the Eq.~\eqref{eq:c1:QCD} is as follows,
\beqa{\label{eq:c1:covSU3}}
{\cal D}^{\alpha}_{\,\Dot{\mu}\,\beta}\eq \partial_{\Dot{\mu}}\,\delta_{\beta}^{\alpha} + i\,\frac{g_{s}}{2}\,\left(G_{\Dot{\mu}}\boldsymbol{\cdot}\lambda\right)^{\alpha}_{\beta},\eeqa 
where, $\lambda$ are the eight generators of $SU(3)_{\mathrm{C}}$ and can be conveniently chosen as Gell-Mann Matrices~\cite{Gell-Mann:1962yej}. The strong couplings, denoted by $g_s$, are universal for all flavours of SM fermions and decrease (or increase) with increasing (or decreasing) energy (cf. Fig.~(\ref{fig:c1:SMgaugerunning})), a phenomenon responsible for asymptotic freedom~\cite{Gross:1973id}.

The Electroweak (EW) part of the Lagrangian, as described in Eq.~\eqref{eq:c1:SMlag}, consists of the following components:
\beqa{\label{eq:c1:EW}}
{\cal L}_{\mathrm {EW}} \eq {\cal L}_{\mathrm {Kinetic}} \,+\, {\cal L}_{\mathrm {Higgs}} \,+\, {\cal L}_{\mathrm {Yukawa}}.
\eeqa
The kinetic Lagrangian, $\cal{L}_{\mathrm{Kinetic}}$, consists of terms: (i) field strength tensors of $SU(2)_L$ and $U(1)_Y$  (ii) interaction of the gauge bosons of $SU(2)_L$ and $U(1)_Y$ with the quarks and leptons in the SM, as shown below;
\beqa{\label{eq:c1:kinetic}}
{\cal L}_{\mathrm{Kinetic}} \eq -\frac{1}{4}\,W_{\Dot{\mu}\Dot{\nu}}\boldsymbol{\cdot}W^{\Dot{\mu}\Dot{\nu}} -\frac{1}{4}\,B_{\Dot{\mu}\Dot{\nu}}\boldsymbol{.}B^{\Dot{\mu}\Dot{\nu}}\nl
\ad \sum_{L}\,\varphi_L^{\dagger}\,i\overline{\sigma}^{\Dot{\mu}}\,\left(\partial_{\Dot{\mu}} + \frac{ig}{2}\, \left(W_{\Dot{\mu}}\boldsymbol{\cdot}\sigma\right) + \frac{ig'}{2}\,\,Y_{\varphi_{L}}\,B_{\Dot{\mu}}\,\right)\,\varphi_{L} \,-\, q^{\dagger}_{\alpha}\,i\overline{\sigma}^{\Dot{\mu}}\,\partial_{\Dot{\mu}\,\beta}^{\alpha}\,q^{\beta}\nl
\ad \sum_{R}\,\chi^{\dagger}_R\,i\overline{\sigma}^{\Dot{\mu}}\,\left(\partial_{\Dot{\mu}} + \frac{i g'}{2}\,Y_{\chi_{R}}\,B_{\Dot{\mu}}\right)\,\chi_R.
\eeqa
$SU(2)_\mathrm{L}$ gauge boson only couples to left-handed quarks and leptons, while the gauge boson associated with the  $U(1)_{\mathrm{Y}}$ couples to both the left and right-handed quarks and leptons. The notation $\sum_{L}$ denotes summing over all left-chiral fermions, while $\sum_{R}$ indicates summing over all left-chiral conjugated fermions. In the second line of Eq.~\eqref{eq:c1:kinetic}, we have removed the contribution of the kinetic term of the left-handed quark fields, which have been already accounted in Eq.~\eqref{eq:c1:QCD}. The notation $Y_{\varphi_{L}}$ ($Y_{\chi_{R}}$) denotes the hypercharge associated with the particular field $\varphi_{L}$ ($\chi_{R}$). The field strength tensors for $W$ and $B$ can be written analogously to that of $SU(3)_C$ given in Eq.~\eqref{eq:c1:FieldStrengthSU3}, except that the structure constant term is absent for $B$ due to its Abelian nature. The gauge couplings $g$ and $g'$, associated with $SU(2)_L$ and $U(1)_Y$ respectively, exhibit differing energy dependencies: $g$ decreases and $g'$ increases with increasing energy (cf. Fig.~(\ref{fig:c1:SMgaugerunning})).

The Higgs field $H$ is the scalar field transforming as a doublet under $SU(2)_{L}$ and is described by the following Lagrangian. 
\beqa{\label{eq:c1:higgs}}
{\cal L}_{\mathrm{Higgs}} \eq \left|\left(\partial_{\Dot{\mu}} + \,\frac{ig}{2}\,\left(W_{\Dot{\mu}}\boldsymbol{\cdot}\sigma\right) + \,\frac{ig^\prime}{2}\,B_{\Dot{\mu}}\right)\,H\right|^2 - V\left(H\right),\eeqa
where, $V\left(H\right)$ is a quadratic function in $H$ and $H^{\dagger}$ respecting the \sm\, gauge symmetry and renormalisability, as shown below:
\beqa{\label{eq:c1:VH}}
V\left(H\right)\eq -\mu^2\,H^{\dagger}H\,+\, \lambda\,\left(H^{\dagger}\,H\right)^2.\eeqa
 The $\mu^2$ and $\lambda$ are not calculable parameters in the SM but need to be fixed from experiments. The Yukawa interaction, written in Eq.~\eqref{eq:c1:EW}, is responsible for mass for different quarks and leptons and can be parameterised as follows, 
\beqa{\label{eq:c1:yukawa}}
-{\cal L}_{\mathrm{Yukawa}}\eq Y_{u\,AB}\,q_A^T\,\cc\,u_B^C\,\tilde{H} \,+\, Y_{d\,AB}\, q_A^T\,\cc\,d_B^C\,H \,+\, Y_{l\,AB}\,l_A^T\,\cc\,e_B^C\,H\nl\hcn,
\eeqa
provided, $\tilde{H}\,\equiv\,i\sigma_2\,H^*$. $\cal{C}$ is the charge conjugation matrix in the Lorentz space and is necessary for the invariance of the bilinear in the Lorentz space.

The SM Lagrangian reduces to only the gauge and the Higgs sector in the limit where all Yukawa couplings vanish. This gauge sector consists of the kinetic terms for quarks and leptons and the kinetic terms for the gauge bosons. In this scenario, we can make the following transformations on the fermion fields without modifying the Lagrangian:
\beqa{\label{eq:c1:accidental}}
q_A\,&\to&\, \big(U_{q}\big)_{AB}\,q_{B},\;\;u^C_{A}\,\to\, \big(U_{u^C}\big)_{AB}\,u^C_{B},\;\;d^C_A\,\to\, \big(U_{d^C}\big)_{AB}\,d^C_{A},\nl l_A\,&\to&\, \big(U_{l}\big)_{AB}\,l_{B},\;\;
e^C_A\,\to\, \big(U_{e^C}\big)_{AB}\,e^C_B,\eeqa
where \( U_{q}^{\dagger} U_q = 1 = \cdots = U_{e^C}^{\dagger} U_{e^C} \). This implies that the SM with $N_{f}$ generations has an extra global symmetry of \([U(N_f)]^5\) in the absence of Yukawa interactions. For three generations, \(N_f = 3\), this symmetry is \([U(3)]^5\). When the Yukawa interactions, given in Eq.~\eqref{eq:c1:yukawa}, are included, the SM Lagrangian is invariant under the following transformations:
\beqa{\label{eq:c1:brokenSM}}
q\,&\to&\,\exp\left(i\,B\,\alpha\right)\,q,\;\;u^C\,\to\,\exp\left(-i\,B\,\alpha\right)\,u^C,\;\;\,d^C \to\,\exp\left(-i\,B\,\alpha\right)\,d^C,\nl
e\,&\to&\,\exp\left(i\,L_{e}\,\alpha_{e}\right)\,e,\;\;e^C\,\to\,\exp\left(-i\,L_{e}\,\alpha_{e}\right)\,e^C,\nl
\mu\,&\to&\,\exp\left(i\,L_{\mu}\,\alpha_{\mu}\right)\,\mu,\;\;\mu^C\,\to\,\exp\left(-i\,L_{\mu}\,\alpha_{\mu}\right)\,\mu^C,\nl
\tau\,&\to&\,\exp\left(i\,L_{\tau}\,\alpha_{\tau}\right)\,\tau,\;\;\tau^C\,\to\,\exp\left(-i\,L_{\tau}\,\alpha_{\tau}\right)\,\tau^C.\eeqa
The extra freedom to rotate in the Lepton sector stems from the masslessness of neutrinos in SM. We identify the $B$ as the Baryon number and $L_{e,\mu,\tau}$ as the generational Lepton numbers. Thus the remnant symmetry is $U(1)_{B}\times U(1)_{L_{e}}\times U(1)_{L_{\mu}}\times U(1)_{\tau}$ $\equiv$ $[U(1)]^4$. Thus inclusion of Yukawa interactions breaks the global $[U(3)]^5\,\to\,[U(1)]^4$ symmetry. Conventionally, $B=\sfrac{1}{3}$ and $L_{e,\mu,\tau}=1$. Consequently, baryon and lepton numbers are the (accidental) symmetry of the Lagrangian.

We now outline the EW symmetry-breaking aspects of the SM. For $\mu^2 \,>\,0$, the potential written in Eq.~\eqref{eq:c1:VH} has the following minimum:
\beqa{\label{eq:c1:minH}}
\langle0\big|H\big|0\rangle\;=\;\frac{1}{\sqrt{2}}\begin{pmatrix}
    0 \\ v
\end{pmatrix}\hspace{1cm}\text{with,\,\,} v=\sqrt{\frac{\mu^2}{\lambda}} \eeqa
The gauge boson mass term can be computed from the first term of the Eq.~\eqref{eq:c1:higgs}, by substituting the vacuum expectation value $(vev)$ of the scalar field mentioned in Eq.~\eqref{eq:c1:minH}.
\beqa{\label{eq:c1:massivegauge}}
\Delta {\cal L}_{\mathrm{Higgs}} \supset \frac{v^2}{8}\left[g^2\,\left(\big(A_1^{\Dot{\mu}}\big)^2 +\big(A_2^{\Dot{\mu}}\big)^2\right) + \big(-g\,A_3^{\Dot{\mu}} + g'\,B^{\Dot{\mu}} \big)^2\right],
\eeqa
The massless gauge bosons absorb the three longitudinal degrees of freedom of Higgs and become massive. One can define specific linear combinations of the gauge bosons that demonstrate their massiveness.
\beqa{\label{eq:c1:transfomedgauges}}
W^{\pm\,\Dot{\mu}}\eq \frac{1}{\sqrt{2}}\,\left(A_1^{\Dot{\mu}}\mp i A^{\Dot{\mu}}_2\right)\hspace{1cm} \text{with mass}\hspace{1cm} m_{W}\,=\,g\frac{v}{2},\nl
Z^{\Dot{\mu}}\eq \frac{1}{\sqrt{g^2 + g'^2}}\,\left(g\,A_3^{\Dot{\mu}}-g'\,B^{\Dot{\mu}}\right)\hspace{1cm}\text{with mass}\hspace{1cm}m_{Z}\,=\,\sqrt{g^2+g'^2}\,\frac{v}{2},\nl
A^{\Dot{\mu}}\eq \frac{1}{\sqrt{g^2 + g'^2}}\,\left(g\,A_3^{\Dot{\mu}}+g'\,B^{\Dot{\mu}}\right)\hspace{1cm}\text{with mass}\hspace{1cm}m_{A}\,=\,0. 
\eeqa
The vector fields $Z$ and $A$ are orthogonal to each other. We identify $A$ as the gauge boson for the unbroken symmetry and is the electromagnetic vector potential. Substituting Eq.~\eqref{eq:c1:transfomedgauges} into the definition of the covariant derivative of the EW group modifies it as follows;
\beqa{\label{eq:c1:modifiedcovariant}}
D^{\Dot{\mu}}\eq \partial^{\Dot{\mu}} - i\,\frac{g}{\sqrt{2}}\,\left(W^{+\,\Dot{\mu}}\,T_+ +  W^{-\,\Dot{\mu}}\,T_-\right) - \frac{i}{\sqrt{g^2+g'^2}}\,Z^{\Dot{\mu}}\,\left(g^2\,T_3-g'^2 Y\right)\nl
\mi \frac{i\,gg'}{\sqrt{g^2+g'^2}}\,A^{\Dot{\mu}}\,\left(T_3 + Y\right),\eeqa
where, $T_{\pm}\equiv\frac{1}{2}\left(\sigma^1 \pm i \sigma^2\right)$. The transformed expressions, given in Eq.~\eqref{eq:c1:modifiedcovariant}, are termed to be in a physical or mass basis.  The expression written in the last line of Eq.~\eqref{eq:c1:modifiedcovariant} indicates that $A$ couples to massless $U(1)$ generator $T+Y$, which is precisely the generator of remnant electromagnetic symmetry. We define the electric charge of any SM particle using the following relation:
\beqa{\label{eq:c1:electricharge}}
Q \equiv T_3 + Y. \eeqa
Electric charge $(Q)$ can be equated to the sum of the third component of weak isospin $(T_3)$ and the weak hypercharge $(Y)$. The weak hypercharge for different SM representations can be determined from their electromagnetic charges. 
Substituting the expression of covariant derivative written in the mass basis, in Eq.~\eqref{eq:c1:modifiedcovariant}, the kinetic term of fermions, written in Eq.~\eqref{eq:c1:kinetic}, modifies as follows:
\beqa{\label{eq:c1:kineticmodi}}
{\cal L} &\supset& \sum_L\,\varphi_{L}^{\dagger}\,i\overline{\sigma}^{\Dot{\mu}}\partial_{\Dot{\mu}}\,\varphi_L + \sum_R\,\chi_{R}^{\dagger}\,i\overline{\sigma}^{\Dot{\mu}}\partial_{\Dot{\mu}}\,\chi_R \nl
\ad g\left(W^+_{\Dot{\mu}}\,J^{\Dot{\mu}\,+} +W^-_{\Dot{\mu}}J^{\Dot{\mu}\,-} + Z_{\Dot{\mu}}J^{\Dot{\mu}}_{Z}\,\right) + e \,A_{\Dot{\mu}}\,J^{\Dot{\mu}}_{\mathrm{EM}},
\eeqa
provided,
\beqa{\label{eq:c1:currents}}
J^{\Dot{\mu}\,+}\eq \frac{1}{\sqrt{2}}\,\left(\nu^{\dagger}\,\overline{\sigma}^{\Dot{\mu}}\,e  + u^{\dagger}\,\overline{\sigma}^{\Dot{\mu}}\,d\right),\nl
J^{\Dot{\mu}\,-}\eq \big(J^{\Dot{\mu}\,+}\big)^{\dagger},\nl
J^{\Dot{\mu}}_{\mathrm{Z}}\eq \frac{1}{\cos\theta_{\mathrm{W}}}\,\left[\sum_{L}\,\varphi^{\dagger}_{L}\,\overline{\sigma}^{\Dot{\mu}}\,\left(T_{3,\phi_L} - Q_{\phi_L}\,\sin^2\theta_{\mathrm{W}}\right)\,\varphi_L \right. \nl \ad \left. \sum_{R}\,\chi^{\dagger}_R\,\overline{\sigma}^{\Dot{\mu}}\,Q_{\chi_{R}}\sin^2\theta_{\mathrm{W}}\,\chi_R\right],\nl
J^{\Dot{\mu}}_{\mathrm{EM}}\eq \sum_{L}\,\varphi^{\dagger}_L\,Q_{\varphi_{L}}\overline{\sigma}^{\Dot{\mu}}\,\varphi_L - \sum_{R}\,\chi^{\dagger}_R\,\overline{\sigma}^{\Dot{\mu}}\,Q_{\chi_R}\,\chi_R,
\eeqa
where, $\theta_{\mathrm{W}}\,=\,\arctan\big(\sfrac{g'}{g}\big)$ and $e\equiv \sfrac{gg'}{\sqrt{g^2 + g'^2}}$.
The neutral current $J_Z$, as defined in Eq.~\eqref{eq:c1:currents}, is a fundamental prediction of the SM. It conserves the charge and flavour of the fermions. It differs from the electromagnetic current because left-chiral particles couple with a different strength than left-chiral conjugated particles, whereas in electromagnetic interactions, particles of both chiralities couple with the same strength. Hence, neutral currents violate the parity in the SM. Neutral currents validate the electroweak theory, and their presence has been verified with great precision in experiments~\cite{GargamelleNeutrino:1973jyy}. 

Now, examining the effect of spontaneous symmetry breaking on the remaining part of the Lagrangian of Eq.~\eqref{eq:c1:SMlag}, i.e. Yukawa Lagrangian. As the Higgs field acquires $vev$, the chiral symmetry of the SM is broken and is shown by the following Lagrangian;
\beqa{\label{eq:c1:fermionmass}}
{\cal L}_{\mathrm{Yukawa}} \eq M_{u\,AB}\,u^T_A\,\cc\,u^C_B + M_{d\,AB}\,d^T_A\,\cc\,d^C_B +M_{l\,AB}\,e^T_A\,\cc\,e^C_B \nl \hcn,\eeqa
where, $M_{f}\,=\,\frac{v}{\sqrt{2}}\,Y_f$ for $f=u,d,e$. In general, the fermion mass matrix \( M^f \) could be a complex square matrix (may not be square sometimes!) in the generation space with dimensions corresponding to the number of fermion generations. Moreover, neutrinos (\( \nu \)) do not acquire mass (Dirac or Majorana) at any perturbative order because their right-handed counterparts (\( \nu^C \)) is absent in the SM. Extending the framework to include \( \nu^C \) would allow neutrinos to gain mass; however, this addition would lead to the lepton number violation if \( \nu^C \) were to obtain a Majorana mass~\cite{Majorana:1937vz,Weinberg:1979sa}.

The mass matrices, in Eq.~\eqref{eq:c1:fermionmass}, are, in general, non-diagonal. One can digonalise them by following unitary transformations:
\beqa{\label{eq:c1:unitary}}
u_A &\to& \big(U_{u}\big)_{AB}\,u_{B},\hspace{1cm}d_A \,\to\, \big(U_{d}\big)_{AB}\,d_{B},\hspace{1cm}e_A \,\to\, \big(U_{e}\big)_{AB}\,e_{B},\nl
u^C_A &\to& \big(U_{u^C}\big)_{AB}\,u^C_{B},\hspace{1cm}d^C_A \,\to\, \big(U_{d^C}\big)_{AB}\,d^C_{B},\hspace{1cm}e^C_A \,\to\, \big(U_{e^C}\big)_{AB}\,e^C_{B}.\nl\eeqa
The abovementioned transformations in Eq.~\eqref{eq:c1:unitary} lead to the following diagonalised mass matrices with real and non-negative eigenvalues:
\beqa{\label{eq:c1:masses}}
U_{u}^{T}\,M_u\,U_{u^C}\eq \mathrm{Diag}\left(m_{u},m_c,m_t\right),\nl
U_{d}^{T}\,M_d\,U_{d^C}\eq \mathrm{Diag}\left(m_{d},m_s,m_b\right),\nl
U_{e}^{T}\,M_e\,U_{e^C}\eq \mathrm{Diag}\left(m_{e},m_{\mu},m_{\tau}\right),
\eeqa
where,  \( m_{i} \) represents the masses of different fermions.

With the above set of unitary transformations given in Eq.~\eqref{eq:c1:unitary}, the charged current interaction in the quark sector becomes the following:
\beqa{\label{eq:c1:VCKM}}
u^{\dagger}_A\,\overline{\sigma}^{\Dot{\mu}}\,d_A &\to& u^{\dagger}_A\,\big(U^{u\dagger}\big)_{BA}\,\overline{\sigma}^{\Dot{\mu}}\,U^d_{AC}\,d_C\eeqa
We observe different flavours of quarks can interact in the mass basis through charged currents. Consequently, these interactions are flavour violating with the strength proportional to the matrix element of the matrix, which is a misalignment between the flavour and mass basis. We call this matrix the Cabibo-Kobayashi-Masakawa $(CKM)$ matrix~\cite{Cabibbo:1963yz,Kobayashi:1973fv}, and it is defined as follows,
\beqa{\label{eq:c1:CKM}}
U^{u\dagger}\,U^d\equiv V_{\mathrm{CKM}}.\eeqa
To extract elements of the CKM matrix, one typically measures the rates of quark flavour-changing processes, such as meson decays and mixing, comparing these experimental results with theoretical predictions involving the CKM elements~\cite{Workman:2022ynf}. Furthermore, neutral currents remain diagonal under the unitary transformation described in Eq.~\eqref{eq:c1:unitary} and flavour violation in them do not occur at the tree level; they only appear at the loop level~\cite{Glashow:1970gm}.

The SM is remarkably successful in explaining almost all observed low-energy data with only a few parameters. Additionally, all particles predicted by the SM have been observed in the experiments, including the $W$ and $Z$~\cite{arnison1983experimental2}, top-quark~\cite{Kobayashi:1973fv,CDF:1995wbb,Czakon:2013goa}, and the latest entry is Higgs~\cite{ATLAS:2012yve,CMS:2012qbp}. 

\section{Limitations of the Standard Model}
{\label{sec:c1:drawbacksSM}}
Despite the success of the SM in predicting phenomena such as the Higgs boson and the existence of neutral currents, it has several drawbacks that are often overlooked. The model faces numerous theoretical challenges on which it is constructed and fails to explain various experimental observations. 

The troublesome theoretical features of SM are as follows:
\begin{enumerate}[label=(\arabic*)]
    \item Gauge structure of the SM\\ 
    In its fundamental descriptions, nature's preference for unitary groups raises a mystery. It is also unclear why the SM is structured around three distinct and seemingly unrelated gauge symmetries. The assignment of particle representations within the SM seems theoretically adhoc.  {\label{pt:c1:theoSM1}}

\item Charge quantisation\\
The prescription to choose the hypercharge of quarks and leptons in the SM is motivated by experimental facts and lacks theoretical understanding. Additionally, the rationale for relating the charges of quarks, leptons, and scalar field remains missing.{\label{pt:c1:theoSM2}}

\item The flavour structure \& incalculable parameters.\\
The SM includes three generations of Weyl fermions, but the reasons behind this specific number remain one of its most compelling mysteries. Several parameters in the SM are incalculable in SM and need to be fixed experimentally. Additionally, there is a significant six-order magnitude difference in the masses between the first and third generations of fermions, which SM does not address. {\label{pt:c1:theoSM5}}

\item Scalar sector and the naturalness problem\\
SM has twelve gauge bosons and fifteen fermions in a single generation. Despite this diversity, only one scalar, the Higgs boson, is present in the model. This raises an intriguing question: why the SM is so economical in the case of scalar?
Also, the Higgs mass parameter \(\mu^2\) is prone to receive large quantum corrections from energy scales higher than the electroweak scale, such as the grand unification scale or the Planck scale. These corrections necessitate precise cancellations to ensure that the resultant \(\mu^2\) is close to the observed mass of the Higgs boson. The greater the hierarchy between the electroweak scale and these higher energy scale, the more fine-tuning is required.{\label{pt:c1:theoSM6}}

\end{enumerate}

The SM also faces several significant challenges on experimental frontier. The matter-antimatter asymmetry is evident from the dominance of matter over antimatter, yet the CP violation incorporated in the SM's weak sector is insufficient to explain the observed cosmological baryon asymmetry~\cite{Sakharov:1967dj,Riotto:1999yt,Dine:2003ax}. Neutrino sector is another area of discrepancy; while the SM predicts massless neutrinos, oscillation data indicates that neutrinos have a small mass~\cite{Super-Kamiokande:1998kpq,SNO:2002tuh,KamLAND:2002uet}. The SM also fails to account for dark matter and dark energy~\cite{Zwicky:1933gu,Planck:2018vyg,SupernovaSearchTeam:1998fmf,SupernovaCosmologyProject:1998vns}, which play crucial roles in the universe's structure and dynamics. Further, SM lacks a comprehensive framework to explain the formation of all experimentally observed bound states. Recent experimental anomalies, such as 
muon's anomalous magnetic moment~\cite{Venanzoni:2023mbe} and tensions in B-meson decay measurements~\cite{London:2019nlu}, further challenge the SM's predictions.

Apart from the various drawbacks of the SM mentioned in pts. (\ref{pt:c1:theoSM1}-\ref{pt:c1:theoSM6}) and experimental inobservations, the SM also fails to incorporate gravity, stabilisation of the electroweak scale \cite{Coleman:1977py,Arnold:1989cb,Kobakhidze:2013tn}, relies only on perturbative analysis.

The issues highlighted above point out that the SM is  incomplete and cannot be considered the ultimate theory. Such a theory, whatever it could be, should \textit{at least} align with experimental observations. Additionally, it might also happen that many of the theoretical shortcomings identified in the SM may not even be present in the ultimate theory, which could offer a more precise and complete description of fundamental forces and particles.

The concept of \textit{unification} relies on finding a single theoretical framework capable of deriving and explaining all observed phenomena. Although this idea of unification may appear \textit{Panglossian}, we pursue it due to historical precedent and human biases toward it. There is no harm in accepting that this approach might be entirely incorrect. While a unifying theory might address some shortcomings of the SM, it could also introduce its own set of complexities. However, the writer remains biased towards and intrigued by the notion of grand unification, a topic that will be explored in the following section.

\section{Grand Unification}

Grand unification is based on the idea of syncretising the known fundamental interactions into a single interaction within a larger symmetry structure. As a result, quarks and leptons in each family are different manifestations of one entity under a unified symmetry group. The distinct strong, weak, and electromagnetic interactions at low energies are unified as aspects of this force under the symmetry group $G$. The simple\footnote{  A simple group is one whose only normal subgroups are a trivial group or the group itself.} group $G$ must contain SM gauge group, \sm, as one of its subgroups. At higher energies, approximately greater than the grand unification scale (\(M_{\mathrm{GUT}}\gg M_{\mathrm{EW}}\)), the distinctions between quarks and leptons, as well as the differences among the strong, weak, and electromagnetic interactions must disappear, consequently allowing conversion of quarks into leptons and vice versa. The differences between quarks and leptons and the distinctions among the three forces are only apparent at low energies $M\,\ll\, M_{\mathrm{GUT}}$. At \(M_{\mathrm{GUT}}\), the grand unified group \(G\) spontaneously breaks\footnote{The unified gauge symmetry can also be broken dynamically~\cite{Napoly:1984wt,Kang:1986ei}.} down into the SM gauge group. The grand unification provides a window to peep processes at extremely short distances $\sim 10^{-32}\,{\mathrm m}$, which are yet to be discovered.

Various Grand Unified Theory (GUT) groups based on unitary groups~\cite{Pati:1973uk,Georgi:1974sy}, orthogonal group~\cite{Fritzsch:1974nn}, and exceptional groups~\cite{Gursey:1975ki,Gursey:1976dn} have been proposed for more than four decades (cf.~\cite{Langacker:1980js,Fukuyama:2004ps,Nath:2006ut} for review). In a GUT like $SU(5)$, quarks and leptons are unified into two irreps, $\mathbf{\overline{5}}$ and $\mathbf{10}$. Further, in an orthogonal group like $SO(10)$, all the SM fermions of a single generation can be unified into a sixteen-dimensional irrep, necessitating the inclusion of a right-handed neutrino. In exceptional groups like $E_{6}$, SM fermions and the right-handed neutrino are unified into a twenty-seven-dimensional irrep, predicting eleven additional fermions per generation. The SM fermions given in Tab.~(\ref{tab:c1:SMcontent})\, can be embedded into irreps of \su and \so\, as shown in  Tab.~(\ref{tab:c1:embeddingSMfermionsintoGUT}).

\begin{table}
    \centering
\scalebox{1.1}{ \begin{tabular}{c|ccc|cc|c}
       \hline\hline Group & $q$ & $u^C$ & $e^C$ & $d^C$ & $l$ & $\nu^C$ \\
        \hline
        SM & $\left(3,2,\sfrac{1}{6}\right)$  & $\left(\overline{3},1,-\sfrac{2}{3}\right)$ & $\left(1,1,1\right)$ & $\left(\overline{3},1,\sfrac{1}{3}\right)$ & $\left(1,2,-\sfrac{1}{2}\right)$ & $\left(1,1,0\right)$ \\ \cline{2-4} \cline{5-6}\cline{6-7}
        $SU(5)$ & \multicolumn{3}{c|}{$\mathbf{10}$}  & \multicolumn{2}{c|}{$\mathbf{\overline{5}}$} & $\mathbf{1}$ \\\cline{2-4} \cline{5-6}\cline{6-7}
        $SO(10)$ & \multicolumn{6}{c}{$\mathbf{16}$}  \\
        \hline\hline
    \end{tabular}}
    \caption{Embedding of SM Fermions in \su\, and \so\, grand unified groups}
    \label{tab:c1:embeddingSMfermionsintoGUT}
\end{table}

\subsection{Grand Unification as a UV Completion of SM}
{\label{ss:c1:aspectsofGUTs}}
Once we assume that GUT can act as a UV completion of the SM and quark and leptons can be unified in the multiplet, some of the theoretical inconsistencies of the SM get readily solved.

\begin{itemize}
    \item Charge quantisation.\\
    The operator representing electric charge is required to be traceless \cite{Slansky:1981yr,Langacker:1980js}. Consequently, the total electric charge of the particles within the five-dimensional representation must balance to zero. This condition necessitates that the combined electric charge of the three down-type antiquarks neutralises the charge of an electron, provided neutrinos are chargeless, resulting in the following relationship:
    \beqa{\label{eq:c1:chargequan}}
    3\,Q_{d^C} + Q_{e} \eq 0. \eeqa
Such a relation, given in Eq.~\eqref{eq:c1:chargequan}, is crucial to ensure the electric neutrality of the matter. It is the unification of symmetries which leads to charge quantisation and suggests that the unrelated symmetries of the SM, the strong and electroweak, are also connected at a deeper level.

\item Strengths of different gauge couplings\\
GUT assume that at high energies, the dynamics of fundamental forces are governed by a single gauge coupling representing the strength of a unified force. After the symmetry breaking of the GUT to the SM gauge groups, the gauge couplings for the respective groups evolve differently from the GUT scale to the EW scale. However, the running of the couplings depends on various parameters like the number of generations, but in a simple scenario, one can conclude that the coupling constants increases for larger gauge groups at lower energies. As a result, the coupling for the strong interaction is greater at the electroweak scale compared to that of the weak interaction, explaining why the strong force dominates at low energies~\cite{Murayama:1998qrc}.

\item The three gauge couplings in the SM are unified into a single gauge coupling in GUTs. Additionally, the masses of quarks and leptons are related within the GUT framework as they reside within the same irreps. Therefore, GUTs reduce arbitrariness in the gauge and Yukawa sectors. 

\item $B$ and $L$ are no longer the accidental symmetries of the SM and hence can address matter-antimatter imbalance. 

\end{itemize}
No additional information is needed beyond the fundamental assumptions of GUTs to account for some of SM's inconsistencies. \\

\subsection{Newer Degrees of Freedom}
{\label{ss:c1:newscalars}}
The unification of SM gauge symmetries into simple gauge groups and the syncretism of quarks and leptons within the same multiplet often necessitates the presence of other particles (scalar and vector) pertinent for elementary essentialities. These essential degrees of freedom are often a part of representations larger than the fermion multiplet and contain newer degrees of freedom which are not part of SM. A GUT has following kinds of irreps:
\begin{enumerate}[label=(\roman*)]
    \item Vector-irreps consisting of gauge bosons, including those of the SM ones, are necessary for maintaining gauge invariance under the larger symmetry group $(SU(5),\,SO(10))$. In the context of $SU(5)$, a typical example of such an irrep is the $24_{\mathrm{V}}$-dimensional irrep. Meanwhile, in $SO(10)$, examples includes the $45_{\mathrm{V}}$-dimensional irrep.
 
\item Scalar irreps, which contain different scalar particles, facilitate the breaking of larger gauge symmetries into the SM gauge symmetry. These irreps include SM singlet scalars that break the larger symmetry upon acquiring a $vev$. For example, in \(SU(5)\), irreps such as \(24_{\hh}\) and \(75_{\hh}\) are involved, while in \(SO(10)\), irreps like \(45_{\hh}\), \(54_{\hh}\), and \(210_{\hh}\) are used. Additionally, scalar irreps are crucial for breaking the SM gauge symmetry into \(SU(3)_C \times U(1)_{\text{EM}}\) and are necessary for reproducing the mass spectrum and mixing angles of SM fermions. Examples include \(5_{\hh}\), \(45_{\hh}\) in \(SU(5)\), and \(10_{\hh}\), $16_{\hh}$,\(120_{\hh}\), and \(\overline{126}_{\hh}\) in \(SO(10)\)~\cite{Langacker:1980js, Slansky:1981yr}.
\item Occasionally, the GUT framework predicts additional fermions not present in the SM. For instance, in \(SO(10)\), the unification of SM fermions in a single sixteen-dimensional irrep necessitates the presence of a singlet field \(\nu^C\), which could be a right-handed neutrino. The presence of right-handed neutrino makes the theory left-right symmetric above the seesaw scale. The natural presence of this singlet makes \(SO(10)\) GUT applaudingly appealing and supports the possibility of neutrino masses through the seesaw mechanism, further enhancing its attractiveness as a unified theory.    
\end{enumerate}

These particles can convert quarks into leptons or antileptons or vice versa, which serve as the basis for new interactions. Such conversions do not occur at the perturbative level\footnote{$B$ and $L$ conversion can occur in the SM due to non-perturbative effects, for example, like sphalleron transition, but the transition rate is too suprressed~\cite{Espinosa:1989qn}.} within the SM. These scalars can also violate flavour and alter the fermions' chirality.
.

\subsection{Scalar Sector of GUTs and their Implications}

New particles reside together with the known SM particles in the irrep. The dimensionality of these irreps is typically larger than the fermion multiplets. These new degrees of freedom can have the following implications.

These new scalars and gauge bosons mediate processes which are not possible in the framework of SM. Hence, these novel phenomena are smoking gun signals of GUTs. The capabilities of new particles challenge the accidental symmetry of the SM, i.e. the conservation of baryon number~\cite{Weyl:1929fm}. The violation of the baryon number has the following consequences.
\begin{itemize}
    \item Nucleon decay\\
No bound states of baryons are less massive than the proton~\cite{ParticleDataGroup:2022pth}; hence, proton decay would violate the principle of baryon number conservation. Proton decay typically happens at higher energy $(\sim M_{\mathrm{GUT}})$, as predicted by GUTs~\cite{Pati:1973uk,PhysRevLett.43.1566,PhysRevLett.43.1571} and the rarity of such processes suggests the amount of baryon number violation is present in very less amount in the universe.

Sakharov first postulated the necessity of baryon number violation to explain the universe's baryon asymmetry~\cite{Sakharov:1967dj}. Proton decay is a fundamental prediction of GUTs, as fundamental as the prediction of neutral currents in weak interactions. 

Various experiments have been set up to detect the proton decay. The Super-Kamiokande (Super-K) detector, commissioned in the 1990s, is the largest proton water Cherenkov detector, most sensitive to the $p\to e^{+}\,\pi^0$ mode~\cite{Super-Kamiokande:2020wjk}. Hyper-Kamiokande, an upgrade over Super-K, will start data collection by 2027~\cite{Hyper-Kamiokande:2018ofw}. The Jiangmen Underground Neutrino Observatory (JUNO), a scintillator detector, will be commissioned by late 2024 and is most sensitive to the $p\to \overline{\nu}\,K^{+}$ channel~\cite{JUNO:2022qgr}.

Proton decay is detected by observing the resulting particles from a proton decay event using detectors like Super-K and Hyper-K, which utilise the Cherenkov effect, and scintillator detectors like JUNO, which detect light produced by charged particles in a scintillating material. So far, there is no statistically significant signal suggesting protons can decay.
    
\item Neutral Nucleon - antinucleon oscillation\\
The neutron-antineutron (\(n - \overline{n}\)) oscillation~\cite{Mohapatra:1980qe,Mohapatra:2009wp}  violates the baryon number by two units and occurs in GUTs at higher order. In contrast to nucleon decay processes, $n-\overline{n}$ happens at relatively smaller energies $10^{4-6}$ GeV.  Additionally, this oscillation effectively renders the neutron a Majorana fermion, albeit with an extremely small Majorana component. These processes are particularly interesting because they necessitate the violation of \(B-L\), which is intrinsically linked to generating neutrino masses in GUTs. Moreover, they provide a more direct probe of the seesaw scale than nucleon decay processes.

 There are two principal experimental approaches for searching $n-\overline{n}$ oscillations: 1) The first utilises free neutrons obtained from nuclear reactors or neutron spallation sources like the Institut Laue-Langevin (ILL) in Grenoble~\cite{Baldo-Ceolin:1994hzw}, 2) the second approach involves monitoring bound neutrons within nuclei at facilities such as Super-K~\cite{Super-Kamiokande:2020bov}. Again, these processes have not yet been detected.

\item Further, these newer degrees of freedom can also violate the lepton number~\cite{Majorana:1937vz,Racah:1937qq,Furry:1939qr} and explain the massiveness of neutrinos~\cite{Pontecorvo:1957qd,Maki:1962mu}. They can also induce flavour-violating processes like neutral meson-antimeson oscillation~\cite{Kobayashi:1973fv} and account for cosmological baryon asymmetry of the universe~\cite{Sakharov:1967dj}.
\end{itemize}

\begin{figure}[t]
    \centering
    \includegraphics[width=0.75\linewidth]{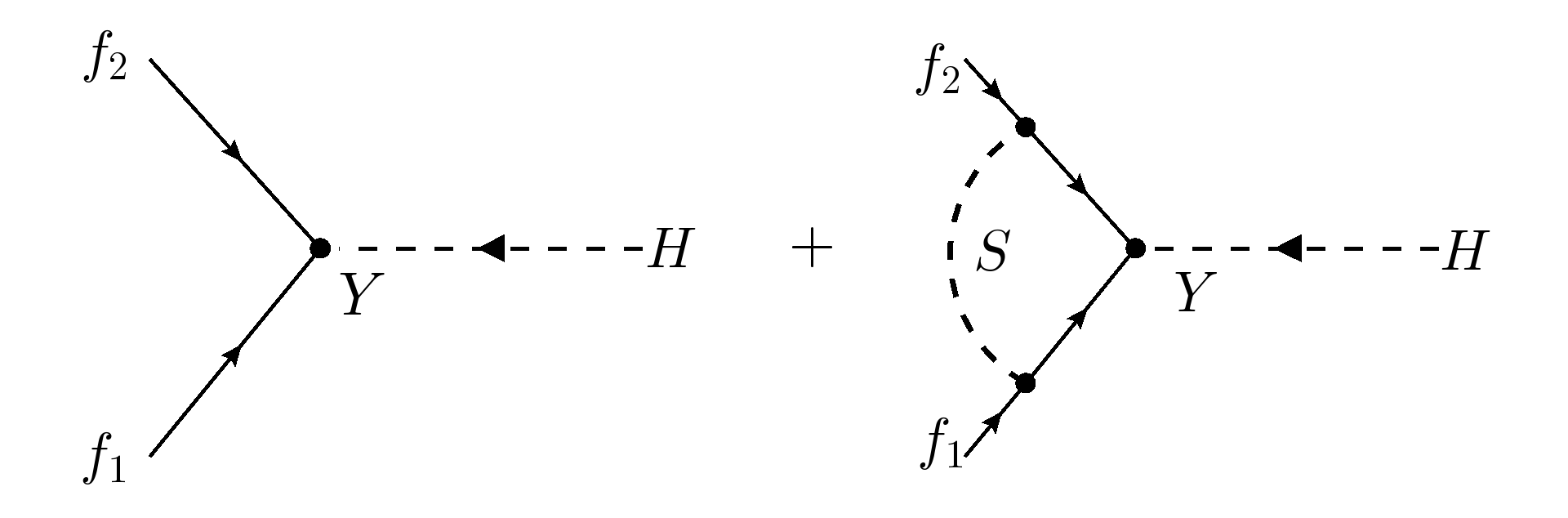}
    \caption{Scalar induced correction to Yukawa coupling.}
    \label{fig:c1:correction}
\end{figure}

These new particles can also affect low-energy observables like gauge couplings and Yukawa couplings. These particles can modify the couplings through quantum corrections, as depicted in Fig~(\ref{fig:c1:correction}). Given that any GUT framework introduces many new particles, they can, in principle, modify every low-energy observable.  These low-energy observables' precise predictions and measurements serve as a signature of GUTs.

As outlined above, introducing new degrees of freedom, including gauge bosons and scalars, can induce phenomena that are not permissible within the SM. Following below, we enumerate the distinction between gauge and scalar bosons and the necessity of focusing on the scalar sector.

\begin{enumerate}[label=(\roman*)]
    \item    Typically, heavy gauge bosons in GUTs obtain masses through the Higgs Mechanism, which is proportional to the GUT breaking scale. This scale is primarily determined by two factors: 1) the convergence of the three SM gauge couplings and 2) the requirement to be consistent with the lower bound from the proton decay rate that aligns with experimental results. Consequently, the masses of the heavy gauge bosons $(M_{X})$ are generally on the order of the GUT scale, which is approximately \(M_{X}\,\sim\,M_{\text{GUT}} \geq 10^{16}\) GeV~\cite{Nath:2006ut}.
    
Although spontaneous symmetry breaking can modify the masses of scalar particles, it does not inherently generate these masses. The exact mechanism for giving distinct masses to different scalar particles within the same irreps in GUTs remains challenging. 
 \item In GUTs, various scalars, including the SM Higgs, coexist within the same irrep. Consequently, scalar-induced interactions are associated with the same couplings that are critical for generating fermion masses in these theories. As a result, these yet-to-be-observed phenomena are deeply connected to the well-studied and tightly constrained sector of the low-energy quark and lepton mass spectrum.
 \item The processes mediated by gauge bosons in GUTs show less sensitivity to the flavour structure of the underlying theory. This is due to the universal nature of the gauge couplings, which couples equally with all generations of fermions.\footnote{Theories in higher spacetime dimensions feature gauge couplings that are not universal in nature~\cite{Randall:1999ee}.} This small sensitivity largely results from the mild effects of rotating to the physical basis. In contrast, processes induced by scalar interactions exhibit high sensitivity to the flavour structure of the underlying GUT, as they depend on the Yukawa couplings, which are preferentially hierarchical in nature.

\item Scalars, in contrast, to gauge bosons, can acquire a $vev$, leading to the breaking of symmetries present in GUTs. For instance,  breaking of the $B-L$ 
symmetry occurs when an SM singlet, charged under the same,
 acquires a $vev$. Such symmetry breaking can contribute to several phenomena, including neutrino mass generation, $B-L$ violating nucleon decays and $n-\overline{n}$ oscillations.

\end{enumerate}

 From the above discussion, it is evident that scalars play a crucial role in probing GUTs, making it essential to study their mass spectrum. One of the direct ways to predict the masses of different scalars is through the analysis of the scalar potential. However, since the scalar potential depends on numerous arbitrary coefficients, certain hypotheses are often adopted to simplify the analysis. A notable hypothesis in this context is the Extended Survival Hypothesis (ESH)~\cite{delAguila:1980qag,Mohapatra:1982aq,Dimopoulos:1984ha}. The ESH suggests that only the scalars necessary for breaking a symmetry at a given energy scale are present at that scale and thus can remain light. This hypothesis reduces the number of fine-tunings to achieve a desired mass scale, with all other scalars assumed to be heavier than that scale.

In this thesis, our aim is to study the phenomenology associated with various scalars and thereby constrain their mass spectrum from phenomenological perspective only. While the ESH provides an economical framework to constrain the mass spectrum of various scalars, it does not fully capture the complexity of scalar dynamics in GUTs. There are cases where the ESH assumptions might oversimplify the scalar spectrum, potentially missing key phenomenological effects from additional scalar fields by making them superheavy. These additional scalars could contribute to low-energy observable effects, indicating the presence of an intermediate scale. Further, it may also require an alternate mechanism that creates substantial hierarchies among different scalar fields within the same multiplet. Nevertheless, we assume that scalar particles within the same irrep can acquire distinct masses and remain significantly below the GUT scale, even if this requires additional fine-tunings and goes beyond the ESH framework.  

We hope to have convinced the reader about the unique importance of the scalars sector and their potential to unveil various subtleties of the structural frameworks of GUTs. The scalar sector is intricately interconnected with the flavour structure of GUTs, highlighting its pivotal role in revealing underlying texture. Further, as they can acquire less mass than the gauge bosons in GUTs, they act as a direct window to validate the GUTs.

\subsection{Underexplored Aspects in the GUTs}

 Scalars have been extensively studied within the bottom-up framework for their roles in low-energy physics, particularly in flavour-changing processes in the quark sector~\cite{Shanker:1981mj,Dorsner:2016wpm}, anomalies in B physics~\cite{Becirevic:2018afm}, and flavour violation in the lepton sector~\cite{Husek:2021isa}. Additionally, their role in generating neutrino masses~\cite{Cai:2017jrq}, addressing issues like CP violation~\cite{Hall:1986fi}, serving as dark matter candidates~\cite{Boehm:2003hm}, and generating cosmological baryon asymmetry of the universe~\cite{Bowes:1996ew,Baldes:2011mh,Fridell:2021gag} is quite well known. Bottom-up approaches typically restrict specific types of couplings by imposing certain symmetries. For example, conserving the baryon number forbids leptoquark couplings of scalars. This freedom is exploited to tune the masses of scalars to reside at lower scales (TeV).

GUTs have also been extensively explored for their capability to yield a realistic fermion mass spectrum across various classes of models~\cite{Babu:1992ia,Bajc:2005zf,Joshipura:2011nn,Altarelli:2013aqa,Dueck:2013gca,Meloni:2014rga,Meloni:2016rnt,Babu:2016bmy,Ohlsson:2018qpt}, achieve gauge coupling unification~\cite{Ellis:1990wk,Mambrini:2015vna,Babu:2015bna}, generate baryogenesis~\cite{Bodeker:2020ghk,Mummidi:2021anm}, and provide insights into the Dirac or Majorana nature of neutrinos~\cite{Strumia:2006db}.

Realistic GUTs, in both renormalisable and non-renormalisable forms, have not yet been fully explored in terms of their capacity to determine the direct implication of various scalars in driving diverse phenomena that signify new physics. Primary among these are processes that violate baryon numbers, such as 1) proton decay, 2) neutron-antineutron oscillation, and 3) baryogenesis. Additionally, these scalars can also induce flavor-violating phenomena. In realistic models, the ability to adjust Yukawa couplings is significantly limited as they are stringently constrained by the need to produce a viable fermion mass spectrum. Consequently, these scalars can be stringently constrained without resorting to arbitrary impositions on the couplings. This exploration will clarify the comprehensive role of scalars in inducing various phenomena and put stringent limits on their masses.

Furthermore, different irreps featuring various types of scalars (and gauge bosons) are inherently present in GUTs in a compact form, and they can potentially influence Yukawa relations at the loop level and modify them in ways that improve the conventional understanding of the Yukawa sector, thereby aiding in the development of minimal\footnote{The notion of minimality is highly subjective. Here, by minimality, we mean the minimum number of parameters in the Higgs Sector.} GUT models. The indirect implications of scalars in correcting conventional tree-level Yukawa relations have not been fully explored, thus warrant a thorough examination.

\section{Thesis: Objectives and Structure}

\lettrine[lines=2, lhang=0.33, loversize=0.15, findent=0.15em]{T}{HIS THESIS AIMS} to investigate and comprehensively analyse the role of the scalar sector in GUTs, exploring both the direct and indirect implications and consequences of scalar fields and constraining them within a realistic GUT framework. The objectives include computing effective couplings using canonically normalised tensors in $SO(10)$ and $SU(5)$ GUTs and evaluating scalar vertices with SM fermions in both renormalisable and non-renormalisable models. Additionally, it seeks to comprehensively analyse the impact of scalars on nucleon stability in realistic renormalisable and non-renormalisable $SO(10)$ GUTs and to constrain the same in realistic \so\, frameworks. Furthermore, the thesis examines the role of other scalars in inducing quark flavour violation and inducing other baryon number violating processes like neutron-antineutron oscillation and the ability to generate the cosmological baryon asymmetry of the universe and constraining the scalars in realistic \so\, framework. Finally, the study also aims to establish the potential of scalars to modify leading-order Yukawa correlations in GUTs, with the motive to minimalise the Yukawa sector of grand unification.

The thesis is organised as follows: In Chapter~(\ref{ch:2}), the methodology for evaluating the effective couplings is detailed by outlining the procedure to obtain an SM invariant coupling from an $SO(10)$ invariant Lagrangian. This chapter also describes obtaining canonically normalised tensors and applying the same analyses to compute effective couplings in renormalsable and non-renormalisable \so\, GUTs. Building on this foundation, Chapter~(\ref{ch:3}) comprehensively classifies scalar capable of inducing proton decays in renormalisable \so\, GUT. Operators for both $B-L$ conserving and violating modes are constructed, and their strengths are computed and examined. The branching patterns of proton decays in a realistic $SO(10)$ model are computed, the distinction between the scalar-induced decays from the conventionally known gauge boson-induced decays is made, and scalars are constrained in the same realistic model. Chapter~(\ref{ch:4}) explores scalar-mediated proton decays in non-renormalisable \so. We particularly focus on the scalar residing in $16_{\rm H}$ and compute their effective couplings with SM fermions. Further, proton decay operators from dimension five interactions are derived, and the different contribution to its strength stemming from various pairs of scalars is computed. In a realistic \so\, model based on $16_{\mathrm{H}}$, stringent limits are put on scalars which affect the proton stability.

Chapter~(\ref{ch:5}) deals with those scalars that do not induce proton decays at leading order in renormalisable \so\,i.e. sextet scalars. Their contribution in inducing quark flavour violation and neutron-antineutron oscillations is computed. Further, constraints coming from the perturbativity of sextet scalars' quartic couplings are derived, and their capability to generate cosmological baryon asymmetry is examined. The chapter concludes with constraints on the masses of sextet scalars within a realistic $SO(10)$ model. In a nutshell, Chapters~(\ref{ch:3}, \ref{ch:4}, and \ref{ch:5}) discusses the direct phenomenological implications of scalars sector of realistic \so\, GUTs.

Expanding the discussion on scalars, Chapter~(\ref{ch:6}) investigates the role of quantum correction in altering the Yukawa relations. It highlights the inconsistency in fermion masses with a single five-dimensional Higgs in the Yukawa sector of \su, expanding the model to include singlet fermions and deriving one-loop matching expressions. Vertex corrections and wave function renormalisation induced by heavy particles are computed. The viability of these one-loop Yukawa relations for achieving realistic fermion masses is tested, showing the necessity of re-examining and minimising Yukawa sectors in conventional GUT models by incorporating quantum corrections. Finally, Chapter~(\ref{ch:7}) concludes the thesis by summarising the key results and highlighting several potential directions for future works. A set of Appendices~(\ref{app:2}-\ref{app:6}) is provided, which contains various supplementary materials, calculations, and additional data supporting the main text.

\clearpage
\begin{savequote}[0.75\linewidth]
	 \qauthor{}
\end{savequote}
\chapter{Methodology for Evaluating Effective Couplings}
\label{ch:2}
\graphicspath{{20_Chapter_2/fig_ch2/}}
\section{Overview}
\label{sec:c2:overview}
\lettrine[lines=2, lhang=0.33, loversize=0.15, findent=0.15em]{W}E START BY outlining the general prerequisites of a GUT group capable of encompassing the SM gauge group in section~(\ref{sec:c2:preGM}), outline some symmetry breaking aspects of \so\, GUT in section~(\ref{sec:c2:symb}). By employing the oscillator expansion technique to decompose the \(SO(10)\) invariant couplings of scalar irreps contributing to the Yukawa Lagrangian, both at the renormalisable and non-renormalisable levels, into the SM invariant form. In the renormalisable sector, we focus on the \(10_{\hh}\) and $45_{\mathrm{V}}$ dimensional irrep, and compute $\fs-\fs-10_{\hh}$ couplings in section~(\ref{sec:c2:RSO}) and $\fs^{\dagger}-\fs-45_{V}$ in section~(\ref{sec:c2:gaugeinteraction}). While in the non-renormalisable sector, we concentrate on the irrep having minimal degrees of freedom, $i.e.$ \(16_{\hh}\) irrep and show its decomposition in section~(\ref{sec:c2:NRSo}). This chapter is summarised in section~(\ref{sec:c2:conclusion}), and the technique outlined in this chapter will be utilised in subsequent chapters to decompose various couplings of various irreps, study various novel phenomena and constrain various scalars within a realistic \(SO(10)\) model.

\section{Prerequisites for a $GUT$ model}
{\label{sec:c2:preGM}}

SM, which describes the known fundamental forces (except gravity), has been elaborated in Chapter~(\ref{ch:1}). The necessity for a grand unified theory has also been pointed out.  Below, we enumerate several prerequisites that any grand unification model must satisfy~\cite{Langacker:1980js, Georgi:1979md}.

\begin{enumerate}[label=(\roman*)]
    \item The gauge symmetry of the SM is a direct product of three mutually commuting groups, denoted as \( SU(3)_{\mathrm C} \times SU(2)_{\mathrm L} \times U(1)_{\mathrm Y} \), collectively known as the SM gauge group. These three distinct symmetries correspond to two, one, and one diagonal generators, summing up to four. The number of diagonal generators is termed as the rank of the group. Therefore, any grand unification group must have at least rank four~\cite{Georgi:1974sy}. Primary group candidates with rank greater than or equal to four are $SU(N)$ and $SO(2N)$ with $N\geq$ $5$~\cite{Slansky:1981yr,Georgi:2000vve,Zee:2016fuk}.
    \item The SM is chiral in nature. As a result, left-handed and right-handed fields have different charges under $SU(2)_{\mathrm{L}}$ gauge symmetry. Hence, any GUT model must have complex representations to accommodate chiral fermions~\cite{Slansky:1981yr}. $SU(N)$ and $SO(4N+2)$, with $N\geq 2$, and  $E_{6}$ exhibits complex representations. 
    
    \item  Anomalies represent a breakdown of classical symmetries at the quantum level, leading to inconsistencies such as non-conservation of currents, violation of gauge invariance, or loss of unitarity. Any class of model based on gauge theories will be renormalisable if it is free of triangle anomaly~\cite{PhysRevD.6.429}. A fermion triangle diagram, when computed, results in infinite and is a source of an anomaly. All irreps of $SO(N)$, with $N\neq6$, are anomaly-safe. Unsafe irreps of $SU(N)$ have been identified in~\cite{Okubo:1977sc,Banks:1976yg}. It turns out the triangle anomalies of $1$, $\overline{5}$ and $10$ irreps of $SU(5)$ add up to zero, which is due to the fact $SU(5)$ is a subgroup of an anomaly-free group $SO(10)$.
    \item A GUT model must exhibit irreps accomodating SM fermions, gauge bosons and Higgs.

\end{enumerate}
Some primary potential and popular GUT candidates are $SU(5)$ and $SO(10)$, with the following characteristics.
\begin{enumerate}[label=(\roman*)]
\item \su\,and \so\, satisfy all the prerequisites to be a GUT group candidate mentioned above.
\item $SU(5)$ and $SO(10)$ are simple Lie groups in which single gauge coupling governs the dynamics above the symmetry-breaking scale.

\item A single sixteen-dimensional irreducible $SO(10)$ spinorial representation can accommodate an entire generation of SM fermions and additionally predicts a singlet fermion (cf. Tab.~(\ref{tab:c1:embeddingSMfermionsintoGUT})). While the SM fermions are partially unified in $\ff$ and $\ft$ dimensional irreps of $SU(5)$.

\item $SO(10)$ is a rank-five group and offers extensive breaking scenarios. It has other rank five subgroups such as \(SU(5) \times U(1)_{\mathrm X}\) and \(SO(6) \times SO(4)\), which could be appropriately broken to yield SM gauge symmetry. $SU(5)$ is a rank four group and is a subgroup of $SO(10).$

\end{enumerate}

 In Appendix~(\ref{app:2}), we review the elementary properties and characteristics of $SO(N)$ and $SU(N)$.

\section{Symmetry breaking}
\label{sec:c2:symb}

 Any imposed symmetry, \(G_{\text{GUT}}\), on the Lagrangian, should be consistent with two elementary demands: 
i) \(G_{\text{GUT}} \supset G_{\text{SM}}\), meaning that the imposed symmetry should contain the SM gauge group as one of its subgroups, ii) it should reproduce the observed fermion masses and mixing angles. Different scalar representations are pertinent to meet these basic demands. 
The conventional way to break a symmetry is via spontaneous symmetry breaking, also called the Higgs Mechanism~\cite{Higgs:1964ia,Fritzsch:1974nn,Buccella:1980qb}. In this mechanism, to break any group, in particular, \so, into one of its subgroups, we need a scalar field that transforms trivially under a certain subgroup. When that scalar field acquires a $vev$, the larger symmetry is broken into that particular subgroup. Since \(SO(10)\) is a rank-five group, various breaking patterns can yield the SM gauge symmetry, highlighting the complexity of the Higgs sector of \(SO(10)\). Different breaking chains can influence the renormalisation group (RG) running of various observables, including the gauge couplings~\cite{Ross:1978wt,Hall:1980kf,Antoniadis:1981gh,Ellis:1990wk}, and can have different phenomenological implications~\cite{Erler:2004in,Mambrini:2015vna}.

\begin{figure}
    \centering
    \includegraphics[width=0.9\linewidth]{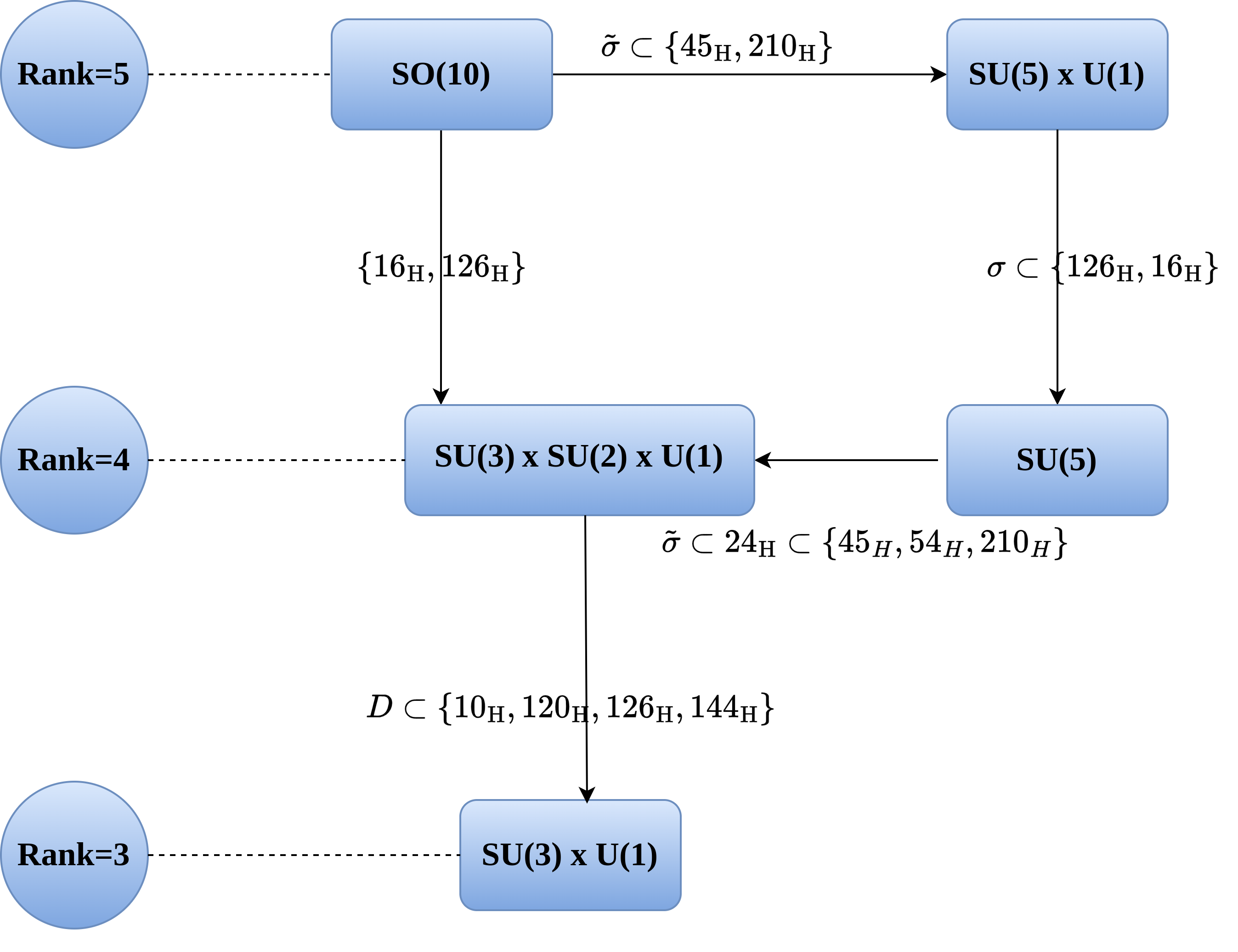}
    \caption{Breaking pattern of \so\, to \smg\, via \su. The SM charges of scalars acquiring $vev$ can be inferred from Tabs.~(\ref{tab:c2:gscalars} and \ref{tab:c2:scalars})}
    \label{fig1:c2:breakingpattern}
\end{figure}

Although many breaking chains are available in $SO(10)$, we shall outline its breaking via only one particular path, $SU(5)\times U(1)_{\mathrm{X}}$ path, which is shown in Fig.~(\ref{fig1:c2:breakingpattern}).
Initially, a $45_{\vv}$-dimensional irrep containing massless gauge bosons is present to preserve the \so\, gauge invariance. Another irrep of scalar fields is required to break the \so\ symmetry into one of its subgroups, $SU(5)\times U(1)_{\mathrm{X}}$. This is achieved using the scalar field irreps $45_{\hh}$ and $210_{\hh}$~\cite{Slansky:1981yr, Fukuyama:2004ps, Nath:2006ut}.   $45_{\hh}$-dimensional irrep  is called an adjoint representation, and when the gauge symmetry is broken using an adjoint representation, the rank of the broken group remains unchanged~\cite{Slansky:1981yr}.

To illustrate the symmetry-breaking mechanism, consider the $45_{\hh}$ irrep. The decomposition of $45_{\hh}$ into irreps of $SU(5)\times U(1)_{\mathrm{X}}$ and the \smg\, is shown below~\cite{Slansky:1981yr, Feger:2019tvk}, where $(...)$ denotes the $SU(5)\times U(1)_{\mathrm{X}}$ irrep, $[(...)]$ denotes the SM charges of fields residing inside it and $c.c.$ denotes complex conjugates.
\beqa{\label{eq:c2:45Hdec}}
45_{\hh}\eq \left(1,0\right)\,\left[\,\left(1,0,0\right)\,\right] \oplus \left(24,0\right)\,\left[\,\left(1,3,0\right)\oplus \left(8,1,0\right) \oplus \left(3,2,-\frac{5}{6}\right) \oplus \left(\overline{3},2,\frac{5}{6}\right)\,\right]\nl
&\oplus& \left\{\left(10,4\right) \left[\,\left(3,2,\frac{1}{6}\right)\oplus \left(\overline{3},1,\frac{2}{3}\right) \oplus \left(1,1,1\right)\right] \oplus \rm{c.c.}\right\}
\eeqa
When \(1_{\hh} \equiv \tilde{\sigma}\) acquires a $vev$, \so\, symmetry is broken into $SU(5)\times U(1)_{\mathrm{X}}$. $20$ gauge bosons residing in $45_{\vv}$ dimensional irrep, which are not part of \(SU(5)\), absorb the degrees of freedom of scalars residing in $45_{\hh}$  and become massive. They obtain masses of the order of the symmetry breaking scale $(\langle\tilde{\sigma}\rangle)$.  Some of the massive gauge bosons are potential proton decay mediators (cf. Chapter~(\ref{ch:3})) and thus dictate the magnitude of the $vev$ to satisfy the proton decay constraints. Typically, \(\langle\tilde{\sigma}\rangle \equiv M_{\text{GUT}} \simeq 10^{16}\,\text{GeV}\).

Another irrep that can induce \so\, symmetry breaking is $210_{\hh}$, whose decomposition into $SU(5)\times U(1)_{\mathrm{X}}$ is as follows:
\beqa{\label{eq:c2:54}}
210_{\hh}\eq \left(1,0\right) \oplus \left\{\left(5,-8\right) + \rm{c.c.}\right\} \oplus \left\{\left(10,4\right) + \rm{c.c.}\right\}\oplus \left(24,0\right)\nl &\oplus&  \left\{\left(40,4\right) + \rm{c.c}\right\} \oplus \left(75,0\right).
\eeqa
There is also \(54_{\text{H}}\) dimensional irrep consisting of scalar fields, which is also used to break the \so\, symmetry~\cite{Babu:1984mz}. However, it lacks $SU(5)\times U(1)_{\mathrm{X}}$ singlet and cannot break \(SO(10)\) to \(SU(5) \times U(1)_{\mathrm{X}}\). As evident from Eqs.~(\ref{eq:c2:45Hdec}, and \ref{eq:c2:54}), upon decomposition under the SM gauge group, these irreps contain multiple multiplets charged under \(B-L\), potentially leading to interactions that violate baryon and/or lepton numbers. The SM and \(B-L\) charges of these fields, along with their multiplicities in \($45$_{\hh}\) and \(210_{\hh}\), are presented in Tab.~(\ref{tab:c2:gscalars}). We adopt the convention where,
\beqa{\label{eq:c2:B-LRel}}
 B-L = \frac{4}{5}Y - \frac{1}{5}X\,,
 \eeqa
where \(X\) and \(Y\) denote the charges under the \(U(1)_{\mathrm{X}}\) and \(U(1)_{\mathrm{Y}}\) subgroups of \(SO(10)\), respectively~\cite{Buchmuller:2019ipg}. The normalisation of Hypercharge and its relation to Electric charge can be inferred from Eq.~\eqref{eq:c1:electricharge}.

\begin{table}[t!]
\begin{center}
\begin{tabular}{lcrcc} 
\hline
\hline
~~SM charges~~&~~Notation~~&~~$B-L$~~&~~$45_{\hh}$~~&~~$210_{\hh}$~~\\
 \hline
$\left(1,1,0\right)$ &$\hat{\sigma}$ &$0$ &$2$ &$3$ \\
$\left(1,1,-1\right)$ &$\hat{\overline{s}}$ &$0$ &$1$ &$1$ \\
$\left(1,1,1\right)$ &$\hat{s}$ &$0$ &$1$ &$1$ \\
$\left(1,2,-\frac{3}{2}\right)$ & $ \overline{{\mathcal D}}^a $ &$-2$ &$0$ &$1$ \\
$\left(1,2,\frac{3}{2}\right)$ & $ {\mathcal{D}}_a $ &$2$ &$0$ &$1$ \\
$\left(1,2,-\frac{1}{2}\right)$ & $\hat{D}^a$ &$2$ &$0$ &$1$ \\
$\left(1,2,\frac{1}{2}\right)$ & $\hat{D}_a$ &$-2$  &0 &$1$\\
$\left(1,3,0\right)$ & $W^a_b$ &$0$  &$1$ &$1$\\
$\left(3,1,-\frac{1}{3}\right)$   &$\hat{T}^{\alpha}$ &$\frac{4}{3}$ &$0$ &$1$ \\
$\left(\overline{3},1,-\frac{1}{3}\right)$ & $\hat{\overline{T}}_{\alpha}$ &$-\frac{4}{3}$ &$0$ &$1$\\ 
$\left(3,1,\frac{2}{3}\right)$   &$\hat{\Theta}_{\alpha\beta}$ &$\frac{4}{3}$ &$1$ &$2$ \\
$\left(\overline{3},1,-\frac{2}{3}\right)$ & $\hat{\overline{\Theta}}^{\alpha\beta}$ &$-\frac{4}{3}$ &$1$ &$2$\\
$\left(\overline{3},1,-\frac{5}{3}\right)$ & ${\overline{\mathcal{U}}}^{\alpha} $ &$-\frac{4}{3}$ &$ 0$ &$1$\\
$\left(3,1,\frac{5}{3}\right)$ & ${\mathcal{U}}_{\alpha} $ &$\frac{4}{3}$ &$ 0$ &$1$\\
$\left(3,2,\frac{1}{6}\right)$ & $Y^{a\,\alpha}_a$ &$-\frac{2}{3}$ &$1$ &$2$\\ 
$\left(\overline{3},2,-\frac{1}{6}\right)$ & $\overline{Y}^a_{\alpha}$ &$\frac{2}{3}$ &$1$ &$2$\\ 
$\left(3,2,-\frac{5}{6}\right)$ & $X^{\alpha}_a$ &$-\frac{2}{3}$ &$1$ &$2$\\ 
$\left(\overline{3},2,\frac{5}{6}\right)$ & $\overline{X}^a_{\alpha}$ &$\frac{2}{3}$ &$1$ &$2$\\
$\left(3,3,\frac{2}{3}\right)$ & $ {\mathcal{V}}_{\alpha\beta\,a}^{b}$ &$\frac{4}{3}$ &$0$ &$1$\\
$\left(\overline{3},3,-\frac{2}{3}\right)$ & $ \overline{\mathcal{V}}^{\alpha\beta\,a}_{b} $ &$-\frac{4}{3}$ &$0$ &$1$\\
$\left(\overline{6},2,-\frac{5}{6}\right)$ & $\overline{{\mathcal{J}}}_{\gamma\,a}^{\alpha\beta} $ &$-\frac{2}{3}$ &$0$ &$1$\\
$\left(6,2,\frac{5}{6}\right)$ & $ {\mathcal{J}}^{\alpha\beta}_{\gamma\,a} $ &$\frac{2}{3}$ &$0$ &$1$\\
$\left(6,2,-\frac{1}{6}\right)$ & $\overline{\mathcal{K}}^{\alpha\beta\,a}_{\gamma} $ &$\frac{2}{3}$ &$0$ &$1$\\
$\left(\overline{6},2,\frac{1}{6}\right)$ & $ {\mathcal{K}}_{\gamma}^{\alpha\beta\,a} $ &$-\frac{2}{3}$ &$0$ &$1$\\
$\left(8,1,0\right)$ & $G^\alpha_\beta$ &$0$ &$1$ &$2$\\ 
$\left(8,1,-1\right)$ & $ {\overline{\mathcal{G}}}^{\alpha}_{\beta} $ &$0$ &$0$ &$1$\\ 
$\left(8,1,1\right)$ & $\mathcal{G}^{\alpha}_{\beta} $ &$0$ &$0$ &$1$\\ 
$\left(8,3,0\right)$ & $\mathbb{G}^{\alpha\,a}_{\beta\,b} $ &$0$ &$0$ &$1$\\
\hline\hline
\end{tabular}
\end{center}
\caption{Classification of different scalar fields residing in $45_{\hh}$ and  $210_{\hh}$ with their charges under the SM gauge group ($SU(3)_{\mathrm{C}}$, $SU(2)_{\mathrm{L}}$, $U(1)_{\mathrm{Y}}$), $B-L$ charge and multiplicities in the given GUT representation. }
\label{tab:c2:gscalars}
\end{table}

Once the \(SO(10)\) gauge symmetry is broken into \(SU(5) \times U(1)_{\mathrm{X}}\), it is further broken into \(SU(5)\) via a singlet charged under \(U(1)_{\mathrm X}\). This further breaking is achieved by the \so\, irreps \(16_{\hh}\) and \(\overline{126}_{\hh}\), $i.e.$ $\sigma$ (cf. Tab.~(\ref{tab:c2:scalars})). The \(\overline{126}_{\hh}\) contributes to the \so\, Yukawa Lagrangian at the renormalisable level, while the \(16_{\hh}\) couples at the non-renormalisable level. Furthermore, to break \(SU(5)\) into the SM gauge group, an \su\, singlet residing in \(24_{\text{H}}\) assumes a $vev$, leading to the breakdown of \su\, into the SM gauge symmetry. Additionally, the SM gauge symmetry is broken into \(SU(3)_{\mathrm C} \times U(1)_{\mathrm{EM}}\) by multiple SM Higgs-like scalar particles residing in the irreps \(10_{\hh}\), \(120_{\hh}\), and \(\overline{126}_{\hh}\) of \(SO(10)\).

So far, we have discussed symmetry breaking in \(SO(10)\), which requires different irreps to reach the electroweak symmetry finally. Once these irreps are present in the theory, they contain different scalar particles and gauge bosons with distinct charges (SM and $B-L$). In further sections, we compute the decompose the \so\, invariant coupling of different irreps to the SM invariant couplings via $SU(5)$ path. We show such explicit decomposition by considering the specific examples of scalar irreps contributing to renormalisable \so\,: $(\fs-\fs-16_{\hh})$, $\fs^{\dagger}-\fs-45_{\vv}$ and non-renormalisable \so\, Yukawa sector $(\fs-\fs-16_{\hh}-16_{\hh})$.
\section{Yukawa Interactions in Renormalisable \so}
\label{sec:c2:RSO}
In the renormalisable class of \(SO(10)\) gauge theory, the Yukawa sector typically involves scalars belonging to the $10_{\hh},\, 120_{\hh} $ and $\overline{126}_{\hh}$ dimensional irreps (refer to Appendix~(\ref{app:2})). Analogous to the case of \(45_{\hh}\) and \(210_{\hh}\), the decomposition of \(10_{\hh}\), \(120_{\hh}\), and \(\overline{126}_{\hh}\) dimensional irreps under SM gauge symmetry results in a plethora of scalar fields.
\begin{table}[t!]
\begin{center}
\begin{tabular}{lcrccc} 
\hline
\hline
~~SM charges~~&~~Notation~~&~~$B-L$~~&~~$10_{\hh}$~~&~~$120_{\hh}$~~&~~$\overline{126}_{\hh}$~~\\
 \hline
$\left(1,1,0\right)$ &$\sigma$ &$-2$ &$0$ &$0$ &$1$ \\
$\left(1,1,1\right)$ & $s$ &$2$ &$0$ &$1$ &0\\
$\left(1,1,-1\right)$ & $\overline{s}$ &$-2$  &0 &1 &1\\
$\left(1,1,-2\right)$   &$X$ &$-2$ &0 &0 &1\\
$\left(1,2,-\frac{1}{2}\right)$ & $\overline{D}_a$ &0 &1 &2 &1\\
$\left(1,2,\frac{1}{2}\right)$ & $D^a$ &0 &1 &2 &1\\
$\left(1,3,1\right)$ & $t^{ab}$ &2 &0 &0 &1\\
$\left(3,1,-\frac{1}{3}\right)$ & $T^{\alpha}$ &$-\frac{2}{3}$ &1 &2 &2\\
$\left(\overline{3},1,\frac{1}{3}\right)$ &$\overline{T}_\alpha$ & $\frac{2}{3}$&1 &2 &1\\
$\left(3,1,\frac{2}{3}\right)$ & $\Theta_{\alpha\beta}$ & $-\frac{2}{3}$ &0 &1 &1\\
$\left(\overline{3},1,-\frac{2}{3}\right)$ &$\overline{\Theta}^{\alpha\beta}$ &$\frac{2}{3}$ &0 &1 &0\\
$\left(3,1,-\frac{4}{3}\right)$ & $\mathcal{T}^\alpha$ & $-\frac{2}{3}$ &0 &1 &1\\
$\left(\overline{3},1,\frac{4}{3}\right)$&$\overline{\mathcal{T}}_\alpha$ &$\frac{2}{3}$ &0 &1 &0\\
$\left(3,2,\frac{1}{6}\right)$ & $\Delta^{\alpha a}$ & $\frac{4}{3}$ &0 &1 &1\\
$\left(\overline{3},2,-\frac{1}{6}\right)$& $\overline{\Delta}_{\alpha a}$&$-\frac{4}{3}$ & 0&1 &1\\
$\left(3,2,\frac{7}{6}\right)$ & $\Omega^a_{\alpha \beta}$ & $\frac{4}{3}$ &0 &1 &1\\
$\left(\overline{3},2,-\frac{7}{6}\right)$&$\overline{\Omega}_a^{\alpha \beta}$ & $-\frac{4}{3}$&0 &1 & 1\\
$\left(3,3,-\frac{1}{3}\right)$ & $\mathbb{T}{^{a\alpha}_{b}}$ & $-\frac{2}{3}$ &0 &1 &0\\
$\left(\overline{3},3,\frac{1}{3}\right)$&$\overline{\mathbb{T}}{^a_{b\alpha}}$ &$\frac{2}{3}$ &0 &1 &1\\
$\left(6,1,\frac{1}{3}\right)$ & $S{^{\alpha}_{\beta\gamma}}$ & $\frac{2}{3}$ &0 &1 &1\\
$\left(\overline{6},1,-\frac{1}{3}\right)$&$\overline{S}{^{\beta\gamma}_{\alpha}}$ &$-\frac{2}{3}$ &0 &1 &0 \\
$\left(6,1,-\frac{2}{3}\right)$ & $ \Sigma^{\alpha\beta}$ & $\frac{2}{3}$ &0 &0

 &1 \\
$\left(6,1,\frac{4}{3}\right)$ & ${\cal S}^{\alpha}_{\beta\gamma}$ & $\frac{2}{3}$ & 0 & 0 & 1 \\
$\left(\overline{6},3,-\frac{1}{3}\right)$ &$\mathbb{S}^{\alpha\beta a}_{\gamma b}$ & $-\frac{2}{3}$ &0 &0 &1\\
$\left(8,2,\frac{1}{2}\right)$ & $O^{\alpha a}_{\beta}$ &0 &0 &1 &1\\
$\left(8,2,-\frac{1}{2}\right)$ & $\overline{O}_{\alpha a}^{\beta}$ &0 &0 &1 &1\\
\hline\hline
\end{tabular}
\end{center}
\caption{Classification of different scalar fields residing in $ 10_{\hh}$, $120_{\hh}$, and $\overline{126}_{\hh}$ with their charges under the SM gauge group ($SU(3)_{\mathrm{C}}$, $SU(2)_{\mathrm L}$, $U(1)_{\mathrm Y}$), $B-L$ charge and multiplicities in the given GUT representation.}
\label{tab:c2:scalars}
\end{table}
Tab.~(\ref{tab:c2:scalars}) provides the SM and \(B-L\) charges of various scalars stemming from the irreps of \(SO(10)\) capable of contributing to the renormalisable Yukawa sector. Some of the scalars listed in the Tabs.~(\ref{tab:c2:gscalars} and \ref{tab:c2:scalars}) may have identical SM charges but differ in \(B-L\) charges; consequently,  scalars with identical SM charges are denoted with a tilde (\(\sim\)) over them in Tab.~(\ref{tab:c2:gscalars}).

In this section, we evaluate the couplings of scalar fields residing in the multiplets that participate in the Yukawa interactions at the renormalisable level. Various techniques have been devised to decompose \(SO(10)\) irreps based on the path ($i.e.$ intermediate symmetry) to reach the SM gauge symmetry~\cite{Wilczek:1979hc,Mohapatra:1979nn,PhysRevD.53.3884,Aulakh:2002zr,Fukuyama:2004ps}. Furthermore, an explicit basis for various irreps of \(SO(10)\) has also been constructed in~\cite{Ozer:2005dwq}. One of the popular techniques to decompose invariants of $SO(10)$ into those of $SU(5)$ is via the oscillator expansion technique~\cite{Mohapatra:1979nn, Nath:2001uw}, which we discuss subsequently.

\subsection{Oscillator Expansion of \so}
\label{ssec:c2:OESO}
Consider $N$ mutually anticommuting creation $(b^{\dagger}_{i})$ and annihilation $b_{i}$ operators, where $1\leq i \leq N$ operators.

\beqa{\label{eq:c2:c&a}}
\big\{b_p,b^{\dagger}_q\big\}\eq \delta_{pq}\hspace{0.5cm}\text{and}\hspace{0.5cm} \big\{b_p,b_q\big\}\,=\,\big\{b^{\dagger}_p,b^{\dagger}_q\big\}\,=\,0
\eeqa
We define $2N$ operators $\gamma_p$ with $1\leq p \leq 2N$ as follows:
\beqa{\label{eq:c2:oe1}}
\gamma_{2p-1} &\equiv& -i\left(b_p - b^{\dagger}_p\right)\hspace{0.5cm}\text{and}\hspace{0.5cm} \gamma_{2p}\,=\, \left(b_p + b^{\dagger}_p\right)
\eeqa
It is straightforward to verify that the defined $\gamma$ operators in Eq.~\eqref{eq:c2:oe1} satisfy the Clifford algebra defined in Eq.~\eqref{eq:c2:gammatrans}. Without the need for the exact structure of the operators, one can yet define their action of the vacuum state $\big|0\rangle$.
\beqa{\label{eq:c2:}}
\big|0\rangle &\equiv& \underbrace{\big|0,0,...,0\rangle}_{\rm{N}\, zeros},\hspace{1cm}b_p\,\big|0\rangle\,=\,0,\,\hspace{0.5cm}\text{and}\hspace{0.5cm}b^{\dagger}_p\big|0\rangle\,=\,\underbrace{\big|0,0,.,1,.,0\rangle}_{\text{1 at p$^{th}$ position}}\nl
\eeqa

Further, defining $T_{pq} \equiv b^{\dagger}_p\,b_q$ which satisfies the $U(N)$ lie algebra. Therefore, any spinor representation $\Phi$ can be written into antisymmetric tensors of $U(N)$, $i.e.$ $\phi$, and a bunch of creation and annihilation operators typically called oscillator basis, as shown below:
\beqa{\label{eq:c2:SOspinor}}
\big|\Phi\rangle \eq \sum_{p=0}^{p=N}\frac{1}{p!}\,b^{\dagger}_{i_1}b^{\dagger}_{i_2}...b^{\dagger}_{i_N}\,\big|0\rangle\, \phi^{i_1i_2...i_p}.\eeqa

The spinorial representation of $\fs_{+}$ (cf. Appendix~(\ref{app:2})) in terms of creation and annihilation operators is as follows~\cite{Mohapatra:1979nn}:
\beqa{\label{eq:c2:16f}}
\big|\fs_{+}\rangle&=&  \big|0\rangle \fn + \frac{1}{2}\,b^{\dagger}_p b^{\dagger}_q\,\big|0\rangle\,\ft^{pq} + \frac{1}{4!}\,\varepsilon^{pqrst}\,b^{\dagger}_pb^{\dagger}_qb^{\dagger}_rb^{\dagger}_s \big|0\rangle\,\ff_{t},
\eeqa
 $\fn$, $\ff$ and $\ft$ are \su\, irreps consisting of SM fermions and singlet fermions.
 
 The decomposition of $\ff$ into \smg\, multiplets is shown below:
\beqa{\label{eq:c2:five}}
\ff &=& \left(\overline{3}, 1, \frac{1}{3}\right) \oplus \left(1,\overline{2},-\frac{1}{2}\right). 
\eeqa
It is evident from the Eq.~\eqref{eq:c2:ff} that the decomposed multiplets have the same quantum numbers as those conjugated down quark and lepton doublet (cf. Tab.~(\ref{tab:c1:SMcontent})). 

Similarly, $\ft$ can be decomposed into SM invariant multiples:
\beqa{\label{eq:c2:10-1}}
\ft &=& \left(3,2,\frac{1}{6}\right) \oplus \left(\overline{3},1,\frac{2}{3}\right) \oplus \left(1,1,1\right).
\eeqa
These decomposed multiplets shown in the Eq.~\eqref{eq:c2:ft} correspond to the SM charges of quark-doublet, conjugated up-quark and conjugated electron (cf. Tab.~\ref{tab:c1:SMcontent}). 

The singlet under \su\, can be associated with the missing ally of SM fermions, $i.e.$ right-handed neutrino, as shown below; 
\beqa {\label{eq:c2:fn}}
\fn &=& \nu^C.
\eeqa

Throughout this thesis, we denote generation indices by uppercase Latin alphabets $\big(1\leq A,B,C...\leq3\big)$, \(SU(5)\) indices by lowercase Latin letters $\big( 1 \leq n , o , p,...,z   \leq 5 \big)$, \(SU(3)\) indices by lowercase Greek letters $\big( 1 \leq \alpha, \beta, \gamma... \leq 3 \big)$, and lowercase Latin alphabets $\big( 4\leq a, b, c,...,m \leq 5\big)$ represent the \(SU(2)\) indices, unless specified explicitly. Moreover, bold letters denote fermion representations, while scalar and gauge representations are denoted with subscript $H$ and $V$, respectively.

At this juncture, we also clarify our definitions of the two-indexed (alias the $SU(2)$) $Levi-Civita$ symbols in alignment with the conventions\footnote{Interested readers can refer to the insightful footnote on page 12 of the review~\cite{Dreiner:2008tw} for other definitions of two-indexed $Levi-Civita$ tensors} for Weyl spinors mentioned in~\cite{Dreiner:2008tw}. The two rank $SU(2)\;Levi-Civita$ tensor used is defined as $\varepsilon_{45}=-\varepsilon_{54}=-\varepsilon^{45}=\varepsilon^{54}=1$, while the three rank $SU(3)\;Levi-Civita$ tensor is defined as $\varepsilon_{123}=\varepsilon^{123}=1$, and similarly for all other cyclic permutations. The transformation rule for $SU(2)$ tensor is consistent with the aforementioned two indexed Levi-Civita tensor definitions, which are as follows:
\beqa{\label{eq:c2:SU2def}}
\psi_a\eq\ \varepsilon_{ab}\,\psi^b, \hspace{1cm}\text{and}\hspace{1cm}\psi^a\,=\,\varepsilon^{ab}\,\psi_b.
\eeqa
Further, $\varepsilon_{ab}\,\varepsilon^{cd}\,=\,-\delta_a^c\,\delta_b^d\,+\, \delta_a^d\,\delta_b^c$, and $\varepsilon_{\alpha\theta\phi}\,\varepsilon^{\beta\theta\phi}\,=\, 2\,\delta_\alpha^\beta$. With the notations established for \(SU(2)\) and \(SU(3)\) indices, along with other essential conventions, we can now define the explicit content of the $\ff$ and $\ft$ irreps in terms of SM fermions. In matrix notation, the contents of $\ff$ dimensional irrep are shown below;
\beqa{\label{eq:c2:ff}}
\ff_{\alpha}&=& d^C_{\alpha},\hspace{0.5cm}\text{and}\hspace{0.5cm} \ff_{a}= \varepsilon_{ab}\,l^b\nl
\ff &=& \begin{pmatrix} d^C_{\alpha}\\ \varepsilon_{ab}\,l^b \end{pmatrix}\;=\; \hspace{0.5cm} \begin{pmatrix} d^C_{1}\\ d^C_{2} \\ d^C_{3} \\ e \\ -\nu \end{pmatrix},\; 
\eeqa
where $d_{1,2,3}$ represents the colour degrees of freedom of the down quark. Further, ${\cal C}$ is the charge conjugation matrix in the Lorentz space as defined in Eq.~\eqref{eq:c1:conjugation}. Similarly, the particle content of $\ft$-dimensional irrep is shown below:
\beqa{\label{eq:c2:ft}}
\ft^{a\alpha}&=& q^{a\,\alpha},\hspace{0.5cm}\;\ft^{\alpha\,\beta}\,=\,\varepsilon^{\alpha\beta\gamma}\,u^C_{\gamma},\;\hspace{0.5cm}\text{and}\hspace{0.5cm} \ft^{ab}\,=\,\varepsilon^{ab}\,e^C.\nl  
\ft &=& \begin{pmatrix}
    0 & u^C_3 & -u^C_2 & u^1 & d^1 \\
    -u^C_3 & 0 & u^C_1 & u^2 & d^2 \\ 
    u^C_2 & -u^C_1 & 0 & u^3 & d^3 \\
    -u^1 & -u^2 & -u^3 & 0 & -e^C\\
    -d^1 & -d^2 & -d^3 & e^C & 0 \\
\end{pmatrix}.
\eeqa

\subsection{$\fs-\fs-10_{\hh}$ couplings}
\label{ssec:c2:161610}
We commence the decomposition of the coupling of \(10_{\text{H}}\) with the \(\fs\)-plet using the technique outlined in section~(\ref{ssec:c2:OESO}).
\beqa{\label{eq:c2:161610op}}
\tilde{H}_{AB}\,\fs^T_A\,\cc\,\fs_B\,10_{\hh} &\to& \tilde{H}_{AB}\,\left<\fs^*_{+\,A}\big|\,\cc\,{\cal B}\, \Gamma_{\mu}\,\phi_{\mu}\,\big|\fs_{+\,B}\right> 
\eeqa
${\mathcal{B}}$ is the charge conjugation operator in \so\, space and is defined in terms of creation and annihilation operators as follows~\cite{Mohapatra:1979nn,Nath:2001uw};
\beqa{\label{eq:c2:Bdef}}
{\cal B}\eq -i\, \prod_{p=1}^{5}\,\left(b_p - b_p^{\dagger}\right).
\eeqa

The action of the charge conjugation operator $({\cal B })$, defined in the Eq.~\eqref{eq:c2:Bdef}, on the $\fs$-spinor, defined in Eq.~\eqref{eq:c2:16f}, can be computed and is given as follows~\cite{Mohapatra:1979nn,Nath:2001uw,Cardoso:2015gfa};
\beqa{\label{eq:c2:16fb}}
\langle \fs^*_{+} \big|\,{\cal B} \eq -i\,\ff_p \langle 0 \big| b_{p}  -i\, \frac{1}{12}\,\ft^{pq}\,\varepsilon_{pqrst}\,\langle 0 \big| b_{r}b_s\,b_t -i\, \fn \langle 0 \big|  b_1b_2b_3b_4b_5.\nl 
\eeqa

Similarly, $10_H$ can also be defined in terms of creation and annihilation operators as follows~\cite{Mohapatra:1979nn,Nath:2001uw};
\beqa{\label{eq:c2:10Hac}}
\Gamma_{\mu}\,10_{\mu}\eq b^{\dagger}_p\,\tilde{5}_{\hh}^p + b_p\,\tilde{\overline{5}}_{p\,\hh},
\eeqa
$\phi_{\mu}$ being a real-scalar, its kinetic term is as follows:
\beqa{\label{eq:c2:ke10}}
{\mathcal{L}}_{\rm{KE}}&\supset& \frac{1}{2}\,\left(\partial_{\Dot{\mu}}10_{\mu}\right)^{\dagger}\,\left(\partial^{\Dot{\mu}}\,10_{\mu}\right)\nl
\eq \left(\partial_{\Dot{\mu}}5_{\hh}\right)^{\dagger}\,\left(\partial^{\Dot{\mu}}5_{\hh}\right) + \left(\partial_{\Dot{\mu}}\overline{5}_{\hh}\right)^{\dagger}\,\left(\partial^{\Dot{\mu}}\overline{5}_{\hh}\right),
\eeqa
where $\Dot{\mu}\,\epsilon\,\{0,1,2,3\}$ and are the Lorentz indices, as defined in Chapter~(\ref{ch:1}). To have canonically normalised kinetic energy term we set, $\tilde{5}_{\hh}\equiv \sqrt{2}\,5_{\hh}$ and $\tilde{\overline{5}}_{\hh}\equiv \sqrt{2}\,\overline{5}_{\hh}$. The fundamental aspect of canonical normalisation of the kinetic term lies in its ability to establish the energy-momentum relation in a conventional form, facilitating standard interpretations. Using Eqs.~(\ref{eq:c2:16f}, \ref{eq:c2:16fb}, and \ref{eq:c2:10Hac}), it is straightforward to derive the expression written in Eq.~\eqref{eq:c2:161610op} and it is shown below, where we have omitted $\cc$ for brevity;
\beqa{\label{eq:c2:1616dec}}
\tilde{H}_{AB}\,\left<\fs^*_{+\,A}\big|B\, \Gamma_{\mu}\,10_{\mu}\,\big|\fs_{+\,B}\right> \eq \tilde{H}_{AB}\,\big<\fs^*_A\big|B\, \left(b^{\dagger}_t\,\tilde{5}_{\hh}^t + b_t\,\tilde{\overline{5}}_{t\,\hh}\right)\nl
& & \times\,\bigg(\big|0\rangle\,\fn_B + \frac{1}{2}\,b^{\dagger}_p b^{\dagger}_q\,\big|0\rangle\,\ft^{pq}_B \nl \ad \frac{1}{4!}\,\varepsilon^{pqrst}\,b^{\dagger}_pb^{\dagger}_qb^{\dagger}_rb^{\dagger}_sb^{\dagger}_t \big|0\rangle\,\ff_B \bigg).
\eeqa
It is worth mentioning that the number of creation and annihilation operators must be equal for a non-zero contribution of the matrix element. The contribution involving $\fn$-plet is as follows:
\beqa{\label{eq:c2:fnanc}}
&\Rightarrow &\ i\,\tilde{H}_{AB}\,\ff_{m\,A}\,\langle 0\big|\,b_m\,b^{\dagger}_t\big|0\rangle\,\fn_B\,\tilde{5}_{\hh}^t\nl \mi \frac{i}{24}\, H_{AB}\,\fn_A \langle 0\big| b_1b_2b_3b_4b_5\, b^{\dagger}_tb^{\dagger}_pb^{\dagger}_qb^{\dagger}_rb^{\dagger}_s\big|0\rangle \varepsilon^{pqrsu}\,\ff_{u\,B}\,\tilde{5}^{t}_{\hh}\nl
\eq -i\tilde{H}_{AB}\,\ff_{m\,A}\,\delta^m_t\,\fn_B\,\tilde{5}^t -H_{AB}\,\frac{i}{24}\,\varepsilon_{12345}\,\varepsilon^{tpqrs}\,\varepsilon^{pqrsu}\,\ff_{u\,B}\,5^{t}_{\hh}\nl
\eq -i\left(\tilde{H}_{AB} + \tilde{H}_{BA}\right)\,\fn_A\,\ff_{p\,B}\,\tilde{5}^p_{\hh},
\eeqa
where, we set $\frac{1}{2}\left(\tilde{H}_{AB}+\tilde{H}_{BA}\right) \equiv H_{AB}$, making $H$ symmetric under generation labels and using canonically normalised definitions of $\tilde{5}_{\hh}\,=\,\sqrt{2}\,5_{\hh}$ and $\tilde{\overline{5}}_{\hh}\,=\,\sqrt{2}\,\overline{5}_{\hh}$. Eq.~(\ref{eq:c2:fnanc}) reduces to the following:
\beqa{\label{eq:c2:fnanc2}}
-i\,2\sqrt{2}\,H_{AB} \,\fn_A\,\ff_{p\,B}\,5^p_{\hh}.
\eeqa
Similarly, for the other term, we have the following.
\beqa{\label{eq:c2:10105}}
\eq -\tilde{H}_{AB}\,\frac{i}{24}\,\varepsilon_{stpqr}\ft^{st}_A\langle 0\big| b_pb_qb_r b^{\dagger}_ub^{\dagger}_vb^{\dagger}_w\big|0\rangle \tilde{5}^u_{\hh}\,\ft^{vw}_B\nl
\eq -\tilde{H}_{AB}\,\frac{i}{24}\,\varepsilon_{stpqr}\,\ft^{st}_A\,\frac{1}{2!}\,\varepsilon^{xypqr}\,\varepsilon_{xyuvw}\,\tilde{5}^t_H\,\ft_B^{vw}\nl
\eq -\tilde{H}_{AB}\,\frac{i}{24}\,\ft^{st}_A\,\frac{3!}{2!}\,\left(-\delta^x_s\delta^y_t + \delta^x_t\delta^y_s \right)\,\tilde{5}^t_{\hh}\,\ft_B^{vw}\nl
\eq \tilde{H}_{AB}\,\frac{i}{4}\,\varepsilon_{stvwu}\,\ft_A^{st}\,\ft_B^{vw}\,\tilde{5}_{\hh}\;=\; \frac{i}{8}\left(\frac{\tilde{H}_{AB}+\tilde{H}_{BA}}{2}\right)\,2\sqrt{2}\,\varepsilon_{stvwu}\,\ft_A^{st}\,\ft_B^{vw}\,5_{\hh}\nl
\eq H_{AB}\,\frac{2\sqrt{2}}{8}\,\varepsilon_{stvwu}\,\ft_A^{st}\,\ft_B^{vw}\,5_{\hh}
\eeqa

Similarly, the derivation of the remaining term is straightforward, and the final result of the interactions of ${\bf 16}$-plet fermions with $10_{\hh}$ becomes the following~\cite{Nath:2001uw}:
\beqa \label{eq:c2:16-16-10}
-{\cal L}^{10_{\hh}}_Y \eq H_{AB}\,{\bf 16}^T_A\,\cc\,{\bf 16}_B\,{\bf 10}_H\,\hc,\nl
 \eq i\, 2\sqrt{2}\, H_{AB}\,\left({\ft^{pq}_A}^T\, \cc\,\ff_{p B}\, \overline{5}_{q\,H}\, +\, \frac{1}{8} \varepsilon_{stuvw}\, \ft^{st\,T}_A\, \cc\,\ft^{uv}_B\, 5^m_{\hh}\right. \nl &-& \fn_A^T\, \cc\,\ff_{iB}\, 5^i_{\hh} \bigg)\hc \eeqa
The expression given above in Eq.~\eqref{eq:c2:16-16-10} is invariant under $SU(5)$. Also, all the scalar \(SU(5)\) irreps appearing in the same expression are defined with canonically normalised kinetic terms.

Further, using the decomposition of $\fn$, $\ff$ and $\ft$ dimensional irreps in terms of SM fermions using Eqs.~(\ref{eq:c2:fn}, \ref{eq:c2:ff}, and \ref{eq:c2:ft}), we can decompose the $SU(5)$ invariant expression given in Eq.~\eqref{eq:c2:16-16-10} to SM invariant. For this, the scalars identified in $5_{\hh}$ and $\overline{5}_{\hh}$ can be decomposed into the following, 
\beqa \label{eq:c2:5dec}
5_{\hh}^\alpha \equiv T^\alpha\,,~~5_{\hh}^a \equiv D^a\,,~~\overline{5}_{\hh\,\alpha} \equiv \overline{T}_\alpha\,,\text{and}~~\overline{5}_{\hh\,a} \equiv \overline{D}_a\,, \eeqa
manner ensuring their kinetic terms are canonically normalised. Substituting Eqs.~(\ref{eq:c2:fn}, \ref{eq:c2:ff}, \ref{eq:c2:ft}, and \ref{eq:c2:5dec}) into Eq.~(\ref{eq:c2:16-16-10}) yields the coupling of SM fermions with the scalars residing in $10_{\hh}$, as shown below:
\beqa \label{eq:c2:10/5}
-{\cal L}^{{10_{\hh}}/5_{\hh}}_Y & \supset & i 2\sqrt{2}\, H_{AB}\, \Bigg[-\bigg( u^{C\,T}_{\gamma A}\, \cc\,e^C_B  +\frac{1}{2} \varepsilon_{\alpha \beta \gamma}\, \varepsilon_{ab}\, q^{a\,\alpha\, T}_A\, \cc\,q^{b\,\beta}_B\nl \ad \nu^{C\,T}_A\,\cc\,d^C_{\gamma\,B}\, \bigg)T^\gamma\,\nl
&-& \left.  \bigg(\varepsilon_{ab}\, q^{a\alpha\,T}_A\,\cc\,u^C_{\alpha\,B}\, - \varepsilon_{ab}\,\nu^{C\,T}_A\,\cc\,l^a_B\,\bigg)\,D^b \right] \hc,\nl
-{\cal L}^{{10_{\hh}}/\overline{5}_{\hh}}_Y & \supset & i 2\sqrt{2}\, H_{AB}\,\Bigg[\bigg(\varepsilon^{\alpha \beta \gamma}\, u^{C\, T}_{\alpha A}\,\cc\,d^C_{\beta B} + \varepsilon_{ab}\, q^{a\,\gamma\,T}_A\,\cc\,l^b_B\bigg) \overline{T}_\gamma \nl \mi q^{a\,\alpha\,T}_A\,\cc\,d^C_B\,\overline{D}_a\Bigg]  \hc.  \eeqa

The following points regarding the different vertices appearing in Eq.~\eqref{eq:c2:10/5} are worth noting.
\begin{enumerate}[label=(\roman*)]
\item In addition to the known couplings of SM-Higgs with SM fermions, the decomposition of $\fs-\fs-10_{\hh}$ yields interactions in which $T$ and $\overline{T}$ have diquark and leptoquark vertices.
    \item The inclusion of only \(10_{\text{H}}\) dimensional irrep in the Yukawa sector results in \(Y_u = Y_d = Y_e^T = Y_{\nu^C} = H\) at the GUT scale, depicting a generational mass degeneracy in all the sector.
    \item \(T^{\gamma}\) couples to \(\nu^C\,d^C\) and \(\nu^C\,l\). We will utilise this specific coupling in Chapter~(\ref{ch:6}) to address the removal of mass degeneracy in the down-quark and charged lepton sector within the context of \(SU(5)\).
\end{enumerate}

Employing a similar procedure, one can also compute the decomposition of $\fs-\fs-120_{\hh}$ and $\fs-\fs-\overline{126}_{\hh}$ which is done in subsequent chapters. Now, we will outline the decomposition of gauge interactions in renormalisable \so.

\section{Gauge Interactions in Renormalisable \so}
{\label{sec:c2:gaugeinteraction}}

The coupling of $45$-dimensional irrep consisting of gauge bosons with $\fs$ can be parameterised as follows:
\beqa{\label{eq:c2:gaugecoupling}}
L_{AB}\,\fs_A^{\dagger}\,\overline{\sigma}^{\Dot{\rho}}\,\fs_B\,45_{\vv\Dot{\rho}} &\to& L_{AB}\,\frac{1}{2}\,\langle \fs_{-A}^{*}\big|\,{\cal B}\,\overline{\sigma}^{\Dot{\rho}}\,\Phi_{\mu\nu}\,\big|\fs_{+B}\rangle\,45_{\vv\,\mu\nu\,\Dot{\rho}},
\eeqa
where $L_{AB}$ is a matrix in generation space and for our case $L_{AB}\,=\,\delta_{AB}$. 

The expansion of $\Phi_{\mu\nu}\,45_{\mu\nu}$ is terms of creation and annihilation operators is as follows~\cite{Nath:2001yj}:
\beqa{\label{eq:c2:45g1}}
\Phi_{\mu\nu}\,45_{\vv\,\mu\nu}\eq \frac{1}{i}\,\left(b_{p}\,b_{q}\,\tilde{10}^{\dagger}_{\vv\,pq} + b^{\dagger}_p\,b^{\dagger}_{q}\,\tilde{10}_{\vv}^{pq} + 2\,b^{\dagger}_p\,b_q\, \tilde{24}^{p}_{\vv\,q} +\big(\frac{2}{5}\,b^{\dagger}_{p}\,b_{q}\delta^p_q - 1\big)\tilde{1}_{\vv}\right).\nl  \eeqa

Following a similar procedure outlined in Eq.~\eqref{eq:c2:ke10}, canonical normalisation of $SU(5)$ irreps residing within $45$ yields the following field redefinition~\cite{Nath:2001yj}.
\beqa{\label{eq:c2:canonical45}}
\tilde{24}_{\vv}\eq \sqrt{2}\,24_{\vv},~~~~\tilde{1}_{\vv}\,=\, \sqrt{10}\,1_{\vv},\nl
\tilde{10}_{\vv}\eq \sqrt{2}\,10_{\vv},~~~~\tilde{10}^{\dagger}_{\vv}\,=\, \sqrt{2}\,10_{\vv}.\eeqa
The action of charge conjugation operator, ${\cal B}$, on $\fs^{\dagger}$ can be computed as follows~\cite{Cardoso:2015gfa}:
\beqa{\label{eq:c2:actionB}}
\langle \fs_{-A}^{*}\big|\,{\cal B} \eq i\fn^{\dagger}_A \langle 0 \big| + \frac{i}{2}\, \ft^{\dagger}_{no\,A} \langle 0 \big|b_{n}b_{o} + \frac{i}{24}\,\varepsilon_{nopqr}\,\ff^{\dagger\,n}_A\,\langle 0 \big|b_{o}b_{p}b_{q}b_{r}.\eeqa
Using Eqs.~(\ref{eq:c2:16f}, \ref{eq:c2:45g1} and \ref{eq:c2:actionB}), one can evaluate the expression mentioned in Eq.~\eqref{eq:c2:gaugecoupling}. We evaluate the same expression by explicitly computing the coupling of $24$-dimensional gauge irrep with fermion-irreps of $SU(5)$, as done below;
\beqa{\label{eq:c2:24actongauge}}
& &L_{AB}\,\frac{1}{2}\,\langle \fs_{-A}^{*}\big|\,{\cal B}\,\overline{\sigma}^{\Dot{\rho}}\,\Phi_{\mu\nu}\,\big|\fs_{+B}\rangle\,45_{\mu\nu\,\Dot{\rho}}\nl
&\supset& L_{AB}\,\frac{1}{2}\,\langle \fs_{-A}^{ *}\big|\,{\cal B}\,\overline{\sigma}^{\Dot{\rho}}\,\left(b^{\dagger}_{s}b_{t}\,\tilde{24}^s_{t}\right)\,b_{p}^{\dagger}b_{q}^{\dagger}\big|0\rangle\,\tilde{10}^{pq}_{\Dot{\rho}}.
\eeqa
It is easy to see that only the second term of Eq.~\eqref{eq:c2:actionB} contributes to the non-zero matrix element. Using,
\beqa{\label{eq:c2:identity}}
\langle 0 \big |b_{i}b_{j}b^{\dagger}_{k}b_{l}b^{\dagger}_{m}b^{\dagger}_{n}\,\big | 0 \rangle\,=\, \delta_{lm}\left(\delta_{jk}\delta_{in}\,-\,\delta_{jn}\delta_{ik}\right)\,-\, \delta_{ln}\left(\delta_{jk}\delta_{im}\,-\,\delta_{jm}\delta_{ik}\right),
\eeqa
the last line given in Eq.~\eqref{eq:c2:24actongauge} reduces to following:
\beqa{\label{eq:c2:10}}
\frac{4\,L_{AB}}{4}\,10^{\dagger}_{pq\,A}\,\overline{\sigma}^{\Dot{\mu}}\,10^{qr}_B\,\tilde{24}^p_{r\,\Dot{\mu}}.\eeqa
Using the canonically normalised $\tilde{24}$ given in Eq.~\eqref{eq:c2:canonical45}, coupling of $\tilde{24}$ modifies as follows:
\beqa{\label{eq:c2:final24}}
L_{AB}\,\sqrt{2}\,10^{\dagger}_{pq}\,\overline{\sigma}^{\Dot{\mu}}\,10^{qr}_B\,\tilde{24}^p_{r\,\Dot{\mu}}.\eeqa

Following the same procedure, the final expression of $\fs-\fs-45_{\vv}$ in terms of canonically normalised $SU(5)$ tensors is shown as follows~\cite{Nath:2001yj}:
\beqa{\label{eq:c2:final45}}
{\cal L}^{(45)}&=&L_{AB}\,\Bigg[\sqrt 5\left(-\frac{3}{5}\,\ff_{A}^{\dagger\,p}\,\overline{\sigma}^{\Dot{\mu}}\,\ff_{B\,p}+
\frac{1}{10}\,\ft^{\dagger}_{A\,pq}\,
\overline{\sigma}^{\Dot{\mu}}\, \ft^{pq} +
\fn^{\dagger}_A\,\overline{\sigma}^{\Dot{\mu}}\, \fn_{B}\right)1_{\vv\,\Dot{\mu}} \nl
&+&{\frac{1} {\sqrt 2}}\left(\fn^{\dagger}_{A} \,\overline{\sigma}^{\Dot{\mu}}\,\ft_{B}^{pq}
+ \frac{1}{2}\,\varepsilon^{rstpq}\,\ft^{\dagger}_{A\,rs}\,\overline{\sigma}^{\Dot{\mu}}\,\ff_{B\,t}\right)\,10^{\dagger}_{\vv\,\Dot{\mu}pq}\nl
&-&\frac{1}{\sqrt 2}\left(
\ft^{\dagger}_{A\,pq}\,\overline{\sigma}^{\Dot{\mu}}\,\fn_{B} 
+ \frac{1}{2}\,\varepsilon_{rstpq}\,
 \ff_{A}^{\dagger\,r}\,\overline{\sigma}^{\Dot{\mu}}\,\ft_{B}^{st}\right)\,10_{\vv\,\Dot{\mu}}^{pq} \nl
&+&\sqrt{2}\left(\ft^{\dagger}_{A\,pr}\,\overline{\sigma}^{\Dot{\mu}}\, \ft_{B}^{rq}+
\ff_{A}^{\dagger\,q}\,\overline{\sigma}^{\Dot{\mu}}\,\ff_{B p}\right)\,24_{\vv\,\Dot{\mu}q}^p\Bigg]. 
\eeqa
The above expression is invariant under $SU(5)$. To decompose the $SU(5)$ invariant expression given in Eq.~(\ref{eq:c2:final45}) into SM invariant, we further decompose $10$-dimensional irrep into canonically normalised tensors, as shown below:
\beqa{\label{eq:c2:10decomposition}}
10^{ab}_{V} \eq \tilde{t}^{ab},\;\;\;10^{a\alpha}_{V}\,=\,\frac{1}{\sqrt{2}}\,Y^{a\alpha},\;\;\;\text{and}\;\;10^{\alpha\beta}_V\,=\,\tilde{\Theta}^{\alpha\beta}.\eeqa

Further, the decomposition of the $24$-dimensional irrep into its canonically normalised fields is shown below;
\beqa{\label{eq:c2:24dec}}
24^{\alpha}_{\vv\,a} &=& X^{\alpha}_a,\hspace{1.8cm} ~~~~~~~~~~24^a_{\vv\,\alpha}\,=\,\overline{X}^{a}_{\alpha},\nl
24^{\alpha}_{\vv\beta}&=& G^{\alpha}_{\beta} + \sqrt{\frac{2}{15}} \delta^{\alpha}_{\beta}\,\hat{\sigma},\hspace{1cm}24^{a}_{\vv\,b} = W^{a}_b -  \frac{3}{2}\sqrt{\frac{2}{15}}\,\delta^{a}_b\,\hat{\sigma}.  
\eeqa
The SM charges of the above gauge fields appearing in Eqs.~(\ref{eq:c2:10decomposition} and \ref{eq:c2:24dec}) can be inferred from Tab.~(\ref{tab:c2:gscalars}). The tracelessness  of \(24_{\hh}\) is the sole reason for the doublet-triplet splitting problem in \(SU(5)\)-GUTs~\cite{Masiero:1982fe,Grinstein:1982um,Babu:2006nf}. Tracelessness indicates that various degrees of freedom of different scalars are interconnected and cannot be chosen arbitrarily.  

Using Eqs.~(\ref{eq:c2:fn}, \ref{eq:c2:ff}, \ref{eq:c2:ft}, \ref{eq:c2:10decomposition}, and \ref{eq:c2:24dec}), the expression given in Eq.~(\ref{eq:c2:final45}) can be decomposed into SM invariant expression consisting of SM fermions. Below, we show the vertices of gauge bosons, $i.e.$ $X$ and $Y$, having diquark and leptoquark couplings that can induce proton decay.
\beqa{\label{eq:c2:XandY}}
{\cal L}^{X\& Y} \eq \,L_{AB}\,\Bigg[\sqrt{2}\,\Big(\varepsilon_{ab} \big(d^C_{\alpha\,A}\big)^{\dagger}\, \overline{\sigma}^{\Dot{\mu}}\, l^b_B\,-\, \varepsilon^{\alpha\beta\gamma}\,\big(q^{a\,\beta}_A\big)^{\dagger}\, \overline{\sigma}^{\Dot{\mu}}\, u^C_{\gamma\,B}\nl \mi\,\varepsilon_{ab}\, \big(e^C_A\big)^{\dagger}\, \overline{\sigma}^{\Dot{\mu}}\, q^{b\alpha}_B \Big)\nl
\ad 2 \Big( \varepsilon_{\alpha\beta\gamma}\,\varepsilon_{ab}\,d^{C\,\dagger}_{\beta\,A}\,\overline{\sigma}^{\Dot{\mu}}\,q^{b\gamma}_B - \big(l^a\big)^{\dagger}\,\overline{\sigma}^{\Dot{\mu}}\,u^C_{\alpha\,B}\Big) \Bigg]\overline{X}^a_{\Dot{\mu}\alpha} \hc.
\eeqa

\section{Yukawa Interactions in Non-Renormalisable \so}

\label{sec:c2:NRSo}

In section~(\ref{sec:c2:RSO}), we have considered irreps, contributing to the Yukawa Lagrangian at the renormalisable level. These tensors, $120_{\hh}$ and $\overline{126}_{\hh}$, have large degree of freedoms and incorporating large tensor representations (LTR) like \(120_{\rm{H}}, \overline{126}_{\rm{H}},\) and \(210_{\rm{H}}\) in the scalar sector affects low-energy phenomenology. However, these LTRs affect the $\beta$-functions, causing low-energy gauge couplings to rise with energy and potentially hitting the Landau pole before the Planck Scale, a phenomenon known as asymptotic slavery in \so\, gauge theories~\cite{Maiani:1977cg,Vaughn:1978st,Rubakov:1999mu,Aulakh:2002ww,Aulakh:2002ph,Aulakh:2003kg,Bajc:2016efj}. To avoid this, smaller tensor representations like \(10_{\rm{H}}, 16_{\rm{H}}, 45_{\rm{H}},\) and \(54_{\rm{H}}\) and non-renormalisable interactions can be used~\cite{Babu:1994dc,Dvali:1996wh,Barr:1997hq,Babu:1998wi,Chang:2004pb,Sayre:2006wf,Preda:2022izo}. These interactions could stem from another theory valid above a certain cutoff scale $(\Lambda)$. Likewise, the SM uses renormalisable interactions for charged fermion masses and non-renormalisable ones for neutral fermions.

$16_H$ is the smallest representation contributing to both charged fermion and neutrino masses, albeit at the non-renormalisable level. As a result, it is the most phenomenologically attractive choice to study the non-renormalisable interactions in \so. The various scalars contained in $16_{\hh}$, along with their SM charge and $B-L$ quantum number, are illustrated in Tab.~(\ref{tab:c2:16scalars}). As evident from Table~(\ref{tab:c2:16scalars}), some of the scalars may possess identical SM charges to those shown in Table~(\ref{tab:c2:scalars}). Consequently, the scalars present in \(16_{\hh}\) are represented with a \(\hat{}\) symbol over them. All the scalars residing in $16_{\hh}$ have a non-zero $B-L$ charge. The canonically normalised decomposition of scalars residing in the $16_{\hh}$-dimensional irrep is as follows~\cite{Patel:2022wya}:
\beqa {\label{eq:c2:SU(5)toSMscalars}}
1_H&=& \hat{\sigma},\hspace{1cm} 5^{\dagger}_{a\,\,H}\;=\; \hat{H}^{\dagger}_a,\hspace{1cm} 5^{\dagger}_{\alpha\,\,H}\;=\;\hat{T}^{\dagger}_{\alpha},\nl 
10^{ab}_H&=& \hat{t}^{ab},\,\hspace{1cm} 10^{a\,\alpha}\;=\; \frac{1}{\sqrt{2}}\,\hat{\Delta}^{a\alpha},\hspace{1cm}10^{\alpha\beta}\;=\;\hat{\Theta}^{\alpha\beta}.
\eeqa

\begin{table}[t!]
    \centering

    \begin{tabular}{ccc}
    \hline\hline
    ~~~~SM Charges~~~~     & ~~~~Notation~~~~ & ~~~~$B-L$ Charges~~~~ \\
    \hline
    $\left(1,1,0\right)$     & $\hat{\sigma}$ & $1$ \\
    $\left(\overline{3},1,\sfrac{1}{3}\right)$     & $\hat{T}^{\dagger}_{\alpha}$ & $\sfrac{-1}{3}$ \\
    $\left(1,2,\sfrac{-1}{2}\right)$     & $\hat{H}^{\dagger}_a$ &$ -1$\\
    $\left(3,2,\sfrac{1}{6}\right)$     & $\hat{\Delta}^{a\alpha}$ &  $\sfrac{1}{3}$\\
    $\left(\overline{3},1,\sfrac{-2}{3}\right)$     & $\hat{\Theta}^{\alpha\beta}$ & $\sfrac{-1}{3}$ \\
    $\left(1,1,-1\right)$ & $\hat{t}^{ab}$   & $1$  \\
    \hline\hline

    \end{tabular}

    \caption{Classification of scalars present in $16_H$ alongwith their SM and $B-L$ charges}
    \label{tab:c2:16scalars}
\end{table}

An $SO(10)$ Lagrangian featuring the interactions of an $\fs$-plet and a scalar $16_{\hh}$ dimensional irrep can be parameterised to first order in $\Lambda$, $\Lambda$ being the typical cutoff scale, as follows; 
\beqa {\label{eq:c2:NRSO10}}
-{\mathcal{L}} &\supset&  \frac{1}{\Lambda} y_{AB}\,\big(\mathbf{16}_{A}\,\mathbf{16}_{B}\;16_{\hh}\,16_{\hh}\big)\,+\, \frac{1}{\Lambda} \bar{y}_{AB}\,\big(\mathbf{16}_{A}\,\mathbf{16}_{B}\;16^{\dagger}_{\hh}\,16^{\dagger}_{\hh}\big) \,\nl \hcn.
\eeqa
where, $y,\,\overline{y}$ are the Yukawa couplings and depicts the strength of non-renormalisable interactions. To decompose down the Lagrangian provided in Eq.~\eqref{eq:c2:NRSO10}, one needs to account for all the Yukawa couplings effectively generated by integrating all possible irreps (scalar or fermions) consistent with the assumed gauge symmetry yielding the interaction of $\fs-16_{\hh}$. The transformation properties of the $16$-dimensional irrep under $SO(10)$, as indicated in Eq.~\eqref{eq:c2:16t16}, suggest that a term like the one in Eq.~\eqref{eq:c2:NRSO10} allows only a limited number of intermediate $SO(10)$ irreps are required to generate the effective coupling. One of the various ways to generate the coupling is specified in Eq.~(\ref{eq:c2:NRSO10}). We represent the coupling as $(...)_{D_1}(...)_{D_2}$, where the representations inside the first (second) bracket are contracted to form $D_1\, (D_2)$, ensuring that $D_1 \times D_2 \supset \mathds{1}$, where $\mathds{1}$ transforms trivially under $SO(10)$. Additionally, $D_1\,(D_2)$ can be either a scalar or a fermion representation;
\beqa {\label{eq:c2:16Fs16Hs}}
      &\bullet &  \frac{h_{AB}}{\Lambda}\big(\fs_A\,\fs_B\big)_{10_H}\,\big(16_H\,16_H\big)_{10_H};\; \frac{\bar{h}_{AB}}{\Lambda}\big(\fs_A\,\fs_B\big)_{10_H}\,\big(16^{\dagger}_H\,16^{\dagger}_H\big)_{10^{\dagger}_H}; \nl  
     & & \frac{\tilde{h}_{AB}}{\Lambda}\big(\fs_A\,16_H\big)_{\mathbf{10}}\,\big(\fs_B\,16_H\big)_{\mathbf{10}}; \text{and}\;\frac{\hat{h}_{AB}}{\Lambda}\Big(\fs_{A}\,16_H\big)_{\ft}\,\Big(\fs_{B}\,16_H\Big)^{*}_{\ft^{\dagger}};\nl 
\eeqa    
where, $ h,\, \bar{h}, \tilde{h},\text{and} \,\hat{h}$ are various Yukawa couplings and $\Lambda$ corresponds to the energy scale where the intermediate \so\, irrep has been integrated out or any other cut-off scale.

We illustrate a specific example that leads to the terms provided in Eq.~\eqref{eq:c2:NRSO10} by integrating out the $10$-dimensional irrep. The relevant Lagrangian for the same is shown below, where we assume that only integration of $10_{\hh}$ can yield the required non-renormalisable vertex;
\beqa{\label{eq:c2:reqNRSO}}
-\cal{L}^{\mathrm h}_{\mathrm{NR}}&\supset& \bigg( H\,\fs^T\,\cc\,\fs\,{10}_{\hh} + \eta\,16_{\hh}\,16_{\hh}\,10_{\hh} \hc \bigg) + M^2_{10_{\hh}}\,10^{\dagger}_{\hh}\,10_{\hh}.\nl\eeqa
The $\fs-\fs-10_{\hh}$ coupling have been already computed in Eq.~\eqref{eq:c2:16-16-10}. To obtain the \(16_{\hh}-16_{\hh}-10_{\hh}\) coupling, we assume its form is analogous to the \(\fs-\fs-10_{\hh}\) coupling, where the \su\, irreps of fermions of \(\fs\) are replaced with the scalars residing in \(16_{\hh}\) with the same \su\, charge. The decomposition of the scalar part of Lagrangian given in Eq.~\eqref{eq:c2:reqNRSO} is as follows:
\beqa{\label{eq:c2:SO10/10H}}
 -{\cal{L}}^{h}_{\mathrm{NR}} &\supset&  \eta\,16_{\hh}\,16_{\hh}\,10_{\hh} + M^2_{10_{\hh}}\,10^{\dagger}_{\hh}\,10_{\hh}\, \nl
&=&  i 2\sqrt{2}\, \eta\,\left(10_{\hh}^{pq}\,5^{\dagger}_{p\,\hh}\, \tilde{5}^{\dagger}_{q\,\hh}\, +\, \frac{1}{8} \varepsilon_{pqrst}\, 10_{\hh}^{pq}\, 5_{\hh}^{rs}\, \tilde{5}^t_{\hh}\, -\, 1_\hh\,5^{\dagger}_{p}\, \tilde{5}^p_{\hh} \right)\nl
&+& M^2_{10_{\hh}}\,\tilde{5}^{\dagger}_{\hh}\,\tilde{5}_{\hh} \hc\,.
\eeqa
To integrate the $10_{\hh}$ irrep, we utilise the equations of motion to solve for the various $SU(5)$ components of $10_H$  shown in Eq.~\eqref{eq:c2:SO10/10H}, and then substitute these results into Eq.~\eqref{eq:c2:16-16-10}~\cite{Nath:2001yj}. We must note that we integrate out those $SU(5)$ irreps which have the same $U(1)_{\mathrm X}$ charge. For example, the $U(1)_{\mathrm X}$ charge of $\tilde{5}^{\dagger}_{\hh} \subset 16_{\hh}$ is $3$, whereas the $U(1)_{\mathrm{X}}$ charge of $5^{\dagger}_{\hh} \subset 10_{\hh}$ is $-2$ ~\cite{Slansky:1981yr,Feger:2019tvk}. Therefore, identical $SU(5)$ representations belonging to different $SO(10)$ representations can have different $U(1)_{\mathrm X}$ charges. Consequently, we denote the $SU(5)$ scalars (and fermions) intended for integration by placing a tilde ($\sim$) over them. 
Further, we also identify $\overline{5}_{\hh}$ as $5^{\dagger}_{\hh}$ in previous expressions of effective couplings in section~(\ref{sec:c2:RSO}).

Integrating out \(\tilde{5}_{\hh}\) from Eq.~\eqref{eq:c2:SO10/10H} and substituting the result into \({\cal L}^{10_{\hh}/5_{\hh}}\), and similarly, integrating out \(\tilde{5}^{\dagger}_{\hh}\) from Eq.~\eqref{eq:c2:SO10/10H} and substituting it into \({\cal L}^{10_{\hh}/\overline{5}_{\hh}}\), where ${\cal L}^{10_{\hh}/5_{\hh}}$ and ${\cal L}^{10_{\hh}/\overline{5}_{\hh}}$ are given in Eq.~\eqref{eq:c2:16-16-10}, allows us to derive the effective non-renormalisable interactions as shown below;
\beqa{\label{eq:c2:h10-1}}
-\lag^{h}_{\rm{NR}} & \supset& \frac{h_{AB}}{\Lambda} \Big( \fs^T_{A}\,\cc\, \fs_B\Big)_{10}\,\Big(16_H\,16_{\hh}\Big)_{10} \nl
&=& \frac{-8\,h_{AB}}{\Lambda}\Big[ \big(\ft^{pq\,T}_A\,\cc\, \ff_{p\,B} \big)\,\big(-1_H\, 5^{\dagger}_{q\,H} + \frac{1}{8} \varepsilon_{qrstu}\,10^{rs}_H\,10^{tu}_H\big)\nl \ad \big(-\fn^T_A\,\cc\,\ff_{p\,B}  \frac{1}{8}\,\varepsilon_{qrstp}\,\ft^{qr\,T}_A\,\cc\ft^{st}_B \big)\,\big(10^{pw}_H\,5^{\dagger}_{w\,H}\big)  \Big] \hc,\nl
\eeqa
where, we set $\frac{\eta\,H_{AB}}{M^2_{10_{\hh}}}\,\,\equiv\,\frac{h_{AB}}{\Lambda}$. The expression given in Eq.~\eqref{eq:c2:h10-1} is invariant under \su\, and can be further decomposed using the dictionary provided in Eqs. (\ref{eq:c2:fn}), (\ref{eq:c2:ff}), (\ref{eq:c2:ft}), and (\ref{eq:c2:SU(5)toSMscalars}) into SM invariant vertices. The resulting expressions is:
\beqa{\label{eq:c2:h10-2}}
-\lag^{h}_{\rm{NR}} & \supset& \frac{h_{AB}}{\Lambda} \Big( \fs^T_{A}\,\cc\, \fs_B\Big)_{10_{\hh}}\,\Big(16_H\,16_{\hh}\Big)_{10_{\hh}}  \hc \nl
&=& \frac{-8\,h_{AB}}{\Lambda}\Big[ \big(\ft^{pq\,T}_A\,\cc\, \ff_{p\,B} \big)\,\big(-1_H\, 5^{\dagger}_{q\,H} + \frac{1}{8} \varepsilon_{qrstu}\,10^{rs}_H\,10^{tu}_H\big) \nl \ad \big(-\fn^T_A\,\cc\,\ff_{p\,B} + \frac{1}{8}\,\varepsilon_{qrstp}\,\ft^{qr\,T}_A\,\cc\ft^{st}_B \big)\,\big(10^{pw}_H\,5^{\dagger}_{w\,H}\big)  \Big] \hc \nl
\eq \frac{-h_{AB}}{\Lambda}\,\bigg[ -\left(4\sqrt{2}\,\varepsilon_{ab}\,q^{a\alpha\,T}_A\,\cc\,u^C_{\alpha\,B}\right)\left(\hat{\Delta}^{b\theta}\,\hat{T}^{\dagger}_{\theta}+ \sqrt{2}\,\hat{t}^{bd}\,\hat{H}^{\dagger}_{d}\right)\nl
\ad \bigg(-4\varepsilon_{ab}\,\varepsilon_{\alpha\beta\gamma}\,q^{a\alpha\,T}_A\,\cc\,q^{b\beta}_B - 8 u^{C\,T}_{\gamma\,A}\,\cc\,e^C_B\bigg)\,\bigg(-\frac{\hat{\Delta}^{d\gamma}}{\sqrt{2}}\,\hat{H}^{\dagger}_d + \hat{\Theta}^{\gamma\lambda}\,\hat{T}^{\dagger}_\lambda \bigg)\nl
\ad \left( q^{a\alpha\,T}_A\,\cc\,d^{c}_B + e^{C\,T}_A\,\cc\,l^a_B\right)\,\left(\sqrt{2}\varepsilon_{ab}\,\varepsilon_{\beta\gamma\lambda}\,\hat{\Delta}^{b\beta}\,\hat{\Theta}^{\gamma\lambda}\right)\nl
\mi \left(\varepsilon^{\alpha\theta\rho}\,u^{C\,T}_{\theta\,A}\,\cc\,d^C_{\rho\,B} + \varepsilon_{ab}\,q^{t\,\alpha}_A\,\cc\,l^b\right)\,2\varepsilon_{\alpha\beta\gamma}\varepsilon_{mn}\left(\hat{\Theta}^{\beta\gamma}\hat{t}^{mn}-\hat{\Delta}^{m\beta}\,\hat{\Delta}^{n\gamma}\right)\nl 
\mi 8 \left( \varepsilon_{ab}\,q^{a\alpha}_A\,\cc\,l^b_B + \varepsilon^{\beta\gamma\alpha}\,u^{C\,T}_{\beta\,A}\,\cc\,d^C_{\gamma\,B}\right)\, \hat{\sigma}\,\hat{T}^{\dagger}_\alpha\nl
\ad \left( e^{C\,T}_A\,\cc\,l^a_B + q^{a\alpha\,T}_B\,\cc\,d^C_{\alpha}\right)\,\hat{\sigma}\,\hat{H}^{\dagger}_A\nl
\mi 8 \bigg(\nu^{C\,T}_A\,\cc\,d^C_\alpha\bigg)\,\bigg(-\frac{\hat{\Delta}^{a\alpha}}{\sqrt{2}} + \hat{\Theta}^{\alpha\beta}\hat{T}^{\dagger}_{\beta}\bigg)\nl
\ad 8 \bigg(\varepsilon_{ab}\,\nu^{C\,T}_A\,\cc\,l^a_B\bigg)\,\bigg(\frac{\hat{\Delta}^{a\alpha}}{\sqrt{2}} + \hat{t}^{bc}\,\hat{H}^{\dagger}_c\bigg)\bigg]\hc
\eeqa
The above expression shows the mass dimension five coupling of SM fermions with the scalars residing in $16_{\hh}$. A simple-looking expression, written in Eq.~\eqref{eq:c2:NRSO10} when decomposed, gives a plethora of interactions.

So far, we have integrated out $10_{\hh}$-irrep to generate the effective coupling mentioned in Eq.~\eqref{eq:c2:NRSO10}. We further provide other possibilities for generating the same effective coupling. Consider the following Lagrangian;
\beqa{\label{eq:c2:reqNRSO2}}
-{\cal{L}}^{\overline{h}}_{\mathrm{NR}}&\supset& \bigg( H\,\fs^T\,\cc\,\fs\,{10}_{\hh} + \left(\eta\,16_{\hh}\,16_{\hh}\,10_{\hh}\right)^{\dagger} \hc \bigg) + M^2_{10_{\hh}}\,10^{\dagger}_{\hh}\,10_{\hh}.\nl\eeqa
This other possibility stems from the fact that $10_{\hh}\times 10^{\dagger}_{\hh}\,\supset\,\mathds{1}$. Following the abovementioned procedure, we can obtain the non-renormalisable interaction emanating from this case;
\beqa{\label{eq:c2:h10b}}
-\lag^{\overline{h}}_{\mathrm{NR}} &\supset & \frac{\bar{h}_{AB}}                 {\Lambda}\Big( \fs^T_{A}\,\cc\, \fs_B\Big)_{10_{\hh}}\,\Big(16^{\dagger}_H\,16^{\dagger}_{\hh}\Big)_{10^{\dagger}_{\hh}} \hc\nl
&=& \frac{8\,\bar{h}_{AB}}{\Lambda}\Big[ \big(\ft^{pq\,T}_A\,\cc\, \ff_{p\,B} \big)\,\big(10^{\dagger}_{qw\,H}\,5^{w}_H\big)\nl \ad  \big(-\fn^T_A\,\cc\,\ff_{p\,B} + \frac{1}{8}\,\varepsilon_{qrstp}\,\ft^{qr\,T}_A\,\cc\ft^{st}_B \big)\nl\,
          & &\times\,\big(-1_H\, 5^{p}_H + \frac{1}{8} \varepsilon^{prstu}\,10^{\dagger}_{rs\,H}\,10^{\dagger}_{tu\,H}\big)  \Big] \hc\nl
\eq \frac{8\bar{h}_{AB}}{\Lambda}\bigg[ \left(\varepsilon_{ab}\,q^{a\alpha\,T}_A\,\cc\,l^b_B + \varepsilon^{\beta\alpha\lambda}u^{C\,T}_{\lambda\,A}\,\cc\,d^C_{\beta\,B}\right)\left(\frac{\hat{\Delta}^{\dagger}_{m\alpha}}{\sqrt{2}}\hat{H}^m + \hat{\Theta}^{\dagger}_{\rho\alpha}\,\hat{T}^{\rho}\right)\nl
\mi \left(q^{m\alpha\,T}_A\,\cc\,d^{C}_{\alpha\,B} + e^C_A\,\cc\,l^m_B\right)\,\left(-\frac{\hat{\Delta}^{\dagger}_{m\beta}}{\sqrt{2}}\,\hat{T}^{\beta} + \hat{t}^{\dagger}_{nm}\,\hat{H}^n\right)\nl
\ad \varepsilon_{ab}\, \left(q^{a\alpha\,T}_A\,\cc\,u^C_{\alpha\,B}\right)\,\hat{H}^b\,\hat{\sigma}\nl
\ad \left(u^{C\,T}_{\alpha\,A}\,\cc\,e^C_B + \frac{1}{2}\,\varepsilon_{ab}\,\varepsilon_{\alpha\beta\gamma}\,q^{a\beta\,T}_A\,\cc\,q^{a\gamma}_B\right)\,\hat{T}^{\alpha}\,\hat{\sigma}\nl
\mi \frac{1}{2\sqrt{2}}\left(q^{m\alpha\,T}_A\,\cc\,u^C_{\alpha\,B}\right)\,\left(\varepsilon^{\beta\gamma\rho}\,\hat{\Delta}^{a\beta}\,\hat{\Theta}^{\dagger}_{\gamma\rho}\right)\nl
\mi 2 \left(u^{C\,T}_{\alpha\,A}\,\cc\,e^C_B\right)\,\varepsilon^{ab}\,\varepsilon^{\alpha\beta\gamma}\,\left(\hat{t}^{\dagger}_{ab}\,\hat{\Theta}^{\dagger}_{\beta\gamma}- \hat{\Delta}^{\dagger}_{a\beta}\,\hat{\Delta}^{\dagger}_{b\gamma}\right)\nl
\mi \varepsilon_{mn}\,\left(\varepsilon_{\lambda\rho\alpha}\,q^{m\lambda\,T}_A\,\cc\, q^{n\rho}_B\right)\,\varepsilon^{\alpha\beta\gamma}\,\varepsilon^{ab}\,\left(\hat{t}^{\dagger}_{ab}\,\hat{\Theta}^{\dagger}_{\beta\gamma}- \hat{\Delta}^{\dagger}_{a\beta}\,\hat{\Delta}^{\dagger}_{b\gamma}\right)\nl
\ad \left(\varepsilon_{ab}\nu^{C\,T}_A\,\cc\,l^a_B\right)\,\left(\frac{1}{4\sqrt{2}}\varepsilon^{ab}\,\varepsilon^{\alpha\beta\gamma}\,\hat{\Delta}^{\dagger}_{b\alpha}\,\hat{\Theta}^{\dagger}_{\beta\gamma} - \hat{\sigma} \,\hat{H}^b\right)\nl
\mi \left(\nu^{C\,T}_A\,\cc\,d^{C}_{\alpha}\right)\left(\frac{1}{8}\,\varepsilon^{\alpha\beta\gamma}\left(\hat{\Theta}^{\dagger}_{\beta\gamma}\,\hat{t}^{\dagger}_{ab} - 2\,\hat{\Delta}^{\dagger}_{a\beta}\,\hat{\Delta}^{\dagger}_{b\gamma} \right) -\hat{\sigma}\,\hat{T}^{\alpha}\right)\bigg]\nl \hcn
\eeqa

We now consider the possibility when the intermediate integrated out irrep is $\ft$-dimensional fermion multiplet. For this, consider the following Lagrangian.
\beqa{\label{eq:c2:fermion10}}
-{\cal L}^{\tilde{h}\&\bar{h}}_{\mathrm{NR}} \eq \tilde{H}\,\fs^{T}\,\cc\,\ft\,16_{\hh} +\bar{H}\,\fs^{T}\,\cc\,\ft^{\dagger}\,16_{\hh} \nl \ad M_{\ft}\,\overline{\ft}\,\ft \hc \eeqa
The decomposition of $\fs$-plet fermion with $\ft$ dimensional fermion, depicted in Eq.~\eqref{eq:c2:fermion10}, can be computed as shown below~\cite{Nath:2001uw}.
\beqa{\label{eq:c2:h10F}}
{\cal L}^{\tilde{h}\&\bar{h}}_{\mathrm{NR}} &\supset & \tilde{H}_{AB}\, \fs^T_{A}\,\cc\,\ft_B\, 16_{\hh}\,+\,M_{\ft}\,\overline{\ft}\,\ft\nl
&=& i 2\sqrt{2}\, \tilde{H}_{AB}\,\Big({\ft^{pq}_A}^T\, \cc\,\tilde{\ff}_{p\,B}\, 5^{\dagger}_{q\,H}\, +\, \frac{1}{8} \varepsilon_{pqrst}\, {\ft^{pq}}^T_A\, \cc\,\big(\tilde{\ff}\big)^{*\,r}_{B}\, 10^{st}_{\hh}\, \nl
&-&\, \fn_A^T\, \cc\,\big(\tilde{\ff}\big)^{*\,p}_{B}\, 5^{\dagger}_{p\,H} \Big) + M_{\ft}\,\overline{\tilde{\ff}}\,\tilde{\ff}\,\hc
\eeqa
where, again the fermion irreps belonging to $\ft$ are shown by putting tilde ($\sim$) over them. Utilising Eq.~\eqref{eq:c2:h10F}, the estimation of higher-dimensional effective couplings can be performed, analogous to the previously mentioned cases when \(10_{\hh}\) was integrated out. We integrate out the fermionic  \su\, irreps of $\ft$ using equations of motion from Eq.~\eqref{eq:c2:h10F} and substitute it in the same equation, resulting in the following expression; 
\beqa{\label{eq:c2:htAB}}
-{\cal L}^{\tilde{h}\&\bar{h}}_{\mathrm{NR}} &\supset &  \frac{\tilde{h}_{AB}}{\Lambda}\Big( \fs^T_{A}\,\cc\, 16_H\Big)_{\ft}\,\Big( \fs^T_{B}\,\cc\, 16_H\Big)_{\ft} \nl &+& \frac{\hat{h}_{AB}}{\Lambda}\Big(\fs^T_{A}\, \cc 16_H\big)_{\ft}\,\Big(\fs^T_{B}\,\cc\,16_H\Big)^{*}_{\ft^{\dagger}}  \hc\nl  
          &=& 
          \Big[\frac{-8\,\tilde{h}_{AB}}{\Lambda} \big(\ft^{pq\,T}_A\,\cc\, 5^{\dagger}_{p\,H} \big)\,\big(-\fn_B\, 5^{\dagger}_{q\,H} + \frac{1}{8} \varepsilon_{qrstu}\,\ft_B^{rs}\,10^{tu}_H\big)\,+ A\leftrightarrow B \Big]\,\nl
          &+& \frac{\hat{h}_{AB}}{\Lambda}\Big[ \big(\ft^{pq\,T}_{A}\cc\,5_{r\,H}^{\dagger}\big)\, \big(\ft^{pq}_{B}\,\,5_{r\,H}^{\dagger}\big)^* + \big(\ff^T_{p\,A}\cc\,10^{qr}_H\big)\, \big(\ff_{p B}\,10^{qr}_H\big)^* \nl
          &+& \big( -\fn^T_A \cc\, 5^{\dagger}_{t\,H} + \frac{1}{8}\varepsilon_{pqrst} \, \ft^{pq\,T}_A\cc 10^{rs}_H\big)\nl\,& & \times\,\big( -\fn_B \, 5^{\dagger}_{t\,H} + \frac{1}{8}\varepsilon_{xywzt} \, \ft^{xy}_B 10^{wz}_H\big)^*\Big] \hc\nl
                    \eq -\frac{8\,\tilde{h}_{AB}}{\Lambda}\Bigg[ \varepsilon^{\rho\beta\alpha}\,\varepsilon_{\alpha\theta\gamma}\left[\left(2 u^{C\,T}_{\rho\,A}\,\cc\,e^C_B\right)\,\left(\hat{T}^{\dagger}_{\beta}\,\hat{\Theta}^{\theta\gamma}\right)\right.\nl
                    \mi \left.\left(\varepsilon_{mn}\,u^{C\,T}_A\,\cc\,q^{m\,\Theta}_B\right)\,\left( \hat{T}^{\dagger}_\beta\,\frac{\hat{\Delta}^{n\gamma}}{\sqrt{2}}\right)\right]\nl
\ad \varepsilon_{mn}\,\left[\left(q^{m\beta\,T}_A\,\cc\,u^C_{\alpha\,B}\right)\left(\hat{T}^{\dagger}_{\beta}\,\frac{\hat{\Delta}^{n\alpha}}{\sqrt{2}}\right)\right.\nl\ad\left.\frac{\varepsilon_{\alpha\gamma\theta}}{2}\left(q^{n\beta\,T}_A\,\cc\,q^{m\alpha}_B\right)\,\left(\hat{T}^{\dagger}_{\beta}\hat{\Theta}^{\gamma\theta}\right)\right]\nl
\mi \left[ \left(e^{C\,T}_A\,\cc\,u^C_{\alpha}\right)\,\left(\hat{H}^{\dagger}_m\frac{\hat{\Delta}^{m\alpha}}{\sqrt{2}}\right) + \frac{\varepsilon_{\alpha\beta\gamma}}{2}\,\left(e^{C\,T}\,\cc\,q^{m\alpha}\right)\,\left(\hat{H}^{\dagger}_m\hat{\Theta}^{\beta\gamma}\right) \right]\nl
\mi \left(\nu^{C\,T}_A\,\cc\,e^{C}_B\right)\,\left(\varepsilon^{mp}\,\hat{H}^{\dagger}_{m}\,\hat{H}^{\dagger}_p\right) + \varepsilon_{\alpha\beta\gamma}\left(\nu^{C\,T}_A\,\cc\,u^C_{\beta\,B}\right)\,\left(\hat{T}^{\dagger}_{\alpha}\hat{T}^{\dagger}_\gamma\right) \bigg] \nl\ad  A\leftrightarrow B  \nl
\ad \frac{8\hat{h}_{AB}}{\Lambda}\bigg[ \left(e^{C\,T}\,\cc\,\big(e^C\big)^*\right)\,\left(\hat{H}^a\,\hat{H}^{\dagger}_a\right)\nl \ad \left(\varepsilon^{\alpha\beta\gamma}\,\varepsilon_{\alpha\theta\phi}\,u^{C\,T}_{\beta\,A}\,\cc\,\big(u^C_{\theta}\big)^*\right)\,\left(\hat{T}^{\dagger}_\gamma\,\hat{T}^{\phi}\right)  \nl
\ad \left(q^{a\alpha\,T}_A\,\cc\,\big(q^{a\beta}\big)^*\right)\,\left(\hat{T}^{\alpha}\,\hat{T}^{\beta}\right) + \left(q^{a\alpha\,T}_A\,\cc\,\big(q^{b\alpha}\big)^*\right)\,\left(\hat{H}^{a}\,\hat{H}^{b}\right)\nl
\mi \left(u^{C\,T}_{\alpha\,A}\,\cc\,\big(u^C_{\beta\,B}\big)^*\right)\,\left(\hat{\Delta}^{b\alpha}\,\hat{\Delta}^{\dagger}_{b\beta}\right) \nl \ad\frac{1}{2}\,\varepsilon_{\alpha\beta\gamma}\,\varepsilon^{\theta\phi\lambda}\,\left(q^{a\alpha\,T}_A\,\cc\,\big(q^{a\theta}_B\big)^*\right)\,\left(\hat{\Theta}^{\beta\gamma}\,\hat{\Theta}^{\dagger}_{\phi\lambda}\right) \nl
\ad \frac{8}{\Lambda}\big(\hat{h}_{AB}\,+\,\hat{h}^*_{BA}\big)\bigg[ \left(\varepsilon^{ab}\,e^{C\,T}\,\cc\,\big(q^{a\alpha}\big)^{*}\right.\nl\ad\left.\varepsilon_{\alpha\beta\gamma}\, q^{b\beta\,T}\,\cc\,\big(u^C_{\gamma\,B}\big)^* \right)\,\left(\hat{H}^{\dagger}_b\,\hat{T}^{\alpha}\right) \nl
\ad \frac{1}{2}\left(u^{C\,T}_{\alpha\,A}\,\cc\,\big(u^{C}_{\alpha\,B}\big)^*\right)\,\left(\varepsilon^{ab}\,\varepsilon_{mn}\,\hat{t}^{\dagger}_{ab}\hat{t}^{mn}\right)\nl \ad \frac{1}{2}\left(\varepsilon_{\alpha\beta\gamma}\,\varepsilon^{\alpha\theta\phi}\,e^{C\,T}_A\,\cc\,\big(e^C\big)^*\right)\,\left(\hat{\Theta}^{\beta\gamma}\,\hat{\Theta}^{\dagger}_{\hat{\Theta}\phi}\right)\nl 
\ad \varepsilon_{ab}\,\varepsilon^{mn}\,\varepsilon_{\alpha\beta\gamma}\,\varepsilon^{\alpha\theta\phi}\,\left(q^{a\beta\,T}_A\,\cc\,\big(q^{m\theta}_B\big)^*\right)\,\left(\hat{\Delta}^{b\gamma}\,\hat{\Delta}^{\dagger}_{n\phi}\right)\nl
\ad \frac{1}{\sqrt{2}}\, \left(\varepsilon_{\alpha\beta\gamma}\,q^{a\alpha\,T}_A\,\cc\,\big(u^C_{\lambda\,B}\big)^*\right)\,\left(\hat{\Theta}^{\beta\gamma}\,\hat{\Delta}^{\dagger}_{b\lambda}\right) \nl
\mi \frac{1}{2} \, \varepsilon_{ab}\varepsilon^{\theta\phi\alpha}\,\left(u^{C\,T}_{\alpha\,A}\,\cc\,\big(e^C_B\big)^*\right)\,\left(\hat{t}^{ab}\,\hat{\Theta}^{\dagger}_{\theta\phi}\right) \nl
\ad \frac{1}{\sqrt{2}} \varepsilon_{ab}\,\varepsilon^{mn}\,\varepsilon^{\eta\theta\phi}\,\left(u^{C\,T}_{\eta\,A}\,\cc\,\big(q^{m\theta}\big)^*\right)\,\left(\hat{t}^{ab}\,\hat{\Delta}^{\dagger}_{n\phi}\right)\nl 
\mi \frac{1}{\sqrt{2}} \, \varepsilon_{\alpha\beta\gamma}\,\varepsilon^{\alpha\theta\phi}\,\varepsilon^{mn}\,\left(e^{C\,T}_A\,\cc\,\big(q^{m\theta}_B\big)^*\right)\,\left(\hat{\Theta}^{\beta\gamma}\hat{\Delta}^{\dagger}_{n\phi}\right) \nl
\ad \left(\nu^{C\,T}\,\cc\,\big(\nu^C_B\big)^*\right)\,\left(\hat{H}^{a}\,\hat{H}^{\dagger}_a + \hat{T}^{\alpha}\,\hat{T}^{\dagger}_{\alpha}\right)\nl \mi \frac{1}{2\sqrt{2}} \varepsilon^{ab}\,\varepsilon^{\alpha\beta\gamma}\left(\nu^{C\,T}_A\,\cc\,\big(u^{C}_{\gamma\,B}\big)^*\right)\,\left(\hat{H}^{\dagger}\,\hat{\Delta}^{\dagger}_{b\gamma}\right)\nl \ad \frac{1}{\sqrt{2}}\left(\nu^{C\,T}_A\,\cc\,\big(q^{b\alpha}_B\big)^*\right)\,\left(\hat{H}^{\dagger}_{a}\,\hat{\Theta}^{\dagger}_{\beta\gamma}\right)\nl
\ad \varepsilon_{ab}\,\varepsilon^{\alpha\beta\gamma}\,\left(\nu^{C\,T}_A\,\cc\,\big(q^{a\beta}_B\big)^*\right)\,\left(\hat{T}^{\dagger}_{\alpha}\,\hat{\Delta}^{\dagger}_{b\gamma}\right)\hc.
\eeqa

The relations provided in (\ref{eq:c2:h10-2}), (\ref{eq:c2:h10b}), and (\ref{eq:c2:htAB}) collectively illustrate the decomposition of non-renormalisable interaction shown in Eq.~\eqref{eq:c2:NRSO10} from the integration of dimension $10$ irrep, which may be either scalar or fermionic. These expressions involve coupling two SM fermions and two scalars and respecting the SM gauge symmetry. These higher dimensional terms contribute not only to mass matrices of charged fermions but also to neutral fermions. Additionally, it can generate the Majorana mass of right-handed neutrinos. Furthermore, some pairs have diquark, leptoquark and dilepton couplings. Consequently, these pairs of scalars could be constrained by different phenomena. On similar lines, one can also decompose the expression given in Eq.~\eqref{eq:c2:NRSO10} using irreps other than $10$-dimensional irreps, which will be shown in Chapter~(\ref{ch:4}).

\section{Conclusion}
\label{sec:c2:conclusion}
\lettrine[lines=2, lhang=0.33, loversize=0.15, findent=0.15em]{I}{N THIS CHAPTER}, we emphasised that \(SU(5)\) and \(SO(10)\) are the potential groups for grand unification and discussed that certain irreps are crucial for meeting elementary requirements. These irreps contribute to the Yukawa Lagrangian at both renormalisable and non-renormalisable levels. These irreps contain diverse scalars that are charged under the \(SU(3)_{\mathrm C} \times SU(2)_{\mathrm L} \times U(1)_{\mathrm Y}\) gauge symmetry, resulting in various types of vertices (leptoquark, diquark and dilepton) with SM fermions. In this chapter, only specific examples of the decomposition of couplings from an \so\, invariant Lagrangian were considered, and the same methodology would be used to evaluate the couplings of other irreps in subsequent chapters. With the knowledge of these various couplings, we can study and analyse different novel predictions and constrain these new scalars from the same.

\clearpage
\graphicspath{{30_Chapter_3/fig_ch3/}}

\chapter{Scalar Induced Proton Decays in Renormalisable GUTs}
{\label{ch:3}}

\section{Overview}
{\label{sec:c3:overview}}
\lettrine[lines=2, lhang=0.33, loversize=0.15, findent=0.15em]{U}NIFIED GAUGE THEORY based on a simple group, as we have seen in the previous Chapter~(\ref{ch:2}), offers a plethora of scalar fields which yield vertices that are absent in the SM. Such vertices often lead to phenomena that serve as the primary basis for the detectability and verifiability of the GUT. One such novel phenomena is the proton decay.

In Chapter~(\ref{ch:2}), we have outlined the technique to compute the SM invariant coupling stemming from \so\, invariant Lagrangian. In this chapter, we begin with computing the resulting effective couplings due to the incorporation of scalar irreps in the Yukawa sector $(i.e.\, 120_{\hh}\,\text{and}\,\overline{126}_{\hh})$ in section~(\ref{sec:c3:couplings}). Consequently, we use these computed couplings to compute the nucleon decay. Before that, we discuss distinct proton decay modes appearing at even and odd mass dimensions in section~(\ref{sec:c3:geninfo}). Based on the same discussion, in sections~(\ref{sec:c3:operators_d6}) and ~(\ref{sec:c3:operators_d7}), we evaluate $D=6$ and $D=7$ mass-dimension operators from different irreps capable of destabilising a proton. We relate the computed operators in terms of quarks and leptons to the decay widths of hadrons in section~(\ref{sec:c3:decay_widths}) and compute the pattern of leading proton decay modes in a realistic \so\, model in section~(\ref{sec:c3:results_model}). We compare our study of scalar-mediated proton decays with the known-conventional gauge boson-mediated proton decays in section~(\ref{sec:c3:gbmpd}) and conclude this chapter by outlining the peculiarities of the scalar-mediated proton decays in section~(\ref{sec:c3:summary}).

\section{Relevant Effective Couplings}
{\label{sec:c3:couplings}}

In the subsection~(\ref{ssec:c2:161610}), we explored the interactions between the $10_{\hh}$ and the $\fs$-fermion multiplet. We found that the $T$ and $\overline{T}$ have leptoquark and diquark couplings capable of destabilising nucleons. The scalar multiplets $120_{\hh}$ and $126_{\hh}$ are also capable of interacting with the $\fs$-plet, as indicated by the Eq.~\eqref{eq:c2:16t16}. We proceed to derive the couplings of the various scalar fields residing in $120_{\hh}$ and $\overline{126}_{\hh}$ with the SM fermions unified in $\fs$ fermion plet.

\subsection{$\fs-\fs-120_{\hh}$ couplings} {\label{ss:c3:1616120}}
We consider the Yukawa interactions of three indexed antisymmetric tensor, $i.e.$ $ 120_{\hh}$, with $\fs$-plet of \so\, whose parameterisation is shown below~\cite{Nath:2001yj,Nath:2006ut}: 
\be \label{eq:c3:16_120}
-{\cal L}^{120_{\hh}}_Y = G_{AB}\,{\bf 16}^T_A\,\cc\,{\bf 16}_B\,{120}_H\,\hc, \ee
where $G_{AB}=-G_{BA}$ is anti-symmetric coupling unlike the case of $10_{\hh}$. The decomposition of $120_{\hh}$, using the syntax provided before Eq.~\eqref{eq:c2:45Hdec}, into $SU(5)\times U(1)_{\mathrm{X}}$ is shown as follows~\cite{Slansky:1981yr},
\beqa{\label{eq:c3:120dec}}
120_{\hh}\eq \left\{\left(5,2\right) \oplus \rm{c.c}\right\} \oplus \left\{\left(10,-6\right) \oplus \rm{c.c}\right\} \oplus \left\{\left(45,2\right) \oplus \rm{c.c}\right\}. 
\eeqa

One can decompose Eq.~(\ref{eq:c3:16_120}) into \(SU(5)\times U(1)_{\mathrm{X}}\) invariant expressions by following a procedure similar to the one outlined in the subsection~(\ref{ssec:c2:OESO}) and the outcome is summarised as follows~\cite{Nath:2001uw}:
\beqa \label{eq:c3:16_120_su5}
-{\cal L}^{120_{\hh}}_Y &=& i \frac{2}{\sqrt{3}}\, G_{AB}\,\Big[- 2\, \fn{^T_A}\,\cc\, \ff_{p B}\, 5^{p}_\hh\, - \ft^{pq\,T}_A\,\cc\,\ff{_{p B}}\,\overline{5}_{q\,\hh}\,\nl 
&+& \sqrt{2}\,\ff{^T_{p A}}\,\cc\,\ff_{q B}\,10^{pq}_\hh - \sqrt{2}\,\fn{^T_{A}}\,\cc\,\ft^{pq}_{B}\,\overline{10}_{pq\,\hh}\,\nl 
&-&   \frac{1}{2\sqrt{2}}\,\varepsilon_{pqrst}\,\ft^{pq\,T}_A\,\cc\,\ft^{tu}_B\,45^{rs}_{u\,\hh}\,+ \sqrt{2}\,\ff{^T_{p A}}\,\cc\, \ft^{qr}_B\,\overline{45}^p_{qr\,\hh}\Big]\,\hc.\nl
\eeqa

The $5_{\hh}$ $(\overline{5}_{\hh})$ and  $45_{\hh}$ $(\overline{45}_{\hh})$ include a scalar field that is a singlet under $SU(2)$ and a triplet under $SU(3)$ $(i.e.\, T)$ with a hypercharge of $-\sfrac{1}{3}$. Consequently, we differentiate them with different subscripts. The \(5_{\hh}\) dimensional \su\, irrep couples with SM fermions in such a way that one of them is a neutral lepton ($i.e.$, neutrino). Utilising the decompositions of $5_{\hh}$ and \(\overline{5}_{\hh}\) mentioned in Eq.~\eqref{eq:c2:5dec}, we obtain the following expression. 
\beqa \label{eq:c3:120/5}
-{\cal L}^{{120_{\hh}}/5_{\hh}}_Y & \supset & -i \frac{4}{\sqrt{3}}\, G_{AB}\, \left(-\varepsilon_{ab}\,\nu_A^{C\,T}\,\cc\,l_B^a\,D^b + \nu^{C\,T}_A\,\cc \, d^C_{\alpha B} T^\alpha \right)\hc, \nl
-{\cal L}^{{120_{\hh}}/\overline{5}_{\hh}}_Y & \supset & -i \frac{2}{\sqrt{3}}\, G_{AB}\, \bigg[\left(\varepsilon^{\alpha \beta \gamma}\, u^{CT}_{\alpha A}\,\cc\, d^C_{\beta B} + \varepsilon_{ab}\, q^{a\gamma\, T}_A\,\cc\, l^b_B \right)\overline{T}_{1\gamma}\nl
&& \left( q^{a\alpha\,T}_A\,\cc\,d^C_B\,+\,e^{C\,T}_A\,\cc\,l_B^a\right)\,\overline{D}_a \bigg] \hc
\eeqa

The canonically normalised decomposition of $10_{\hh}$ is analogous to the decomposition of $10_{\vv}$ done in Eq.~\eqref{eq:c2:10decomposition}.
When decomposed, the coupling of $10_{\hh}$ and $\overline{10}_{\hh}$ of \su\, irreps yields the following expression, where the latter again couples to neutrinos.
\beqa \label{eq:c3:120/10}
-{\cal L}^{{120_{\hh}}/10_{\hh}}_Y & \supset & -i 2\sqrt{\frac{2}{3}}\, G_{AB}\, \Big(\sqrt{2}\,\varepsilon_{ab}\, l^{a\,T}_A\,\cc\, d^C_{\alpha\,A}\, \Delta^{b\alpha}\,- d^{CT}_{\alpha A}\,\cc\,d^{C}_{\beta B}\,\overline{\Theta}^{\alpha \beta} \nl
&+& \varepsilon_{ab}\,l^{a\,T}_A\,\cc\,l^{b}_B\Big)\hc,\nl
-{\cal L}^{{120_{\hh}}/\overline{10}_{\hh}}_Y & \supset & -i 2\sqrt{\frac{2}{3}}\,G_{AB}\,\Big[-2\,\nu^{C\,T}_A\,\cc\,e^C_B\,s + \sqrt{2}\, \nu^C_A\,\cc\,q^{a\,\alpha}_B\,\overline{\Delta}_{a\alpha}\nl 
&+&\varepsilon_{\alpha\beta\gamma}\,\nu^{C\,T}_A\,\cc\,u^C_{\alpha\,B}\,\Theta_{\beta\gamma} \Big] \hc
\eeqa

We provide a procedure for the decomposition and canonical normalisation of \(\overline{45}_{\hh}\) in \(SU(5)\). The tensor \(\overline{45}\) is completely antisymmetric in its lower two indices. Below, we write all possible combinations in which one can place $SU(2)$ and $SU(3)$ indices in $\overline{45}$;
\beqa{\label{eq:c3:45decc}}
\overline{45}^{p}_{qr} \eq \overline{45}^a_{bc} \oplus \overline{45}^a_{\alpha b} \oplus \overline{45}^{\alpha}_{ab} \oplus \overline{45}^a_{\alpha\beta} \oplus \overline{45}^\alpha_{\beta\gamma}.
\eeqa

To begin with, we can write the following tensor as a sum of another traceless tensor and its trace part, as shown below;
\beqa{\label{eq:c3:45p1}}
\overline{45}^{a}_{bc}\eq c_1 \left(\delta_a^b \overline{D}_c-\delta^a_c\overline{D}_b\right)\nl
\overline{45}^{\alpha}_{\beta\,a}\eq O^\alpha_{\beta\,a} + c_2\,\delta^\alpha_\beta\,\overline{D}_a .
\eeqa
with $O$ being the traceless tensor by definition. The tracelessness condition yields the following;
\beqa
& &\overline{45}^a_{bc}\,\delta^b_a + \overline{45}^\alpha_{\beta\,a}\,\delta^{\beta}_{\alpha} = 0 \hspace{1cm}
\Rightarrow  c_1 = -3c_2.\nonumber
\eeqa
For canonically normalised $\overline{45}$ kinetic term we have, 
\beqa{\label{eq:c3:45ke}}
\left(\partial_{\Dot{\mu}}\overline{45}^{a}_{bc}\right)^{\dagger}\,\left(\partial^{\Dot{\mu}}\overline{45}^{a}_{bc}\right) + 2 \left(\partial_{\Dot{\mu}}\overline{45}^{\alpha}_{\beta\,a}\right)^{\dagger}\,\left(\partial^{\Dot{\mu}}\overline{45}^{\alpha}_{\beta\,a}\right).
\eeqa
The additional factor of $2$ in the above equation is due to the symmetry in the exchange of $SU(3)$ and $SU(2)$ labels, $i.e.$ $\beta$ and $a$.
\beqa{\label{eq:c3:45ke2}}
2\left(\partial_{\Dot{\mu}}O^{\alpha}_{\beta\,a}\right)^{\dagger}\,\left(\partial^{\Dot{\mu}}O^{\alpha}_{\beta\,a}\right) + 24\,c_2^2 \left(\partial_{\Dot{\mu}}\overline{D}_a\right)^{\dagger}\,\left(\partial^{\Dot{\mu}}\overline{D}_a\right)
\eeqa
Hence $O\to \frac{1}{\sqrt{2}}O$ and $c_2\to \frac{1}{2\sqrt{6}}$ for a canonically normalised kinetic term.

Similarly, other decomposed irreps can be canonically normalised, giving us the following result~\cite{Patel:2022wya}.
\beqa \label{eq:c3:45dec}
\overline{45}^{\gamma}_{\alpha \beta} &\equiv& S^{\gamma}_{\alpha \beta} + \frac{1}{2 \sqrt{2}} \left( \delta^\gamma_{\alpha} \overline{T}_\beta - \delta^\gamma_\beta \overline{T}_\alpha \right)\,,~~\overline{45}^a_{\alpha \beta}  \equiv   \Omega^a_{\alpha \beta}\,\,, \nonumber \\
\overline{45}^\beta_{\alpha a} &\equiv& \frac{1}{\sqrt{2}} \overline{O}^\beta_{\alpha a} + \frac{1}{2\sqrt{6}} \delta^\beta_\alpha \overline{D}_a\,,~~\overline{45}^\beta_{ab} \equiv \frac{1}{\sqrt{2}} \varepsilon_{ab} {\cal T}^\beta\,,\nonumber \\
\overline{45}^a_{b \alpha} & \equiv & \frac{1}{\sqrt{2}} \overline{\mathbb{T}}^a_{b \alpha} - \frac{1}{2 \sqrt{2}} \delta^a_b \overline{T}_\alpha\,,~~ \overline{45}^{c}_{ab} \equiv -\frac{\sqrt{3}}{2\sqrt{2}} \left( \delta^c_a \overline{D}_b - \delta^c_b \overline{D}_a\right)\,. \eeqa

The couplings of the remaining fields present in the \(45_{\hh}\) and \(\overline{45}_{\hh}\)-plets can be straightforwardly computed using the decomposition provided in Eq.~\eqref{eq:c3:45dec}. We obtain:
\beqa \label{eq:c3:120/45}
-{\cal L}^{{120_{\hh}}/45_{\hh}}_Y & \supset & i \frac{2}{\sqrt{3}}\, G_{AB}\, \bigg[2\, u^{C\,T}_{\alpha A}\,\cc\,e^C_B\,T_1^\alpha\,+\,\varepsilon^{\alpha \beta \gamma}\,u^{CT}_{\alpha A}\,\cc\,u^C_{\beta B}\,\overline{{\cal T}}_\gamma \Big. \nonumber\\ 
& + & \Big. \varepsilon_{\alpha \beta \gamma}\,  \varepsilon_{ab}\, q^{\alpha a T}_A\,\cc\,q^{\beta c}_B\, \mathbb{T}^{b \gamma}_c\,-\,\varepsilon_{\alpha \beta \gamma}\,e^{C  T}_A\,\cc\,q^{\gamma a}_B\,\overline{\Omega}^{\alpha \beta}_a \nl
&+& 2\,\varepsilon_{ab}\,u^{C\,T}_{\alpha\,A}\,\cc\,q^{b\beta}_B\,O^{\alpha\,a}_{\beta} \bigg] \hc\nl
\eeqa
\beqa \label{eq:c3:120/45b}
-{\cal L}^{{120_{\hh}}/\overline{45}_{\hh}}_Y & \supset & -i \frac{2}{\sqrt{3}}\, G_{AB}\, \bigg[\left(\varepsilon^{\alpha \beta \gamma}\, u^{CT}_{\alpha A}\,\cc\,d^C_{\beta B} + \varepsilon_{ab}\, q^{\gamma a T}_A\,\cc\,l^b_B\right) \overline{T}_{2 \gamma}\,  \nonumber \\ 
& + &  2\, e^{CT}_A\,\cc\, d^C_{\alpha B}\, {\cal T}^\alpha\,+\, 2\, \varepsilon_{bc}\,q^{\alpha a T}_A\,\cc\,l^b_B\,\overline{\mathbb{T}}^c_{a \alpha} \nonumber \\
& - &  \varepsilon^{\alpha \beta \gamma}\,\varepsilon_{ab}\,l^{bT}_A\,\cc\,u^C_{\gamma B}\,\Omega^a_{\alpha\beta} - d^{C\,T}_A\,\cc\,q^{a\,\alpha}_B\,\overline{O}^{\beta}_{a\alpha} \nl
&-&  \frac{1}{\sqrt{3}}\,d^C_{\alpha\,A}\,\cc\,q^{a\alpha}_B\,\overline{D}_b + \sqrt{3}\,l^a_A\,\cc\,e^C_B\,\overline{D}_a \bigg]
\eeqa

The following comments are in order regarding the couplings of various scalars residing in $120_{\hh}$-dimensional \so\,irrep.

\begin{enumerate}[label=(\roman*)]
    \item The effective couplings given in Eqs.~(\ref{eq:c3:120/5}, \ref{eq:c3:120/10}, \ref{eq:c3:120/45}, and \ref{eq:c3:120/45b}) determine the various leptoquark, diquark and dilepton vertices stemming from the coupling of $120_{\hh}$ with $\fs$-plet. 
    \item The couplings of $\overline{5}_{\hh}$, given in Eq.~\eqref{eq:c3:120/5} yields, $Y_d\,=-\,Y_e^T\,=\,G$ at the GUT scale.
    \item  From the Eq.~\eqref{eq:c3:120/10}, we notice $\overline{10}_{\hh}$ couples to neutrinos, while $10_{\hh}$ has three scalars, namely $\Theta$, $\Delta$ and $t$ having diquark, leptoquark and dilepton vertices respectively.

    \item \(45_{\text{H}}\) and \(\overline{45}_{\text{H}}\) each contain seven scalars. Among them, \(T\), \({\mathcal{T}}\), and \({\mathbb T}\) possess leptoquark vertices, while \(\overline{T}\), \(\overline{{\mathcal{T}}}\), and \({\mathbb T}\) couple to diquarks. Additionally,  $O$ and $\overline{O}$ couple solely to quarks, while \(\Omega\) and \(\overline{\Omega}\) exclusively feature leptoquark vertices.
    \item Interestingly, if the lightest Higgs stems from $\overline{45}_{\hh}$, then $3\,Y_d=\,Y_e^T$, as evident from Eq.~(\ref{eq:c3:120/45b}).

\end{enumerate}

\subsection{$\fs-\fs-\overline{126}_{\hh}$  couplings}
{\label{ss:c3:1616126}}
The \(\overline{ 126}_H\) possesses a diverse scalar spectrum comprising of  $22$ distinct scalar fields that transform as irrep under SM gauge symmetry~(cf.~Tab.~(\ref{tab:c2:scalars})), and the Yukawa Lagrangian involving this field is expressed as:

\be \label{eq:c3:16-16-126}
-{\cal L}^{\overline{126}_{\hh}}_Y = F_{AB}\,{\bf 16}^T_A\,\cc\,{\bf 16}_B\,\overline{126}_{\hh}\,\hc, \ee
where, $F_{AB}$ is constrained by its symmetricity in generation space. The decomposition of $\overline{126}_{\hh}$ into $SU(5)\times U(1)_{\mathrm{X}}$ is shown below~\cite{Slansky:1981yr};
\beqa{\label{eq:c2:126dec}}
126_{\hh}\eq \left(1,10\right) \oplus \left(5,2\right) \oplus   \left(\overline{10},6\right) \oplus \left(15,-6\right) \oplus \left(\overline{45},2\right) \oplus \left(50,2\right). 
\eeqa

Again, the above Lagrangian given in Eq.~\eqref{eq:c3:16-16-126} can be decomposed respecting the $SU(5)$ gauge symmetry, as shown below~\cite{Nath:2001uw};
\beqa \label{eq:c3:16_126_su5}
-{\cal L}^{\overline{ 126}_{\hh}}_Y &=& i \frac{2}{\sqrt{15}}\, F_{AB}\,\Big[-\fn{^T_A}\, \cc\, \fn_B\, 1_0 + \fn{^T_A}\,\cc\, \ft_B^{pq}\, \overline{10}_{pq\,\hh} \, \Big. \nonumber \\
&-& \sqrt{\frac{3}{2}}\,\left(\fn{^T_A}\, \cc\, \ff_{p B}\, 5^{p}_\hh\, + \frac{1}{24}\varepsilon_{pqrst}\, \ft^{pq\,T}_A\,\cc\, \ft^{rs}_B\,5^t_\hh \right)\,\nonumber \\
& + & \Big. \ft^{pq\,T}_A\, \cc\,  \ff_{r B}\, \overline{45}^r_{pq\,\hh} - \frac{1}{4\sqrt{3}}\,\varepsilon_{pqrst}\, \ft^{pq\,T}_A\,\cc\, \ft^{uv}_B\,50^{rst}_{uv\,\hh} \nl
\mi \ff{_{p A} ^T}\, \cc\, \ff_{q B}\, 15^{pq}_\hh\Big] \hc .\eeqa

We proceed analogous to the previous subsection~(\ref{ss:c3:1616120}) and decompose the $SU(5)\times U(1)_{\mathrm{X}}$ invariant expression written in Eq.~\eqref{eq:c3:16_126_su5} and starting with the coupling of the singlet. The singlet, alias the \(\sigma\), present in \(\overline{126}_{\hh}\) couples exclusively to right-handed neutrinos. This term generates Majorana masses for right-handed neutrinos when the singlet acquires a $vev$, as depicted below;
\beqa{\label{eq:c3:126-1-1}}
-{\cal L}^{\overline{126}_{\hh}/1_{\hh}}_Y & \supset & -i\,\frac{2}{\sqrt{15}}\,F_{AB}\,\nu^{C\,T}_A\,\cc\,\nu^C_B\,\sigma \hc.
\eeqa

Moreover, \(\overline{126}_{\hh}\) also includes a \(5_{\hh}\) of \(SU(5)\), whose decomposition is provided earlier in Eq.~\eqref{eq:c2:5dec}, and it couples similarly to the standard \(5_{\text{H}}\), as illustrated earlier in Eq.~\eqref{eq:c2:10/5}.
\beqa \label{eq:c3:126/5}
-{\cal L}^{\overline{126}_{\hh}/5_{\hh}}_Y & \supset & i \sqrt{\frac{2}{5}}\, F_{AB}\, \Big[-\left( \frac{1}{3} u{^{CT}_{\gamma A}}\, \cc\, e^C_B +\frac{1}{6} \varepsilon_{\alpha \beta \gamma}\, \varepsilon_{ab}\, q^{a\alpha T}_A\, \cc\, q^{b\beta}_B \right)T_1^\gamma\,\nl
&-& \frac{1}{3}\,\varepsilon_{ab}\,q^{a\alpha\,T}_A\,\cc\,u^C_{\alpha\,B}\,H^b\Big] \hc
\eeqa
Since \(\overline{ 126}_{\hh}\) contains more than one colour triplet field residing in $5_{\hh}$ and $50_{\hh}$, consequently, we distinguish them by assigning a subscript. Furthermore, utilising the canonically normalised decomposition of \(10_{\hh}\) computed in Eq.~\eqref{eq:c2:10decomposition}, one can deduce its couplings with SM fermions as shown below;
\beqa{\label{eq:c3:126-10-10}}
-{\cal L}^{\overline{ 126}_{\hh}/10_{\hh}}_Y & \supset & i\,\frac{4}{\sqrt{15}}\,F_{AB}\,
\Big( \nu^{C\,T}_A\,\cc\,q^{a\alpha}_B\,\Delta_{a\alpha} - \nu^{C\,T}_A\,\cc\,e^C_B\,s\nl \ad \frac{1}{2}\,\varepsilon^{\alpha\beta\gamma}\,\nu^{C\,T}_A\,\cc\,u^C_{\alpha}\,\Theta_{\beta\gamma}\Big)\hc.
\eeqa

Moving on to the decomposition of two indexed symmetric tensor $15_{\hh}$ into the normalised irreps of $SU(5)$ and is given by:
\beqa \label{eq:c2:15dec}
15^{\alpha \beta} = \Sigma^{\alpha \beta},\,~~15^{\alpha a} = \frac{1}{\sqrt{2}} \Delta^{\alpha a},\,~~\text{and}~~15^{ab} = t^{ab}\,.\eeqa
Allowed vertices of scalars stemming from $15_{\hh}$ with SM fermions are shown below; 
\beqa \label{eq:c3:126/15}
-{\cal L}^{\overline{ 126}_{\hh}/15_{\hh}}_Y & \supset & i\,\frac{2}{\sqrt{15}}\, F_{AB}\, \Big[\sqrt{2}\,\varepsilon_{ab}\, d{^{CT}_{\alpha A}}\, \cc\, l^a_B\, \Delta^{\alpha b} -\varepsilon_{ae}\,\varepsilon_{bf}\,l^{e\,T}_B\,\cc\,l^f_B\,t^{ab}\nl &-& d^{C\,T}_{\alpha\,A}\,\cc\,d^C_{\beta}\,\Sigma^{\alpha\beta} \Big]\hc. \eeqa
 
Further, using the canonically normalised decomposed $45_{\hh}$ given in Eq.~\eqref{eq:c3:45dec}, the coupling of various $B-L$ charged scalars with SM fermions is shown below; 
\beqa \label{eq:c3:126/45:2}
-{\cal L}^{\overline{126}_{\hh}/\overline{45}_{\hh}}_Y & \supset & i \sqrt{\frac{2}{15}}\, F_{AB}\, \bigg[\left(\varepsilon^{\alpha \beta \gamma}\, u^{CT}_{\alpha A}\,\cc\,\,d^C_{\beta B} + \varepsilon_{ab}\, q^{\gamma a T}_A\,\cc\,l^b_B\right) \overline{T}_\gamma\,  \nonumber \\ 
& + &  2\, e^{CT}_A\,\cc\, d^C_{\alpha B}\, {\cal T}^\alpha\,+\, 2\, \varepsilon_{bc}\,q^{\alpha a T}_A\,\cc\,l^b_B\,\overline{\mathbb{T}}^c_{a \alpha} \nl
&+& 2\,q^{a\alpha\,T}_A\,\cc\,d^C_{\beta\,B}\,\overline{O}_{a\alpha}^{\beta} + \frac{1}{\sqrt{3}}\,q^{a\alpha\,T}_A\,\cc\,d^C_{\alpha\,B}\,\overline{D}_a \nl \ad \sqrt{3}\,e^{C\,T}_A\,\cc\,l_B^a\,\overline{D}_a\bigg]\hc.
\eeqa

Finally, considering the \(50_{\hh}\) of \su\, dimensional irrep is totally antisymmetric in both lower and upper indices. The coupling with the \(50_{\hh}\)-plet can be computed from its following decomposition~\cite{Patel:2022wya}:
\beqa \label{eq:c3:50dec}
50^{\alpha \beta \gamma}_{\sigma \rho} & \equiv & \frac{1}{6} \left[ \left(\delta^\alpha_\sigma \delta^\beta_\rho -  \delta^\alpha_\rho \delta^\beta_\sigma \right) T^\gamma +  \left(\delta^\alpha_\rho \delta^\gamma_\sigma -  \delta^\alpha_\sigma \delta^\gamma_\rho \right) T^\beta  +  \left(\delta^\beta_\sigma \delta^\gamma_\rho -  \delta^\beta_\rho \delta^\gamma_\sigma \right) T^\alpha \right]\,,\nonumber \\
50^{\alpha \beta \gamma}_{\sigma a} & \equiv & \frac{1}{2\sqrt{3}} \left( \delta^\alpha_\sigma \overline{\Omega}^{\beta \gamma}_a + \delta^\beta_\sigma \overline{\Omega}^{\gamma \alpha}_a + \delta^\gamma_\sigma \overline{\Omega}^{\alpha \beta}_a\right)\,, \nonumber \\
50^{\alpha \beta a}_{\sigma \rho} & \equiv & \frac{1}{2\sqrt{6}} \left(\delta^\alpha_\sigma O^{\beta a}_\rho - \delta^\alpha_\rho O^{\beta a}_\sigma +\delta^\beta_\rho O^{\alpha a}_\sigma - \delta^\beta_\sigma O^{\alpha a}_\rho \right)\,,\nonumber\\
50^{\alpha \beta \gamma}_{ab} & \equiv & \frac{1}{2 \sqrt{3}} \varepsilon^{\alpha \beta \gamma} \varepsilon_{ab} X\,,~~50^{\alpha \beta a}_{\sigma b} \equiv \frac{1}{\sqrt{6}} \mathbb{S}^{\alpha \beta a}_{\sigma b} + \frac{1}{12} \delta^a_b \left( \delta^\alpha_\sigma T^\beta - \delta^\beta_\sigma T^\alpha \right)\,, \nonumber \\
50^{\alpha a b}_{\sigma \rho} & \equiv & \frac{1}{2\sqrt{3}} \varepsilon^{ab} {\cal S}^\alpha_{\sigma \rho}\,,~~50^{\alpha \beta a}_{bc}  \equiv  \frac{1}{2\sqrt{3}} \left(\delta^a_c \overline{\Omega}^{\alpha \beta}_b - \delta^a_b \overline{\Omega}^{\alpha \beta}_c \right)\,, \nonumber \\
50^{\alpha a b}_{\sigma c} & \equiv & \frac{1}{2\sqrt{6}} \left(\delta^b_c O^{\alpha a}_\sigma - \delta^a_c O^{\alpha b}_\sigma \right)\,,~~50^{\alpha ab}_{cd} = \frac{1}{6} \left(\delta^a_c \delta^b_d - \delta^a_d \delta^b_c \right) T^\alpha\,. \eeqa

\beqa \label{eq:c3:126/50}
-{\cal L}^{\overline{126}_{\hh}/50_{\hh}}_Y & \supset & -i \frac{2}{3\sqrt{5}}\, F_{AB}\, \bigg[-\left(u^{CT}_{\gamma A}\,\cc\, e^C_B +\frac{1}{2} \varepsilon_{\alpha \beta \gamma}\, \varepsilon_{ab}\, q^{ a\alpha\, T}_A\,\cc\, q^{b\,\beta }_B \right)T_2^\gamma\nl
&+& \sqrt{3}\,\varepsilon_{\alpha\beta\gamma}\,q^{a\alpha\,T}_A\,\cc\,e^C_B\,\overline{\Omega}^{\beta\gamma}_b - \sqrt{6}\,\varepsilon_{ab}\,u^{C\,T}_{\alpha\,A}\,\cc\,q^{a\beta}_B\,O^{\alpha\,b}_\beta \nl &-& \frac{24}{\sqrt{15}}\,e^{C\,T}_A\,\cc\,e^C_B\,X + \frac{3\sqrt{15}}{2}\, \varepsilon_{mn}\,\varepsilon_{\alpha\beta\gamma}\,q^{m\alpha\,T}_A\,\cc\,q^{b\sigma}_B\,{\mathbb S}^{\beta\gamma\,n}_{\sigma\,b} \bigg]\nl\hcn
\eeqa
In the following points, we summarise the notable features of couplings of \(126_{\text{H}}\) with the \(\fs\)-plet as follows:
\begin{enumerate}[label=(\roman*)]
    \item $\overline{126}_{\hh}$ features a singlet, $\sigma$, charged under $B-L$ and can generate Majorana mass of right-handed neutrinos as mentioned in Eq.~\eqref{eq:c3:126-1-1}.
    \item Three scalars are present in \(\overline{10}_{\hh}\), namely \(\Theta\), \(s\), and \(\Delta\), which respectively have leptoquark, dilepton, and diquark couplings, analogous to the \(10_{\hh}\) present in \(120_{\hh}\).
    \item The two-indexed symmetric tensor, $i.e.$, \(15_{\hh}\), also comprises three scalars, namely \(\Delta\), \(\Sigma\), and \(t\), which possess leptoquark, diquark, and dilepton vertices respectively, as evident from Eq.~\eqref{eq:c3:126/15}.
    \item The \(\overline{45}_{\hh}\) residing in \(\overline{126}_{\hh}\) couples analogously to the \(\overline{45}_{\hh}\) residing in \(120_{\hh}\), but the Yukawa couplings are symmetric in the former case.
    \item The irrep \(50_{\hh}\) stemming from \(\overline{126}_{\hh}\) comprises six scalars and lacks any SM-like Higgs. These six scalars are \(X\), which couples only to charged leptons, \(T_2\), having leptoquark and diquark couplings, \(\overline{\Omega}\), featuring only leptoquark coupling, and \({\mathcal {S}}\), ${\mathbb S}$, and \(O\) exhibiting only diquark couplings. 
    \item SM like Higgs residing in $\overline{5}_{\hh}$ yields $Y_{d}=Y_{e}^T$, while that residing in $\overline{45}$ yields $Y_d=3 Y_e^T$ at the GUT scale, provided considering the contribution of one doublet at a time.
\end{enumerate}

\begin{table}[t!]
\begin{center}
\begin{tabular}{cccc} 
\hline
\hline
~~Coupling~~&~~$10_{\hh}$~~&~~$120_{\hh}$~~&~~$\overline{126}_{\hh}~~$\\
\hline
 &  & $T_1$, $T_{2}$, $\overline{T}_1$, $\overline{T}_2$ & $T_{1}$, $T_2$, $\overline{T}$ \\
Leptoquark& $T$, $\overline{T}$ & $\Delta$, $\overline{\Delta}$, $\Omega$, $\overline{\Omega}$  & $\Delta$, $\Omega$\\
& & $\Theta$, ${\cal T}$, $\overline{\mathbb{T}}$ & $\Theta$, ${\mathcal T}$, $\overline{{\mathbb T}}$ \\
\hline
 &  & $T_1$, $T_{2}$, $\overline{T}_1$, $\overline{T}_2$ & $T_{1}$, $T_2$, $\overline{T}$ \\
Diquark & $T$, $\overline{T}$ & $O$, $\overline{O}$, $S$, $\overline{S}$ & $O$, $\overline{O}$\\
 & & $\overline{\Theta}$, $\overline{{\mathbb T}}$, ${\mathcal T}$ & ${\mathcal{S}}$, $\mathbb{S}$ \\ 
 \hline\hline
\end{tabular}
\end{center}
\caption{Various scalars exhibiting leptoquark and diquark couplings within the $10_{\hh}$, $120_{\hh}$, and $\overline{126}_{\hh}$ multiplets. }
\label{tab:c3:scalarsummary}
\end{table}

Tab.~(\ref{tab:c3:scalarsummary}) summarises the scalar fields yield diquark and leptoquark vertices. These scalars are charged under $B-L$ and could potentially induce proton decay. 
Consequently, we utilise these evaluated couplings to compute the proton decay spectrum.

\section{$B-L:$ Conservation vs. Violation}
\label{sec:c3:geninfo}

 Proton decay is a baryon number violating phenomenon, and $B$ and $L$ are the accidental symmetries of the SM as discussed in Chapter~(\ref{ch:1}) and can only be violated in the SM through non-perturbative effects like sphalleron transition~\cite{Espinosa:1989qn}. These accidental symmetries are preserved by the tree-level Lagrangian but are violated by quantum corrections and hence are called \textit{anomalous}.
 
The anomaly of these discrete symmetries can be inferred from the computation of the triangle diagrams, where the individual currents for the baryon and lepton numbers are typically expected to be zero. However, \(B-L\) (baryon number minus lepton number) can be defined such that this current is conserved, establishing \(B-L\) as an exact global symmetry of the SM. Furthermore, the \(B-L\) symmetry can be gauged and this global symmetry can be promoted to a local one, with one additional gauge boson introduced for local \(B-L\) gauge invariance. This also requires the addition of one right-handed neutrino per generation for cancellation of $(B-L)^3$ anomaly. The gauging of the \(B-L\) symmetry and the relating of its breaking scale to the Majorana masses of the right-handed neutrinos are essential for the implementation of the typical seesaw (Type I) mechanism in the conventional GUTs~\cite{Minkowski:1977sc, Yanagida:1979as, Mohapatra:1979ia, Schechter:1980gr}. Once \(B-L\) is gauged, it can also be related to the subgroups of \(SO(10)\), ultimately relating it to the electric charges of fermions (cf.~\cite{Mohapatra:2014yla} for an impactful discussion on \(B-L\)).

As proton decay is a baryon number violating process, its different decay modes can be classified into two distinct categories: 1) $B-L$ conserving or $B+L$ violating and 2) $B-L$ violating or $B+L$ conserving.

In modes that conserve $B-L$, the change in baryon number is equal to the change in lepton number, such that $\Delta (B-L) = 0$ or $\Delta (B+L) = 2$, where $\Delta$ denotes the change. Additionally, any change in $B-L$ must always be an even number~\cite{Kobach:2016ami,Helset:2019eyc}. Operators involving only SM fermions with $B-L = 0, 4, 8, ...$ typically appear at even mass dimensions in the Lagrangian, $i.e.$, $D=6, 8, 10, ...$~\cite{Kobach:2016ami,Helset:2019eyc}. In $B-L$ conserving modes, the proton decays into an antilepton and is accompanied by one or more mesons.

\begin{figure}[t]
    \centering
    \includegraphics[scale=0.1]{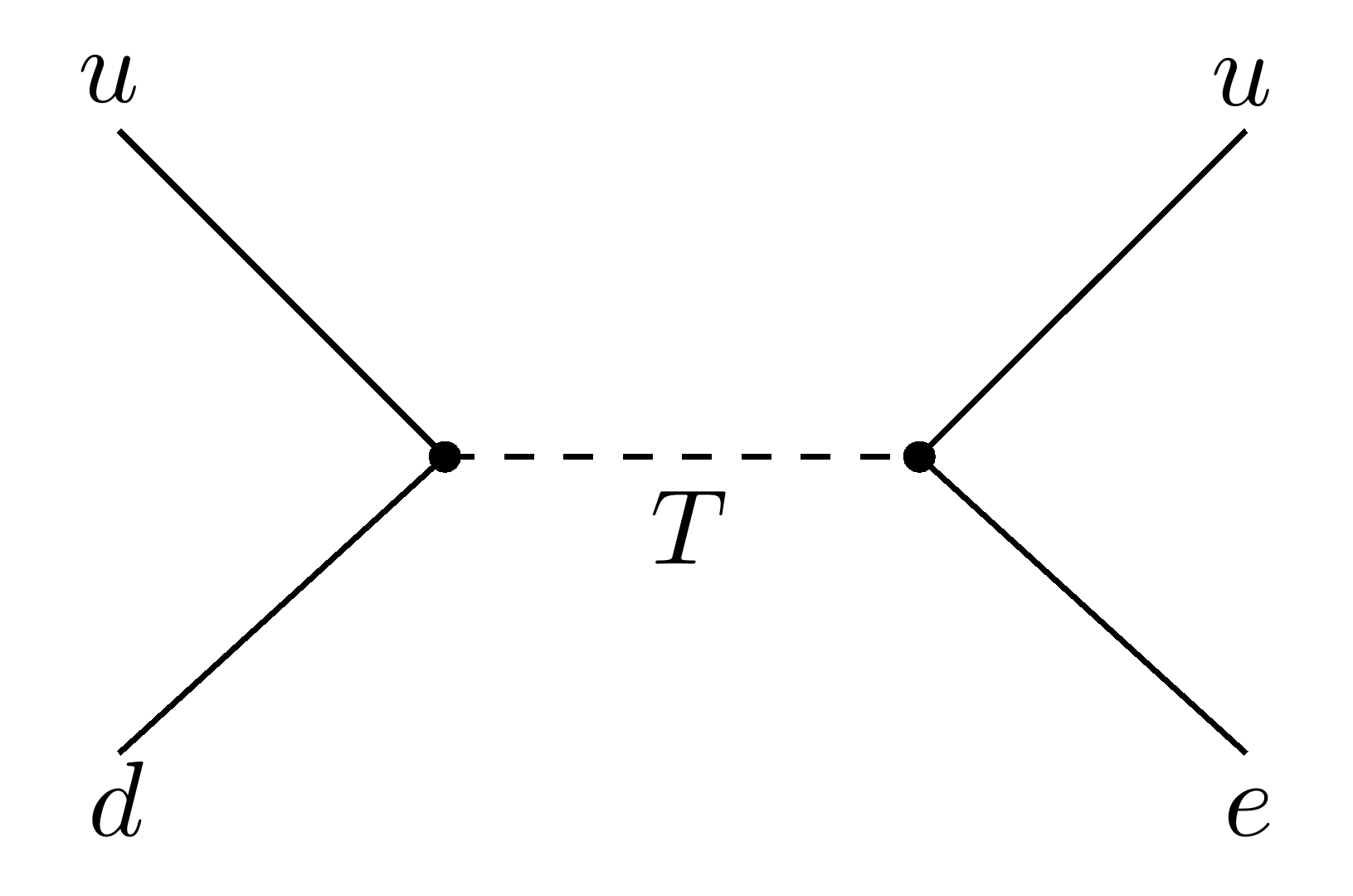} \hspace{0.4cm} \includegraphics[scale=0.12]{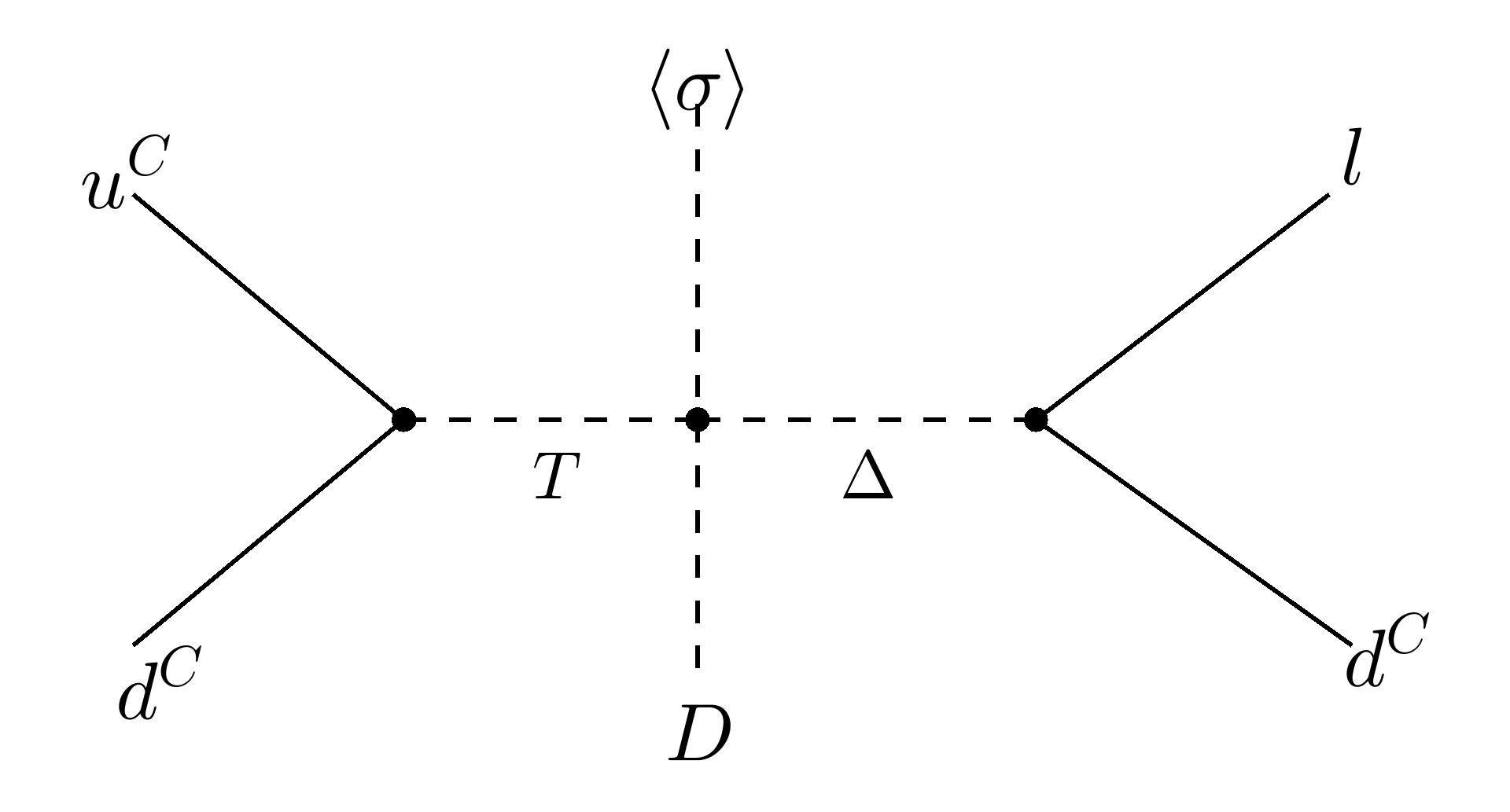}
    \caption{Topologies of $B-L$ conserving (left Feynman graph) and $B-L$ violating (right Feynman graph) proton decay, drawn using~\cite{Harlander:2020cyh}. $B-L$ violating diagram is drawn in the unbroken Electroweak phase.}
    \label{fig:c4:pdecay:b-l}
\end{figure}

In contrast to $B-L$ conserving modes, in $B-L$ violating modes, the change in $\Delta (B-L)$ is 2 or $\Delta (B+L) = 0$. Operators with $B-L = 2, 6, 10, ...$ typically appear at odd mass dimensions in the Lagrangian. The strength of such modes is proportional to the $B-L$ breaking scale. In such modes, proton decay results in a lepton accompanied by two or more mesons.

In the upcoming sections, we will consider the effective operators capable of inducing $B-L$ conserving and violating modes at $D=6$ and $D=7$, respectively. Additionally, we will focus only on two-body proton decay modes, as decays involving three or more bodies are inherently phase-space suppressed.

\section{Dimension-Six Effective Operators}
\label{sec:c3:operators_d6}
We now calculate the leading mass dimension $D=6$ effective operators essential for the $B$ and $L$ violating decays of baryons by integrating out the effects of heavy scalar fields present in the model. We analyse each case involving $10_{\hh}$, $120_{\hh}$ and $\overline{126}_H$ independently and discuss also the potential of mixing of scalar fields with the same SM charges among themselves.

\subsection{Effective operators from $10_{\hh}$}
\label{ss:c3:op10}
The contribution to nucleon decay in \so\, models featuring only ${10}_H$ stems from the interaction between a pair of colour triplet and anti-triplet, $T$ and $\overline{T}$, as illustrated in Eq.~\eqref{eq:c2:10/5}. The most general mass terms for these triplets (and antitriplets) are expressed as follows:
\be \label{eq:c3:triplet_mix}
m_T^2\, T^\dagger T + m_{\overline{T}}^2\, \overline{T}^\dagger \overline{T} + \left( \mu^2\, T\, \overline{T}+\hc\right).\,\ee
The initial two terms may arise from the $SO(10)$ invariant combination $10_{\hh}^\dagger 10_{\hh}$, and the final term from $ 10_{\hh}^2$ \cite{Slansky:1981yr}. Generally, $m_T \neq m_{\overline{T}}$ because mass splitting can occur if the model includes $45_{\hh}$ and/or $54_{\hh}$. Due to the mixing term, the physical states differ and can be identified through the following substitutions:
\be \label{eq:c3:T_redf}
T \to T \cos\theta_T + \overline{T}^* \sin\theta_T\,,~~ \overline{T} \to \overline{T} \cos\theta_T - T^* \sin\theta_T\,. \ee
Here, the mixing angle $\theta_T$ is obtained from Eq.~\eqref{eq:c3:triplet_mix} as; 
\be \label{eq:c3:mixing_angle}
\tan 2\theta_T = \frac{2 \mu^2}{m_{\overline{T}}^2 - m_T^2}\,. \ee 
With the above replacements, Eq.~\eqref{eq:c3:triplet_mix} becomes
\be \label{eq:c3:mass_term_T}
M_T^2\, T^\dagger T + M_{\overline{T}}^2\, \overline{T}^\dagger \overline{T}\,.\ee

By substituting Eq.~\eqref{eq:c3:T_redf} into Eq.~\eqref{eq:c2:10/5} and integrating out the triplet $(T)$ and anti-triplet $(\overline{T})$ in the mass basis, we derive the effective Lagrangian relevant for $B$-violating baryon decays as shown below;
\beqa \label{eq:c3:op_10_0}
{\cal L}^{{10_{\hh}}}_{\rm eff} &=& 8 \left(\frac{c_T^2}{M_T^2}+\frac{s_T^2}{M_{\overline{T}}^2}\right) H_{AB} H^\dagger_{CD}\,\, \varepsilon_{\alpha \beta \gamma}\, \left(u^{\alpha\,T}_{A}\, \cc\,d^{\beta}_{ B}\right)\,\left( {e^C_C}^\dagger\,\cc\,{u^C_{\gamma D}}^*\right) \nonumber \\
& + &8 \left(\frac{s_T^2}{M_T^2}+\frac{c_T^2}{M_{\overline{T}}^2}\right) H_{AB} H^\dagger_{CD}\,\, \varepsilon^{\alpha \beta \gamma}\, \left(u{^{CT}_{\alpha A}}\,\cc\, d_{\beta B}^C\right)\, \left(e_C^\dagger \,\cc\, u_{\gamma D}^*\right)  \nonumber \\
& - &8 \left(\frac{s_T^2}{M_T^2}+\frac{c_T^2}{M_{\overline{T}}^2}\right) H_{AB} H^\dagger_{CD}\,\, \varepsilon^{\alpha \beta \gamma}\, \left(u{^{CT}_{\alpha A}}\,\cc\, d_{\beta B}^C\right)\, \left( \nu_C^\dagger \,\cc\, d_{\gamma D}^*\right)  \nonumber \\
& + &  8\, s_T c_T \left(\frac{1}{M_T^2}-\frac{1}{M_{\overline{T}}^2} \right) H_{AB} H^T_{CD}\,\varepsilon^{\alpha \beta \gamma}\,\left( u{^{CT}_{\alpha A}}\,\cc\, d^C_{\beta B}\right)\, \left({e^{CT}_C}\,\cc\,{u^C_{\gamma D}}\right)\nonumber \\
& + & 8\, s_T c_T \left(\frac{1}{M_T^2}-\frac{1}{M_{\overline{T}}^2} \right) H_{AB} H^T_{CD}\,\, \varepsilon_{\alpha \beta \gamma}\,\left( u^{\alpha\,T} _{A} \,\cc\, d^{\beta}_{ B}\right)\, \left(e_C^T\,\cc\, u^{\gamma}_{D} \right)\nonumber \\
& - & 8\, s_T c_T \left(\frac{1}{M_T^2}-\frac{1}{M_{\overline{T}}^2} \right) H_{AB} H^T_{CD}\,\, \varepsilon_{\alpha \beta \gamma}\,\left( u^{\alpha\,T}_{ A} \,\cc\, d_{\beta B}\right)\, \left(\nu_C^T \,\cc\, d^{\gamma}_{D} \right)\nl \hcn,\eeqa
where $c_T\equiv\cos\theta_T$ and $s_T\equiv\sin\theta_T$ and used the properties of the charge conjugation matrix given in Eq.~\eqref{eq:c1:conjugation}.

The effective operators in the physical basis of fermions can be obtained by replacing \( f \to U_f f \) in the aforementioned \({\cal L}^{{10}_{\hh}}_{\rm eff}\). The unitary matrices \( U^f \) and \( U_{f^C} \) (for \( f=u, d, e, \nu \)) can be explicitly computed from their respective fermion mass matrices. In the physical basis, the resulting expression is as follows:
\beqa \label{eq:c3:op_10}
{\cal L}^{{ 10_{\hh}}}_{\rm eff} &=& h[u_A,d_B,e^C_C,u^C_D]\,\, \varepsilon_{\alpha \beta \gamma}\, \left(u^{\alpha\,T}_{A}\, \cc\,d^{\beta}_{ B}\right)\,\left( {e^C_C}^\dagger\,\cc\,{u^C_{\gamma D}}^*\right) \nonumber \\
& + & h[u^C_A,d^C_B,e_C,u_D]\,\, \varepsilon^{\alpha \beta \gamma}\, \left(u{{^{CT}_{\alpha A}}}\,\cc\, d_{\beta B}^C\right)\, \left(e_C^\dagger \,\cc\, u_{\gamma D}^*\right)\, \nonumber \\
& + & h[u^C_A,d^C_B,\nu_C,d_D]\,\, \varepsilon^{\alpha \beta \gamma}\, \left(u{^{CT}_{\alpha A}}\, \cc\, d_{\beta B}^C\,\right)\,\left( \nu_C^\dagger\,\cc\,d_{\gamma D}^*\right)\, \nonumber\\
& + & h^\prime[u^C_A,d^C_B,e^C_C,u^C_D]\,\, \varepsilon^{\alpha \beta \gamma}\, \left(u{^{CT}_{\alpha A}}\,\cc\, d^C_{\beta B}\right)\, \left(e{^{CT}_C}\,\cc\,u^C_{\gamma D}\right) \nonumber \\
& + & h^\prime[u_A,d_B,e_C,u_D]\,\, \varepsilon_{\alpha \beta \gamma}\, \left(u^{\alpha\,T}_A\,\cc\,d^{\beta}_{ B}\right)\, \left(e_C^T\,\cc\,u^{\gamma}_{D}\right)\, \nonumber \\
& + & h^\prime[u_A,d_B,\nu_C,d_D]\,\, \varepsilon_{\alpha \beta \gamma}\, \left(u^{\alpha\,T }_{A}\,\cc\,d^{\beta}_{ B}\right)
\, \left(\nu_C^T\,\cc\,d^{\gamma}_{D}\right)\,\,\nl \hcn,\eeqa
together with,
\beqa \label{eq:c3:coff_op_10}
h[u_A,d_B,e^C_C,u^C_D] &=& 8 \left(\frac{c_T^2}{M_T^2}+\frac{s_T^2}{M_{\overline{T}}^2}\right)  \left(U_u^T H U_d \right)_{AB}\,\left(U_{e^C}^\dagger H^\dagger U_{u^C}^*\right)_{CD}\,, \nonumber \\
h[u^C_A,d^C_B,e_C,u_D] &=& 8 \left(\frac{s_T^2}{M_T^2}+\frac{c_T^2}{M_{\overline{T}}^2}\right)  \left(U_{u^C}^T H U_{d^C}\right){_{AB}}\, \left(U_{e}^\dagger H^\dagger U_{u}^* \right){_{CD}}\,, \nonumber \\
h[u^C_A,d^C_B,\nu_C,d_D] &=& - 8 \left(\frac{s_T^2}{M_T^2}+\frac{c_T^2}{M_{\overline{T}}^2}\right) \left(U_{u^C}^T H U_{d^C} \right){_{AB}}\,\left(U_{\nu}^\dagger H^\dagger U_{d}^*\right){_{CD}}\,, \nonumber \\
h^\prime[u^C_A,d^C_B,e^C_C,u^C_D] &=& 8 s_T c_T \left(\frac{1}{M_T^2}-\frac{1}{M_{\overline{T}}^2} \right)  \left(U_{u^C}^T H U_{d^C} \right){_{AB}}\,\left(U_{e^C}^T H^T U_{e^D}\right){_{CD}}\,, \nonumber \\
h^\prime[u_A,d_B,e_C,u_D] &=& 8\, s_T c_T \left(\frac{1}{M_T^2}-\frac{1}{M_{\overline{T}}^2} \right) \left(U_u^T H U_d \right)_{AB}\,\left(U_e^T H^T U_u\right)_{CD}\,, \nonumber \\
h^\prime[u_A,d_B,\nu_C,d_D] &=&-8\, s_T c_T \left(\frac{1}{M_T^2}-\frac{1}{M_{\overline{T}}^2} \right) \left(U_u^T H U_d \right)_{AB}\,\left(U_\nu^T H^T U_d \right)_{CD}\,. \nl \eeqa
 When there is no mixing term between \( T \) and \( \overline{T} \) ($i.e.$, \( \theta_T = 0 \)), all \( h' \), in Eq.~\eqref{eq:c3:op_10}, coefficients are zero. Furthermore, the coefficients of the first operator and the subsequent two operators become independent of each other.

\subsection{Effective operators from $120_{\hh}$}
\label{ss:c3:op120}

Now, we compute effective mass dimension six $B-L$ conserving operators stemming from three indexed antisymmetric tensor $i.e.$ $120_{\hh}$. 
Since $120_{\hh}\times 120_{\hh}\,\supset \mathds{1} $, where $\mathds{1}$ transforms trivially under \so\, transformations, consqeuently $T$ $\in 45_{\hh}$ and $\overline{T}_2 \in \overline{45}_{\hh}$ can have a mixing term. Similarly, ${\cal T}$ and $\mathbb{T}^{a \alpha}_b$ can mix with their conjugate partners appearing in Eqs.~(\ref{eq:c3:120/45}, and \ref{eq:c3:120/45b}). Integrating out these heavy scalars from Eqs. (\ref{eq:c3:120/45} and \ref{eq:c3:120/45b}), we obtain the following four fermion interactions.
\beqa \label{eq:c3:op_120_0}
{\cal L}^{{ 120_{\hh}}}_{\rm eff} &=& \frac{4}{3}\left(\frac{1}{M_{\overline{T}_1}^2}-\frac{c_T^2}{M_{\overline{T}_2}^2} - \frac{s_T^2}{M_T^2}\right)\,G_{AB} G^\dagger_{CD}\,\nl 
& &\times\,\varepsilon^{\alpha \beta \gamma}\, \left(u{^{CT}_{\alpha A}} \,\cc^{-1}\,d_{\beta B}^C\right)\, \left(e_C^\dagger\,\cc\,u_{\gamma D}^* \right)\, \nonumber \\
\mi \frac{4}{3}\left(\frac{1}{M_{\overline{T}_1}^2}-\frac{c_T^2}{M_{\overline{T}_2}^2} - \frac{s_T^2}{M_T^2}\right)\,G_{AB} G^\dagger_{CD}\,\nl & & \times\, \varepsilon^{\alpha \beta \gamma}\, \left(u{^{CT}_{\alpha A}} \,\cc\,d_{\beta B}^C\right)\, \left(\nu_C^\dagger\,\cc\, d_{\gamma D}^*\right)\, \nonumber \\
& + & \frac{8}{3}s_T c_T \left(\frac{1}{M{^2_T}}-\frac{1}{M_{\overline{T}_2}^2} \right) G_{AB} G_{CD}\,\nl
& & \times\,\varepsilon^{\alpha \beta \gamma}\, \left(u{^{CT}_{\alpha A}}\,\cc\,d^C_{\beta B}\,\right)\,\left( e{^{CT}_C}\,\cc\,{u^C_{\gamma D}}\right) \nonumber \\
& -& \frac{8}{3} s_{\cal T} c_{\cal T} \left(\frac{1}{M_{\cal T}^2}-\frac{1}{M_{\overline{\cal T}}^2} \right) G_{AB} G_{CD}\,\nl 
& & \times\,\varepsilon^{\alpha \beta \gamma}\, \left(u{^{CT}_{\alpha A}}\,\cc\,u^C_{\beta B}\right)\,\left(e{^{CT}_C}\,\cc\,{d^C_{\gamma D}}\right)\, \nonumber \\
& -& \frac{8}{3} s_{\mathbb{T}} c_{\mathbb{T}} \left(\frac{1}{M_\mathbb{T}^2}-\frac{1}{M_{\overline{\mathbb{T}}}^2} \right) G_{AB} G_{CD}\,\nl
& & \times\,\varepsilon_{\alpha \beta \gamma}\, \varepsilon_{bc} \varepsilon_{da}\,\left(q_{ A}^{\alpha\,a T}\,\cc\,l^b_B\right)\, \left(q_{ C}^{\beta\,d T}\,\cc\, q^{\gamma\,c}_{ D}\right) \hc\eeqa
Here, $\theta_T$, $\theta_{\cal T}$ and the angles \(\theta_{\mathbb{T}}\) parametrise the mixing between \(T\)-\(\overline{T}\), \({\cal T}\)-\(\overline{\cal T}\), and \(\mathbb{T}\)-\(\overline{\mathbb{T}}\), respectively, similar to the method described in Eq.~(\ref{eq:c3:mixing_angle}). Since \(T\)  couples solely to leptons and quarks, its contribution to nucleon decay occurs exclusively through the mixing term.

Using the Fierz rearrangement for two-component Weyl spinors~\cite{Dreiner:2008tw},
\beqa \label{eq:c3:fierz}
\left(\psi_1^T \,\cc\, \psi_2 \right) \left(\psi_3^T \,\cc\, \psi_4 \right) \eq - \left(\psi_1^T\,\cc\,\psi_3 \right) \left(\psi_4^T\,\cc\,\psi_2 \right)\nl \mi \left(\psi_1^T\,\cc\,\psi_4 \right) \left(\psi_2^T\,\cc\, \psi_3 \right)\,,\eeqa
we amend the operator given in the eighth line of Eq.~\eqref{eq:c3:op_120_0} as,
\beqa \label{eq:c3:3op}
 & & G_{AB} G_{CD}\,\varepsilon^{\alpha \beta \gamma}\, \left(u{^{CT}_{\alpha A}}\,\cc\,\, u^C_{\beta B}\right)\, \left(e{^{CT}_C}\,\cc\,d^C_{\gamma D}\right)\nl \eq 2 G_{AD} G_{CB}\,\,\varepsilon^{\alpha \beta \gamma}\, \left(u{^{CT}_{\alpha A}}\,\cc\,d^C_{\beta B}\right)\, \left(e{^{CT}_C}\,\cc\,u^C_{\gamma D}\right)\,,\eeqa
where we have also used the antisymmetricity of $G$. Further, using $\varepsilon_{bc} \varepsilon_{da} = \delta_{bd} \delta_{ca} - \delta_{ba} \delta_{cd}$ and Eq.~\eqref{eq:c3:fierz}, the term given in Eq.~\eqref{eq:c3:op_120_0}  can be reduced as follows;
\beqa \label{eq:c3:4op}
& & G_{AB} G_{CD} \varepsilon_{\alpha \beta \gamma}\, \varepsilon_{bc} \varepsilon_{da}\,\left(q_{ A}^{\alpha\,a T}\,\cc\,l^b_B\right)\, \left(q_{ C}^{\beta\,d T}\,\cc\, q^{\gamma\,c}_{ D}\right)\nonumber \\
&=& - \left(G_{AB} G_{CD} - 2 G_{AD} G_{CB}\right)\, \varepsilon_{\alpha \beta \gamma}\, \left(u_{ A}^{\alpha\,T}\,\cc\,d_{ B}^{\beta}\right)\, \left(e_C^T\,\cc\,u^{\gamma}_{ D}\right) \nonumber \\
&-& \left(G_{AB} G_{CD} - 2 G_{AC} G_{BD}\right)\, \varepsilon_{\alpha \beta \gamma}\, \left(u_{ A}^{\alpha\,T}\,\cc\,d_{ B}^{\beta}\right)\, \left(\nu_C^T\,\cc\,d^{\gamma}_{D}\right)\,. \eeqa

Using Eqs.~(\ref{eq:c3:3op} and \ref{eq:c3:4op}), we modify Eq.~\eqref{eq:c3:op_120_0} and transforming into the physical basis of fermions, we get
\beqa \label{eq:c3:op_120}
{\cal L}^{{ 120_{\hh}}}_{\rm eff} &=& g[u^C_A,d^C_B,e_C,u_D]\,\, \varepsilon^{\alpha \beta \gamma}\, \left(u{^{CT}_{\alpha A}}\,\cc\,d_{\beta B}^C\right)\, \left(e_C^\dagger\,\cc\, u_{\gamma D}^*\right)\, \nonumber \\
& + & g[u^C_A,d^C_B,\nu_C,d_D]\,\, \varepsilon^{\alpha \beta \gamma}\, \left(u{^{CT}_{\alpha A}}\,\cc\,d_{\beta B}^C\right)\, \left(\nu_C^\dagger\,\cc\,d_{\gamma D}^*\right)\, \nonumber\\
& + & g^\prime[u^C_A,d^C_B,e^C_C,u^C_D]\,\, \varepsilon^{\alpha \beta \gamma}\, \left(u{^{CT}_{\alpha A}}\,\cc\,d_{\beta B}^C\right)\, \left(e{^{CT}_C}\,\cc\,u^C_{\gamma D}\right)\,\nonumber \\
& + & g^\prime[u_A,d_B,e_C,u_D]\,\, \varepsilon_{\alpha \beta \gamma}\, \left(u_{A}^{\alpha\,T}\,\cc\,d_{ B}^{\beta}\right)\, \left(e_C^T\,\cc\,u_{ D}^{\gamma}\right)\, \nonumber \\
& + & g^\prime[u_A,d_B,\nu_C,d_D]\,\, \varepsilon_{\alpha \beta \gamma}\,\left( u_{A}^{\alpha\,T}\,\cc\,d_{ B}^{\beta}\right)\, \left(\nu_C^T\,\cc\,d_{D}^{\gamma}\right)\,\, \hc\,.\nl\eeqa
The strengths of these operators in terms of different unitary matrices, along with the Yukawa coupling, are given below;
\beqa \label{eq:c3:coff_op_120}
g[u^C_A,d^C_B,e_C,u_D] &=& \frac{4}{3}\left(\frac{1}{M_{\overline{T}_1}^2}-\frac{c_T^2}{M_{\overline{T}_2}^2}- \frac{s_T^2}{M_T^2}\right)\, \left(U_{u^C}^T G U_{d^C}\right){_{AB}}\, \left(U_{e}^\dagger G^\dagger U_{u}^* \right){_{CD}}\,, \nonumber \\
g[u^C_A,d^C_B,\nu_C,d_D] &=& -\frac{4}{3}\left(\frac{1}{M_{\overline{T}_1}^2}-\frac{c_T^2}{M_{\overline{T}_2}^2}- \frac{s_T^2}{M_T^2}\right)\, \left(U_{u^C}^T G U_{d^C} \right){_{AB}}\,\left(U_{\nu}^\dagger G^\dagger U_{d}^*\right){_{CD}}\,,\nonumber \\
g^\prime[u^C_A,d^C_B,e^C_C,u^C_D] &=& \frac{8}{3} s_T c_T \left(\frac{1}{M_T^2}-\frac{1}{M_{\overline{T}_2}^2} \right)\, \left(U_{u^C}^T G U_{d^C} \right){_{AB}}\,\left(U_{e^C}^T G U_{u^C}\right){_{CD}}\, \nonumber \\
& - & \frac{16}{3} s_{\cal T} c_{\cal T} \left(\frac{1}{M_{\cal T}^2}-\frac{1}{M_{\overline{\cal T}}^2} \right) \left(U_{u^C}^T G U_{u^C} \right)_{AD}\,\left(U_{e^c}^T G U_{d^C}\right)_{CB}\,, \nonumber \\
g^\prime[u_A,d_B,e_C,u_D] &=& \frac{8}{3} s_{\mathbb{T}} c_{\mathbb{T}} \left(\frac{1}{M_\mathbb{T}^2}-\frac{1}{M_{\overline{\mathbb{T}}}^2} \right) \Big[\left(U_u^T G U_d \right)_{AB}\,\left(U_e^T G U_u\right)_{CD} \big. \nonumber \\ 
&-& \Big. 2 \left(U_u^T G U_u \right)_{AD}\,\left(U_e^T G U_d\right)_{CB}\Big]\,, \nonumber \\
g^\prime[u_A,d_B,\nu_C,d_D] &=& \frac{8}{3} s_{\mathbb{T}} c_{\mathbb{T}} \left(\frac{1}{M_\mathbb{T}^2}-\frac{1}{M_{\overline{\mathbb{T}}}^2} \right) \Big[\left(U_u^T G U_d \right)_{AB}\,\left(U_\nu^T G U_d\right)_{CD} \big. \nonumber \\ 
&-& \Big. 2 \left(U_u^T G U_\nu \right)_{AC}\,\left(U_d^T G U_d\right)_{BD}\Big],\, \eeqa
the obtained strengths of the operators, $i.e.$ $g^{\prime} $, are very different from $h^\prime$ due to additional contributions supplied by the new scalar fields residing in $120_{\hh}$.

\subsection{Effective operators from $\overline{126}_{\hh}$}
\label{ss:c3:op126}
At last, we conclude this section by giving effective operators capable of destabilising a proton originating from the integration of scalars residing in $\overline{126}_{\hh}$. Unlike the situation with \(10_{\hh}\), the colour triplets and  complex conjugate of anti-triplets in \(\overline{126}_{\hh}\) do not mix due to the \(SO(10)\) gauge symmetry, which forbids \((\overline{126}_{\hh})^2\)~\cite{Slansky:1981yr}. However, the triplets within \(\overline{5}_{\hh}\) and \(50_{\hh}\) from \(\overline{126}_{\hh}\) generally can mix, though such mixing depends on gauge-invariant interactions with other fields and is thus model-specific. Additionally, as evident from Tab.~(\ref{tab:c3:scalarsummary}), \({\cal T}^\alpha\) and \(\overline{\mathbb{T}}^c_{a \alpha}\) possess only leptoquark couplings. As a result, the effective operators critical for nucleon decay are simplified in the case of \(\overline{126}_{\hh}\). This simplification occurs upon integrating the triplets and anti-triplets from the Eqs.~(\ref{eq:c3:126/5}, \ref{eq:c3:126/45:2}, and \ref{eq:c3:126/50}), we find 
\beqa \label{eq:c3:op_126_0}
{\cal L}^{\overline{126}_{\hh}}_{\rm eff} &=& \frac{2}{45}\left(\frac{1}{M_{T_1}^2}-\frac{2}{M_{T_2}^2}\right)\,F_{AB} F^\dagger_{CD}\,\, \varepsilon_{\alpha \beta \gamma}\, \left(u_{A}^{\alpha\,T}\,\cc\, d_{ B}^{\beta}\right)\, \left({e^C_C}^\dagger\,\cc\,{u^C_{\gamma D}}^*\right) \nonumber \\
& - &\frac{2}{15}\frac{1}{M_{\overline{T}}^2}\,F_{AB} F^\dagger_{CD}\,\, \varepsilon^{\alpha \beta \gamma}\, \left(u{^{CT}_{\alpha A}}\,\cc\,d_{\beta B}^C\right)\, \left(e_C^\dagger\,\cc\,u_{\gamma D}^* \right)\,\nl
\ad \frac{2}{15}\frac{1}{M_{\overline{T}}^2}\,F_{AB} F^\dagger_{CD}\,\, \varepsilon^{\alpha \beta \gamma}\, \left(u{^{CT}_{\alpha A}}\,\cc\,d_{\beta B}^C\right)\, \left(\nu_C^\dagger C^{-1} d_{\gamma D}^*\right) \hc,\, \eeqa
where $M_{T_1}$, $M_{T_2}$, and $M_{\overline{T}}$ are the masses of triplets $T_1$, $T_{2}$ and anti-triplet $\overline{T}$, respectively. 

In the physical basis, the above Eq.~\eqref{eq:c3:op_126_0} can be manipulated as follows:
\beqa \label{eq:c3:op_126}
{\cal L}^{\overline{ 126}_{\hh}}_{\rm eff} &=& f[u_A,d_B,e^C_C,u^C_D]\,\, \varepsilon_{\alpha \beta \gamma}\, \left(u_{A}^{\alpha\,T}\,\cc\,d_{ B}^{\beta}\right)\,\left( {e^C_C}^\dagger\,\cc\,u{^{C*}_{\gamma D}}\right) \nonumber \\
& + & f[u^C_A,d^C_B,e_C,u_D]\,\, \varepsilon^{\alpha \beta \gamma}\, \left(u{^{CT}_{\alpha A}}\,\cc\,d_{\beta B}^C\right)\, \left(e_C^\dagger\,\cc\,u_{\gamma D}^*\right)\, \nonumber \\
& + & f[u^C_A,d^C_B,\nu_C,d_D]\,\, \varepsilon^{\alpha \beta \gamma}\, \left(u{^{CT}_{\alpha A}}\,\cc\,d_{\beta B}^C\right)\, \left(\nu_C^\dagger\,\cc\,d_{\gamma D}^*\right)\, \hc\,,\nl\eeqa
provided,
\beqa \label{eq:c3:coff_op_126}
f[u_A,d_B,e^C_C,u^C_D] &=& \frac{2}{45}\left(\frac{1}{M_{T_1}^2}-\frac{2}{M_{T_2}^2}\right) \left(U_u^T F U_d \right)_{AB}\,\left(U_{e^C}^\dagger F^\dagger U_{u^C}^*\right)_{CD}\,, \nonumber \\
f[u^C_A,d^C_B,e_C,u_D] &=& -\frac{2}{15}\frac{1}{M_{\overline{T}}^2} \left(U_{u^C}^T F U_{d^C}\right){_{AB}}\, \left(U_{e}^\dagger F^\dagger U_{u}^* \right){_{CD}}\,, \nonumber \\
f[u^C_A,d^C_B,\nu_C,d_D]&=&  \frac{2}{15}\frac{1}{M_{\overline{T}}^2} \left(U_{u^C}^T F U_{d^C} \right){_{AB}}\,\left(U_{\nu}^\dagger F^\dagger U_{d}^*\right){_{CD}}\,. \eeqa
The structure of these operators is similar to those derived from $10_{\hh}$, with the significant distinction being the relative factor between contributions mediated by colour triplets and anti-triplets.

The operators provided in Eqs.~(\ref{eq:c3:op_10}, \ref{eq:c3:op_120}, and \ref{eq:c3:op_126}) encapsulate $B$ and $L$ violating, yet $B-L$ conserving, baryon decays mediated by scalars from the $10_{\hh}$, $120_{\hh}$, and $\overline{126}_{\hh}$, respectively. This analysis includes the potential for mixing among various scalars within the same representations. Generally, scalar fields with the same SM charges but residing in different GUT representations may also mix. For instance, in models containing at least two of the aforementioned scalar irreps, various triplets and anti-triplets can intermix, with the lightest pair often being a specific linear combination of these. The contribution from this lightest pair turns out to be the most significant. The operators resulting from integrating out this pair are crucial in assessing proton decay. These operators generally have coefficients that are linear combinations of relevant $h$, $g$, and $f$ values, among others. However, the precise determination of this mixing relies not only on the specifications of the Yukawa sector but also on the complete scalar potential of the model, making this analysis highly dependent on the model. Nonetheless, the findings provided can be directly applied to compute proton decay in specific models where the mixing among various triplets and anti-triplets is deterministically known.

\section{Dimension-Seven Effective Operators}
\label{sec:c3:operators_d7}
In the previous section~(\ref{sec:c3:operators_d6}), we discussed the leading $D=6$ operators, which simultaneously violate $B$ and $L$ by equal amounts, thereby conserving $B-L$. As noted in section~(\ref{sec:c3:geninfo}), another category of operators that violate $B-L$ by two units emerges at $D=7$ ~\cite{Weinberg:1980bf,Weldon:1980gi}. These operators result from quartic couplings that involve SM singlet field $(\sigma)$, an electroweak doublet $(D\,\text{and}\,\overline{D})$, and two colour triplet fields. Utilising the fields listed in Tab.~(\ref{tab:c2:scalars}), the following combinations, which are invariant under \smg\, can be formulated for such quartic terms:
\beqa {\label{eq:c3:d7invariants}}
\sigma\, D^a\,T^\alpha\,\overline{\Delta}_{\alpha a}\, &,&~~\sigma\, \overline{D}_a\,\overline{T}_\alpha\,\Delta^{\alpha a}\,; \label{eq:c3:q1} \\
\sigma\, D^a\,\Theta_{\alpha \beta}\,\overline{\Omega}^{\alpha \beta}_a\, &,&~~\sigma\, \overline{D}_a\,\overline{\Theta}^{\alpha \beta}\,\Omega_{\alpha \beta}^a\,; \label{eq:c3:q2} \\
\sigma\, D^a\,\overline{\Theta}^{\alpha \beta}\,\Delta^{\gamma b}\,\epsilon_{\alpha \beta \gamma}\,\epsilon_{ab}\, &,&~~\sigma\, \overline{D}_a\,\Theta_{\alpha \beta}\,\overline{\Delta}_{\gamma b}\,\epsilon^{\alpha \beta \gamma}\,\epsilon^{ab}\,; \label{eq:c3:q3} \\
\sigma\, D^a\,\overline{\Delta}_{\alpha b}\,\mathbb{T}^{\alpha b}_a\, &,&~~\sigma\, \overline{D}_a\,\Delta^{\alpha b}\,\overline{\mathbb{T}}_{\alpha b}^a\,. \label{eq:c3:q4} \eeqa
When the field $\sigma$ acquires a $vev$, it breaks the $B-L$ symmetry. The above-mentioned quartic terms lead to the formation of $D=7$ operators involving four fermions and a Higgs doublet. Following electroweak symmetry breaking, these interactions produce effective four-fermion $B-L = 2$ violating operators~\cite{Babu:2012iv,Babu:2012vb}.

The quartic terms given in Eqs.~(\ref{eq:c3:q1}$-$\ref{eq:c3:q4}) could originate from various $SO(10)$ invariant combinations of scalar irreps, such as  $(\overline{126}_{\hh})^2\, ({120}_{\hh})^2$ and $(\overline{126}_{\hh})^4$. At least one $\overline{126}_{\hh}$ (or ${126}_{\hh}$) is necessary in these terms as a source of $\sigma$, while the electroweak doublets may stem from ${10}_{\hh}$, $\overline{126}_{\hh}$, or ${120}_{\hh}$. The triplets that appear in Eqs.~(\ref{eq:c3:q1}$-$\ref{eq:c3:q4}) could stem from ${10}_{\hh}$,  ${120}_{\hh}$, and $\overline{126}_{\hh}$ or even scalars like ${126}_{\hh}$, which do not directly participate in the Yukawa sector. While the $126_{\hh}$ do not directly interact with the $\fs$-plet, they can still contribute to nucleon decay through their interactions with scalar sub-multiplets in ${10}_{\hh}$, $\overline{126}_{\hh}$, and ${120}_{\hh}$. Therefore, pinpointing the exact operators in this context demands a full specification of the scalar sector of the model. As we remain ignorant of the complete model, our subsequent analysis will proceed under the assumption that the colour triplet fields in Eqs.~(\ref{eq:c3:q1}$-$\ref{eq:c3:q4}) originate from either ${10}_{\hh}$, ${120}_{\hh}$ or $\overline{126}_{\hh}$,.

\subsection{Effective Operators From $ 10_{\hh}$}
The formation of $D=7$ operators requires at least two distinct types of colour triplets, as evident from Eqs.~(\ref{eq:c3:q1}-\ref{eq:c3:q4}). Given that ${10}_{\hh}$ contains only one type of colour triplet, it alone cannot generate $B+L$ conserving operators at the leading order.

\subsection{Effective Operators From $120_{\hh}$}
In contrast to ${10}_{\hh}$, ${120}_H$ is capable of generating $D=7$ operators due to the allowed mixing between various scalar sub-multiplets and their conjugates. Using Eqs.~(\ref{eq:c3:q1}$-$\ref{eq:c3:q4}) and utilising the evaluated couplings for different triplet fields discussed in section (\ref{sec:c3:couplings}), we derive the following leading order operators;
\beqa \label{eq:c3:op_120_0_d7}
{\cal L}^{120_{\hh}}_{\rm eff} &=& -\frac{8 \lambda v_\sigma}{3}\left[\left( \frac{1}{M{^2_{\Delta}}\,M{^2_{\overline{T}_1}}}+\frac{c_\Delta c_T}{M_\Delta^2 M^2_{\overline{T}_2}}\right) + \frac{s_\Delta s_T}{M_{\overline{\Delta}}^2 M_{T}^2}\right]\,G_{AB}^*G_{CD}^*\,\nonumber\\
& & \hspace{10 pt}\,\times\, \,\varepsilon^{\alpha \beta \gamma}\,\epsilon^{ab}\,\overline{D}_a \,\left(u^{C \dagger}_{\alpha A}\,\cc\,{d^C_{\beta B}}^*\right)\,\left(l^\dagger_{bC}\,\cc\, {d^{C}_{\gamma D}}^*\right)\, \nonumber \\
& - & \frac{4\sqrt{2} \lambda v_\sigma}{3}\left(\frac{c_\Theta c_\Omega}{M_{\overline{\Theta}}^2 M_\Omega^2} + \frac{s_\Theta s_\Omega}{M_\Theta^2 M_{\overline{\Omega}}^2} \right)\,G_{AB}^*G_{CD}^*\nl 
& &  \hspace{10 pt}\,\times\, \,\varepsilon^{\alpha \beta \gamma}\, \varepsilon^{ab}\,\overline{D}_a\,\left(d^{C \dagger}_{\alpha A}\,\cc\,{d^C_{\beta B}}^*\right)\, \left(l^\dagger_{b C}\,\cc\,{u^C_{\gamma D}}^*\right)\, \nonumber \\
& - & \frac{4\sqrt{2} \lambda v_\sigma}{3}\left(\frac{c_\Theta s_\Omega}{M_{\overline{\Theta}}^2 M_\Omega^2} - \frac{s_\Theta c_\Omega}{M_\Theta^2 M_{\overline{\Omega}}^2}\right)\, G_{AB}^*G_{CD}\, \nl 
& & \hspace{10pt}\, \times\,\varepsilon_{\alpha \beta \gamma}\,\overline{D}_a\, \left(d^{C \dagger}_{\alpha A}\,\cc\,{d^C_{\beta B}}^*\right)\, \left(e^{C T}_C\,\cc\,{q^{a\gamma}_{D}}\right)\, \nonumber \\
& + & \frac{8\sqrt{2} \lambda v_\sigma}{3} \left(\frac{c_\Theta c_\Delta}{M_{\overline{\Theta}}^2 M_\Delta^2} + \frac{s_\Theta s_\Delta}{M_{\Theta}^2 M_{\overline{\Delta}^2}}\right)\, G_{AB}^*G_{CD}^*\,
\nl
& & \hspace{10pt}\,\times \, \epsilon^{\alpha \beta \gamma}\,D^a\,\left(d^{C \dagger}_{\alpha A}\,\cc\,{d^C_{\beta B}}^*\right)\, \left(l^\dagger_{a C}\,\cc\,{d^C_{\gamma D}}^*\right)\, \nonumber \\
& - & \frac{8 \lambda v_\sigma}{3} \left(\frac{c_{\mathbb{T}} s_\Delta}{M_{\overline{\Delta}}^2 M_{\mathbb{T}}^2}- \frac{s_{\mathbb{T}} c_\Delta}{M_\Delta^2 M_{\overline{\mathbb{T}}}^2} \right)\, G_{AB}^*G_{CD}\nl
& & \hspace{10pt}\,\times\,
\varepsilon^{\alpha \beta \gamma}\, D^a\,\left(q^{\dagger}_{\alpha b A}\,\cc\,{q^*_{\beta a B}}\right)\, \left(l^{bT}_C\,\cc\,d^{C}_{\gamma D}\right)\, \hc. \nonumber \\
\eeqa
In the above expressions, Eq.~(\ref{eq:c3:op_120_0_d7}), \(c_\chi \equiv \cos \theta_\chi\), \(s_\chi \equiv \sin \theta_\chi\), where \(\theta_\chi\) represents the angle that denotes the mixing between the \(\chi\)-\(\overline{\chi}\) fields, similar to the one previously defined in Eqs.~(\ref{eq:c3:T_redf} and \ref{eq:c3:mixing_angle}). Additionally,  $\lambda$ is a quartic coupling, and $v_\sigma$ is the $vev$ that breaks $B-L$. The operator in the first line of Eq.~\eqref{eq:c3:op_120_0_d7} corresponds to \(\tilde{O}_1\), the second to \(\tilde{O}_2\), the third to \(\tilde{O}_5\), the fourth to \(\tilde{O}_6\), and the last to \(\tilde{O}_4\) as listed in~\cite{Babu:2012vb}. It is important to note that the operator \(\tilde{O}_3\) does not emerge because the coupling of \(\mathbb{T}\) with a pair of quark doublets is prohibited by the flavour anti-symmetry of \(G\).

Using the Fierz rearrangement as outlined in Eq.~\eqref{eq:c3:fierz}, the operator described in the fourth line of Eq.~\eqref{eq:c3:op_120_0_d7} can be reformulated in terms of the one in the first line. Once the electroweak doublets acquire $vev$, the resulting four-fermion operator from Eq.~\eqref{eq:c3:op_120_0_d7} in the physical basis can be parameterised as follows:
\beqa \label{eq:c3:op_120_d7}
{\cal L}^{120_{\hh}}_{\rm eff} &=&\tilde{g}[u^C_A,d^C_A,\nu_C,d^C_D]\,\varepsilon^{\alpha \beta \gamma}\, \left(u^{C T}_{\alpha A}\,\cc\,{d^C_{\beta B}}\right)\,\left(\nu_C^T\,\cc\, d^{C}_{\gamma D}\right)\, \nonumber \\
& + & \tilde{g}^\prime[d^C_A,d^C_B,e^C_C,d_D]\,\varepsilon_{\alpha \beta \gamma}\, \left(d^{C \dagger}_{\alpha A}\,\cc\,d^{C *}_{\beta B}\right)\, \left(e^{C T}_C\,C^{-1}\,d^{\gamma}_{D}\right)\, \nonumber \\
& + & \tilde{g}[d^C_A,d^C_B,e_C,d^C_D]\,\varepsilon^{\alpha \beta \gamma}\, \left(d^{C T}_{\alpha A}\,\cc\,d^C_{\beta B}\right)\, \left(e^T_C\,\cc\,d^C_{\gamma D}\right)\, \nonumber \\
& + & \tilde{g}^\prime[u_A,d_B,\nu_C,d^C_D]\,\varepsilon^{\alpha \beta \gamma}\, \left(u^{\dagger}_{\alpha A}\,\cc\,{d^*_{\beta B}}\right)\, \left(\nu^{T}_C\,\cc\,d^{C}_{\gamma D}\right)\, \nonumber \\
& + & \tilde{g}^\prime[d_A,d_B,e_C,d^C_D]\,\varepsilon^{\alpha \beta \gamma}\, \left(d^{\dagger}_{\alpha A}\,\cc\,{d^*_{\beta B}}\right)\, \left(e^{T}_C\,\cc\,d^{C}_{\gamma D}\right)\,\hc,\nl\eeqa
such that,
\beqa \label{eq:c3:coff_op_120_d7}
 \tilde{g}[u^C_A,d^C_B,\nu_C,d^C_D] &=&\frac{8 \lambda v_\sigma v_{\overline{D}} }{3}\left[\left( \frac{1}{M{^2_{\Delta}}\,M{^2_{\overline{T}_1}}}+\frac{c_\Delta c_T}{M_\Delta^2 M_{\overline{T}_2}^2}\right) + \frac{s_\Delta s_T}{M_{\overline{\Delta}}^2 M_{T}^2}\right]\nl
 & & \hspace{14pt}\,\times\,
 \left(U_{u^C}^T G U_{d^C} \right)_{AB}\,\left(U_\nu^T G U_{d^C}\right)_{CD} \nonumber \\
 & + &   \frac{8\sqrt{2} \lambda v_\sigma v_{\overline{D}} }{3}\left(\frac{c_\Theta c_\Omega}{M_{\overline{\Theta}}^2 M_\Omega^2} + \frac{s_\Theta s_\Omega}{M_\Theta^2 M_{\overline{\Omega}}^2} \right)\,\nl
 & &\hspace{14pt}\,\times\,\left(U{_{\nu}^T} G U_{u^C} \right)_{CA}\,\left(U{_{d^C}^T} G U_{d^C} \right)_{BD}\,,\nonumber\\
\tilde{g}^\prime[d^C_A,d^C_B,e^C_C,d_D] & = &- \frac{4\sqrt{2} \lambda v_\sigma v_{\overline{D}} }{3}\left(\frac{c_\Theta s_\Omega}{M_{\overline{\Theta}}^2 M_\Omega^2}  - \frac{s_\Theta c_\Omega}{M_\Theta^2 M_{\overline{\Omega}}^2}  \right)\,\nl
& & \hspace{14pt}\,\times\,\left(U_{d^C}^\dagger G^* U_{d^C}^*\right)_{AB} \,\left(U_{e^C}^T G U_{d}\right)_{CD} \,, \nonumber \\
\tilde{g}[d^C_A,d^C_B,e_C,d^C_D]& = & \frac{8\sqrt{2} \lambda v_\sigma v_D}{3} \left(\frac{c_\Theta c_\Delta}{M_{\overline{\Theta}}^2 M_\Delta^2} + \frac{s_\Theta s_\Delta}{M_{\Theta}^2 M_{\overline{\Delta}^2}} \right)\,\nl
& & \hspace{14pt}\,\times\,\left(U_{d^C}^T G U_{d^C} \right)_{AB}\,\left(U{_ e^T} G U_{d^C}\right)_{CD}\,, \nonumber \\ 
\tilde{g}^\prime[u_A,d_B,\nu_C,d^C_D]& =& -\frac{8 \lambda v_\sigma v_D}{3} \left(\frac{c_{\mathbb{T}} s_\Delta}{M_{\overline{\Delta}}^2 M_{\mathbb{T}}^2} - \frac{s_{\mathbb{T}} c_\Delta}{M_\Delta^2 M_{\overline{\mathbb{T}}}^2} \right)\,\nl
& & \hspace{14pt}\,\times\,\left(U_u^\dagger G^* U_d^* \right)_{AB}\, \left(U_\nu^T G U_{d^C} \right)_{CD} \,,\nonumber \\ 
\tilde{g}^\prime[d_A,d_B,e_C,d^C_D] & = & -\frac{8 \lambda v_\sigma v_D}{3} \left(\frac{c_{\mathbb{T}} s_\Delta}{M_{\overline{\Delta}}^2 M_{\mathbb{T}}^2} - \frac{s_{\mathbb{T}} c_\Delta}{M_\Delta^2 M_{\overline{\mathbb{T}}}^2} \right)\,\nl
& & \hspace{14pt}\,\times\,\left(U_d^\dagger G^* U_d^* \right)_{AB}\, \left(U_e^T G U_{d^C} \right)_{CD}\,. \eeqa
In the above Eq.~\eqref{eq:c3:coff_op_120_d7}, the primed and unprimed coefficients are defined such that all the \(\tilde{g}^\prime = 0\), indicating that there is no mixing between the different scalars and their conjugate partners.

\subsection{Effective Operators From $\overline{126}_{\hh}$}
As indicated in Tab.~(\ref{tab:c2:scalars}), $\overline{126}_{\hh}$ does not include the fields $\overline{\Theta}^{\alpha \beta}$ and $\mathbb{T}^{a \alpha}_b$. Among the other fields that contribute to generating $D=7$ operators, $\Theta_{\alpha \beta}$ and $\overline{\Delta}_{\alpha a}$ reside in $\overline{10}_{\hh}$ of $SU(5)$, which interact exclusively with the RH neutrinos, as detailed in Eq.~\eqref{eq:c3:126-10-10}. Additionally, $\Delta^{\alpha a}$, $\overline{\Omega}^{\alpha \beta}_a$, and $\overline{\mathbb{T}}^a_{\alpha b}$ exhibits only in lepto-quark couplings. Thus, only the second term in Eq.~\eqref{eq:c3:q1} can induce the desired operator. By applying Eqs.~(\ref{eq:c3:126/15} and \ref{eq:c3:126/45:2}) and integrating out $\Delta^{\alpha a}$ and $\overline{T}_\alpha$, we derive the following result:
\beqa \label{eq:c3:op_126_0_d7}
{\cal L}^{\overline{ 126}_{\hh}}_{\rm eff} &=& \frac{4 \lambda v_\sigma}{15 M_\Delta^2 M_{\overline{T}}^2}\,F_{AB}^*F_{CD}^*\,\varepsilon^{\alpha \beta \gamma}\, \varepsilon^{ab}\overline{D}_a\,\left(u^{C \dagger}_{\alpha A}\,\cc\,{d^C_{\beta B}}^*\right)\, \left(l_{b C}^\dagger\,C^{-1}\,{d^C_{\gamma D}}^*\right)\,\nl\hcn,\eeqa
The resulting operator can be identified with operator \(\tilde{O}_1\) as given in~\cite{Babu:2012vb}\footnote{It should be noted that we have assumed the absence of \({126}_{\hh}\), which forbids \(T\)-\(\overline{T}\) mixing. If such mixing were permitted, one might also encounter an operator similar to \(\tilde{O}_3\) as described in~\cite{Babu:2012vb}}. After the electroweak symmetry breaking, the above operator in Eq.~\eqref{eq:c3:op_126_0_d7}, reduces to $D=6$ operator, containing three quarks and a neutrino field. Transforming to the physical basis, the operator defined in Eq.~\eqref{eq:c3:op_126_0_d7} converted to the following form.
\beqa \label{eq:c3:op_126_d7}
{\cal L}^{\overline{ 126}_{\hh}}_{\rm eff} &=& \tilde{f}[u^C_A,d^C_B,\nu_C,d^C_D]\,\varepsilon^{\alpha \beta \gamma}\,\left(u^{C T}_{\alpha A}\,\cc\,{d^C_{\beta B}}\right)\, \left(\nu_C^T\,\cc\,{d^C_{\gamma D}}\right),\,\hc \nl\,\eeqa
provided, 
\beqa \label{eq:c3:coff_op_126_d7}
 \tilde{f}[u^C_A,d^C_B,\nu_C,d^C_D] &=& -\frac{4 \lambda v_\sigma v_{\overline{D}}}{15 M_\Delta^2 M_{\overline{T}}^2}\, \left(U_{u^C}^T F U_{d^C} \right)_{AB}\,\left(U_\nu^T F U_{d^C}\right)_{CD}\,, \eeqa
 where $v_{\overline{D}}$ is a $vev$ of $\overline{D}$ residing in $\overline{126}_{\hh}$ as described earlier.

In summary, the operators written in Eqs.~(\ref{eq:c3:op_120_d7} and \ref{eq:c3:op_126_d7}) quantify the $B$, $L$, and $B-L$ violating baryon decays mediated by scalars residing in $120_{\hh}$ and $\overline{126}_{\hh}$ respectively, at the tree level. Analogous to the previous section~(\ref{sec:c3:operators_d6}), we have also considered the potential for mixing among various scalars from these representations. It is important to note that the masses of $T$, $\overline{T}$, $\mathbb{T}$, and $\overline{\mathbb{T}}$ are already constrained by the leading order $d=6$ operators discussed previously. Consequently, the $B-L$ violating nucleon decays could provide lower bounds on the masses of $\Delta$ and $\overline{\Delta}$ in $\overline{126}_{\hh}$ as well as $\Delta$, $\overline{\Delta}$, $\Theta$, $\overline{\Theta}$, $\Omega$, and $\overline{\Omega}$ in ${120}_{\hh}$, depending on their interactions with quarks and leptons and the scale at which $B-L$ is broken.

\section{Nucleon Partial Decay Widths}
\label{sec:c3:decay_widths}
In the previous sections~(\ref{sec:c3:operators_d6} and \ref{sec:c3:operators_d7}), we have derived the most general $B-L$ conserving and violating operators able to induce nucleon decay. Further, we compute the proton's two body leading decay modes using these operators.

\subsection{$B-L$ conserving decays}
\label{ss:c3:B-Lcd}
The $B+L$ violating mass dimension six effective operators listed previously in Eqs.~(\ref{eq:c3:op_10}, \ref{eq:c3:op_120}, and \ref{eq:c3:op_126}) can be parametrised in terms of the following six independent operators respecting \smg\, symmetry:
\beqa \label{eq:c3:gen_op}
{\cal L}_{\rm eff} & = & y[u_A,d_B,e^C_C,u^C_D]\,\, \varepsilon_{\alpha \beta \gamma}\, \left(u_{ A}^{\alpha\,T}\,\cc\,d_{B}^{\beta}\right)\, \left({e^C_C}^\dagger\,\cc\,{u^C_{\gamma D}}^*\right) \nonumber \\
& + & y[u^C_A,d^C_B,e_C,u_D]\,\, \varepsilon^{\alpha \beta \gamma}\, \left(u{^{CT}_{\alpha A}}\,\cc\,d_{\beta B}^C\right)\, \left(e_C^\dagger\,\cc\,u_{\gamma D}^*\right)\, \nonumber \\
& + & y[u^C_A,d^C_B,\nu_C,d_D]\,\, \varepsilon^{\alpha \beta \gamma}\, \left(u{^{CT}_{\alpha A}}\,\cc\,d_{\beta B}^C\right)\, \left(\nu_C^\dagger\,\cc\,d_{\gamma D}^*\right)\, \nonumber\\
& + & y^\prime[u^C_A,d^C_B,e^C_C,u^C_D]\,\, \varepsilon^{\alpha \beta \gamma}\, \left(u{^{CT}_{\alpha A}}\,\cc\, d^C_{\beta B}\right)\, \left(e{^{CT}_C}\,\cc\,u^C_{\gamma D}\right) \nonumber \\
& + & y^\prime[u_A,d_B,e_C,u_D]\,\, \varepsilon_{\alpha \beta \gamma}\, \left(u_{A}^{\alpha\,T}\,\cc\,d_{ B}^{\beta}\right)\, \left(e_C^T\,\cc\,u_{D}^{\gamma}\right)\, \nonumber \\
& + & y^\prime[u_A,d_B,\nu_C,d_D]\,\, \varepsilon_{\alpha \beta \gamma}\, \left(u_{ A}^{\alpha\,T}\,\cc\,d_{ B}^{\beta}\right)\, \left(\nu_C^T\,\cc\,d_{ D}^{\gamma}\right)\,\hc\,,\nl\eeqa
In cases where only a single GUT scalar representation is involved, \(y\) corresponds to \(h\), \(g\), or \(f\). If multiple scalar fields are taken into account, as outlined in the previous sections, \(y\) may represent a linear combination of \(h\), \(f\), and \(g\). These operators are consistent with the most general dimension six operators derived from effective field theory, as listed in ~\cite{PhysRevLett.43.1566,PhysRevLett.43.1571}. However, in this context, the coefficients of these operators can be explicitly calculated using the fundamental Yukawa couplings specified within a particular \so\, GUT model. 

The operators mentioned in Eq.~\eqref{eq:c3:gen_op} can be rewritten using the following transformation properties given in Eq.~\eqref{eq:c1:conjugation} and properties of charge conjugation matrix leads to,
\be \label{eq:c3:identity}
\psi^T\,\cc\,\chi = \overline{(\psi_L)^C}\,\chi_L\,,\hspace{0.5cm}\text{and}\hspace{0.5cm}~~{\psi^C}^\dagger\,\cc\,{\chi^C}^* = \overline{(\psi_R)^C}\,\chi_R\,,\ee
where $\overline{\psi^C} = \psi^T \cc$. Using the above identities and $\overline{\psi^C}\, \chi = \overline{\chi^C}\,\psi$, the baryon number violating operators  listed in Eq.~\eqref{eq:c3:gen_op} can be rewritten into the following form:
\beqa \label{eq:c3:gen_op_2}
{\cal L}_{\rm eff} & = & y[u_A,d_B,e^C_C,u^C_D]\,\, \varepsilon_{\alpha \beta \gamma}\, \left(\overline{(d^{\beta}_{ B L})^C}\, u^{\alpha}_{ A L}\right)\,\, \left(\overline{(u^{\gamma}_{ D R})^C}\, e_{C R}\right) \nonumber \\
& + & y^*[u^C_A,d^C_B,e_C,u_D]\,\, \varepsilon_{\alpha \beta \gamma}\, \left(\overline{(d^{\beta}_{ B R})^C}\, u^{\alpha}_{A R}\right)\,\, \left(\overline{(u^{\gamma}_{D L})^C}\, e_{C L}\right) \nonumber \\
& + & y^*[u^C_A,d^C_B,\nu_C,d_D]\,\, \varepsilon_{\alpha \beta \gamma}\,\left(\overline{(d^{\beta}_{B R})^C}\, u^{\alpha}_{A R}\right)\,\, \left(\overline{(d^{\gamma}_{D L})^C}\, \nu_{C L}\right) \nonumber \\
& + & y^{\prime *}[u^C_A,d^C_B,e^C_C,u^C_D]\,\, \varepsilon_{\alpha \beta \gamma}\, \left(\overline{(d^{\beta}_{ B R})^C}\, u^{\alpha}_{ A R}\right)\,\, \left(\overline{(u^{\gamma}_{ D R})^C}\, e_{C R}\right) \nonumber \\
& + & y^\prime[u_A,d_B,e_C,u_D]\,\, \varepsilon_{\alpha \beta \gamma}\, \left(\overline{(d^{\beta}_{B L})^C}\, u^{\alpha}_{A L}\right)\,\, \left(\overline{(u^{\gamma}_{ D L})^C}\, e_{C L}\right) \nonumber \\
& + & y^\prime[u_A,d_B,\nu_C,d_D]\,\, \varepsilon_{\alpha \beta \gamma}\,\left(\overline{(d^{\beta}_{B L})^C}\, u^{\alpha}_{A L}\right)\,\, \left(\overline{(d^{\gamma}_{D L})^C}\, \nu_{C L}\right).\,\nl \hcn\eeqa
where $y^*[A,B,C,D]\,=\, \left(y[A,B,C,D]\right)^*$ and the same for $y^\prime$.

In Eq.~\eqref{eq:c3:gen_op_2}, all quarks can be assigned a baryon number \(B=\frac{1}{3}\) and a lepton number \(L=0\). The conjugation operation inverts the baryon number, while a simultaneous $bar\, (\overline{\phantom{x}})$ operation nullifies the conjugation's effect. Similarly, all leptons are assigned a lepton number of \(1\) and a baryon number \(B=0\). Consequently, the operators specified in Eq.~\eqref{eq:c3:gen_op_2} exhibits \(B-L\,=\,0\) and \(B+L=2\). Moreover, the \(B+L\) violating operators mentioned in Eq.~\eqref{eq:c3:gen_op_2} are invariant under \(SU(3)_C \times U(1)_{\mathrm{EM}}\) gauge symmetry. As noted earlier, they preserve \(B-L\), in which a proton decays into an antilepton accompanied by a meson.

The operators given in the Eq.~\eqref{eq:c3:gen_op_2} involve three quarks and one lepton field. These quarks must be appropriately matched with the known hadronic states observed at low energies. For instance, a typical \({\mathcal{O}} = (uud)(e^+)\) contributing to the decay \(p \to e^+ \, \pi^0\), one would like to study the transition of quarks within a proton $(uud)$ to the final states (positron and pion). This would involve the computation of $\langle \pi^0 \big| uud \big| p \rangle$, which is typically called the hadronic matrix element and is a non-perturbative object. Various techniques have been devised to compute the hadronic matrix elements like i) Based on approximate spin-flavour symmetry of partons~\cite{Buras:1977yy,Jarlskog:1978uu,Goldman:1980ah}, ii) MIT bag model~\cite{Din:1979bz,Donoghue:1979pr,Goldman:1980ah}, iii) Chiral Perturbation Theory~\cite{Claudson:1981gh,Isgur:1982iz,Kaymakcalan:1983uc,Chadha:1983sj},  iv) Lattice QCD~\cite{Gavela:1988cp,JLQCD:1999dld}, and v) Light Cone Sum Rules~\cite{Haisch:2021hvj}.

The proton decay width expression for the mode $p\to e^{+}_j\,\pi^0$, calculated using chiral perturbation theory, is given in the expression below~\cite{Nath:2006ut}. The parameters involved in this expression can be inferred from the Appendix~(\ref{app:3}), which are taken from~\cite{JLQCD:1999dld}.
\beqa{\label{eq:c3:pepi}}
\Gamma[p \to e_j^+\pi^0] &=& \frac{(m_p^2 - m_{\pi^0}^2)^2}{32\, \pi\, m_p^3 f_\pi^2} A^2 \left( \frac{1+\tilde{D}+\tilde{F}}{\sqrt{2}}\right)^2\nl
& & \,\times\, \bigg(\left| \alpha\, y[u_1,d_1,e^C_i,u^C_1] + \beta\, y^{\prime *}[u^C_1,d^C_1,e^C_i,u^C_1] \right|^2 \Big. \nl
& + & \Big. \left| \alpha\, y^*[u^C_1,d^C_1,e_i,u_1] + \beta\, y^\prime[u_1,d_1,e_i,u_1]\right|^2 \bigg), 
\eeqa 
where, $e^+_{1}\,=\,e^{+}$ and $e^+_2\,=\,\mu^+$.

From Eq.~\eqref{eq:c3:pepi}, it is evident that the couplings relevant to proton decay into charged leptons include \(y[u_1, d_i, e^C_j, u^C_1]\), \(y[u^C_1, d^C_i, e_j, u_1]\), \(y^\prime[u^C_1, d^C_i, e^C_j, u^C_1]\), and \(y[u_1, d_i, e_j, u_1]\). For proton decay into neutrinos, the pertinent couplings are \(y[u_1, d_i, \nu_j, d_k]\) and \(y[u^C_1, d^C_i, \nu_j, d_k]\) (cf. Eq.~\eqref{eq:app:c3:decay_width}). We sum over all neutrino flavours while computing proton decaying into neutrinos, as the individual flavours are undetected. Further, the other leading two body proton decay modes have been relegated to the Appendix~(\ref{app:c3:decaywidths}).

 Additionally, the tree-level contribution mediated by ${\cal T}$ and $\overline{\cal T}$ (as evident from Eq.~\eqref{eq:c3:coff_op_120}) vanishes due to the anti-symmetric nature of $G$, as shown:
\beqa \label{eq:c3:cancel_G}
\left(U_{u^C}^T G U_{u^C} \right)_{11} = 0\,.\eeqa
These fields, however, can induce proton decay through dimension-six operators that appear at the loop level. Similar findings were reported in the context of $SU(5)$ GUTs in~\cite{Dorsner:2012nq}, where 1-loop diagrams involving additional $W$-boson exchange were also analysed. Consequently, in models incorporating ${ 120}_{\hh}$, proton decays through dimension-six operators at tree level are solely mediated by $T$, $\overline{T}$, and $\mathbb{T}$, $\overline{\mathbb{T}}$.

\subsection{$B-L$ violating decays}
{\label{ss:c3:B-Lvd}}
The $B-L$ violating or $B+L$ conserving decays of nucleons arise from the operators derived in Eqs.~(\ref{eq:c3:op_120_d7} and \ref{eq:c3:op_126_d7}). They can be further generalised as follows: 
\beqa \label{eq:c3:gen_op_d7}
{\cal L}_{\rm eff} &=& \tilde{y}[u^C_A,d^C_A,\nu_C,d^C_D]\,\varepsilon^{\alpha \beta \gamma}\, \left(u^{C T}_{\alpha A}\,\cc\,{d^C_{\beta B}}\right)\,\left(\nu_C^T\,\cc\, d^{C}_{\gamma D}\right)\, \nonumber \\
& + & \tilde{y}^\prime[u_A,d_B,\nu_C,d^C_D]\,\varepsilon^{\alpha \beta \gamma}\, \left(u^{\dagger}_{\alpha A}\,\cc\,{d^*_{\beta B}}\right)\, \left(\nu^{T}_C\,C^{-1}\,d^{C}_{\gamma D}\right)\, \nonumber \\
& + & \tilde{y}[d^C_A,d^C_B,e_C,d^C_D]\,\varepsilon^{\alpha \beta \gamma}\, \left(d^{C T}_{\alpha A}\,\cc\,d^C_{\beta B}\right)\, \left(e^T_C\,\cc\,d^C_{\gamma D}\right)\, \nonumber \\
& + & \tilde{y}^\prime[d_A,d_B,e_C,d^C_D]\,\varepsilon^{\alpha \beta \gamma}\, \left(d^{\dagger}_{\alpha A}\,\cc\,{d^*_{\beta B}}\right)\, \left(e^{T}_C\,\cc\,d^{C}_{\gamma D}\right)\, \nonumber \\
& + & \tilde{y}^\prime[d^C_A,d^C_B,e^C_C,d_D]\,\varepsilon_{\alpha \beta \gamma}\, \left(d^{C \dagger}_{\alpha A}\,\cc\,d^{C *}_{\beta B}\right)\, \left(e^{C T}_C\,\cc\,d^{\gamma}_{D}\right)\, \hc\,,\nl\eeqa
where $y = g$ or $f$. Using the identities for two-component Weyl spinors given in Eq.~\eqref{eq:c1:conjugation}, the operators written in Eq.~\eqref{eq:c3:gen_op_d7} can be rewritten in the following convenient form;  
\beqa \label{eq:c3:gen_op_2_d7}
{\cal L}_{\rm eff} &=& \tilde{y}^*[u^C_A,d^C_A,\nu_C,d^C_D]\,\varepsilon_{\alpha \beta \gamma}\, \left(\overline{(d^{\beta}_{ B R})^C}\,u^{\alpha}_{ A R}\right)\,\, \left(\overline{\nu_{C\,L}}\,d^{\gamma}_{ D R}\right)\, \nonumber \\
& + & \tilde{y}^{\prime *}[u_A,d_B,\nu_C,d^C_D]\,\varepsilon_{\alpha \beta \gamma}\, \left( \overline{(d^{\beta}_{B L})^C}\,u^{\alpha}_{A L}\right)\,\, \left(\overline{\nu_{C\,L}}\,d^{\gamma}_{ D R}\right)\, \nonumber \\
& + & \tilde{y}^*[d^C_A,d^C_B,e_C,d^C_D]\,\varepsilon_{\alpha \beta \gamma}\, \left(\overline{(d^{\beta}_{ B R})^C}\,d^{\alpha}_{A R}\right)\,\, \left(\overline{e_{CL}}\,d^{\gamma}_{ D R}\right)\, \nonumber \\
& + & \tilde{y}^{\prime *}[d_A,d_B,e_C,d^C_D]\,\varepsilon_{\alpha \beta \gamma}\, \left(\overline{(d^{\beta}_{ B L})^C}\,d^{\alpha}_{ A L}\right)\,\, \left(\overline{e_{C L}}\,d^{\gamma}_{ D R}\right)\, \nonumber \\
& + & \tilde{y}^\prime[d^C_A,d^C_B,e^C_C,d_D]\,\varepsilon_{\alpha \beta \gamma}\, \left(\overline{(d^{\beta}_{ B R})^C}\,d^{\alpha}_{ A R}\right)\,\, \left(\overline{e_{CR}}\,d^{\gamma}_{ D L}\right)\,\hc.\nl\eeqa

Analogous to the previous subsection~({\ref{ss:c3:B-Lcd}}) in which we have derived $B+L$ violating decays, we can assign similar baryon and lepton numbers to the quarks and lepton and obtain that these sets of operators have $B-L=2$ or $B+L=0$. Another evident difference from the previous set of operators in Eq.~\eqref{eq:c3:gen_op_2} is that the outgoing charged fermion is a lepton. Thus, the operators obtained above violate $B-L$ by two units and lead to processes in which a nucleon decays into lepton along with a meson. Moreover, these operators also respect only $SU(3)_C\,\times\,U(1)_{\mathrm{Y}}$ gauge symmetry.

The first two operators in Eq.~\eqref{eq:c3:gen_op_2_d7} induce the proton decay through channels $p \to \nu\, \pi^+$ and $p \to \nu\, K^+$ and their decay widths are shown in the following expressions;
\beqa \label{eq:c3:decay_width_d7_1}
\Gamma[p \to \nu \pi^+] &=& \frac{(m_p^2 - m_{\pi^+}^2)^2}{32\, \pi\, m_p^3 f_\pi^2} A^2\,\left(1+\tilde{D}+\tilde{F} \right)^2\nl  & & \,\times\,\sum_{i=1}^3 \left| \alpha\, \tilde{y}^*[u^C_1,d^C_1,\nu_i, d^C_1] \beta\, \tilde{y}^{\prime *}[u_1,d_1,\nu_i, d^C_1] \right|^2\,, \nonumber\\
\Gamma[p \to \nu K^+] &=& \frac{(m_p^2 - m_{K^+}^2)^2}{32\, \pi\, m_p^3 f_\pi^2}\, A^2\,\nl 
& & \,\times\,\sum_{i=1}^3 \Big| \frac{2\tilde{D}}{3} \frac{m_p}{m_B} \tilde{C}^\nu_{L1i} + \left(1+\frac{\tilde{D}+3\tilde{F}}{3} \frac{m_p}{m_B}\right) \tilde{C}^\nu_{L2i} \Big|^2 \,,\eeqa 
together with,
\beqa \label{eq:c3:C_dw_d7}
\tilde{C}^\nu_{L1i} &=& \alpha\, \tilde{y}^*[u_1^C,d_2^C,\nu_i,d^C_1] + \beta\, \tilde{y}^{\prime *}[u_1,d_2,\nu_i,d^C_1]\,,\nonumber \\
\tilde{C}^\nu_{L2i} &=& \alpha\, \tilde{y}^*[u_1^C,d_1^C,\nu_i,d^C_2] + \beta\, \tilde{y}^{\prime *}[u_1,d_1,\nu_i,d^C_2]\,. \eeqa
Even in this case, we sum over all flavours of neutrinos to compute the decay width of the proton. 
The other three operators listed in Eq.~\eqref{eq:c3:gen_op_2_d7} do not contribute to proton decay but do lead to $B-L$ violating neutron decays. The decay width for this process is given as follows~\cite{Nath:2006ut}: 
\beqa \label{eq:c3:decay_width_d7_2}
\Gamma[n \to e_i^-\pi^+] &=& \frac{(m_n^2 - m_{\pi^+}^2)^2}{32\, \pi\, m_n^3 f_\pi^2} A^2 \left( \frac{1+\tilde{D}+\tilde{F}}{\sqrt{2}}\right)^2  \Big(\left| \beta\, \tilde{y}^{\prime *}[d^C_1,d^C_1,e^C_i,d_1] \right|^2 \Big. \nonumber \\
& + & \Big. \left| \alpha\, \tilde{y}^*[d^C_1,d^C_1,e_i,d^C_1] + \beta\, \tilde{y}^{\prime *}[d_1,d_1,e_i,d_1^C]\right|^2 \Big).\,\eeqa
It is evident that when \( y=g \), implying the operators are induced only by the triplets in \( {120}_{\hh} \), the decay width for \( n \to e_i^-\pi^+ \) at the leading order is zero due to the anti-symmetric nature of \( G \), as shown in Eq.~\eqref{eq:c3:coff_op_120_d7}.

\section{Decay Width Estimation in Realistic \so\, Model}
\label{sec:c3:results_model}
Using the general findings regarding nucleon decays detailed in section~({\ref{sec:c3:decay_widths}}), we now calculate the partial decay widths of the proton within a specific $SO(10)$ model. We use a known-minimal and realistic \so\, model whose Yukawa sector includes complex ${10}_{\hh}$\footnote{ Although $10_{\hh}$ is a real representation implying $10_{\hh} = 10^{\dagger}_{\hh}$, which leads to $\sfrac{\langle 5_{\hh} \rangle}{\langle 5^{\dagger}_{\hh} \rangle} \sim \frac{m_{t}}{m_b} = 1$, this is in contradiction with the experimentally observed mass relation. This situation necessitates complexifying the $10_{\hh}$ by adding another $10_{\hh}$ dimensional irrep~\cite{Bajc:2005zf}}.
 and $\overline{ 126}_{\hh}$ scalar irreps, each comprising a pair of colour singlets and $SU(2)$ doublets with $Y=\pm 1/2$, as listed in Tab.~(\ref{tab:c2:scalars}). A pair of linear combinations of these doublets, $h_u$ and $h_d$, is presumed to remain significantly lighter than the GUT scale, and $vevs$ of $h_u$ and $h_d$ contribute to generating masses for all charged fermions. The $\overline{126}_{\hh}$ includes an SM singlet with $B-L$ charge, whose $vev$ imparts masses to the heavy RH neutrinos. Together with the Dirac masses from the $vevs$ of $h_{u,d}$, this \so\, framework enables a naturally suppressed mass for the SM left-handed neutrinos through the type I seesaw mechanism. 

The effectiveness of this framework in accurately reproducing the observed spectrum of fermion masses and mixing parameters has been thoroughly examined in multiple studies, such as~\cite{Babu:1992ia,Bajc:2005zf,Joshipura:2011nn,Altarelli:2013aqa,Dueck:2013gca,Meloni:2014rga,Meloni:2016rnt,Babu:2016bmy,Ohlsson:2018qpt,Mummidi:2021anm}. In a recent study~\cite{Mummidi:2021anm}, several viable solutions within this model that successfully reproduce the known fermion mass spectrum and also explain the observed baryon asymmetry through Leptogenesis.

In this model, the absence of \(120_{\hh}\) implies that proton decay is induced only by colour triplet scalars with \(Y = \pm 1/3\), as discussed in section~(\ref{sec:c3:couplings}). Additionally, this minimal model incorporates a \(U(1)\) Peccei-Quinn symmetry~\cite{Peccei:1977hh}, where the fermion field \(\fs\) carries a charge of \(+1\) and both \(10_{\hh}\) and \(\overline{126}_{\hh}\) carry a charge of \(-2\). This setup allows for Yukawa couplings involving \(10_{\hh}\) but prevents those involving \(10_{\hh}^\dagger\), resulting in only two Yukawa coupling matrices and enhancing the model's predictability~\cite{Joshipura:2011nn}. Due to the Peccei-Quinn symmetry, the gauge-invariant term \(10_{\hh}^2\) is forbidden, thus preventing any mixing between the components \(T\) and \(\overline{T}\) of \({10}_{\hh}\). As a result, \(\theta\) is set to $0$ in Eq.~\eqref{eq:c3:mixing_angle} and all \(h^\prime\) coefficients vanish in Eq.~\eqref{eq:c3:op_10}. Consequently, proton decay in this scenario is determined by only three independent operators, as listed in the first three entries of Eq.~\eqref{eq:c3:gen_op_2}.

Generally, the interaction terms in the scalar potential that cause the mixing of electroweak doublets can also lead to mixing between different \(T\) and \(\overline{T}\) from \(10_{\hh}\) and \(\overline{ 126}_{\hh}\). In this scenario, the coefficients of the effective operators specified in Eq.~\eqref{eq:c3:gen_op_2} are linear combinations of the corresponding \(h\) and \(f\). However, the exact combination depends on the entire scalar potential, which is rather very specific to the underlying models. To compute the proton lifetime, we use a simplified approach, assuming that the lightest pair of triplets primarily stems from either \(10_{\hh}\) or \(\overline{126}_{\hh}\). Hence, the coefficients \(y\) are either \(h\) or \(f\), and these can be precisely determined from the fundamental Yukawa coupling matrices \(H\) and \(F\) and the flavour rotation matrices \(U_f\). The relevant information of fitting and parameters for the best-fit solution are listed in~\cite{Mummidi:2021anm} and have been relegated in the Appendix~(\ref{app:c3:fit}) for reference.

\subsection{ Proton Decay Pattern}
{\label{ss:c3:pdecaypatters}}
In previous sections~(\ref{sec:c3:operators_d6}, \ref{sec:c3:operators_d7}, and \ref{sec:c3:decay_widths}), we discussed various operators that contributed to proton decay and provided the decay width expressions for the tree level two-body decay modes. Now, we calculate the branching patterns for the leading decay modes of the proton. The results for various branching ratios, assuming that the lightest \(T\) and \(\overline{T}\) primarily originate from \({10}_{\hh}\), are presented in Tab.~(\ref{tab:c3:10_res}). Additionally, the branching ratios computed under the assumption that the lightest pair of triplets come from \(\overline{126}_{\hh}\) are shown in Tab.~(\ref{tab:c3:126_res}).
\begin{table}[t]
\begin{center}
\begin{tabular}{lccc} 
\hline
\hline
Branching Ratio [\%]& ~~~$M_T \ll M_{\overline{T}}$~~~  & ~~~$M_T \gg M_{\overline{T}}$~~~ & ~~~$M_T = M_{\overline{T}}$~~~\\
\hline
${\rm BR}[p\to e^+ \pi^0]$ & $< 1$ & $<1$ & $< 1$\\
${\rm BR}[p\to \mu^+ \pi^0]$  & $7$ & $< 1$ & $< 1$\\
${\rm BR}[p\to \bar{\nu} \pi^+]$  & $0$ & $15$ & $14$\\
${\rm BR}[p\to e^+ K^0]$ & $< 1$ & $< 1$ & $< 1$\\
${\rm BR}[p\to \mu^+ K^0 ]$  & $93$ & $<1$ & $2$\\
${\rm BR}[p\to \bar{\nu} K^+]$  & $0$ & $84$ & $83$\\
${\rm BR}[p\to e^+ \eta]$ & $<1$  &  $<1$ & $<1$\\
${\rm BR}[p\to \mu^+ \eta]$  & $<1$ &  $<1$ & $< 1$\\
\hline\hline
\end{tabular}
\end{center}
\caption{Estimates of proton decay branching fractions based on the best-fit solution, considering different mass hierarchies between the \(T\) and \(\overline{T}\) scalars within \({10}_{\hh}\).}
\label{tab:c3:10_res}
\end{table}
\begin{table}[t]
\begin{center}
\begin{tabular}{lccc} 
\hline
\hline
Branching Ratio [\%]& ~~~$M_{T_{1,2}} \ll M_{\overline{T}}$~~~  & ~~~$M_{T_{1,2}} \gg M_{\overline{T}}$~~~ & ~~~$M_{T_{1,2}} = M_{\overline{T}}$~~~\\
\hline
${\rm BR}[p\to e^+ \pi^0]$ & $< 1$ & $<1$ & $< 1$\\
${\rm BR}[p\to \mu^+ \pi^0]$  & $11$ & $< 1$ & $< 1$\\
${\rm BR}[p\to \bar{\nu} \pi^+]$  & $0$ & $11$ & $11$\\
${\rm BR}[p\to e^+ K^0]$ & $< 1$ & $< 1$ & $< 1$\\
${\rm BR}[p\to \mu^+ K^0 ]$  & $88$ & $10$ & $11$\\
${\rm BR}[p\to \bar{\nu} K^+]$  & $0$ & $78$ & $77$\\
${\rm BR}[p\to e^+ \eta]$ & $<1$  &  $<1$ & $<1$\\
${\rm BR}[p\to \mu^+ \eta]$  & $<1$ &  $<1$ & $< 1$\\
\hline\hline
\end{tabular}
\end{center}
\caption{Estimates of proton decay branching fractions calculated for the best-fit solution, considering various mass hierarchies between the \(T_{1,2}\) and \(\overline{T}\) scalars residing in \(\overline{126}_{\hh}\).}
\label{tab:c3:126_res}
\end{table}

The crucial inferences drawn from Tabs.~(\ref{tab:c3:10_res}, and \ref{tab:c3:126_res}) are listed as follows:
\begin{enumerate}[label=(\roman*)]
\item In scalar-induced proton decay, the proton predominantly decays into channels involving \(K^0\mu^+\) or \(K^+\overline{\nu}\).\label{aa}
\item Additionally, the branching ratio \( {\rm BR}[p \to e^+ \pi^0] \) is significantly less than \( {\rm BR}[p \to \mu^+ \pi^0] \).\label{ab}
\item When \(\overline{T}\) is lighter than \(T\), the proton mainly decays through channels that involve charged mesons and neutrinos. Since \(T\) does not couple to neutrinos, decays mediated by \(T\) typically result in neutral mesons and charged leptons.\label{ac}
\item A comparison between Tabs.~(\ref{tab:c3:10_res} and \ref{tab:c3:126_res}) indicates that the decay patterns of the proton are not significantly influenced by whether the lightest \(T\) and \(\overline{T}\) resides in \({10}_{\hh}\) or \(\overline{126}_{\hh}\). This suggests that the qualitative results would remain consistent even if the lightest pairs, \(T\) and \(\overline{T}\), are general linear combinations of triplets and anti-triplets respectively, from \({10}_{\hh}\) and \(\overline{126}_{\hh}\).\label{ad}
\end{enumerate}

\begin{figure}[t]
\centering
\includegraphics[scale=0.38]{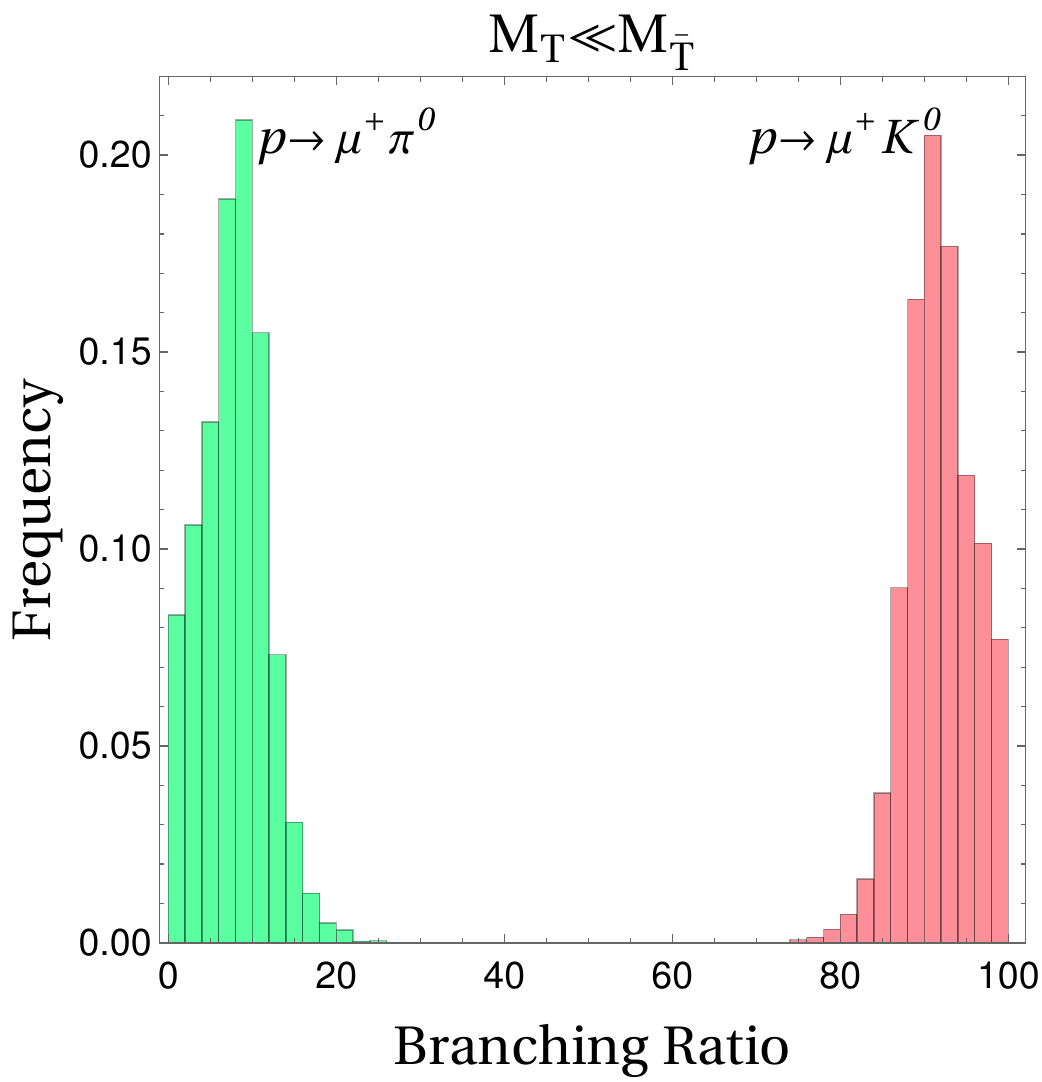}\hspace{0.2cm}
\includegraphics[scale=0.38]{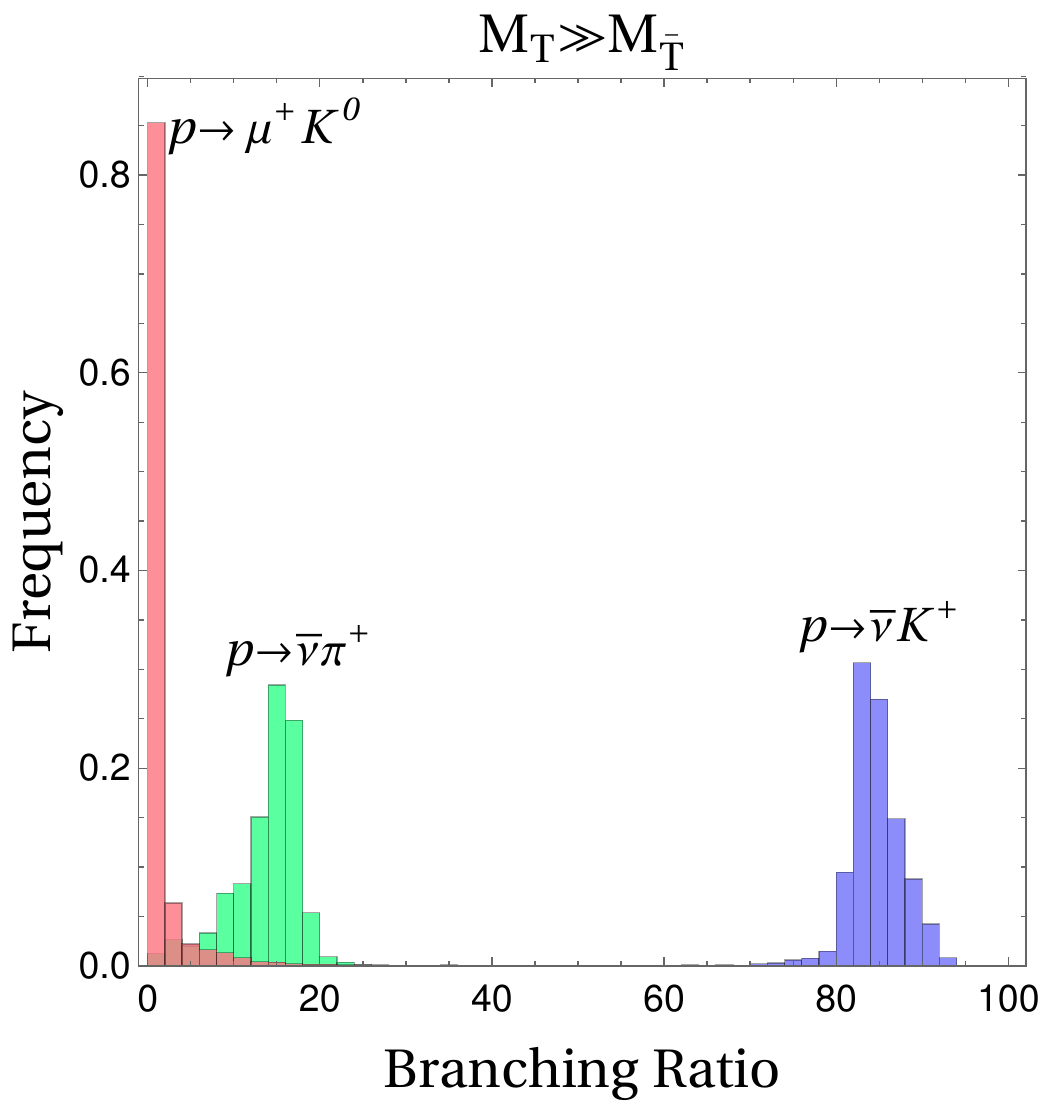}\hspace{0.2cm}
\includegraphics[scale=0.38]{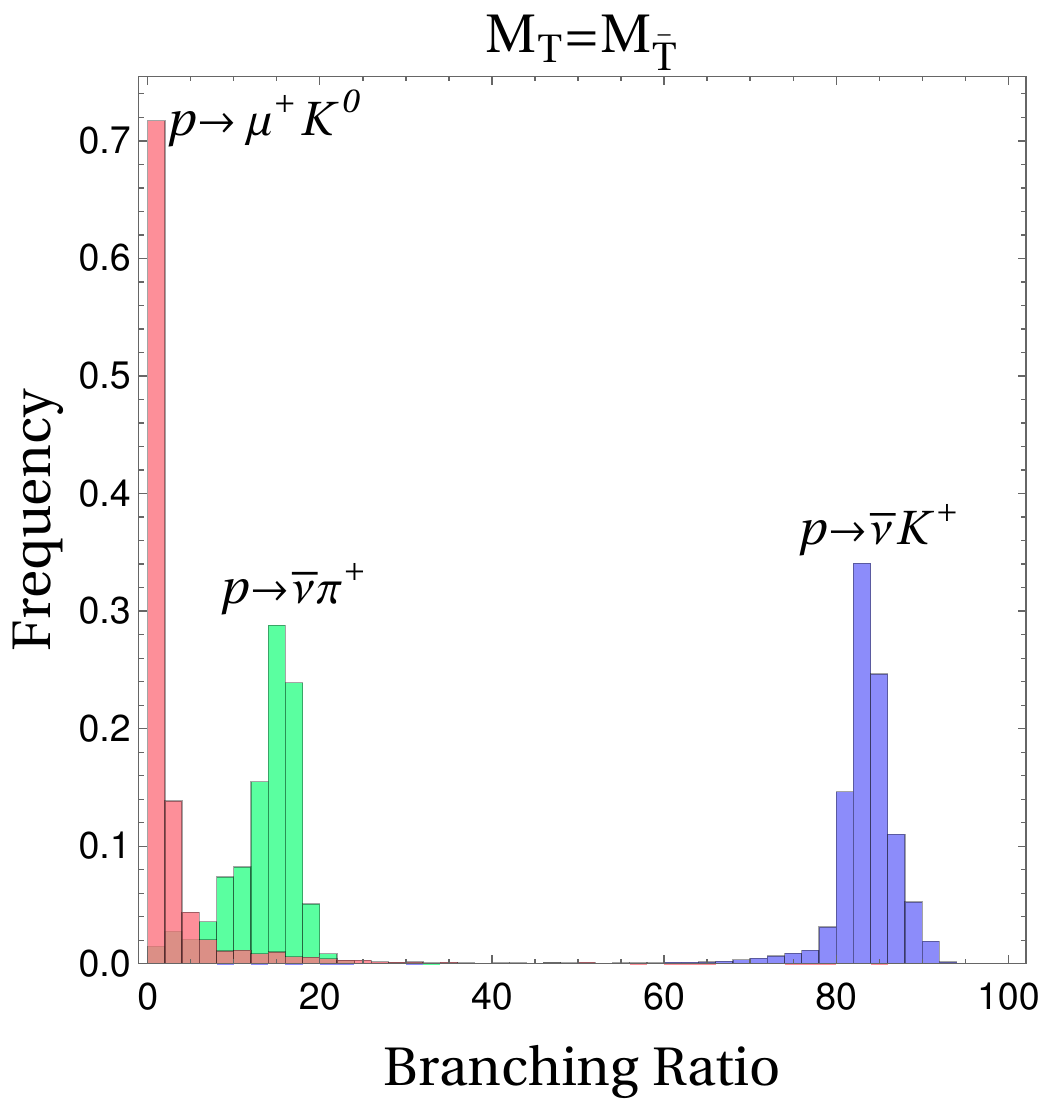}
\caption{The scalar-mediated proton decay spectrum within the framework of the minimal renormalisable non-supersymmetric \(SO(10)\) model. Various branching ratios have been calculated under the assumption that the lightest pair of colour triplets, \(T\) and \(\overline{T}\), originate from \(10_{\hh}\).}
\label{fig:c3:fig1}
\end{figure}

\begin{figure}[t]
\centering
\includegraphics[scale=0.38]{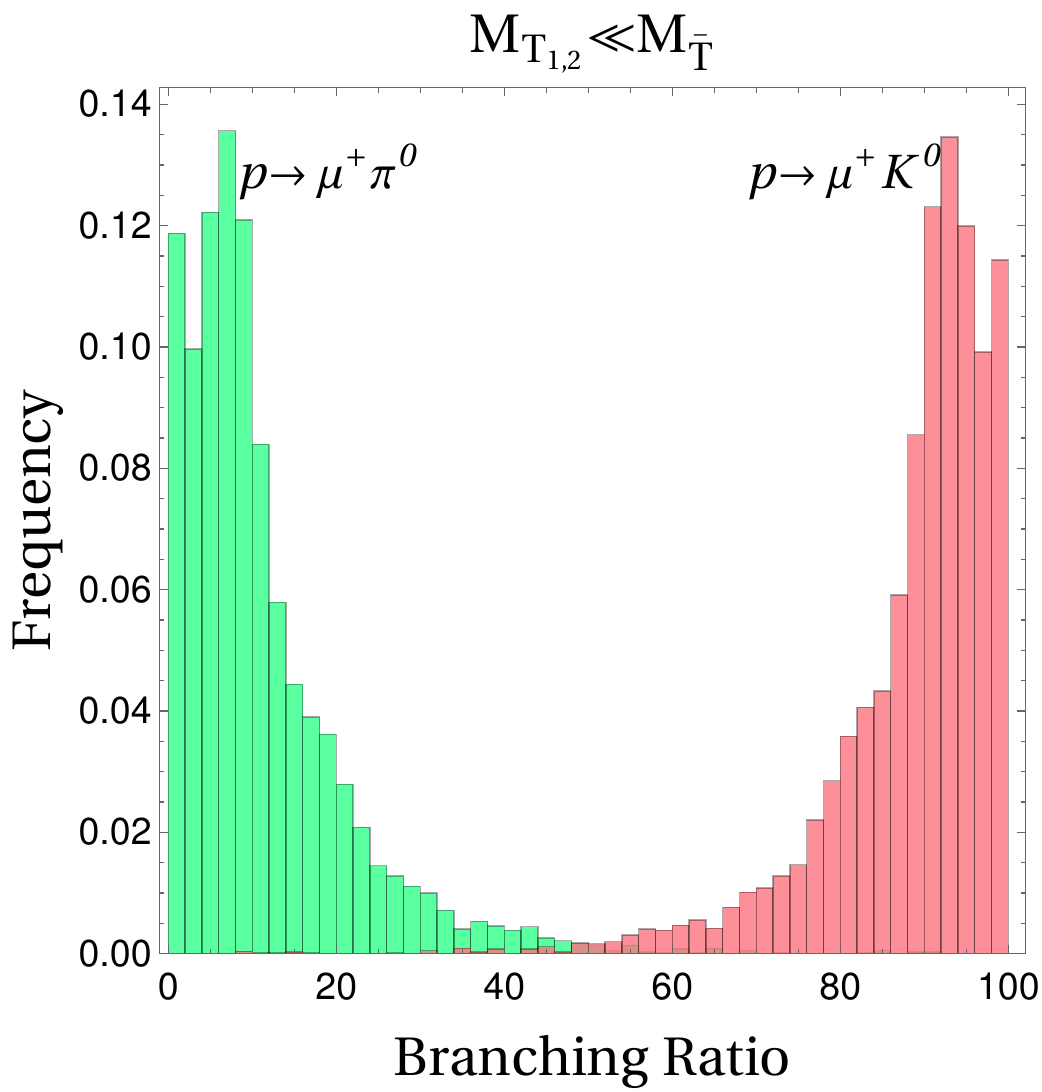}\hspace{0.2cm}
\includegraphics[scale=0.38]{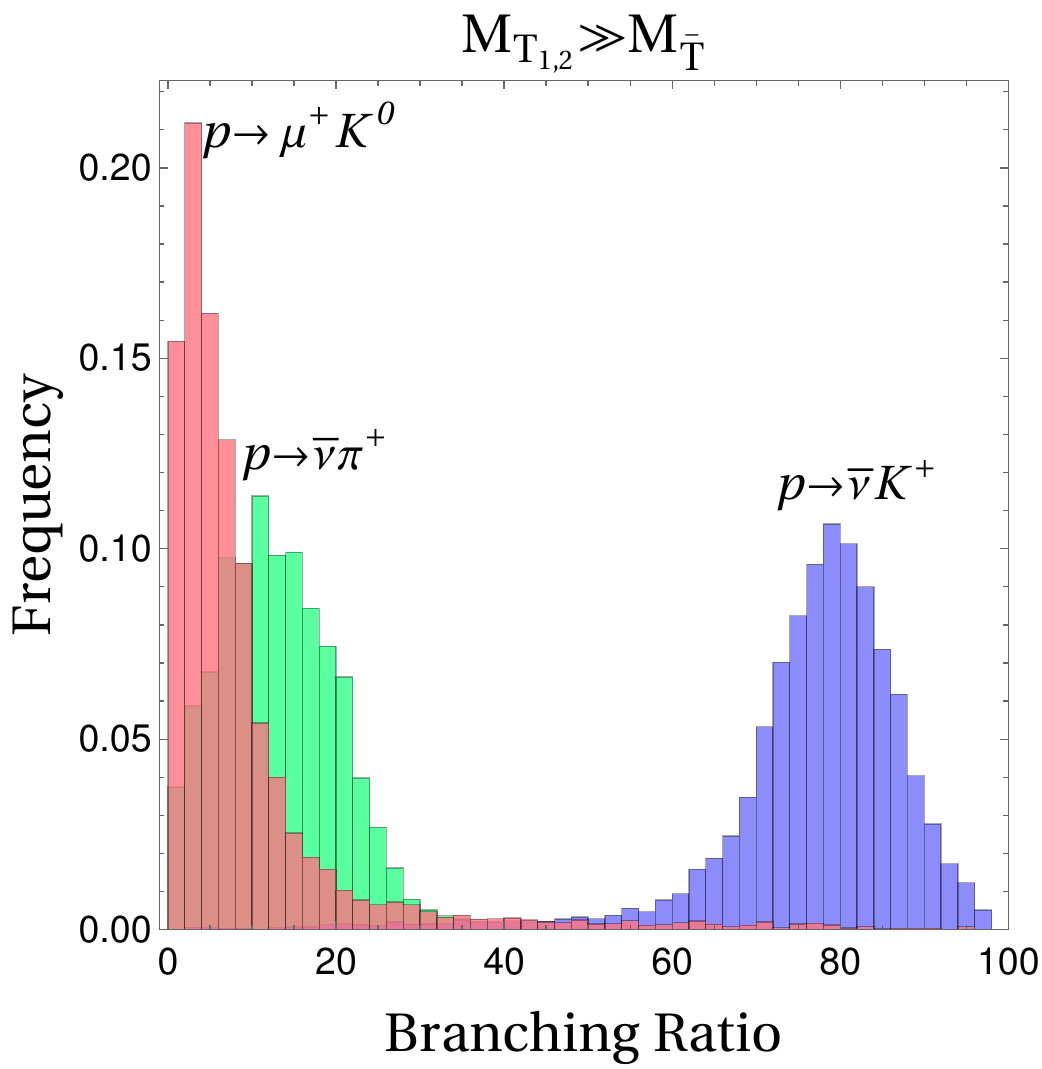}\hspace{0.2cm}
\includegraphics[scale=0.38]{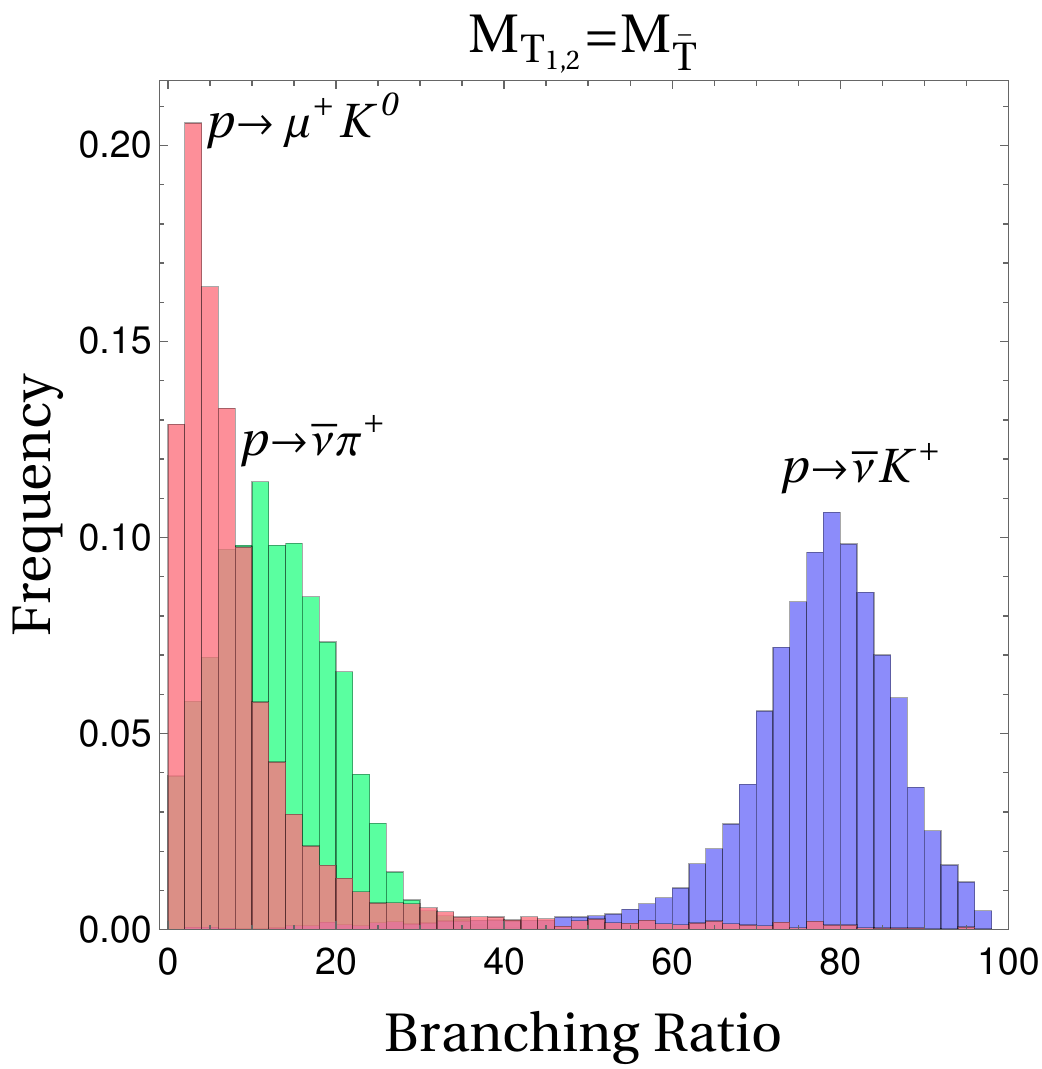}
\caption{Similar to Fig. (\ref{fig:c3:fig1}), but with the assumption that the lightest triplets predominantly originate from \(\overline{ 126}_{\hh}\).}
\label{fig:c3:fig2}
\end{figure}

To check the robustness of the above observations, we evaluate the proton decay branching ratios for not only the best-fit solution but also for several other solutions with acceptable $\chi^2$ values at the minimum, determined in~\cite{Mummidi:2021anm}. The results are displayed in Figs.~(\ref{fig:c3:fig1} and (\ref{fig:c3:fig2}).
From the Figs.~(\ref{fig:c3:fig1} and \ref{fig:c3:fig2}), it is evident that the predictions for various branching ratios as listed in Tabs.~(\ref{tab:c3:10_res} and (\ref{tab:c3:126_res}) remain consistent across other viable solutions as well. In Figs.~~(\ref{fig:c3:fig1} and \ref{fig:c3:fig2}), frequency represents the probability of the event in the solution space. The key characteristic of the proton decay spectrum, as referenced in (\ref{aa}, \ref{ab}, \ref{ac}, and \ref{ad}), can be explained by the flavour structure of the model. The realistic fermion mass spectrum in the underlying model results in hierarchical structures for the matrices \(H\) and \(F\), which are given as follows:
\be \label{eq:c3:HF_form}
H \sim \frac{\lambda^4}{\alpha_1} \left(\ba{ccc} \lambda^7 & 0 & 0\\
0 & \lambda^3 & 0\\
0 & 0 & 1 \ea\right)\,\hspace{0.5cm}\text{and}\,~~
F \sim \frac{\lambda^4}{\alpha_2} \left(\ba{ccc} \lambda^5 & \lambda^4 & \lambda^3\\
\lambda^4 & \lambda^3 & \lambda^2\\
\lambda^3 & \lambda^2 & \lambda \ea\right)\,,\ee
 where, \(\lambda = 0.23\) represents the Cabibbo angle, and we have omitted coefficients of order one for simplicity in front of each element. It is important to mention the branching ratios listed in both the Tabs. (\ref{tab:c3:10_res} and \ref{tab:c3:126_res}) are independent of the uncertain parameter \(\alpha_{1,2}\) mentioned in Eq.~\eqref{eq:c3:HF_form}. The Yukawa matrices for the charged fermions—\(Y_u\), \(Y_d\), and \(Y_e\)—are derived from linear combinations of \(H\) and \(F\). The unitary matrices that diagonalise these Yukawa matrices and the neutrino mass matrix typically exhibit the following generic form:
\be \label{eq:c3:U_form}
U_u \sim U_d \sim U_e \sim\left(\ba{ccc} 1 & \lambda & \lambda^3\\
\lambda & 1 & \lambda^2 \\
\lambda^3 & \lambda^2 & 1 \ea\right)\,,\hspace{0.5cm}\text{and}\,~~U_\nu \sim \left( \ba{ccc} &  &  \\  & {\cal O}(1) & \\  &  &  \ea \right)\,\ee
where $U_\nu$ is matrix having all the elements of ${\cal O}(1)$.  $U_{u,d,e}$ could taken as $CKM$-type, leading to small quark mixing while the large mixing in the PMNS matrix arises through $U_\nu$.  From Eqs.~(\ref{eq:c3:HF_form} and \ref{eq:c3:U_form}), we find
\beqa \label{eq:c3:UHU}
U_f^T\,H\,U_{f^\prime} &\sim & \frac{\lambda^4}{\alpha_1} \left(\ba{ccc} \lambda^5 & \lambda^4 & \lambda^3\\ \lambda^4 & \lambda^3 & \lambda^2\\
\lambda^3 & \lambda^2 & 1 \ea\right)\,,~~U_f^T\,F\,U_{f^\prime} \sim  \frac{\lambda^4}{\alpha_2} \left(\ba{ccc} \lambda^5 & \lambda^4 & \lambda^3\\ \lambda^4 & \lambda^3 & \lambda^2\\
\lambda^3 & \lambda^2 & \lambda \ea\right)\,, \nonumber \\
U_\nu^T\,H\,U_d &\sim & \frac{\lambda^4}{\alpha_1} \left(\ba{ccc} \lambda^3 & \lambda^2 & 1\\ \lambda^3 & \lambda^2 & 1\\
\lambda^3 & \lambda^2 & 1 \ea\right)\,,~~U_\nu^T\,F\,U_d \sim  \frac{\lambda^4}{\alpha_2} \left(\ba{ccc} \lambda^3 & \lambda^2 & \lambda\\ \lambda^3 & \lambda^2 & \lambda\\
\lambda^3 & \lambda^2 & \lambda \ea\right)\,.\eeqa
for every $f,f^\prime = u,d,e$.

Incorporating these results into the Eqs.~(\ref{eq:c3:coff_op_10}, \ref{eq:c3:coff_op_126}, and \ref{eq:app:c3:decay_width}), we arrive at the following
\beqa \label{eq:c3:pattern_a}
\frac{\Gamma[p \to e_i^+\,\pi^0]}{\Gamma[p \to e_i^+\,K^0]} & \simeq & \frac{(m_p^2 - m{^2_{\pi^0}})^2}{(m_p^2 - m{^2_{K^0}})^2}\,\frac{(1+\tilde{D}+\tilde{F})^2}{2 \left(1+\frac{m_p}{m_B}(\tilde{D}-\tilde{F})\right)^2}\, {\lambda}^2 \simeq 3 \lambda^2\,, \nonumber \\
\frac{\Gamma[p \to \overline{\nu}\,\pi^+]}{\Gamma[p \to \overline{\nu}\,K^+]} & \simeq & \frac{(m_p^2 - m{^2_{\pi^+}})^2}{(m_p^2 - m{^2_{K^+}})^2}\,\frac{(1+\tilde{D}+\tilde{F})^2}{\left(1+\frac{m_p}{m_B}(\tilde{D}+\tilde{F})\right)^2}\,  {\lambda}^2 \simeq 2 \lambda^2\,.\eeqa
which explains the observation listed as point \ref{aa} above. Further, point \ref{ab} can be explained as
\be \label{eq:c3:pattern_b}
\frac{\Gamma[p \to e^+\,\pi^0]}{\Gamma[p \to \mu^+\,\pi^0]} \simeq \frac{\Gamma[p \to e^+\,K^0]}{\Gamma[p \to \mu^+\,K^0]} \simeq \lambda^2\,, \ee
 Furthermore, point \ref{ac} can be explained from the following.
\be \label{eq:c3:decaylambda}
\frac{\Gamma[p \to \mu^+\,K^0]}{\Gamma[p \to \overline{\nu}\,K^+]}  \simeq  \frac{(m_p^2 - m{^2_{K^0}})^2}{(m_p^2 - m{^2_{K^+}})^2}\,\frac{\left(1+\frac{m_p}{m_B}(\tilde{D}-\tilde{F})\right)^2}{\left(1+\frac{m_p}{m_B}(\tilde{D}+\tilde{F})\right)^2}\,  \frac{4 {\lambda}^2}{9} \simeq 0.2 \lambda^2\,.\ee
Additionally, it is evident from Eq.~\eqref{eq:c3:UHU} that the flavour structure of the couplings relevant to proton decay is similar for both \(10_{\hh}\) and \(\overline{126}_{\hh}\). This similarity supports the observations noted in \ref{ad}.

\subsection{Constraint on the Masses of $T$ and $\overline{T}$}
{\label{ss:c3:constraintsonscalars}}
Finally, by utilising the current experimental limits on the proton decay lifetime for various channels, we derive the stringent limits on the masses of \(T\) and \(\overline{T}\) within the realistic model being considered. When the lightest \(T\) primarily originates from \(10_{\hh}\), the stringent limit on its mass is imposed by the decay channels that involve \(\mu^+\). We find:
\beqa \label{eq:c3:limit_T_10}
\tau/{\rm BR}[p \to \mu^+\, K^0] &=& 1.6 \times 10^{33}\,{\rm yrs}\,\times\left(\frac{\alpha_1}{0.1}\right)^4 \times \left(\frac{M_T}{1.4 \times 10^{11}\,{\rm GeV}} \right)^4\,, \nonumber \\
\tau/{\rm BR}[p \to \mu^+\, \pi^0] &=& 1.6 \times 10^{34}\,{\rm yrs}\,\times\left(\frac{\alpha_1}{0.1}\right)^4 \times \left(\frac{M_T}{1.3 \times 10^{11}\,{\rm GeV}} \right)^4\,,\nl \eeqa
provided, the first factor on the right-hand side represents the current experimental lower bounds on the lifetime of proton decay in the respective channels. Similarly, when \(\overline{T}\) predominantly originates from \( 10_{\hh}\), the most stringent upper bound on its mass is derived from the proton decay channels into neutrinos and \(K^+\):
\beqa \label{eq:c3:limit_Tbar_10}
\tau/{\rm BR}[p \to \overline{\nu}\, K^+] &=& 5.9 \times 10^{33}\,{\rm yrs}\,\times\left(\frac{\alpha_1}{0.1}\right)^4 \times \left(\frac{M_{\overline{T}}}{6.4 \times 10^{11}\,{\rm GeV}} \right)^4.\,
\eeqa
The experimental limits on the lifetimes for various proton decay channels referenced in the above calculations are taken from~\cite{Super-Kamiokande:2020wjk, Super-Kamiokande:2012zik, Super-Kamiokande:2014otb}. These studies provide the current lower bounds on the lifetime of the proton at $90\,\%$ confidence level.

If the lightest $T$ and $\overline{T}$ arise dominantly from $\overline{126}_{\hh}$, we find 
\beqa \label{eq:c3:limit_T_126}
\tau/{\rm BR}[p \to \mu^+\, K^0] &=& 1.6 \times 10^{33}\,{\rm yrs}\,\times\left(\frac{\alpha_2}{0.1}\right)^4 \times \left(\frac{M_{T_i}}{2.5 \times 10^{10}\,{\rm GeV}} \right)^4\,, \nonumber \\
\tau/{\rm BR}[p \to \mu^+\, \pi^0] &=& 1.6 \times 10^{34}\,{\rm yrs}\,\times\left(\frac{\alpha_2}{0.1}\right)^4 \times \left(\frac{M_{T_i}}{2.7 \times 10^{10}\,{\rm GeV}} \right)^4,\,
\eeqa
and 
\beqa \label{eq:c3:limit_Tbar_126_1}
\tau/{\rm BR}[p \to \overline{\nu}\, K^+] &=& 5.9 \times 10^{33}\,{\rm yrs}\,\times\left(\frac{\alpha_2}{0.1}\right)^4 \times \left(\frac{M_{\overline{T}}}{1.1 \times 10^{11}\,{\rm GeV}} \right)^4.\,\eeqa
It can be noticed that the limits on the masses of $T$ and $\overline{T}$ are slightly lower than the ones given in Eqs.~(\ref{eq:c3:limit_T_10} and \ref{eq:c3:limit_Tbar_10}). This is due to the fact that the magnitude of Yukawa couplings with $\overline{ 126}_{\hh}$ are somewhat smaller than those with $10_{\hh}$. 

 \subsection{Constraint on the Mass of $\Delta$}
In addition to the \(D=6\) operators previously discussed in section~(\ref{sec:c3:operators_d6}), a \(D=7\) operator emerges within the model due to \(B-L\) violation and mixing between \(\overline{T}\) and \(\Delta\), as outlined in section~(\ref{sec:c3:operators_d7}). This operator induces proton decay modes such as \(p \to \nu\,\pi^+\) and \(p \to \nu\,K^+\). Given that the mass of \(\overline{T}\) is already constrained by the \(D=6\) operators, the lower bound on the mass of \(\Delta\) can be deduced. To calculate this, we assume the quartic coupling \(\lambda = 1\), \(\alpha_2 = 0.1\), \(v_{\overline{D}} = 174\) GeV, and \(M_{\overline{T}} = 1.1 \times 10^{11}\) GeV, as specified in Eq.~\eqref{eq:c3:limit_Tbar_126}. Substituting these parameters into Eqs.~(\ref{eq:c3:coff_op_126_d7} and \ref{eq:c3:decay_width_d7_1}), we determine:
\beqa \label{eq:c3:limit_Tbar_126}
\tau/{\rm BR}[p \to \nu\, K^+] &=& 5.9 \times 10^{33}\,{\rm yrs}\,\times\left(\frac{10^{11}\,{\rm GeV}}{v_\sigma}\right)^2 \times \left(\frac{M_{\Delta}}{7.0 \times 10^{6}\,{\rm GeV}} \right)^2.\nl\eeqa
The calculated lower bound on \(M_\Delta\) turns out to be two orders of magnitude lower than that reported in~\cite{Babu:2012vb} for the same \(B-L\) breaking scale. This difference is due to the values of the Yukawa couplings used, as well as an additional factor of \((4/15)^2\) given in Eq.~\eqref{eq:c3:coff_op_126_d7}, which stems from the decomposition.

\section{Gauge Boson Mediated Proton Decay}
\label{sec:c3:gbmpd}
To clearly distinguish between scalar-mediated and gauge boson-mediated proton decay, we calculate the gauge-mediated proton decay partial widths for a quantitative comparison. The currents associated with \(B-L\) charged gauge bosons, \(X \sim (3,2,-5/6)\) and \(Y \sim (3,2,1/6)\), which belong to the adjoint representation of \(SO(10)\), are detailed as follows\footnote{There is also a gauge boson with SM charges \(Z\sim (3,1,2/3)\) having a \(B-L\) charge of \(4/3\). However, it does not mediate proton decay by itself \cite{Buchmuller:2019ipg}.}~\cite{Machacek:1979tx}.
The interactions of the \(X\) and \(Y\) gauge bosons with SM fermions can be inferred from the Eq.~\eqref{eq:c2:XandY}, where $L_{AB}\,=\,g_{10}\,\delta_{AB}$;
\beqa \label{eq:c3:LXY}
-{\cal L}^{X\&Y} & = & i\,g_{10}\,\Big[\sqrt{2}\,\overline{X}_{\Dot{\mu}} \left(q^{\dagger}_A\, \overline{\sigma}^{\Dot{\mu}}\, u^C_A + e^{C\,\dagger}_A\, \overline{\sigma}^{\Dot{\mu}}\, q_A - d^{C\,\dagger}_A\, \overline{\sigma}^{\Dot{\mu}}\, l_A \right) \Big. \nonumber \\
& + &  \Big. 2\,\overline{Y}_{\Dot{\mu}} \left(q^{\dagger}_A\, \overline{\sigma}^{\Dot{\mu}}\, d^C_A + \nu^{C\,\dagger}_A\, \overline{\sigma}^{\Dot{\mu}}\, q_A - u^{C\,\dagger}_A\, \overline{\sigma}^{\Dot{\mu}}\, l_A \right) \Big]\,\hc\,,\eeqa
where $g_{10}$ is unified gauge coupling and we have supressed the $SU(3)$ and $SU(2)$ indices for brevity. Integrating out the degrees of freedom of $X$ and $Y$ bosons using classical equations of motion, the relevant effective operators for the proton decay are as follows:
\beqa \label{eq:c3:LXY_2}
{\cal L}_{\rm eff} & = & \frac{g_{10}^2}{2 M_X^2} \left(u^{C\,\dagger}_A\, \overline{\sigma}^{\Dot{\mu}}\, q_A\right)\,\, \left(e^{C\,\dagger}_B\, \overline{\sigma}_{\Dot{\mu}}\, q_B - d^{C\,\dagger}_B \overline{\sigma}_{\Dot{\mu}} l_B\right) \nonumber \\
& - & \frac{g_{10}^2}{2 M_Y^2} \left(d^{C\,\dagger}_A \,\overline{\sigma}^{\Dot{\mu}}\, q_A\right)\,\, \left(u^{C\,\dagger}_B\, \overline{\sigma}_{\Dot{\mu}}\, l_B\right)\,\hc\,.\eeqa

Using Fierz relations for two-component Weyl spinors~\cite{Dreiner:2008tw}  and transforming into the physical basis, $f \to U_{f} f$, we obtain, (cf.~\cite{Buchmuller:2019ipg} for details)
\beqa \label{eq:c3:LXY_3}
{\cal L}_{\rm eff} & = & k[u_A,d_B,e^C_C,u^C_D]\left(\overline{e^C}_C\, \overline{u^C}_D\, u_A\, d_B\right) \nonumber \\
& + & k[u^C_A,d^C_B,e_C,u_D]\left(\overline{d^C}_B\, \overline{u^C}_A\, u_D\, e_C\right) \nonumber \\
& + & k[u^C_A,d^C_B,\nu_C,d_D]\left(\overline{d^C}_B\, \overline{u^C}_A\, d_D\, \nu_C\right)\, \hc,,\eeqa
with
\beqa \label{eq:c3:coff_gauge}
k[u_A,d_B,e^C_C,u^C_D] &=& \frac{g_{10}^2}{M_X^2}\bigg[\left(U_{e^C}^\dagger U_d\right)_{CB} \left(U_{u^C}^\dagger U_{u}\right)_{DA}\nl
\ad \;\; \left(U_{e^C}^\dagger U_u\right)_{CA} \left(U_{u^C}^\dagger U_d \right)_{DB} \bigg]\,,\nonumber \\
k[u^C_A,d^C_B,e_C,u_D] &=& -\frac{g_{10}^2}{M_X^2} \left(U_{d^C}^\dagger U_e\right)_{BC} \left(U_{u^C}^\dagger U_{u}\right)_{AD}\nl
\mi\;\frac{g_{10}^2}{M_Y^2} \left(U_{d^C}^\dagger U_u\right)_{BD} \left(U_{u^C}^\dagger U_e \right)_{AC}\,,\nonumber \\
k[u^C_A,d^C_B,\nu_C,d_D] &=& \frac{g_{10}^2}{M_X^2} \left(U_{d^C}^\dagger U_\nu\right)_{BC} \left(U_{u^C}^\dagger U_d\right)_{AD}\nl
\ad \;\; \frac{g_{10}^2}{M_Y^2} \left(U_{d^C}^\dagger U_d\right)_{BD} \left(U_{u^C}^\dagger U_\nu \right)_{AC}.\,\nl\eeqa

\begin{figure}[h!]
\centering
\includegraphics[scale=0.38]{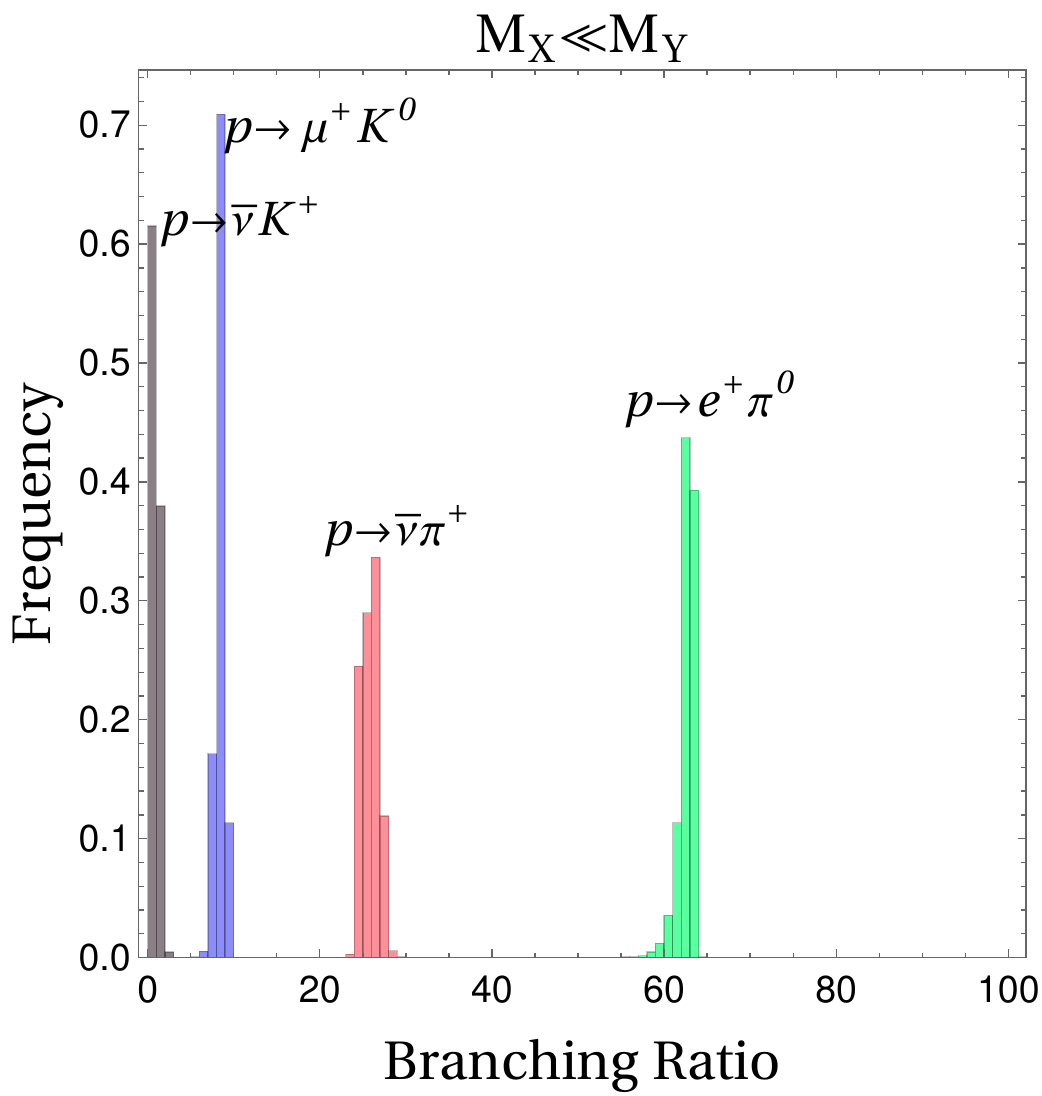}\hspace{0.2cm}
\includegraphics[scale=0.38]{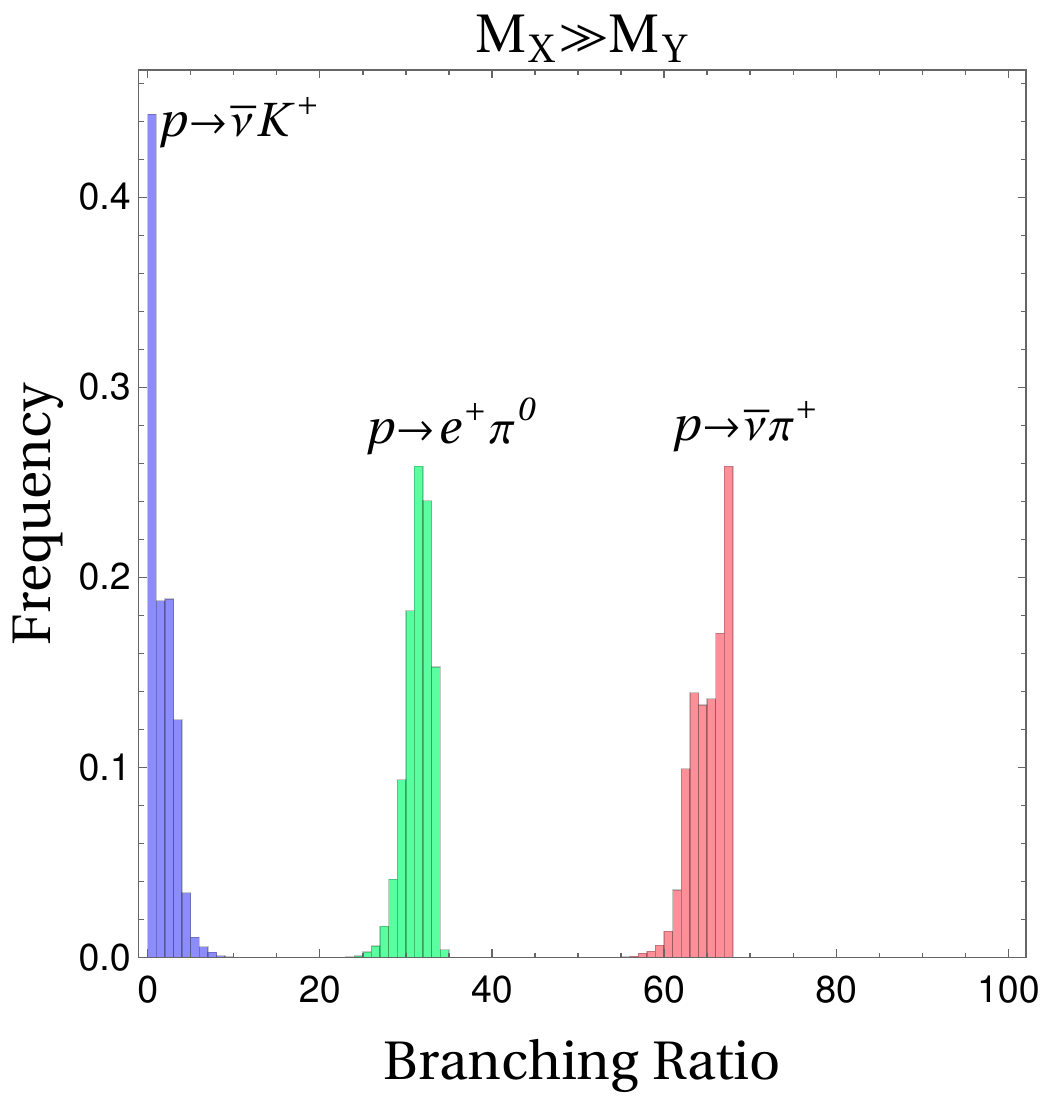}\hspace{0.2cm}
\includegraphics[scale=0.38]{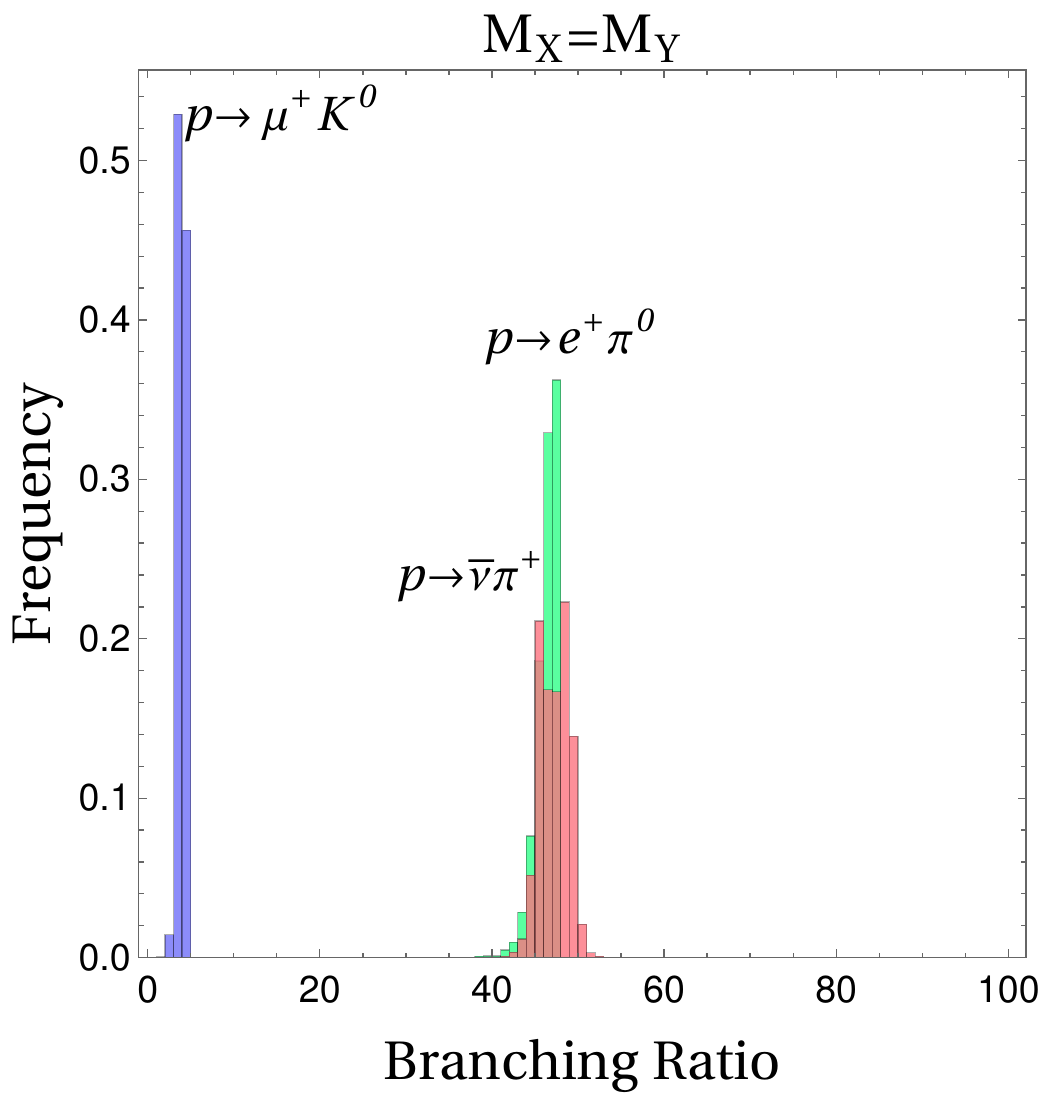}
\caption{The spectrum of proton decay induced by gauge bosons for several solutions within the same \(SO(10)\) model as in Figs.~(\ref{fig:c3:fig1} and \ref{fig:c3:fig2}).}
\label{fig:c3:fig3}
\end{figure}

The decay widths of the proton into various channels can be calculated by setting \(y[f_1, f_2, f_3, f_4] = k[f_1, f_2, f_3, f_4]\) and \(y^\prime[f_1, f_2, f_3, f_4] = 0\), using the expressions given in the Eqs.~(\ref{eq:c3:pepi}, and \ref{eq:app:c3:decay_width}). The branching pattern for the leading decay modes of the proton has been computed from the best-fit solution, assuming different mass hierarchies among the \(X\) and \(Y\) gauge bosons, is shown in Tab.~(\ref{tab:c3:gauge_res}). Additionally, the spectrum of branching ratios for several solutions with acceptable \(\chi^2\) values is displayed in Fig.~(\ref{fig:c3:fig3}).
\begin{table}[t]
\begin{center}
\begin{tabular}{lccc} 
\hline
\hline
Branching Ratio [\%]& ~~~$M_X \ll M_{Y}$~~~  & ~~~$M_X \gg M_{Y}$~~~ & ~~~$M_X = M_{Y}$~~~\\
\hline
${\rm BR}[p\to e^+ \pi^0]$ & $63$ & $33$ & $47$\\
${\rm BR}[p\to \mu^+ \pi^0]$  & $1$ & $1$ & $< 1$\\
${\rm BR}[p\to \bar{\nu} \pi^+]$  & $26$ & $55$ & $46$\\
${\rm BR}[p\to e^+ K^0]$ & $< 1$ & $< 1$ & $< 1$\\
${\rm BR}[p\to \mu^+ K^0 ]$  & $9$ & $<1$ & $4$\\
${\rm BR}[p\to \bar{\nu} K^+]$ & $3$ & $15$ & $1$\\
${\rm BR}[p\to e^+ \eta]$ & $<1$ & $<1$ & $<1$\\
${\rm BR}[p\to \mu^+ \eta]$  & $<1$ &  $<1$ & $< 1$\\
\hline\hline
\end{tabular}
\end{center}
\caption{Proton decay branching fractions estimated for the best fit solution for various hierarchies among the masses of $X$ and $Y$ gauge bosons in case of the minimal renormalizable non-supersymmetric $SO(10)$ model.}
\label{tab:c3:gauge_res}
\end{table}
It is evident from Tab.~(\ref{tab:c3:gauge_res}) and Fig.~(\ref{fig:c3:fig3}) that the branching pattern of gauge boson-mediated proton decay is entirely different from the scalar field-induced proton decay. The important differences are outlined as follows:
\begin{enumerate}[label=(\roman*)]
\item In the case of gauge boson mediation the proton dominantly decays into $\pi^0\,e^+$ or $\pi^+ \overline{\nu}$, while the scalar-induced proton decay favours the channels involving $K^0\,\mu^+$ or $K^+\,\overline{\nu}$.\label{ba}
\item  Typically, for the case of scalar mediated proton decay ${\rm BR}[p\to e^+ \pi^0] \ll {\rm BR}[p\to \mu^+ \pi^0]$. This is in contrast to the gauge boson-mediated decays in which one typically finds ${\rm BR}[p\to e^+ \pi^0] \gg {\rm BR}[p\to \mu^+ \pi^0]$. This difference is primarily due to the hierarchical nature of the Yukawa couplings, which are typically larger for the second generation fermions than the first. In contrast, the gauge parameter couples uniformly across all generations of fermions. {\label{bb}}
\end{enumerate}

\section{Summary}
\label{sec:c3:summary}
\lettrine[lines=2, lhang=0.33, loversize=0.15, findent=0.15em]{P}ROTON DECAY IS one of the crucial phenomena which can validify GUTs and offers a glimpse into high-energy phenomena from a low-energy. Observing proton decay would definitively discard the conservation of the baryon number. Additionally, it could shed light on the nature of the theories valid at ultra-violet scales, where violations of baryon number and lepton number originate. This understanding necessitates calculating nucleon decay widths within specific models, considering all potential sources of its decay.

In this chapter, we consider the most general scalar spectrum possible in the renormalisable \so\, Yukawa sector and proceed to explicitly compute the couplings of various scalars that can lead to baryon and lepton number-violating decays of hadrons at the tree level. We calculate effective operators by integrating out the scalar fields while also considering the possibility of mixing between the scalars within the same GUT multiplet. We evaluated  \(D=6\) operators that conserve baryon-minus-lepton number (\(B-L\)) and \(D=7\) operators which violate \(B-L\). Thereafter, we quantify the proton decay widths using these operators and offer a detailed analysis of scalar-mediated proton decay within a realistic GUT model based on $10_{\hh}$ and $\overline{126}_{\hh}$.

Key highlights from our comprehensive analysis are outlined as follows:
\begin{itemize}
\item The irreps capable of contributing to the renormalisable \so\, Yukawa sector, $i.e.$ \(10_{\hh}\), \(120_{\hh}\), and \(\overline{126}_{\hh}\), consists of different scalar fields which are charged under \(B-L\), but not all of them can induce proton decay. The \(D=6\) operators induced baryon decays originate from only three pairs of colour triplet fields: \(T(3,1,-\frac{1}{3})\), \({\cal T}(3,1,-\frac{4}{3})\), and \(\mathbb{T}(3,3,-\frac{1}{3})\) along with their conjugate partners.
\item In the class of realistic models based on ${10}_{\hh}$ and/or $\overline{126}_{\hh}$, only $T$ and $\overline{T}$ induce the nucleon decay at the tree level. Additionally, $\overline{126}_{\hh}$ contains ${\cal T}$ and $\mathbb{T}$ fields, it only exhibits leptoquark vertex.
\item When the Yukawa sector has $120_{\hh}$, the contribution to proton decay from \({\cal T}\)-\(\overline{\cal T}\) vanishes at the tree level due to the anti-symmetric nature of the Yukawa couplings. Consequently, these fields can only contribute to proton decay at the loop level.

\item The $B+L$ conserving nucleon decays is induced by $D=7$ operators at the leading order, and is mediated by $\Theta(3,1,\frac{2}{3})$, $\Delta(3,2,\frac{1}{6})$, $\Omega(3,2,\frac{7}{6})$ and their conjugate partners. In the \so\, models devoid of ${ 120}_{\hh}$, only $\Delta$ can induce such decays.
\end{itemize}

In a minimal model based on \({10}_{\hh}\) and \(\overline{126}_{\hh}\), the proton decay spectrum observations are different when scalar-mediated contributions are dominant and have been discussed in the section~(\ref{sec:c3:results_model}). The primary source of distinction stems from the nature of Yukawa and gauge couplings. Yukawa couplings are more sensitive to the flavour structure of the underlying GUT model. Consequently, proton decay primarily results in \(\overline{\nu}\, K^+\) or \(\mu^+\, K^0\) if \(\overline{T}\) or \(T\) is lighter, respectively. Additionally, it is noted that \({\rm BR}[p\to \mu^+ \pi^0]\) is significantly greater than \({\rm BR}[p\to e^+ \pi^0]\). These patterns contrast with those of gauge boson-mediated proton decays, where protons typically decay into \(\overline{\nu}\, \pi^+\) or \(e^+\, \pi^0\) and the branching ratio for \(\mu^+\) is much lower than for \(e^+\). Therefore, if scalar-mediated contributions predominate, they can offer valuable insights into the Yukawa structure of the theory, significantly influencing the predicted proton decay patterns.

\clearpage
\thispagestyle{empty}\newpage
\vspace*{\fill}
\begin{Huge}
\begin{center}
    \textbf{This page is intentionally left blank.}
\end{center}
\end{Huge}\vspace*{\fill}
\newpage
\thispagestyle{empty}
\chapter{Scalar Induced Proton Decays in Non-Renormalisable  GUTs}
{\label{ch:4}}
\graphicspath{{40_Chapter_4/fig_ch4/}}
\section{Overview}
{\label{sec:c4:overview}}

\lettrine[lines=2, lhang=0.33, loversize=0.15, findent=0.15em]{N}ON-RENORMALISABLE GRAND unified model with irreps contributing to the Yukawa sector having fewer degrees of freedom offer alternative solutions to the inconsistencies found in renormalisable GUT models in the UV regime, as discussed in Chapter~(\ref{ch:2}). In this chapter, our focus is to study non-renormalisable \so\, Yukawa interactions in which LTR are absent. We consider \so\, models constructed from irreps with less degree of freedoms and couples at the non-renormalisable level. $16_{\hh}$ is one such irrep and we evaluated the coupling of \(\fs\) with \(16_{\hh}\) in Chapter~(\ref{ch:2}). As many $B-L$ charged scalars reside in $16_{\hh}$ (cf. Tab.~(\ref{tab:c2:16scalars})), they can induce proton decays. Analogous to Chapter~(\ref{ch:3}), we constrain the scalars residing in $16_{\hh}$ from proton decays in this chapter.


In Section~(\ref{sec:c4:couplings}), we continue the analysis of the couplings between $16_{\hh}$ and $\fs$, initiated in Chapter~(\ref{ch:2}). We compute the contributions from various leptoquark and diquark couplings, generated by different scalar pairs, to $B-L$ conserving two-body proton decay modes in section~(\ref{sec:c4:pdeff}). In the same section, we also discuss some general characteristics observed from the evaluated coefficients. We argue that $16_{\hh}$ can assume a role similar to that of $\overline{126}_{\hh}$ in renormalisable \so\, GUTs, as detailed in section~(\ref{sec:c4:pds}), and utilise the evaluated strengths of various operators to compute the branching pattern of proton decay in a realistic \so\, model based on $16_{\hh}$. Key aspects of proton decay, including the leading decay mode and constraints on various pairs of tree-level proton decay mediators, are highlighted, and the study is concluded in section~(\ref{sec:c4:conc}).

\section{Couplings of $16_\hh$ with $\fs$-plet}
\label{sec:c4:couplings}
In the section~(\ref{sec:c2:NRSo}), we have considered the coupling of $16_{\hh}$ with $\fs$ and computed its decomposition when the intermediate integrated out irrep in $10$-dimensional, considering both the possibility $10$-dimensional irrep to be a fermion or a scalar. Further, in this section, we consider other possible irreps that yield the coupling mentioned in Eq.~\eqref{eq:c2:NRSO10} when other irreps are integrated out. We again show the coupling of $16_{\hh}$ with $\fs$-fermion plet as shown below:
-\beqa {\label{eq:c4:NRSO10}}
{\mathcal{L}}_{\mathrm{NR}} &\supset&  \frac{1}{\Lambda} y_{AB}\,\big(\mathbf{16}_{A}\,\mathbf{16}_{B}\;16_{\hh}\,16_{\hh}\big)\,+\, \frac{1}{\Lambda} \bar{y}_{AB}\,\big(\mathbf{16}_{A}\,\mathbf{16}_{B}\;16^{\dagger}_{\hh}\,16^{\dagger}_{\hh}\big) \,\hc,\nl
\eeqa
where $\Lambda$ is the cut-off scale. The non-renormalisable coupling mentioned in Eq.~\eqref{eq:c4:NRSO10} can be decomposed through various ways, and one particular possibility has been discussed in section~(\ref{sec:c2:NRSo}). Below, we provide the generalised ways of evaluating the effective couplings given in Eq.~\eqref{eq:c4:NRSO10}. The convention followed in depicting the following invariants is also outlined in the section~(\ref{sec:c2:NRSo});
\beqa {\label{eq:c4:16Fs16Hs}}
      & \bullet &  \frac{g_{AB}}{\Lambda}\big(\fs_A\,\fs_B\big)_{120_\hh}\,\big(16_\hh\,16_\hh\big)_{120_\hh};\; \frac{\bar{g}_{AB}}{\Lambda}\big(\fs_A\,\fs_B\big)_{120_\hh}\,\big(16^{\dagger}_\hh\,16^{\dagger}_\hh\big)_{120^{\dagger}_\hh}; \nl
     & &  \frac{\tilde{g}_{AB}}{\Lambda}\big(\fs_A\,16_\hh\big)_{\mathbf{120}}\,\big(\fs_B\,16_\hh\big)_{\mathbf{120}}; \;\text{and}\; \frac{\hat{g}_{AB}}{\Lambda}\big(\fs_A\,16_\hh\big)_{\mathbf{120}}\,\big(\fs_B\,16_\hh\big)^*_{\mathbf{120^{\dagger}}} \nl
     &\bullet &   \frac{\bar{f}_{AB}}{\Lambda}\big(\fs_A\,\,\fs_B\big)_{126^{\dagger}_\hh}\,\big(16^{\dagger}_\hh\,16^{\dagger}_\hh\big)_{126_\hh};\;\text{and} \;\frac{\hat{f}_{AB}}{\Lambda}\big(\fs_A\,\,16_\hh\big)_{\mathbf{126}^{\dagger}}\,\big(\fs_B\,16_\hh\big)^*_{\mathbf{126}} \nl
     & \bullet & \frac{k_{AB}}{\Lambda}\, \big(\fs_A\,16^{\dagger}_\hh\big)_{\mathbf{1}}\,\big(\fs_B\,16^{\dagger}_\hh\big)_{\mathbf{1}};\;\frac{\tilde{k}_{AB}}{\Lambda}\, \big(\fs_A\,16^{\dagger}_\hh\big)_{\mathbf{45}}\,\big(\fs_B\,16^{\dagger}_\hh\big)_{\mathbf{45}};\nl 
     & & \text{and}\; \frac{\bar{k}_{AB}}{\Lambda}\, \big(\fs_A\,16^{\dagger}_\hh\big)_{\mathbf{210}}\,\big(\fs_B\,16^{\dagger}_\hh\big)_{\mathbf{210}}.
\eeqa    
Here, \( g, \bar{g}, \tilde{g}, \hat{g}, \bar{f}, \hat{f}, k, \tilde{k}, \) and \( \bar{k} \) represent various Yukawa couplings, and \( \Lambda \) denotes the energy scale at which the intermediate \so\, irrep has been integrated out. It is to be noted that in Eq.~\eqref{eq:c4:16Fs16Hs}, we have not considered the combination $\big(\fs-\fs^{\dagger}\big)_{1,\,45,\,210}\,$\,$\big(...\big)_{1,\,45,\,210}$ as the coupling $\big(\fs-\fs^{\dagger}\big)$ is conventionally a part of gauge sector and are related to term mentioned in the last line of Eq.~(\ref{eq:c4:16Fs16Hs}) through Fierz transformations at $SO(10)$ level.

We begin with the decomposition of effective terms resulting from integrating out the $120$-dimensional irrep. Following the methodology used in Chapter~(\ref{ch:2}), we derive the expression below when $120$-dimensional irrep is integrated out;
\beqa {\label{eq:c4:gAB}}
-\lag^{120}_{\mathrm{NR}} &\supset & \frac{g_{AB}}{\Lambda}\Big( \fs^T_{A}\,\cc\, \fs_B\Big)_{120}\,\Big(16_\hh\,16_{\hh}\Big)_{120}\nl
\ad \frac{\bar{g}_{AB}}{\Lambda}            \Big( \fs^T_{A}\,\cc\, \fs_B\Big)_{120}\,\Big(16^{\dagger}_\hh\,16^{\dagger}_{\hh}\Big)_{120^{\dagger}}\,\nl   
            &+& \frac{\tilde{g}_{AB}}{\Lambda}\Big( \fs^T_{A}\,\cc\, 16_\hh\Big)_{\mathbf{120}}\,\Big( \fs_{B}\,\cc\, 16_\hh\Big)_{\mathbf{120}} \nl
            &+& \frac{\hat{g}_{AB}}{\Lambda}\Big(\fs^T_{A}\, \cc 16_\hh\big)_{\mathbf{120}}\,\Big(\fs_{B}\,\cc\,16_\hh\Big)^{*}_{\mathbf{120}^{\dagger}}
          \hc \nl 
          &=& \frac{-4\,g_{AB}}{3\Lambda}\,\Big[ 2 \big(\fn_A\,\cc\,\ff_{p\,B}\big)\,\big(10^{pq}_\hh\,5^{\dagger}_{q\,\hh}\big) - 4 \big(\ft_A^{pq\,T}\,\cc\,\ff_{p\,B}\big)\,\big(1_\hh\,5^{\dagger\,q}_\hh\big) \nl
          &+& 2 \big(\ff^T_{p\,A}\,\cc \ff_{q\,B}\big)\,\big(1_\hh\,10^{pq}_\hh\big)
          + 2 \big(\fn_A\,\cc\,10^{pq}_B\big)\,\big(5^{\dagger}_{p\,\hh}\,5^{\dagger}_{q\,\hh}\big)\nl
          &-&\frac{1}{2} \big(\varepsilon_{pqrst}\,\ft^{pq\,T}_{A}\,\cc\,\ft^{rn}_B\big)\,\big(5^{\dagger}_{n\,\hh}\,10^{st}_\hh\big)
          - \frac{1}{2} \big(\ff_{A\,n}\,\ft^{pq}_B\big)\,\big(\varepsilon_{pqrst}\,10^{rs}_\hh\,10^{tn}_\hh\big)\Big] \,\nl 
          &+& \frac{4\bar{g}_{AB}}{3\Lambda}\,\Big[ 4\big( \fn_{p\,A}^{T}\cc\,\ff_B\big)\,\big(1_\hh\,5^{p}_\hh\big)+ \big(\ft^{pq\,T}_A\cc\,\ff_{r\,B}\big)\,\big(10^{\dagger}_{pq\,\hh}\,5^{r}_\hh\big) \nl
          &+& 2 \big(\ff_{p\,A}\cc\,\ff_{q\,B}\big)\,\big(5^p_\hh\,5^q_\hh\big) + 2 \big(\fn_A^T\cc\,\ft^{pq}_{B}\big)\,\big(1_\hh\,10^{\dagger}_{pq\,\hh}\big)\nl \ad 2 \big(\ff_{p\,A}\cc\,\ft_B^{qr}\big)\,\big(5^{p}_\hh\,10^{\dagger}_{qr}\big)\nl
          &+& \frac{1}{8} \big(\varepsilon_{pqrst}\ft_A^{pq\,T}\cc\,\ft_B^{rn}\big)\,\big(\varepsilon^{xyzst}10^{\dagger}_{xy\,\hh}\,10^{\dagger}_{zn\,\hh}\big)\Big]\nl
          &-& \frac{4\tilde{g}_{AB}}{3\,\Lambda}\Big[ -2\big( \fn_A^T\,\cc 5^{\dagger}_{p\,\hh}\big)\,\big(\ft^{pq}_B\,5^{\dagger}_{q\,\hh}\big)-2 \big( \ff^T_{p\,A}\cc\,5^{\dagger}_{q\,\hh}\big)\,\big(\fn_B\,10^{pq}_\hh\big) \nl
          &-& \frac{1}{2}\big(\varepsilon_{pqrst}\,\ft_A^{pq\,T}\,\cc\,10^{rn}_\hh\big)\,\big(\ff_{n\,B}\,10^{st}_\hh\big)\Big]\nl
          &+& \frac{4\hat{g}_{AB}}{3\,\Lambda}\Big[ \big(\ft^{pq\,T}_A\cc\,5^{\dagger}_{p\,\hh}\big)\,(\ft_B^{sq}\,5^{\dagger}_{s\,\hh}\big)^* + 2\big(\ff^T_{p\,A}\cc\,10^{qr}_{\hh}\big)\,\big(\ff_{p\,B}\,\,10^{qr}_{\hh}\big)^* \nl
          &+& \big(\ff^T_{p\,A}\cc\,5^{\dagger}_{q\,\hh}\big)\,\big(\ff_{p\,B}\,\,5^{\dagger}_{q\,\hh}\big)^*\nl \ad \frac{1}{8}\,\big(\varepsilon_{pqrst}\,\ft^{pq\,T}_A\,\cc\,10^{rn}_\hh\big)\,\big(\varepsilon_{xyzst}\,\ft^{xy}_B\,\,\,10^{zn}_\hh\big)^*\Big] \hc
\eeqa
The above expression consists of various couplings between two $SU(5)$ fermion multiplets and two scalar multiplets, a result of integrating out the $120$ irrep. This equation depends on the Yukawa couplings $g$, $\bar{g}$, $\tilde{g}$, and $\hat{g}$, with the first two being antisymmetric in nature. In Eq.~\eqref{eq:c4:gAB}, the expression is invariant under \su\, and can be decomposed into effective couplings respecting \smg\,symmetry using Eqs.~(\ref{eq:c2:fn}, \ref{eq:c2:ff}, \ref{eq:c2:ft}, and \ref{eq:c2:SU(5)toSMscalars}).

We next address the computation of effective non-renormalisable coupling resulting from integrating out the degrees of freedom of $126$-dimensional irrep. Unlike the previous scenarios involving the $10_{\hh}$ and $120_\hh$ irreps, the $126$ allows fewer decomposition possibilities resulting in terms given in Eq.~\eqref{eq:c4:NRSO10}. This limitation stems from the transformation properties of the $126$-dimensional irrep~\cite{Slansky:1981yr} and is also used in Chapter~(\ref{ch:3}). The decomposition of Eq.~\eqref{eq:c4:NRSO10}, considering the possibility of \(\overline{126}\) serving both as a scalar and a fermion irrep, is shown as follows:
\beqa {\label{eq:c4:fAB}}
-\lag^{\overline{126}}_{\mathrm{NR}}     & \supset &  \frac{\bar{f}_{AB}}{\Lambda}\big(\fs^T_A\,\cc\,\fs_B\big)_{126^{\dagger}_\hh}\,\big(16^{\dagger}_\hh\,16^{\dagger}                               _\hh\big)_{126_\hh} \nl
              &+& \frac{\hat{f}_{AB}}{\Lambda}\big(\fs_A^T\,\cc\,16_\hh\big)_{\mathbf{126}^{\dagger}}\,\big(\fs_B\,     16_\hh\big)^*_{\mathbf{126}}  \hc\nl
             &=& \frac{-4\,\bar{f}_{AB}}{15\,\Lambda}\,\Big[ \big(\fn_A^T\,\cc\,\ft_B^{pq}\big)\,\big(1_\hh\,10^{\dagger}_{pq\,\hh}\big) - \big(\ff_{p\,A}\cc\,\ff_{q\,B}\big)\,\big(5^p_\hh\,5^q_\hh\big) \nl
             &+& \frac{3}{2} \big(\fn_A^T\,\cc\,\ff_{p\,B} +\frac{1}{24} \varepsilon_{qrstp}\,\ft^{qr}_A\,\cc\,\ft^{st}_B\big)\nl \, & & \hspace{6cm}\times\,\big(1_\hh\,5^p_\hh + \frac{1}{24} \varepsilon^{uvwxp} 10^{\dagger}_{uv\,\hh}\,10^{\dagger}_{wx\,\hh}\big) \nl
             &+& \big(\ft_A^{pq\,T}\,\cc\,\ff_{r\,B}\big)\,\big(10^{\dagger}_{pq\,\hh}\,5^k_\hh\big) + \big(\ff^T_{p\,A}\cc\,10^{qr}_{\hh}\big)\,\big(\ff_{p\,B}\,\,10^{qr}_{\hh}\big)^* \nl 
             \mi  \frac{1}{4\sqrt{3}}\,\big(\varepsilon_{pqnuv}\, \ft_{A}^{pq\,T}\,\cc\,\ft_B^{rs}\big)\,\big(\varepsilon^{xynuv}\,10^{\dagger}_{xy\,\hh}\,10^{\dagger}_{rs\,\hh}\big) \Big]\nl
             &+& \frac{4\hat{f}_{AB}}{15\,\Lambda}\,\Big[ \big(\ff^T_{p\,A}\,\cc\,5^{\dagger}_q\big)\,\big(\ff_{p\,B}\,\cc\,5^{\dagger}_q\big)^* + \big(\ft^{pq\,T}_A\cc\,5^{\dagger}_{p\,\hh}\big)\,(\ft_B^{sq}\,5^{\dagger}_{s\,\hh}\big)^* \nl
 \ad \frac{1}{288}\,\big( \varepsilon_{pqtuv}\,\ft^{pq\,T}_A\,\cc\,10_\hh^{rs}\big)\,\big( \varepsilon_{wxtuv}\,\ft^{wx}_B\,10_\hh^{rs}\big)^* \nl
             &+&  \frac{1}{288}\,\big( \varepsilon_{pqtuv}\,\ft^{rs\,T}_A\,\cc\,10_\hh^{pq}\big)\,\big( \varepsilon_{wxtuv}\,\ft^{rs}_B\,10_\hh^{wx}\big)^* \Big], \hc, 
\eeqa
where, $\bar{f}$ and $\tilde{f}$ are the Yukawa couplings. Further, the \su\, invariant expressions mentioned in the above Eq.~\eqref{eq:c4:fAB} can be decomposed further into the couplings involving SM fermions and scalars of Tab.~(\ref{tab:c2:16scalars}).

So far, we have considered the cases where two identical $16$-plets couples. Now we focus on the scenarios where one $16$-plet couples to its conjugate. This interaction leads to limited possible outcomes, as specified in Eq.~\eqref{eq:c2:16t16}.  In contrast to the scenarios involving $10$, $120$, and $126$ irreps, only the fermionic adjoint irreps can mediate this case when $16$ couples to its conjugate and the limited possibilities are shown in the last two lines of Eq.~\eqref{eq:c4:NRSO10}. The interaction between the $\fs$-plet and $\mathbf{45}$ can be parameterised as follows~\cite{Nath:2001yj};
\beqa{\label{eq:c4:1645}}
-\lag^{\mathbf{45}}_{Y} &\supset & \tilde{k}_{AB}\, \fs^T_A\,\cc\,\mathbf{45}_B\,16^{\dagger}_\hh \nl
&=& \frac{i}{\sqrt{2}}\,\tilde{k}_{AB}\, \Bigg[\sqrt{5}\Big( \frac{3}{5} \ff_{p\,A}^T\,\cc\,5^p_\hh + \frac{1}{10}\, \ft_A^{pq\,T}\,\cc\, 10^{\dagger}_{pq\,\hh} - \fn_A^T\,\cc\,1^{\dagger}_\hh\Big) \tilde{\fn}_{B} \nl
&+& \Big(-\ft_A^{st\,T}\,\cc\,1_\hh + \frac{1}{2} \varepsilon^{pqrst} \ff_{p\,A}^T\,\cc\,10^{\dagger}_{qr\,\hh}\Big) \big(\tilde{\ft}\big)^{*}_{st\,B}\nl
&+& \Big( -\fn^T\,\cc\,10^{\dagger}_{st\,\hh} + \frac{1}{2} \varepsilon_{pqrst}\, \ft^{pq\,T}_A\,\cc\,5^r_{\hh} \Big) \tilde{\ft}^{st}_{B} \nl
&+& 2 \Big( \ft^{pr\,T}_A\,\cc\,10^{\dagger}_{rq} - \ff^T_{q\,A}\,\cc\,5^{p}\big)\,\tilde{\mathbf{24}}^q_{p\,B}\Bigg]. \hc
\eeqa

Using the above $SU(5)$ invariant decomposition, we integrate out the degrees of freedom of ${\mathbf{45}}$ and substitute it back in the same equation, yielding the required effective non-renormalisable coupling and the resulting expression is shown below:
\beqa {\label{eq:c4:ktab}}
-\lag^{\mathbf{45}}_{\rm{NR}} &\supset & -\frac{\tilde{k}_{AB}}{\Lambda}\, \big(\fs^T_A\,\cc\,16^{\dagger}_\hh\big)_{\mathbf{45}}\,\big(\fs_B\,16^{\dagger}_\hh\big)_{\mathbf{45}} \nl
     &=& -\frac{\tilde{k}_{AB}}{2\Lambda}\Big[5\,\big( -\fn_A^T\,\cc\,1_\hh + \frac{1}{10}\,\ft_A^{pq\,T}\,\cc\,10^{\dagger}_{pq\,\hh} + \frac{3}{5}\, \ff^T_{p\,A}\,\cc\,5^{p}_\hh\big)\,\nl
     & & \hspace{4cm} \times\,\big( -\fn_B\,\,1_\hh  \frac{1}{10}\,\ft_B^{st}\,\,\,10^{\dagger}_{st\,\hh} +\frac{3}{5} \,\ff_{t\,B}\,\,\,5^{t}_\hh\big)\nl\ad
4\big(\ft_A^{pr\,T}\,\cc10^{\dagger}_{rq\,\hh}\big)\,\big(\ft^{qn}_B\,10^{\dagger}_{np\,\hh}\big)\nl 
      & - & 4 \big(\ft_A^{pr\,T}\,\cc10^{\dagger}_{rq\,\hh}\big)\,\big(\ff_{p\,B}\,5^q_\hh\big) -4  \big(\ff^T_{p\,A}\,\cc\,5^q_\hh\big)\,\big(\ft_B^{pr}\,\cc10^{\dagger}_{rq\,\hh}\big)\nl &+& 4 \big(\ff^T_{p\,A}\,\cc\,5^q_\hh\big)\,\big(\ff_{q\,B}\,\cc\,5^p_\hh\big) \nl
      &+& \big(-\ft_A^{pq\,T}\,\cc\,1_\hh + \frac{1}{2}\,\varepsilon^{rstpq}\,\ff_{A\,r}^{T}\,\cc\,10^{\dagger}_{st\,\hh}\big)\,\nl & & \hspace{4cm}\times\,\big(-\fn_B\,10^{\dagger}_{pq\,\hh} + \frac{1}{2} \varepsilon_{uvwpq}\ft^{uv}\,5^w_\hh\big) \Big]\nl \hcn 
\eeqa

Analogous to Eq.~\eqref{eq:c4:ktab}, we can compute the coupling of $\fs$ and $16^{\dagger}_{\hh}$ with $\mathbf{1}$ and $\mathbf{210}$. The effectively generated couplings mediated by these irreps are shown below; 
\beqa{\label{eq:c4:kAB}}
-\lag^{\mathbf{1}\&\mathbf{210}}_{\mathrm{NR}} &\supset& \frac{k_{AB}}{\Lambda}\, \big(\fs^T_A\,\cc\,16^{\dagger}_\hh\big)_{\mathbf{1}}\,\big(\fs_B\,16^{\dagger}_\hh\big)_{\mathbf{1}} 
\nl \ad   \frac{\bar{k}_{AB}}{\Lambda}\, \big(\fs^T_A\,\cc\,16^{\dagger}_\hh\big)_{\mathbf{210}}\,\big(\fs_B\,16^{\dagger}_\hh\big)_{\mathbf{210}} \hc\nl
     &=& -\frac{k_{AB}}{\Lambda}\,\Big[\big( \fn_A^T\,\cc\,1_\hh + \frac{1}{2}\,\ft_A^{pq\,T}\,\cc\,10^{\dagger}_{pq\,\hh} - \ff_{p\,A}\,\cc\,5^{p}_\hh\big)\,\nl & & \hspace{4cm}\,\times\, \big( \fn_A\,\,1_\hh 
       \frac{1}{2}\,\ft_B^{pq}\,\,10^{\dagger}_{pq\,\hh} - \ff_{p\,B}\,\,\,5^{p}_\hh\big)\Big] \nl
      &-& \frac{2\bar{k}_{AB}}{3}\,\Big[\frac{3}{2}\big(\ff^T_{p\,A}\,\cc\,5^q+\frac{1}{3}\ft^{qr\,T}_A\,\cc\,10^{\dagger}_{rp\,\hh}\big)\,\big(\ff_{q\,B}\,5^p+\frac{1}{3}\ft^{pn}_B\,10^{\dagger}_{nq\,\hh}\big)\nl
      &+& \frac{5}{8}\big(\fn_A^T\,\cc\,1_\hh \frac{1}{10}\ft^{pq\,T}\,\cc\,10^{\dagger}_{pq\,\hh}+\frac{1}{5}\ff^T_{p\,A}\,\cc\,5^p_\hh\big)\,\nl& & \hspace{4cm}\times\,\big(\fn_B\,1_\hh +\frac{1}{10}\ft^{st}_B\,10^{\dagger}_{st\,\hh}+\frac{1}{5}\ff_{s\,B}\,5^s_\hh\big)\nl
      &+& \frac{3}{4}  \big(\ft_A^{pq\,T}\,\cc\,1_\hh + \frac{1}{6}\,\varepsilon^{rstpq}\,\ff_{A\,r}^{T}\,\cc\,10^{\dagger}_{st\,\hh}\big)\,\nl & & \hspace{4cm}\times\,\big(\fn_B\,10^{\dagger}_{pq\,\hh} + \frac{1}{6} \varepsilon_{uvwpq}\ft^{uv}\,5^w_\hh\big)\nl
      &+& \frac{1}{36}\,\big(\varepsilon_{pqrst}\,\ft^{in\,T}_{A}\,\cc\,5^q_\hh\big)\,\big(\varepsilon^{uvrst}\,\ff_{u\,B}\,10^{\dagger}_{vn\,\hh}\big)\nl \ad\frac{1}{16}\big(\ft^{pq\,T}_A\,\cc\,10^{\dagger}_{st\,\hh}\big)\,\big(\ft^{st}_B\,10^{\dagger}_{pq\,\hh}\big)\nl
      &+& \big(\ff_A^T\cc\,1_\hh\big)\,\big(\fn_B\,5^{i}_\hh\big) \Big] \hc
\eeqa

The $SU(5)$ invariant expressions given in Eqs. (\ref{eq:c2:h10-1}, \ref{eq:c2:h10b}, \ref{eq:c2:htAB}, \ref{eq:c4:gAB}, \ref{eq:c4:fAB}, \ref{eq:c4:ktab}, and \ref{eq:c4:kAB}) illustrate various $SU(5)$ invariant expressions resulting from the decomposition of the term given in Eq.~\eqref{eq:c4:NRSO10}. This demonstrates that an $SO(10)$ invariant combination—comprising two $\fs$-plets and $16_{\hh}$, along with its conjugate—can be broken down at the leading non-renormalisable level into multiple $SU(5)$ invariant forms. These forms stem from different methods of contracting the same $SO(10)$ invariant term. In section~(\ref{sec:c2:NRSo}), we have further decomposed the evaluated \(SU(5)\) invariant expressions into expressions adhering to SM gauge symmetry. Here, we omit all the further decomposition of above \su\, decomposition into expression respecting \sm\, gauge symmetry. As evident from various decomposed expressions, all the effective couplings are mass dimension five, and these can have significant phenomenological implications, from affecting the masses of fermions (both charged and neutral) to inducing baryon number-violating processes, such as nucleon decay.

As we have done in Chapter~(\ref{ch:3}), we aim to constrain the spectrum of scalar residing in $16_{\hh}$ (cf. Tab.~(\ref{tab:c2:16scalars})), through proton decays. As we have learnt from Chapter~(\ref{ch:3}), the relevant scalar (here, pair of scalar) must possess both leptoquark and diquark couplings to induce proton decay. In Tab.~(\ref{tab:c4:scalarclassification}), we gather the couplings of respective scalar pairs with SM fermions, derived from the decomposition of Eqs.~(\ref{eq:c2:h10-1}, \ref{eq:c2:h10b}, \ref{eq:c2:htAB}, \ref{eq:c4:gAB}, \ref{eq:c4:fAB}, \ref{eq:c4:ktab}, and \ref{eq:c4:kAB}). For brevity, we suppress the $SU(3)$, $SU(2)$, and family indices while mentioning the couplings in Tab.~(\ref{tab:c4:scalarclassification}).
\begin{table}[t]
    \centering
    \begin{tabular}{ccc}
    \hline
        ~~~Pair of Scalar~~~ & ~~~Diquark~~~ & ~~~Leptoquark~~~ \\
    \hline\hline
     $\hat{\sigma} - \hat{\sigma}$     & \xmark & \xmark \\
     $\hat{\sigma} - \hat{T}$ & $q\,q; u^C\,d^C$  & $q\,l; u^C\,e^C$  \\
     $\hat{\sigma} - \hat{H}$& $q\,u^C;\,q\,d^C$ & \xmark \\
     $\hat{\sigma} - \hat{\Delta}$& \xmark & $d^C\,l$ \\
     $\hat{\sigma} - \hat{\Theta}$ & $d^C\,d^C$  & \xmark \\
     $\hat{\sigma} - \hat{t}$& \xmark & \xmark \\
     $\hat{T} - \hat{T}$& $d^C\,d^C$ & \xmark \\
     $\hat{T} - \hat{H}$& $q\,u^{C\,*}$ & $q\,e^{C\,*}$  \\
     $\hat{T} - \hat{\Delta}$& $q\,d^C$ & \xmark \\
     $\hat{T} - \hat{\Theta}$& $q\,q; u^C\,d^C$  & $q\,l; u^C\,e^C$\\
     $\hat{T} - \hat{t} $& \xmark & $d^C\,e^C$  \\
     $\hat{H} - \hat{H}$& $q\,q$;\, $d^C\,d^C$ & \xmark \\
     $\hat{H} - \hat{\Delta}$& $q\,q; u^C\,d^C$  & $q\,l; u^C\,e^C$\\
     $\hat{H} - \hat{\Theta}$& \xmark & $l\,u^C$ \\
     $\hat{H} - \hat{t}$& $q\,d^C$ & \xmark \\
     $\hat{\Delta} - \hat{\Delta}$& $q\,q;\;u^C\,d^C$ & $q\,l;\;u^C\,e^C$ \\
     $\hat{\Delta} - \hat{\Theta}$& $q\,u^{C\,*}$ & $q\,e^{C\,*}$ \\
     $\hat{\Delta} - \hat{t}$& $q\,u^{C\,*}$ & $q\,e^{C\,*}$   \\
     $\hat{\Theta} - \hat{\Theta}$& $u^C\,d^C$;\, $d^C\,d^C$ & \xmark  \\
     $\hat{\Theta} - \hat{t}$& $q\,q; u^C\,d^C$  & $q\,l; u^C\,e^C$\\
     $\hat{t} - \hat{t}$& \xmark & \xmark \\
     \hline
    \end{tabular}
    \caption{The coupling of different pairs of scalars residing in $16_{\hh}$ with SM fermions. The symbol “\xmark” indicates the absence of the respective coupling (diquark or leptoquark).}
    \label{tab:c4:scalarclassification}
\end{table}
Tab. (\ref{tab:c4:scalarclassification}) summarises the various couplings of pair of scalars with the SM fermions and some notable features regarding it, are as follows:
\begin{itemize}
    \item The depiction of different scalar pairs in Tab.~(\ref{tab:c4:scalarclassification}) represents all possible configurations appearing in the Eqs~. (\ref{eq:c2:h10-1}, \ref{eq:c2:h10b}, \ref{eq:c2:htAB}, \ref{eq:c4:gAB}, \ref{eq:c4:fAB}, \ref{eq:c4:ktab}, and \ref{eq:c4:kAB}). For instance, $\hat{\sigma} - \hat{T}$ denotes $\hat{\sigma} - \hat{T}$, $\hat{\sigma} - \hat{T}^*$, $\hat{\sigma}^* - \hat{T}$, and $\hat{\sigma}^* - \hat{T}^*$.
    \item Scalar coupling to its conjugate partner can interact with any SM fermion and its conjugate. Consequently, such interactions are not listed in Tab.~(\ref{tab:c4:scalarclassification}).
    \item The couplings involving $\nu^C$ are excluded from Tab.~(\ref{tab:c4:scalarclassification}) as they do not contribute to the tree-level proton decay at dimension six.
    \item Among the twenty scalar pairings identified, only eight possess the necessary leptoquark and diquark couplings capable of inducing proton decay. These pairs are $\hat{\sigma} - \hat{T}$, $\hat{T} - \hat{H}$, $\hat{T} - \hat{\Delta}$, $\hat{H} - \hat{\Delta}$, $\hat{\Delta} - \hat{\Delta}$, $\hat{\Delta} - \hat{\Theta}$, $\hat{\Delta} - \hat{t}$, and $\hat{\Theta} - \hat{t}$.
    \item It is evident that $\hat{\Delta}$ can couple with any scalar listed in Tab.~(\ref{tab:c2:16scalars}), and can generate both diquark and leptoquark vertices.
\end{itemize}

Having summarised the diquark and leptoquark interactions of various scalar pairs derived from the decomposed $SU(5)$ invariant expressions, we will compute their specific contributions to the various effective nucleon decay operators through their Wilson coefficients.

\section{Dimension Six Effective Operators} {\label{sec:c4:pdeff}}

The SM gauge symmetry-preserving and $B-L$ conserving ( or $B+L$ violating) effective mass dimension-six ($D=6$) four-fermion operators, which can induce nucleon decay (whether of a free proton or a bound neutron), are given below~\cite{PhysRevLett.43.1566,PhysRevLett.43.1571,Sakai:1981pk}. As discussed in section~(\ref{sec:c3:geninfo}), these operators mediate two-body nucleon decays, where the nucleon decays into either charged anti-leptons or neutral ones, along with mesons. The vertices inducing these decays are prohibited in the SM, as SM preserves the Baryon number. Thus, they emerge from integrating out non-SM particle degrees of freedom, either scalar or vector.
\beqa \label{eq:c4:bnc}
\mathcal{O}_{1} &=& c_1[A,B,C,D]\,\varepsilon_{\alpha\beta\gamma}\,\varepsilon_{gh}\,\varepsilon_{ab}\,\Big(q^{g\alpha\,T}_{A}\,\cc q_B^{h\,\beta}\Big)\,\Big( q^{a\,\gamma\,T}_C\,\cc l^b_D\Big),\nonumber\\
\mathcal{O}_2 &=& c_2[A,B,C,D]\,\varepsilon_{\alpha\beta\gamma}\,\varepsilon_{gh}\,\Big(q^{g\alpha\,T}_{A}\,\cc\, q_B^{h\,\beta}\Big)\,\Big( u^{C\,T}_{\,\gamma\,C}\,\cc e^{C}_D\Big)^{*},\nonumber\\
\mathcal{O}_3 &=& c_3[A,B,C,D]\,\varepsilon^{\alpha\beta\gamma}\,\varepsilon^{ab}\, \Big(u^{C\,T}_{\alpha\,A}\cc\, d^C_{\beta\,B}\Big)\,\Big(q^{a\,\gamma\,T}_C\cc\,l^b_D\Big)^*,\nonumber\\
\mathcal{O}_4 &=& c_4[A,B,C,D]\,\varepsilon^{\alpha\beta\gamma}\, \Big(u^{C\,T}_{\alpha\,A}\cc\, d^C_{\beta\,B}\Big)\,\Big( u^{C\,T}_{\,\gamma\,C}\,\cc e^{C}_D\Big).
\eeqa
The coefficients $c_{1,2,3,4}$ represent various Wilson coefficients associated with newly introduced degrees of freedom, reflecting the effective strength of each respective operator. Typically, these coefficients depend on the Yukawa couplings and are inversely proportional to the mass of the mediating particle(s). $A$, $B$, $C$, and $D$ are the generation indices for the SM fermions involved in these processes. Additionally, only up-quarks can contribute to proton decay, while down-quarks and charged leptons can participate up to the first two generations, and all three generations of neutral fermions are permitted.

\begin{figure}[t]
    \centering
    \includegraphics[width=0.85\linewidth]{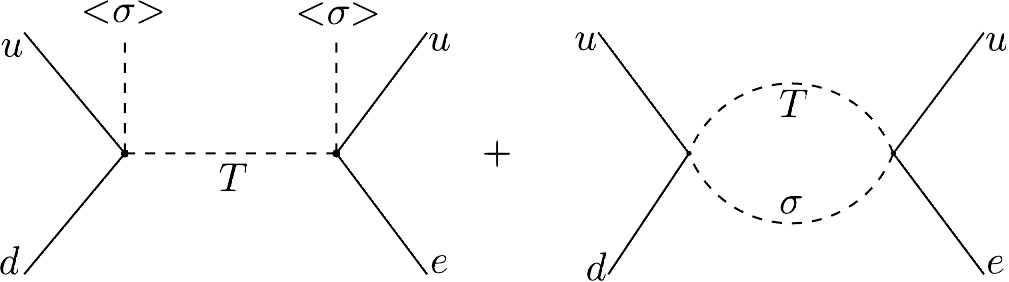}
    \caption{Tree and loop level proton decay topologies generated by the pair $\sigma-T$, drawn using \cite{Harlander:2020cyh}.}
    \label{fig:c4:pdecayconfig}
\end{figure}

The Lagrangian, given in Eqs.~(\ref{eq:c2:h10-1}, \ref{eq:c2:h10b}, \ref{eq:c2:htAB}, \ref{eq:c4:gAB}, \ref{eq:c4:fAB}, \ref{eq:c4:ktab}, and \ref{eq:c4:kAB}), consists of non-renormalisable effective couplings which are capable of giving rise to the $D=6$ operators listed in Eq.~(\ref{eq:c4:bnc}) through two primary topologies: 1) Integrating out a pair of scalars which leads to the formation of $D=8$ operators, while another pair of scalars simultaneously acquires a $vev$, effectively reducing it to a $D=6$ operator (as shown in the left diagram of Fig.~(\ref{fig:c4:pdecayconfig})); 2) The two pairs of scalars producing $D=6$ operators propagating inside the loop (as depicted in the right diagram of Fig.~(\ref{fig:c4:pdecayconfig})). These $D=6$ effective operators might originate from specific terms in Eqs.~(\ref{eq:c2:h10-1}, \ref{eq:c2:h10b}, \ref{eq:c2:htAB}, \ref{eq:c4:gAB}, \ref{eq:c4:fAB}, \ref{eq:c4:ktab}, and \ref{eq:c4:kAB}), or from a combination of them. Integrating out various irreps will not necessarily yield all four fermion operators of Eq.~\eqref{eq:c4:bnc}. Other $B-L$ conserving $D=8$ operators provided in Eq.~\eqref{eq:c4:bnc} might induce proton decays involving multi-meson modes, which are phase space suppressed compared to two-body decays and are not discussed in this analysis~\cite{Helo:2019yqp,Heeck:2019kgr}.

We compute the strengths of various $D=6$ nucleon decay operators. While deriving these contributions, we have taken the contribution of one specific \so\, irrep mediating the decomposition of the term given in Eq.~\eqref{eq:c4:NRSO10} at a time. However, in principle, similar operators generated by the same scalar pair from different \so-mediated irreps could mix with each other, a possibility that we have not considered.


We start by examining the contribution of the \(\hat{\sigma} - \hat{T}\) pair, which is capable of inducing proton decay as it couples to both leptoquark and diquark, as outlined in Tab.~(\ref{tab:c4:scalarclassification}). This scalar pair can induce proton decay at both tree and loop levels, as depicted in Fig. (\ref{fig:c4:pdecayconfig}). Integrating out the degrees of freedom of $\hat{T}$ generates a \(D=8\) operator. Further, \(\hat{\sigma}\), which is singlet under SM gauge symmetry, acquiring a $vev$ reduces this \(D=8\) effective operator to a \(D=6\) operator, given in Eq.~\eqref{eq:c4:bnc}. Additionally, this pair also mediates proton decay at the loop level, with \(\hat{\sigma}\) and \(\hat{T}\) propagating within the loop. The contributions to the effective strengths of the operators, \(\mathcal{O}_{1,2,3,4}\), in the flavour basis, mediated by the \(\hat{\sigma} - \hat{T}\) pair, are shown below;
\beqa \label{eq:c4:p1}
c_1\big[A,B,C,D\big] &=& \bigg(\frac{\langle\hat{\sigma}\rangle ^2}{\Lambda^2\,M_{\hat{T}}^2} + \frac{D\big(M^2_{\hat{T}},M^2_{\hat{\sigma}}\big)}{\Lambda^2}\bigg)\,                           \Big[32 \bar{h}_{AB}\,h_{CD}   \Big],\nonumber\\
c_2\big[A,B,C,D\big] &=& \bigg(\frac{\langle\hat{\sigma}\rangle^2}{\Lambda^2\,M_{\hat{T}}^2} + \frac{D\big(M^2_{\hat{T}},M^2_{\hat{\sigma}}\big)}{\Lambda^2}\bigg)\,                           \Big[ 32 \, \bar{h}_{AB}\, \bar{h}^*_{CD} - \frac{8}{2025} f_{AB}\, f^*_{CD}                                   \nonumber\\
                     &+&  
                      2\, \tilde{k}_{AB}\,\tilde{k}^*_{CD} + \frac{16}{9} \bar{k}_{AB}\,\big(\bar{k}^*_{CD}+\bar{k}^*_{DC}\big)\Big],\nonumber\\
c_3 \big[A,B,C,D\big] &+& \bigg(\frac{\langle\hat{\sigma}\rangle^2}{\Lambda^2\,M_{\hat{T}}^2} + \frac{D\big(M^2_{\hat{T}},M^2_{\hat{\sigma}}\big)}{\Lambda^2}\bigg)\,                           \Big[ 64\,h_{AB}\,h^*_{CD} + 36\, \bar{h}_{AB}\, \bar{h}^*_{CD} \nl \ad \frac{128}{9}\, g_{AB}\,g_{CD} \Big],\nonumber\\
c_4 \big[A,B,C,D\big] &=& \bigg(\frac{\langle\hat{\sigma}\rangle^2}{\Lambda^2\,M_{\hat{T}}^2} + \frac{D\big(M^2_{\hat{T}},M^2_{\hat{\sigma}}\big)}{\Lambda^2}\bigg)\,                            \big( 64\ h_{AB}\,\bar{h}_{CD}  \big),
\eeqa
where, \(M_{\hat{T}}\) and \(M_{\hat{\sigma}}\) represent the masses of the Triplet and \(\hat{\sigma}\), respectively. The terms within the parentheses in Eq.~(\ref{eq:c4:p1}) comprise two distinct contributions: the first term is due to the tree-level contribution, while the second term arises from loop-level processes. Typically,  \(\langle\hat{\sigma}\rangle \neq M_{\hat{\sigma}}\), and the precise relationship between \(\hat{\sigma}\) and its mass depend on the details of the scalar potential. Furthermore, the effective strengths of these operators are influenced by the Yukawa couplings and the masses of the mediators involved. From the coefficients shown in Eq.~\eqref{eq:c4:p1}, it is evident that the \(\hat{\sigma} - \hat{T}\) pair can generate the generation of all four operators listed in Eq.~\eqref{eq:c4:bnc}. Moreover, although \(\hat{\sigma}\) carries a \(B-L\) charge (cf. Tab.~(\ref{tab:c2:16scalars})), the operators resulting from \(\hat{\sigma}\) acquiring a $vev$ conserve \(B-L\) and appear at mass-dimension six. 

 The loop contribution of the pair $\hat{\sigma}-\hat{T}$, given in Eq.~\eqref{eq:c4:p1} depend upon the loop integral, which is shown below;
\beqa \label{eq:c4:lf}
D\big(M_1^2,M_2^2\big) &\equiv& -\frac{1}{16\pi^2}\Bigg( \frac{M_1^2\log\Big(\frac{M_1^2}{\mu^2}\Big)-M_2^2\log\Big(\frac{M_2^2}{\mu^2}\Big)}{M_1^2-M_2^2}-1\Bigg). 
\eeqa
The above loop function is symmetric under the exchange of $M_1$ and $M_2$, where $M_1$ and $M_2$ are the masses of scalars propagating inside the loop. Furthermore, the renormalisation scale, $\mu$, can be chosen as $1$ GeV, corresponding to the scale at which the proton at rest can decay. Ideally, all parameters contributing to proton decay should be renormalised down to this scale ($1$ GeV).

Next, we consider the pair \(\hat{T}-\hat{\Theta}\), which also possesses both diquark and leptoquark couplings. The contribution of this pair to the coefficients of the effective \(D=6\) operators is shown below; 
\beqa \label{eq:c4:p2}
c_1 \big[A,B,C,D\big] &=& \frac{1}{\Lambda^2}\, D\big(M_{\hat{T}}^2,M^2_{\hat{\Theta}}\big)\,\big( -32\,h_{AB}\,h_{CD} \big),\nonumber\\
c_2 \big[A,B,C,D\big] &=& \frac{1}{\Lambda^2}D\big(M_{\hat{T}}^2,M^2_{\hat{\Theta}}\big)\,\Big[ -64\,h_{AB}\,\bar{h}^*_{CD} -8\,\tilde{h}_{AB}\,\tilde{h}^*_{CD} \Big], \nonumber\\
c_3 \big[A,B,C,D\big] &=& \frac{1}{\Lambda^2}D\big(M_{\hat{T}}^2,M^2_{\hat{\Theta}}\big)\,\Big[ \frac{-64}{9}\,\bar{g}_{AB}\,\bar{g}^*_{CD}\nl \mi 2\big(\tilde{k}_{AB}+ \frac{3}{10}\tilde{k}_{BA}\big)\, \big( \frac{1}{2}\tilde{k}^*_{CD} \tilde{k}^*_{DC}\big)\nl &-& \,\frac{2}{3}\, \left(\bar{k}_{AB}-\frac{5}{4}\bar{k}_{BA}\right)\,\left(\bar{k}^*_{CD}-4\bar{k}_{DC}\right) \Big], \nl
c_4 \big[A,B,C,D\big] &=& \frac{1}{\Lambda^2}D\big(M_{\hat{T}}^2,M^2_{\hat{\Theta}}\big)\,\big( -\frac{64}{9}\, \bar{g}_{AB}\,g_{CD} \big), 
\eeqa
where \(M_{\hat{T}}\) and \(M_{\hat{\Theta}}\) represent the masses of \(\hat{T}\) and \(\hat{\Theta}\), respectively. Neither of the scalars $\hat{T}-\hat{\Theta}$ can acquire a $vev$, it would violate the \(SU(3)_{\rm{C}} \times U(1)_{\mathrm{EM}}\) symmetry. Consequently, this pair can generate proton decay only at the loop level and contribute to the coefficients of all the four operators mentioned in Eq.~\eqref{eq:c4:bnc}.

Next, we consider when the pair $\hat{T} - \hat{H}$ mediates the proton decay;
\beqa {\label{eq:c4:p3}}
c_2[A,B,C,D] &=&  -64\,\bigg(\frac{\langle \hat{H}\rangle^2}{\Lambda^2\,M_{\hat{T}}^2} + \frac{D\big(M^2_{\hat{H}},M^2_{\hat{T}}\big)}{\Lambda^2}\bigg)\,\big(\hat{h}_{AC}+\hat{h}^*_{CA}\big)\,\big(\hat{h}_{BD}+\hat{h}^*_{DB}\big).\nl
\eeqa
\(M_{\hat{H}}\) represents the mass of \(\hat{H}\). Although \(\hat{H}\) typically transforms like the SM Higgs, it is charged under \(B-L\) (cf. Tab.~(\ref{tab:c2:gscalars})), which restricts \(\hat{H}\) from acquiring a large $vev$. This scalar pair can mediate proton decay at tree and loop levels, contributing to effective \(D=6\) operators. However, it specifically generates only one \(D=6\) effective operator, \({\mathcal{O}}_2\). The derivation of the expression in Eq.~\eqref{eq:c4:p3} utilises the Fierz transformation rules for two-component spinors, as outlined in Eq.~\eqref{eq:c3:fierz}~\cite{Dreiner:2008tw}.

We further consider when the $D=6$ operators are generated by the pair $\hat{H} - \hat{\Delta}$. The contribution of these pair of scalars to the strength of effective $D=6$ operators is shown below,
\beqa \label{eq:c4:p4}
c_1 \big[A,B,C,D\big] &=& \left( \frac{\langle \hat{H}\rangle^2}{\Lambda^2\,M^2_{\hat{{\Delta}}}} + \frac{D\big(\mh,\md\big)}                   {\Lambda^2}\right)\,\left( 32\,h_{AB}\,\bar{h}_{CD} -           \frac{64\sqrt{2}}{9}\,g_{AB}\,\bar{g}_{CD}   \right),\nonumber\\
c_2 \big[A,B,C,D\big] &=& \Big( \frac{\langle \hat{H}\rangle^2}{\Lambda^2\,M^2_{\hat{\Delta}}} + \frac{D\big(\mh,\md\big)}{\Lambda^2}\Big)\,\Bigg( 16\,h_{AB}\,h^*_{CD} \nl \ad \left.\frac{16}{9} g_{AB}\,g^*_{CD} + 32\,\tilde{h}_{AB}\,\tilde{h}^*_{CD} \right), \nonumber\\
c_3 \big[A,B,C,D\big] &=& \Big( \frac{\langle \hat{H}\rangle^2}{\Lambda^2\,M^2_{\hat{\Delta}}} + \frac{D\big(\mh,\md\big)}{\Lambda^2}\Big)\,\Bigg[ 32\,\bar{h}_{AB}\,\bar{h}^*_{CD} + \frac{64}{9}\,\bar{g}_{AB}\,g^*_{CD} \big) \nl
                        &+&  \left(\tilde{k}_{AB}-2\sqrt{2}\tilde{k}_{BA}\right)\,\left( \left(\frac{1}{\sqrt{2}}-\frac{3}{20}\right)\tilde{k}^*_{CD}\right. \nl \ad \left. \left(\sqrt{2}-\frac{3}{20}\right)\tilde{k}^*_{DC}\right)                         -  2\left(\frac{1}{\sqrt{2}}\bar{k}_{AB}-\frac{2\sqrt{2}}{3}\bar{k}_{BA}\right)\nl 
                        & & \hspace{1cm}\times\;\left( \sqrt{2}\big(k_{CD}+2 k_{DC}\big) - \big( \frac{\sqrt{2}}{48} \bar{k}_{DC}+\frac{4\sqrt{2}}{3}\bar{k}_{CD}\big)\right) \Bigg],\nl
c_4 \big[A,B,C,D\big] &=& \Big( \frac{\langle \hat{H}\rangle^2}{\Lambda^2\,M^2_{\hat{\Delta}}} + \frac{D\big(\mh,\md\big)}{\Lambda^2}\Big)\,\left( 32 \bar{h}_{AB}\,h_{CD} -\frac{32}{9}\,g_{AB}\,g_{CD} \right). \nl
\eeqa
$M_{\hat{H}}$, and $M_{\hat{\Delta}}$ are the masses of $\hat{H}$ and $\hat{\Delta}$ respectively. This pair mediates proton decays at tree and loop levels and contributes to all four effective $D=6$ proton decay operators. 

Further, the pair $\hat{\Delta} - \hat{\Delta}$ can only induce loop-level proton decay, as shown below;
\beqa \label{eq:c4:p5}
c_1 \big[A,B,C,D\big] &=& \frac{1}{\Lambda^2}\, D\big(M_{\hat{\Delta}}^2,M^2_{\hat{\Delta}}\big)\,\left( - 64                                \bar{h}_{AB}\,h_{CD} + \frac{10}{9}\,\bar{g}_{AB}\,g_{CD}  \right),\nl
c_2 \big[A,B,C,D\big] &=& \frac{1}{\Lambda^2}D\big(M_{\hat{\Delta}}^2,M^2_{\hat{\Delta}}\big)\,\Big[                                       -16\sqrt{2}\,\bar{h}_{AB}\,\bar{h}^*_{CD} + \frac{10}{9}\, \bar{g}                              _{AB}\,g^*_{CD} \nl \ad \frac{8}{405}\,f_{AB}\,f^*_{CD} + \frac{36\sqrt{2}}{5}\, \tilde{k}_{AB}\,\tilde{k}^*_{DC}\nl \ad 64 \left(\frac{1}{\sqrt{2}}-\frac{16}{3}\right)\bar{k}_{AB}\,\bar{k}^*_{DC} \bigg],\nl
c_3 \big[A,B,C,D\big] &=& \frac{1}{\Lambda^2}D\big(M_{\hat{\Delta}}^2,M^2_{\hat{\Delta}}\big)\,\left( 64 h_{AB}                            \,h^*_{CD} + \frac{8}{3}\, g_{AB}\,\bar{g}_{CD} -8 g_{AB}\,g^*_{CD}\right. \nl \ad \left.\frac{8}{9}\,\tilde{g}_{AB}\,\tilde{g}^*_{CD}\right),\nonumber\\
c_4 \big[A,B,C,D\big] &=& \frac{8}{\Lambda^2}D\left(M_{\hat{\Delta}}^2,M^2_{\hat{\Delta}}\right)\,\left( 16 h_{AB}\,                           \bar{h}_{CD} - \frac{8}{3}\, g_{AB}\,\bar{g}_{CD} \right). 
\eeqa
This particular pair generates all the effective $D=6$ operators mentioned in Eq.~\eqref{eq:c4:bnc}. 

Considering the case of the pair $\Delta-\Theta$, which mediates loop level proton decay and generates only operator $\mathcal{O}_2$, as shown below;
\beqa {\label{eq:c4:p6}}
c_2[A,B,C,D] &=&  \frac{D\big(M^2_{\hat{\Delta}},M^2_{\hat{\Theta}}\big)}{\Lambda^2}\,\Big[\big(\hat{h}_{AD}+\hat{h}^*_{DA}\big)\,\big(\hat{h}_{CB}+\hat{h}^*_{BC}\big)\nl \ad \frac{16\sqrt{2}}{9}\hat{g}_{AC}\big(\hat{g}_{BD} + \hat{g}^*_{BD}\big) + \frac{16\sqrt{2}}{45}\,\hat{f}_{AD}\,\hat{f}_{BC}\Big].
\eeqa

Along a similar line, the pair $\hat{\Delta} - \hat{t}$ induces proton decay at one-loop and contributes to the effective strength of the operator ${\mathcal{O}_2}$; 
\beqa {\label{eq:c4:p7}}
c_2[A,B,C,D] &=&  \frac{D\big(M^2_{\hat{{\Delta}}},M^2_{\hat{t}}\big)}{\Lambda^2}\,\Big[ \frac{8\sqrt{2}}{9}\hat{g}_{AC}\big(\hat{g}_{BD} + \hat{g}^*_{BD}\big) - \frac{8\sqrt{2}}{45}\,\hat{f}^*_{DA}\,\hat{f}^*_{BC}\Big].\nl
\eeqa
where, $M_{\hat{t}}$ denotes the mass of the scalar $\hat{t}$.

To conclude, we evaluated the contribution to operators, $\mathcal{O}_{1,2,3,4}$  generated by the pair $\Theta-t$, as shown below; 
\beqa \label{eq:c4:p8}
c_1 \big[A,B,C,D\big] &=& \frac{1}{\Lambda^2}\, D\big(M_{\hat{\Theta}}^2,M^2_{\hat{t}}\big)\,\left(16 \bar{h}                              _{AB}\,h_{CD} + \frac{10}{9}\,\bar{g}_{AB}\,g_{CD}  \right),\nonumber\\
c_2 \big[A,B,C,D\big] &=& \frac{1}{\Lambda^2}D\big(M_{\hat{\Theta}}^2,M^2_{\hat{t}}\big)\,\bigg[ 16\,\bar{h}                               _{AB}\,\bar{h}^*_{CD} + \frac{10}{9}\, \bar{g}                                                _{AB}\,g^*_{CD} \nl \ad \frac{8}{405}\,f_{AB}\,f^*_{CD}
                      -   \frac{8}{20}\ \tilde{k}_{AB}\,\big(\tilde{k}^*_{CD}+\tilde{k}^*_{DC}\big) \nl \mi \frac{16}{15} \bar{k}_{AB}\,\left(\frac{29}{4}\,\bar{k}^*_{CD} + \bar{k}^*_{DC}\right)\bigg],\nl
c_3 \big[A,B,C,D\big] &=& \frac{1}{\Lambda^2}D\big(M_{\hat{\Theta}}^2,M^2_{\hat{t}}\big)\,\left( 64 h_{AB}                                 \,h^*_{CD}  + \frac{8}{3}\, g_{AB}\,\bar{g}^*_{CD} + 4 g_{AB}                                       \,g^*_{CD}\right.\nl \mi \left. \frac{16}{9}\tilde{g}_{AB}\,\tilde{g}^*_{CD}\right),\nonumber\\
c_4 \big[A,B,C,D\big] &=& \frac{1}{\Lambda^2}D\big(M_{\Theta}^2,M^2_t\big)\,\left( 32 h_{AB}\,                               \bar{h}_{CD} - \frac{8}{3}\, g_{AB}\,\bar{g}_{CD} \right). 
\eeqa

The expressions derived in Eqs.~(\ref{eq:c4:p1}, \ref{eq:c4:p2}, \ref{eq:c4:p3}, \ref{eq:c4:p4}, \ref{eq:c4:p5}, \ref{eq:c4:p6}, \ref{eq:c4:p7}, and \ref{eq:c4:p8}) quantify the contributions of various scalar pairs capable of mediating nucleon decays that violate $B$ and $L$ numbers while conserving \(B-L\). These contributions are influenced by the scalar masses and the associated Yukawa couplings. The following points summarise, mentioned in (\ref{pt:c4:ia}-\ref{pt:c4:ig}), the key inferences of these evaluated strengths of the operators mentioned in Eq.~\eqref{eq:c4:bnc}:
\begin{enumerate}[label=(\roman*)]
    \item Only three scalar pairs can mediate nucleon decay at the tree level: $\hat{\sigma}-\hat{T}$, $\hat{H}-\hat{T}$, and $\hat{H}-\hat{\Delta}$. Both $\hat{\sigma} - \hat{T}$ and $\hat{H} - \hat{\Delta}$ can contribute to all the $D=6$ operators, whereas $\hat{H}-\hat{T}$ exclusively contributes to $\mathcal{O}_2$. Consequently, it only induces proton decay modes involving charged antileptons accompanied with neutral mesons, such as $p\to e^+ \pi^0$ and $p\to \mu^+\,K^0$, among others. \label{pt:c4:ia}
    \item Five additional pairs are capable of inducing nucleon decay only at the loop level: $\hat{T} - \hat{\Theta}$, $\hat{\Delta} - \hat{\Delta}$, $\hat{\Delta} - \hat{\Theta}$, $\hat{\Delta} - \hat{t}$, and $\hat{\Theta} - \hat{t}$. Of these, $\hat{T} - \hat{\Theta}$, $\hat{\Delta} - \hat{\Delta}$, and $\hat{\Theta} - \hat{t}$ contribute to all four $B-L$ conserving operators, potentially resulting in proton decay into charged and neutral antileptons. Additionally, $\hat{\Delta} - \hat{\Theta}$ and $\hat{\Delta} - \hat{t}$ only generate $\mathcal{O}_2$. \label{pt:c4:ib}
    \item Terms associated with the Yukawa couplings $\hat{h}$, $\hat{g}$, and $\hat{f}$ only contribute to $\mathcal{O}_2$, thus can only contribute to the modes in which the proton decays into charged anti-leptons together with mesons. Hence, it offers a unique signature of its detectability. \label{pt:c4:ic}
    \item Expressions with $\tilde{h}$ and $\tilde{g}$ Yukawa couplings contribute to both $\mathcal{O}_{2}$ and $\mathcal{O}_{3}$. \label{pt:c4:id}
    \item Expressions where the intermediate integrated out irreps are $10$ and $120$ dimensional, it generates all the four operators given in Eq.~\eqref{eq:c4:bnc}. In contrast, those involving ${\bf 45}$ and ${\bf 210}$ induces only $\mathcal{O}_2$ and $\mathcal{O}_3$, and the integration of $\overline{126}$-dimensional tensor results in only $\mathcal{O}_{2}$. \label{pt:c4:ie}
    \item The term with Yukawa coupling $k$ does not contribute to nucleon decay, as it solely gives rise to diquark or dilepton vertices. \label{pt:c4:if}
    \item As we have studied in Chapter~(\ref{ch:3}), in renormalisable class \so\, GUTs only two scalars with SM charges $(3,1,\frac{1}{3})$ and $(3,3,-\frac{1}{3})$ together with their conjugate partners can induce $B-L$ conserving tree-level proton decay. Additionally, $B-L$ violating modes at the tree level are mediated by the pair $(3,2,\frac{1}{6}) - (3,1,\frac{1}{3})$ and its conjugate partner~\cite{Patel:2022wya}. In non-renormalisable GUTs, however, the lowest dimension for tree-level $B-L$ conserving proton decay is $D=8$ and is always mediated by a pair of scalars, as previously discussed. \label{pt:c4:ig}
\end{enumerate}

Nevertheless, some scalars from these eight different pairs may also contribute to \(B-L\) violating two-body proton decays, as also discussed in Chapter~(\ref{ch:3}). However, if \(B-L\) conserving proton decays typically appear at dimension \(D=8\), we anticipate that \(B-L\) violating two-body proton decay modes to appear at dimension \(D=9\) or higher, specifically at odd dimensions. However, our analysis primarily focuses on constraining the mediators based on \(B-L\) conserving two-body proton decay modes.

\section{Constraints from Proton Decay} 
{\label{sec:c4:pds}}
In previous section~(\ref{sec:c4:pdeff}) we have identified various pairs of scalars capable of mediating proton decays at both tree and loop levels and have analysed their contributions, in terms of Wilson coefficients, to the effective \(D=6\) $B-L$ conserving proton decay operators. These coefficients depend on the specific Yukawa couplings. These Yukawa couplings also contribute to the masses of charged and neutral fermions in a bottom-up scenario.

The class of renormalisable \so\, theories featuring the Yukawa sector with complex  $10_\hh$ and $\overline{126}_{\hh}$ is known for reproducing the observed fermion mass spectrum and mixing angles~\cite{Joshipura:2011nn,Mummidi:2021anm}. Such a particular model has also been used in the previous Chapter~(\ref{ch:3}) to study scalar-induced proton decay spectrum in realistic \so\, GUTs. The Lagrangian of such realistic \so\, model based on $10_{\hh}$ and $\overline{126}_{\hh}$ is shown below,
\beqa {\label{eq:c4:realso10}}
-\mathcal{L}_{R} &=&  \fs_A\,\Big( H_{AB}\,10_\hh + F_{AB}\,\overline{126}_{\hh} \Big)\fs_B.  
\eeqa 
As mentioned in the previous Chapter~(\ref{ch:2}), $H$ and $F$ are the symmetric undergeneration indices. From Eq.~\eqref{eq:c4:realso10}, one can forbid the coupling of $\overline{126}_{\hh}$ with $\fs$-plet fermion and instead incorporate $16_{\hh}$ which couples to $\fs$-plet at non-renormalisable level. The Lagrangian of such a setup is shown below;
\beqa{\label{eq:c4:SO10RNR}}
-{\cal{L}}_{\mathrm{NR}} &=& H_{AB}\,\fs_A\,\fs_B\,10_\hh + \frac{y_{AB}}{\Lambda} \left(\fs_A\,\fs_B\,16_\hh\,16_\hh\right) \nl
\ad\, \frac{\bar{y}_{AB}}{\Lambda} \,\big(\mathbf{16}_{A}\,\mathbf{16}_{B}\;16^{\dagger}_{\hh}\,16^{\dagger}_{\hh}\big)\hc.
\eeqa

In the absence of \(\overline{126}_{\hh}\), its role is solely assumed to acquired by \(16_{\hh}\) for a realistic \so\, model. Consequently, the non-renormalisable term mentioned in Eq.~\eqref{eq:c4:SO10RNR} can only be decomposed by the \(10_{\hh}\)-dimensional \so\, irrep, as the coupling of different irreps with the \(\fs\)-plet is restricted. This restriction could be imposed by introducing a discrete symmetry that permits the coupling of the \(16\)-plet with \(10_{\hh}\) at the renormalisable level and with \(16_{\hh}\) at the non-renormalisable level. As a result, only the \(h\) and \(\bar{h}\) couplings mentioned in Eq.~\eqref{eq:c2:16Fs16Hs} are allowed and could assume the magnitude of the Yukawa coupling that \(126_{\hh}\) would typically have with the \(\fs\)-plet fermion to yield a realistic fermion mass spectrum.
\beqa {\label{eq:c4:hNR}}
h\frac{\langle\hat{\sigma}\rangle}{\Lambda} &= & \bar{h}\frac{\langle\hat{\sigma}\rangle}{\Lambda} \hspace{0.5cm}\sim \hspace{0.5cm}  F.
\eeqa
Furthermore, we impose the condition that the magnitudes of \(h\) and \(\bar{h}\) remain within perturbative limits, $i.e.$ \(|h|\) and \(|\bar{h}|\) \(\leq 4\pi\). This requirement for perturbativity sets a lower bound on the parameter \(\hat{\sigma}\) as \(10^{-4.5}\Lambda\), while there is no upper bound imposed on \(\hat{\sigma}\). We set \(\Lambda\) as the reduced Planck scale (\(\sim 10^{18.5}\) GeV), making the \(B-L\) scale closer to the conventional GUT scale ($\sim 10^{16}\,\rm{GeV}$). Choosing a \(B-L\) scale lower than this threshold would render the couplings \(h\) and $\bar{h}$ non-perturbative, thereby making the \so\, model unrealistic. The best-fit values of the entries of \(F\) can be found in Appendix~(\ref{app:2}), which have also been used in Chapter~(\ref{ch:3}).

The breaking of \so\, gauge invariance into the SM gauge symmetry leads to the following expressions of the effective Yukawa couplings resulting from Eq.~(\ref{eq:c4:SO10RNR}); 
\beqa {\label{eq:c4:yukrel}}
Y_{u,d,e,\nu} &=& c_1^{u,d,e,\nu} H + \frac{\langle\hat{\sigma}\rangle}{\Lambda} \big(c_2^{u,d,e,\nu}\, h +  c_3^{u,d,e,\nu}\,\bar{h} \big), \nl
M_{R} &=&  r\frac{\langle\hat{\sigma}\rangle^2}{\Lambda} \big( h + \bar{h}\big) , 
\eeqa
where, \(c_{1}^{u,d,e,\nu}\) represent the various \({\mathcal{O}}(1)\) Clebsch-Gordan coefficients \cite{Mummidi:2021anm}. The coefficients \(c_{2,3}^{u,d,e,\nu}\) and \(r\) are also \({\mathcal{O}}(1)\) Clebsch-Gordan coefficients. When \(\hat{\sigma}\), the \(B-L\) charged singlet-scalar found in \(16_{\hh}\), acquires a $vev$, it generates a Majorana mass for the right-handed neutrinos, similar to the scenario in a renormalisable \so\, model. Light neutrino masses can be generated using the Type-I seesaw mechanism.

\subsection{Branching Pattern}
\label{ss:c4:brpattern}
In our scenario, we consider the contribution from the non-renormalisable part of the Yukawa Lagrangian, as provided by Eq.~\eqref{eq:c4:SO10RNR}, to be comparable to that of the renormalisable part. Together, both contributions must comply with the constraints imposed by proton decay. As discussed in the previous section in our framework, only the Yukawa couplings \(h\) and \(\bar{h}\) can contribute to proton decay. Further, it is evident from the Eq.~\eqref{eq:c4:p1} that only the scalar pairs \(\hat{\sigma}-\hat{T}\) and \(\hat{H}-\hat{\Delta}\) are capable of inducing proton decay at the tree level~\cite{Shukla:2024owy}.

We have elaborated the method to compute the proton decay in section~(\ref{sec:c3:decay_widths}). The expressions of partial decay widths of leading proton decay modes are given in Eq.~(\ref{app:c3:decaywidths}), along with slightly modified definitions of the coefficients for the present case, which are provided in Appendix~(\ref{app:3}). Using the decay width expressions from Eq.~\eqref{eq:app:c3:decay_width}, together with the definitions of coefficients in Eq.~\eqref{eq:app4:C_dw}, and determining the explicit forms of \(h\) and \(\bar{h}\) based on the form of \(F\) given in Eq.~\eqref{input_prm}, we can calculate the branching pattern of proton decay. The Tab.~(\ref{tab:c4:pdecayvariation}) shows the branching patterns for leading proton decay modes, assuming the involvement of scalar pairs capable of inducing proton decay at the tree level.
\begin{table}[t]
\begin{center}
\begin{tabular}{lccc} 
\hline\hline
Branching ratio [\%]& ~~~$\sigma-T$~~~  & ~~~$H-\Delta$~~~ & ~~~$\sigma-T\,+ H-\Delta $~~~\\
& $M_\Delta\to\infty$ & $M_{T} \to \infty$  &~~~with $M_T=M_{\Delta}$\\
\hline
${\rm BR}[p\to e^+ \pi^0]$ & $< 1$ & $<1$ & $< 1$\\
${\rm BR}[p\to \mu^+ \pi^0]$  & $2$ & $ 2$ & $2$\\
${\rm BR}[p\to \bar{\nu} \pi^+]$  & $12$ & $13$ & $13$\\
${\rm BR}[p\to e^+ K^0]$ & $< 1$ & $< 1$ & $< 1$\\
${\rm BR}[p\to \mu^+ K^0 ]$  & $3$ & $2$ & $2$\\
${\rm BR}[p\to \bar{\nu} K^+]$  & $83$ & $84$ & $82$\\
${\rm BR}[p\to e^+ \eta]$ & $<1$  &  $<1$ & $<1$\\
${\rm BR}[p\to \mu^+ \eta]$  & $<1$ &  $<1$ & $< 1$\\
\hline\hline
\end{tabular}
\end{center}
\caption{Proton decay branching fractions estimated for different hierarchies, assuming $\big\langle \sigma\big\rangle = 10^{14}$ GeV and $\big\langle H \big\rangle = 246$ GeV.}
\label{tab:c4:pdecayvariation}
\end{table}
The inferences drawn from Tab.~(\ref{tab:c4:pdecayvariation}) are as follows:
\begin{enumerate} [label=(\roman*)]
    \item In non-renormalisable \so\, models, where proton decay is mediated by pairs of scalars, the decay into second-generation mesons is favoured over the first generation. This preference stems from the hierarchical nature of Yukawa couplings, as also explained previously discussed in Chapter~(\ref{ch:3})~\cite{Patel:2022wya}. {\label{pt:c4:aa}}
    \item  For realistic \so\ scenario, the exact values of the $B-L$ scale and $\Lambda$ are irrelevant for the tree-level branching pattern of the proton decay induced by $\hat{\sigma}-\hat{T}$. {\label{pt:c4:bb}}
    \item  The loop contribution of the scalar pair to proton decay becomes significantly suppressed when the \(B-L\) scale is set lower than \(10^{11}\) GeV. Consequently, this loop contribution has been disregarded in the calculations for the branching pattern of the proton. \label{p:c4:c} 
  \end{enumerate}

The above-mentioned feature of the proton decay spectrum in pt.~\ref{pt:c4:aa} spectrum can be understood from the flavour structure, as elaborated in section~(\ref{ss:c3:pdecaypatters}). The structure of $h$ and $\bar{h}$ can be computed from the structure of $F$, given in Eq.~\eqref{eq:c3:HF_form}. From this structure, one can compute the leading order Unitary matrices, $U_{u,d,e}$, which diagonalises these Yukawa matrices, again which could be inferred from Eq.~\eqref{eq:c3:U_form}.
All the strengths of the operators, appearing in Eqs.~(\ref{eq:c4:p1}, \ref{eq:c4:p2}, \ref{eq:c4:p3}, \ref{eq:c4:p4}, \ref{eq:c4:p5}, \ref{eq:c4:p6}, \ref{eq:c4:p7}, and \ref{eq:c4:p8}), are in the flavour basis. To compute proton decay, one has to transform from the flavour basis to the physical basis, as done in Eq.~\eqref{eq:c3:HF_form}. However, such unitary transformations slightly differ from those mentioned in Eq.~\eqref{eq:c3:HF_form}, as mentioned below;
\beqa \label{eq:c4:UFU}
~~U_f^T\,f\,U_{f^\prime} \sim  \frac{\Lambda}{\hat{\sigma}} \,\lambda^4\, \left(\ba{ccc} \lambda^5 & \lambda^4 & \lambda^3\\ \lambda^4 & \lambda^3 & \lambda^2\\
\lambda^3 & \lambda^2 & \lambda \ea\right)\,=U_d^T\,f\,U_\nu \eeqa
Substituting the results from Eq.~\eqref{eq:c4:UFU} into the expressions for the partial decay widths of the proton, given in Eqs.~(\ref{eq:app:c3:decay_width} and \ref{eq:app4:C_dw}), we 
get the similar ratio evaluated in Eqs.~(\ref{eq:c3:pattern_a}, \ref{eq:c3:pattern_b}, and \ref{eq:c3:decaylambda}) explaining
 the branching pattern observed in Tab.~(\ref{tab:c4:pdecayvariation}), which is mediated by \(\hat{\sigma}-\hat{T}\). Similarly, the branching pattern for proton decay mediated by \(\hat{H}-\hat{T}\) results in similar ratios.

The observation noted in pt.~\ref{pt:c4:bb} can be explained as follows: As evident from Eq.~\eqref{eq:c4:p1}, the tree-level contribution of the coefficient for a specific scalar pair scales as \( c \sim \frac{\langle\hat{\sigma}\rangle^2}{\Lambda^2} |h|^2\), where \(h\) represents a specific Yukawa coupling. From Eq.~\eqref{eq:c4:hNR}, in a realistic \so\, scenario, we have \(h \sim \frac{\Lambda}{\langle\hat{\sigma}\rangle}\), which makes the coefficient $c$  effectively independent of both \(\langle\hat{\sigma}\rangle\) and \(\Lambda\). This configuration of \(h\) is essential to accurately reproducing the observed fermion mass spectrum. However, choosing a different value for \(h\) that does not depend on \(\hat{\sigma}\) and \(\Lambda\) would introduce a dependence of the coefficient on these parameters, potentially affecting the mass scale of the mediator.

Claim~\ref{p:c4:c} can be understood by examining the loop contribution of the \(\hat{\sigma}-\hat{T}\) pair to proton decay. By substituting the results from Eqs.~(\ref{eq:c4:p1} and \ref{eq:c4:UFU}) into the relation for the partial decay width of proton decay given in Eq.~\eqref{eq:app:c3:decay_width}, we calculate the resulting decay into \(\overline{\nu}\, K^+\). The expression derived from these substitutions is outlined below;
\beqa{\label{eq:c4:ptc}}
\Gamma[p\to \overline{\nu}\,K^+] &=& \frac{\big(D\big(M_{\hat{T}}^2,M_{\hat{\sigma}}^2\big)\big)^2}{\big\langle\hat{\sigma}\big\rangle^4} \frac{(m_p^2 - m_{K^\pm}^2)^2}{32\, \pi\, m_p^3 f_\pi^2} \nl & & \times\,\left(1+\frac{m_N}{m_S}(\tilde{D}-\tilde{F})+\frac{m_N}{m_{\Lambda}}\big(\tilde{D}+3\tilde{F}\big)\right)^2\,32^2\,\lambda^{30},\nl
\eeqa
here, various symbols have their usual meanings as given in the Appendix~(\ref{app:3}), and \(D(M_{\hat{T}}^2, M_{\hat{\sigma}}^2)\) refers to the loop function defined in Eq.~\eqref{eq:c4:lf}. The expression provided in Eq.~\eqref{eq:c4:ptc} can be modified to calculate $\langle\hat{\sigma}\rangle$ as follows:
\beqa {\label{eq:ptcc}}
\big\langle\hat{\sigma}\big\rangle &=& \Bigg[\tau_{p\to \overline{\nu}\,K^+}\,\times\,\frac{\big(D\big(M_{\hat{T}}^2,M_{\hat{\sigma}}^2\big)\big)^2}{\big\langle\hat{\sigma}\big\rangle^4} \frac{(m_p^2 - m_{K^\pm}^2)^2}{32\, \pi\, m_p^3 f_\pi^2}\nl
& &\left(1+\frac{m_N}{m_S}(\tilde{D}-\tilde{F})+\frac{m_N}{m_{\Lambda}}\big(\tilde{D}+3\tilde{F}\big)\right)^2\,32^2\,\lambda^{30}\Bigg]^{0.25} \nonumber
\eeqa
From the current lower bound on the lifetime of $p\to \overline{\nu}\,K^+$~\cite{Super-Kamiokande:2014otb}, we arrive at the following: 
\beqa{\label{eq:c4:ptccc}}
\big\langle \hat{\sigma} \big \rangle &>& \Big[D\big(M^2_T,M^2_{\sigma}\big)\Big]^{0.5}\, 2.5*10^{11}\; \rm{GeV}
\eeqa
The loop function in the above Eq.~\eqref{eq:c4:ptccc} reaches its maximum value when the masses of the involved scalars are degenerate, and its value is always less than $1$, given that the maximum mass attained by any scalar propagating inside the loop is \(M_p\), the Planck mass. As a consequence, \(\hat{\sigma}\) must exceed \(10^{11}\) GeV to meet the constraints imposed by proton decay experiments. In realistic \so\, scenarios, \(\hat{\sigma}\) is typically around \(10^{14}\) GeV. Reducing \(\hat{\sigma}\) below \(10^{14}\) GeV would compromise the perturbativity of the Yukawa couplings. Therefore, the scalar pair that mediates proton decay at the loop level cannot be stringently constrained under these conditions.

\subsection{Estimation of bound on the mediator}
Having established the branching pattern for the primary proton decay modes mediated by scalar pairs capable of inducing tree-level proton decays, we proceed to constrain their masses based on the current lower bounds on proton lifetime provided by experimental results~\cite{Super-Kamiokande:2014otb}. The lower bound on the mass of pair $\hat{\sigma} - \hat{T}$ from its tree-level contribution is as follows:
\beqa \label{eq:c4:limit_T}
\tau/{\rm BR}[p \to \overline{\nu}\, K^+] &=& 5.9 \times 10^{33}\,{\mathrm{yrs}}\,\times \left(\frac{M_{\hat{T}}}{2.8 \times 10^{11}\,{\mathrm{ GeV}}} \right)^4\,,\eeqa
which is independent from the precise value of $\hat{\sigma}$.

Similarly, the following is the lower bound on the mass of $\hat{\Delta}$ from its tree-level contribution.
\beqa \label{eq:c4:limit_D}
\tau/{\rm BR}[p \to \overline{\nu}\, K^+] &=& 5.9 \times 10^{33}\,{\rm yrs}\,\times \left(\frac{M_{\hat{\Delta}}\,\big\langle\hat{\sigma}\big\rangle}{4.9 \times 10^{13}\,\mathrm{GeV}^2} \right)^4\,\left(\frac{ 246 }{\big\langle \hat{H}\big\rangle}\right)^4. \nl\eeqa
Eqs.~(\ref{eq:c4:limit_T} and \ref{eq:c4:limit_D}) show the constraints on the masses of \(M_{\hat{T}}\) and \(M_{\hat{\Delta}}\) respectively. The lower bound on \(M_{\hat{T}}\) is notably independent of the \(B-L\) scale. In contrast, the bound on \(M_{\hat{\Delta}}\) is dependent on the ratio \(\frac{\langle\hat{\sigma}\rangle}{\langle \hat{H}\rangle}\)
Both mediators, \(\hat{T}\) and \(\hat{\Delta}\), demonstrate independence from the specific value of the cutoff scale (\(\Lambda\)). Importantly, the mass of \(\hat{\Delta}\) depends on the ratio given in Eq.~(\ref{eq:c4:limit_D}), which can be adjusted to achieve a relatively lighter mass for \(\hat{\Delta}\), around the TeV scale. This adjustment opens exciting possibilities for the experimental probing of \(\hat{\Delta}\) at future collider experiments~\cite{Dorsner:2016wpm, Raj:2016aky, Padhan:2019dcp}. Furthermore, in renormalisable \so\, models, \(\hat{\Delta}\) plays a role in mediating \(B-L\) violating proton decay modes, as shown in the previous Chapter~(\ref{ch:3}) and induce the decay channel \(p \to \nu K^+\) which imposes a lower mass bound on \(\Delta\) exceeding \(10^6\) GeV (cf. Eq.~\eqref{eq:c3:limit_Tbar_126}. However, the mass bound on $\hat{\Delta}$ is significantly reduced in the scenario considered here.

\section{Summary} {\label{sec:c4:conc}}

This study deals with a variant of the non-renormalisable \(SO(10)\) model, which lacks LTRs. Within this framework, the \(16_{\hh}\) representation emerges as the smallest irrep coupling with the \(\fs\)-plet, capable of generating masses for charged and neutral fermions at the non-renormalisable level. We begin with decomposing the coupling of \(16_{\hh}\) with the \(\fs\)-fermions, exploring all the permitted decompositions allowed by integrating out allowed \(SO(10)\) irreps and further decomposed it into \(SU(5)\)-invariant expressions. Further, we analyse the contributions of different pairs of scalars, at both tree and loop levels, to the effective \(D=6\) \(B-L\) conserving operators. We argued that the \(16_{\hh}\) irrep could effectively perform the role typically played by \(\overline{126}_{\hh}\) in renormalisable \(SO(10)\) GUTs, thereby assigning its Yukawa coupling to be similar to that of \(\overline{126}_{\hh}\). Finally, we impose constraints on these pairs of scalars residing in the \(16_{\hh}\) representation through proton decays.

The key findings of this chapter are summarised as follows:
\begin{itemize}
    \item Proton decay in the non-renormalisable \(SO(10)\) model is always mediated by pairs of scalars, occurring in two distinct topologies that result in effective \(D=6\) operators. Firstly, at the tree level, one scalar is integrated out while the other scalar acquires a $vev$. Secondly, both scalars contribute to proton decay through loops.
    
    \item In total, eight different pairs are capable of inducing proton decay. Among them, only three can mediate proton decay at the tree level: \(\hat{\sigma}-\hat{T}\), \(\hat{H}-\hat{\Delta}\), and \(\hat{H}-\hat{T}\), while the other pairs induce proton decay at the loop level.
    
    \item  Due to the hierarchical Yukawa couplings, protons favours to decay into second-generation mesons accompanied by charged or neutral antileptons. For a realistic scenario, the lower bound on \(M_{\hat{T}}\) is greater than \(10^{11}\) GeV, while \(M_{\hat{\Delta}}\) is constrained to be at least $1$ TeV. The bound on \(M_{\hat{\Delta}}\) arises purely from low energy searches.
\end{itemize}

As we have seen in Chapter~(\ref{ch:3}), in renormalisable versions of \(SO(10)\) GUTs, proton decay at the lowest mass dimension ($i.e.$, six) is mediated by a single scalar particle, which can typically remain lighter than the conventional GUT scale. However, in the non-renormalisable version, proton decay at the lowest dimension is always induced by a pair of scalars. In certain scenarios, these mediators may remain closer to the electroweak scale than the GUT scale. These scalars can be light enough to lead to visible effects in the below energy phenomenology. For instance, as we have seen in this Chapter, $\hat{\Delta}$ can simultaneously acquire low mass and contribute to proton decay. This goes beyond the conventional understanding that proton decay mediators should necessarily be super-heavy. This happens due to the consideration of non-renormalisable couplings. 

It is worth to mention that introducing non-renormalisable interactions allows all possible operators, including four-fermion interactions appearing at mass dimension \(D=6\), potentially resulting from integrating out (other) heavy scalar(s). These operators could significantly affect any process. However, we have focussed on non-renormalisable operators containing fields of \(16_{\hh}\). The non-renormalisable effective couplings stemming from the decomposition of \(\fs-\fs-16_{\hh}-16_{\hh}\) are of mass dimension five; hence, the four-fermion operators are comparatively more suppressed for the same cut-off scale. Consequently, we have ignored such four fermion operators and other such operators appearing at higher dimensions. 

We summarise the results of the previous Chapter~(\ref{ch:3}) and this Chapter in the following Table.
\begin{table}[h!]
 \centering
    \begin{tabular}{c|ccrcc}
    \hline\hline
        \so\,Model & Scalar & SM Charge & $B-L$ & Mode & Lower Bound \\
        \hline
        \multirow{2}{*}{$10_{\hh} + \overline{126}_{\hh}$} & $T$ & $\left(3,1,-\sfrac{1}{3}\right)$ & $-\sfrac{2}{3}$ & Conserving & $M_{T}\,\geq\, 10^{11}\,\rm{GeV}$ \\
        & $\Delta$ & $\left(3,2,\sfrac{1}{6}\right)$ & $\sfrac{4}{3}$ & Violating & $M_{\Delta}\,\geq\,10^{6}\,\rm{GeV}$ \\
        \hline
        \multirow{2}{*}{$10_{\hh} + 16_{\hh}$} & $\hat{\sigma}-\hat{T}$ & $\left(3,1,-\sfrac{1}{3}\right)$ & $\sfrac{2}{3}$ & Conserving & $M_{\hat{T}}\geq 10^{11}\,\rm{GeV}$ \\
        & $\hat{H}-\hat{\Delta}$ & $\left(3,2,\sfrac{1}{6}\right)$ & $\sfrac{2}{3}$ & Conserving & $M_{\hat{\Delta}}\,\geq\,10^{3}\,\rm{GeV}$ \\
        \hline\hline
    \end{tabular}
   \caption{Scalars contributing to tree-level nucleon decays in realistic renormalisable and non-renormalisable \so\, model, provided $\langle \hat{H}\rangle \,=\, 246$ GeV.}
    \label{tab:c4:scalarssummary}
\end{table}

\clearpage
\chapter{Scalar Induced Matter-Antimatter Oscillations $\&$ Asymmetry }
\graphicspath{{50_Chapter_5/fig_ch5/}}
{\label{ch:5}}
\section{Overview}
\label{sec:c5:intro}
\lettrine[lines=2, lhang=0.33, loversize=0.15, findent=0.15em]{R}OLE OF SCALAR fields in affecting the stability of the nucleon, stemming from renormalisable and non-renormalisable \so\, interactions have been discussed in Chapter~(\ref{ch:3} and \ref{ch:4}) respectively. In this chapter, we particularly focus on the scalar fields that only couple with quarks, reside in irreps of \(10_{\hh}\), \(120_{\hh}\), and \(\overline{126}_{\hh}\), have non-zero \(B-L\) charge, and do not contribute to proton decay at leading order. A class of such fields is the colour sextet scalar.

Sextet scalar transforms as $six$-dimensional irrep under $SU(3)_{\mathrm{C}}$, can be either singlet or triplet under $SU(2)_{\mathrm{L}}$ and couples solely to quarks. These sextet fields can induce various interesting phenomena and thus are valid candidates that demand focused study. This chapter again begins with the effective couplings of these sextet fields with SM fermions in section~(\ref{sec:c5:couplings}). Using these effective vertices, we compute their contribution in quark flavour violation and neutral baryon-antibaryon oscillation in sections~(\ref{sec:c5:qfv} and  \ref{sec:c5:nnbar}), respectively. In section~(\ref{sec:c5:quartic}), we discuss the constraints on sextet scalars from the effectively generated quartic coupling. We discuss an interesting possibility to generate the cosmological baryon asymmetry of the universe using the sextet scalars in section~(\ref{sec:c5:baryo}). We constrain the spectrum of the sextets using the aforementioned phenomena in section~(\ref{sec:c5:results}) and conclude this chapter in section~(\ref{sec:c5:summary}).  

\section{Effective Couplings of Sextet Scalars}
\label{sec:c5:couplings}

As discussed in the Chapter~(\ref{ch:2}), the Yukawa sector of renormalisable $SO(10)$ GUTs consists of $10_{\hh}$, $120_{\hh}$, and $\overline{126}_{\hh}$ dimensional irreps. In the Tab.~(\ref{tab:c5:sextetscalars}), we have extracted out the relevant sextet fields from the Tab.~(\ref{tab:c2:scalars}), which arises only from $120_{\hh}$ and $\overline{126}_{\hh}$. Incidentally, no sextet scalar field resides in $10_{\hh}$-irrep. The Tab.~(\ref{tab:c5:sextetscalars}) also shows the $B-L$ charges and their multiplicity in their respective parent multiplet.  

\begin{table}[t]
\begin{center}
\begin{tabular}{lcrccc} 
\hline
\hline
~~SM charges~~&~~Notation~~&~~$B-L$~~&~~$120_{\hh}$~~&~~$\overline{ 126}_{\hh}$~~\\
 \hline
$\left(6,1,\frac{1}{3}\right)$ & $S^{\alpha}_{\beta\gamma}$ & $\frac{2}{3}$  &1&1\\
$\left(\overline{6},1,-\frac{1}{3}\right)$&$\overline{S}{^{\beta\gamma}_{\alpha}}$ &$-\frac{2}{3}$  &1&0 \\
$\left(6,1,-\frac{2}{3}\right)$ & $ \Sigma^{\alpha\beta}$ & $\frac{2}{3}$  &0&1 \\
$\left(6,1,\frac{4}{3}\right)$ & ${\cal S}^{\alpha}_{\beta\gamma}$ & $\frac{2}{3}$  & 0 & 1 \\
$\left(\overline{6},3,-\frac{1}{3}\right)$ &$\mathbb{S}^{\alpha\beta a}_{\gamma b}$ & $-\frac{2}{3}$ &0&1\\
\hline\hline
\end{tabular}
\end{center}
\caption{Different coloured sextet fields, their charges according to the SM gauge groups $SU(3)_C$, $SU(2)_L$, $U(1)_{\mathrm{Y}}$), their $B-L$ charges, and multiplicity in the $120_{\hh}$ and $\overline{126}_{\hh}$ scalar representations of $SO(10)$.}
\label{tab:c5:sextetscalars}
\end{table}

The interactions of the coloured sextet fields with the SM quarks can be calculated using the method discussed in section~(\ref{ssec:c2:OESO}), along with the decomposition of canonically normalised tensors described in Chapters~(\ref{ch:2}, and \ref{ch:3}). Additionally, as evident from Tab.~(\ref{tab:c5:sextetscalars}), the sextet field \(S^{\alpha}_{\beta\gamma}\) appears in both \(120_{\hh}\) and \(\overline{126}_{\hh}\) and is degenerate. Therefore, we refer to \(S^{\alpha}_{\beta\gamma}\) in \(120_{\hh}\) as \(\tilde{S}^{\alpha}_{\beta\gamma}\).

We begin with the couplings of sextets residing in $120_{\hh}$, which are not given in the section~(\ref{ss:c3:1616120});
\beqa \label{eq:c5:sextet_120}
-{\cal L}^{{120}_{\hh}}_Y &=& G_{AB}\,\fs^T_A\,\cc\,\fs_B\,120_{\hh}\,\hc\, \nonumber \\
& \supset & -\frac{2i}{\sqrt{3}}\,G_{AB}\,\Big(\varepsilon^{\alpha \beta \gamma}\, u^{C T}_{\gamma A}\,\cc\,d^C_{\sigma B}\, \tilde{S}^\sigma_{\alpha \beta} - \varepsilon_{\alpha \beta \gamma}\, u^{\gamma T}_A\,\cc\,d^\sigma_B\, \overline{S}^{\alpha \beta}_\sigma \Big)\,\nl\hcn,\eeqa
where we have used the same dictionary to decompose into SM fields mentioned in section~(\ref{ssec:c2:OESO}). Additionally,  as section~(\ref{ss:c3:1616120}) mentions, $G$ is anti-symmetric in flavour indices. Similarly, the vertices of sextets with SM fermions residing in $\overline{126}_{\hh}$ are as follows:
\beqa \label{eq:c5:sextet_126}
-{\cal L}^{\overline{126}_{\hh}}_Y &=& F_{AB}\,\fs^T_A\,\cc\,\fs_B\,\overline{126}_{\hh}\,\hc\, \nonumber \\
& \supset & -\frac{i}{\sqrt{15}}\,F_{AB}\,\Big(2\, d^{C T}_{\alpha A}\,\cc\,d^C_{\beta B}\, \Sigma^{\alpha \beta}\,-\, \sqrt{2}\, \varepsilon^{\alpha \beta \gamma}\, u^{C T}_{\gamma A}\,\cc\,d^C_{\sigma B}\, {S}^\sigma_{\alpha \beta} \Big. \nonumber \\
& & \Big. + \varepsilon^{\alpha \beta \gamma}\, u^{C T}_{\sigma A}\,\cc\,u^C_{\gamma B}\,{\cal S}^\sigma_{\alpha \beta}\,+\,\sqrt{2}\, \epsilon_{\alpha \beta \gamma}\, \varepsilon_{ab}\, q^{a \alpha T}_{A}\,\cc\,q^{\sigma c}_B\,\mathbb{S}^{\beta \gamma b}_{\sigma c}\Big)\,\nl \hcn\,.\eeqa
Here, \(F\) is symmetric in the flavour space. As it is evident, \(\overline{S}\) and \(\mathbb{S}\) interact with left-chiral quark fields, while the other coloured sextets couple with right-chiral quarks.

In the \so\, models having irreps  $120_{\hh}$ and \(\overline{ 126}_{\hh}\) together, the fields \(S\) and \(\tilde{S}\) can mix with each other through gauge-invariant terms such as  $\left(\,120_{\hh}\right.$ $\,\left.\overline{{126}}^{\dagger}_{\hh}\,45_{\hh}\,\right)$ or $\left(\,120_{\hh}\,\overline{126}^{\dagger}_{\hh}\, 210_{\hh}\,\right)$. The resulting physical states are combinations of \(S\) and \(\tilde{S}\). Further, we assume that this mixing is characterised by real parameters;
\be \label{eq:c5:S12}
S_1 = S\,\cos\theta \,+\,\tilde{S}\,\sin\theta\,,~~ S_2 = -S\,\sin\theta\, \,+\,\tilde{S}\,\cos\theta\,\,.\ee
We define $c_{\theta}\equiv \cos\theta$ and $s_{\theta}\equiv \sin\theta$ which will be used in further sections. Additionally, the linear combination $S_1$ is assumed to be the lighter mass eigenstate, $i.e.$ $M_{S_1} < M_{S_2}$. 

After substituting \(S\) and \(\tilde{S}\) with the physical states \(S_{1,2}\) as described in Eq.~\eqref{eq:c5:S12}, and transforming the quark fields into their physical basis by transforming \(f \to U_f f\), the Yukawa couplings between the various sextet fields \(\Phi=\Sigma, S_i, {\cal S}, \mathbb{S}, \overline{S}\), and the quarks can be rewritten as follows:
\beqa \label{eq:c5:sextet_all}
-{\cal L}_\Phi & = & Y^\Sigma_{AB}\,d^{C T}_{\alpha A}\,\cc\,d^C_{\beta B}\, \Sigma^{\alpha \beta}\,+\, Y^{S_i}_{AB}\,\varepsilon^{\alpha \beta \gamma}\, u^{C T}_{\gamma A}\,\cc\,d^C_{\sigma B}\, {S_i}^\sigma_{\alpha \beta}\, \nonumber \\
& + & Y^{\cal S}_{AB}\,\varepsilon^{\alpha \beta \gamma}\, u^{C T}_{\sigma A}\,\cc\,u^C_{\gamma B}\,{\cal S}^\sigma_{\alpha \beta}\,+\,Y^{\mathbb{S}}_{AB}\, \varepsilon_{\alpha \beta \gamma}\, \epsilon_{ab}\, q^{a \alpha T}_{A}\,\cc\,q^{\sigma c}_B\,\mathbb{S}^{\beta \gamma b}_{\sigma c} \nonumber \\
& + & Y^{\overline{S}}_{AB}\,\varepsilon_{\alpha \beta \gamma}\, u^{\gamma T}_A\,\cc\,d^\sigma_B\, \overline{S}^{\alpha \beta}_\sigma\,\hc,\eeqa
where $i=1,2$. The $3 \times 3$ matrices $Y^\Phi$ obtained from Eqs.~(\ref{eq:c5:sextet_120} and \ref{eq:c5:sextet_126}) are given as:
\beqa \label{eq:c5:YPhi}
Y^\Sigma &=& -\frac{2 i}{\sqrt{15}}\,U_{d^C}^T\, F\, U_{d^C}\,,\nl
Y^{S_1} \eq \frac{\sqrt{2}i}{\sqrt{3}}\,\left(\frac{1}{\sqrt{5}} c_\theta\, U_{u^C}^T\,F\,U_{d^C}\, -\, \sqrt{2}s_\theta\, U_{u^C}^T\,G\,U_{d^C}\right)\,, \nl
Y^{S_2} &=& -\frac{\sqrt{2}i}{\sqrt{3}}\,\left(\frac{1}{\sqrt{5}} s_\theta\, U_{u^C}^T\,F\,U_{d^C}\, +\, \sqrt{2}c_\theta\, U_{u^C}^T\,G\,U_{d^C}\right)\,,
\nl Y^{\cal S}  \eq  -\frac{i}{\sqrt{15}}\,U_{u^C}^T\,F\,U_{u^C}\,, \nonumber \\
Y^{\mathbb{S}} & = & -\frac{\sqrt{2} i}{\sqrt{15}}\,U_{q}^T\,F\,U_{q^\prime}\,,\nl
Y^{\overline{S}}  \eq \frac{2 i}{\sqrt{3}}\,U_u^T\,G\,U_d \,.\eeqa
Here, \(q, q' = u, d\). The matrices \(Y^\Sigma\) and \(Y^{\cal S}\) are symmetric in the flavor space, while \(Y^{\mathbb{S}}\) is symmetric when \(q = q'\). It is important to note that the matrices \(U_u\) and \(U_d\) determine the quark mixing matrix, defined as \(U_u^\dagger U_d \equiv V_{\rm CKM}\). Thus, assuming \(U_u \approx U_d\) is a reasonable approximation, and at the leading order, \(Y^{\mathbb{S}}\) can be considered symmetric.

The advantage of formulating the expressions in Eq.~\eqref{eq:c5:YPhi} is that all the \(Y^\Phi\) matrices can be explicitly calculated within realistic \(SO(10)\) models where the fundamental couplings \(F\), \(G\), and the diagonalising matrices \(U_f\) are derived from fits to the fermion mass data as mentioned in section~(\ref{sec:c3:results_model}). In subsequent sections, we will employ these couplings to analyse various phenomenologically significant processes that only involve sextet scalars: quark flavour violation, neutron-antineutron oscillation and baryogenesis.

\section{Quark flavour violation}
\label{sec:c5:qfv}

The \(Y^\Phi\) shown in Eq.~(\ref{eq:c5:YPhi}) depends on the Yukawa couplings $G$ and $F$. Generally, $G$ and $F$ are not diagonal in generation space, allowing sextet scalars to introduce a new source of flavour violation in the quark sector. Additionally, $Y^{\Phi}$ depends on unitary matrices $(U_{u,u^C,d,d^C})$ that diagonalise charged fermion mass matrices. This rotation from flavour to physical basis can also act as a flavour violation, making a diagonal Yukawa matrix non-diagonal.  The most stringent constraints on this type of new physics come from \(|\Delta F|=2\) processes, which involve neutral meson-antimeson oscillations~\cite{Buras:2003jf,Buras:2005xt,Buras:2020xsm,Antonelli:2009ws}. Following effective theory analysis, we assess the contributions from sextets to \(K^0-\overline{K}^0\), $B^{0}_d-\overline{B}^0_d$, $B^{0}_s-\overline{B}^0_s$, and \(D^0-\overline{D}^0\) mixing at both tree and one-loop levels.

Integrating out various sextets scalar fields from Eq.~\eqref{eq:c5:sextet_all}, the effective Lagrangian can be written as follows:
\beqa \label{eq:c5:L_F2}
{\cal{L}}^{\Delta F=2}_{\rm eff} &=& \sum _{q=d,u} \left(c_q\, {\cal{O}}_q\;+\; \tilde{c}_q\, {\cal{\tilde{O}}}_q \right)\,\hc\,,\eeqa
we find the following independent operators
\beqa \label{eq:c5:O_F2}
{\cal{\tilde{O}}}_{d}\;&=&\;\Big(\overline{d{_{R}^{\alpha}}}_{A} \gamma^{\Dot{\mu}}\,d{_{R\,C}^{\alpha}}\Big)\,\left(\overline{d{_{R}^{\beta}}}_{B}\,\gamma_{\Dot{\mu}}\,d{_{R\,D}^{\beta}}\right)\,, \nonumber \\
{\cal{\tilde{O}}}_{u}\;&=&\;\Big(\overline{u{_{R}^{\alpha}}}_{A} \gamma^{\Dot{\mu}}\,u{_{R\,C}^{\alpha}}\Big)\,\left(\overline{u{_{R}^{\beta}}}_{B}\,\gamma_{\Dot{\mu}}\,u{_{R\,D}^{\beta}}\right)\,, \nonumber \\
{\cal O}_{d}\;&=&\;\Big(\overline{d{_{L}^{\alpha}}}_{A} \gamma^{\Dot{\mu}}\,d{_{L\,C}^{\alpha}}\Big)\,\left(\overline{d{_{L}^{\beta}}}_{B}\,\gamma_{\Dot{\mu}}\,d{_{L\,D}^{\beta}}\right)\,, \nonumber \\
{\cal O}_{u}\;&=&\;\Big(\overline{u{_{L}^{\alpha}}}_{A} \gamma^{\Dot{\mu}}\,u{_{L\,C}^{\alpha}}\Big)\,\left(\overline{u{_{L}^{\beta}}}_{B}\,\gamma_{\Dot{\mu}}\,u{_{L\,D}^{\beta}}\right)\,.\eeqa
Here, we use $q = q_L$ and $q^C = \cc\, q_R^*$, for $q=u,d$, to obtain the above operators in the usual left- and right-chiral notations as also done in Chapter~(\ref{ch:3}). We have also used two components Fierz transformation rules, given in Eq.~\eqref{eq:c3:fierz}, to match the integrated out operators with the conventional ones given in Eq.~\eqref{eq:c5:O_F2}. Further, ${\cal O}_{u,d}$ and ${\cal \tilde{O}}_{u,d}$ are related by the interchange of $L \leftrightarrow R$ chirality. These operators change the flavour by two units and induce flavour-changing neutral meson-antimeson oscillations in the up-type and down-type quark sectors, respectively.

\begin{figure}[t]
    \centering
    \includegraphics[scale=0.4]{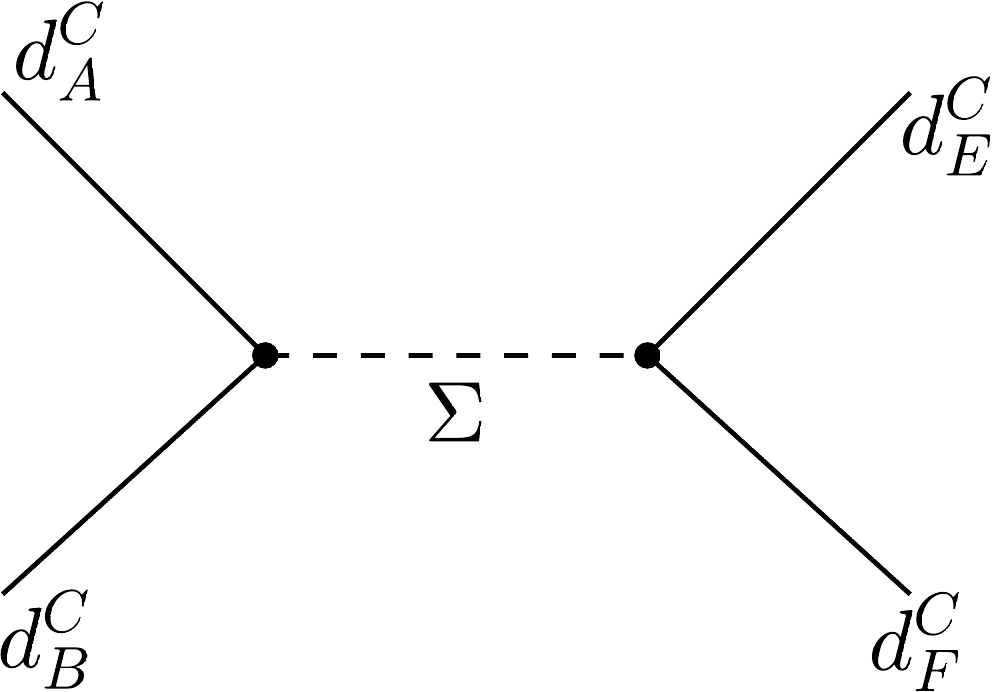}
    \caption{Feynman graph of flavour violation process induced by $\Sigma$, drawn using~\cite{Harlander:2020cyh}}
    \label{fig:c5:QFV}
\end{figure}

We integrate out the heavy degrees of freedom of sextet scalars using the equations of motion in Eq.~\eqref{eq:c5:sextet_all}, resulting in four-fermion operators. These operators are then matched with those given in Eq.~\eqref{eq:c5:L_F2}, yielding the strength of the operators, as shown below;
\beqa \label{eq:c5:cd1}
\tilde{c}_d [A,B,C,D] & = & \frac{1}{M_{\Sigma}^2}\bigg[\frac{1}{2} Y^\Sigma_{AB} (Y^{\Sigma}_{CD})^* \nl
\mi \frac{6}{64\pi^2}\left[\left(Y_2^\Sigma\right)_{AC}\left(Y_2^\Sigma\right)_{BD} + \left(Y_2^\Sigma\right)_{AD}\left(Y_2^\Sigma\right)_{BC} \right]\bigg]\nonumber\\
& - & \frac{1}{M^2_{S_1}}\,\frac{6}{16\pi^2} \left[\left(\bar{Y}_2^{S_1}\right)_{AC}\left(\bar{Y}_2^{S_1}\right)_{BD} + \left(\bar{Y}_2^{S_1}\right)_{AD}\left(\bar{Y}_2^{S_1}\right)_{BC}\right].\,\nl \eeqa
In the above expression, Eq.~\eqref{eq:c5:cd1}, \(Y_2^\Phi = Y^\Phi Y^{\Phi \dagger}\) and \(\bar{Y}_2^\Phi = Y^{\Phi \dagger} Y^\Phi\) represent hermitian matrices. The first term indicates the tree-level contribution mediated by \(\Sigma\), while the second and third terms represent contributions from 1-loop processes involving the scalars \(\Sigma\) and \(S_1\), respectively. It is important to note that the contribution from \(S_2\) has not been included due to the implicit assumption that  \(M_{S_2} > M_{S_1}\).

Analogously, we find the following effective strengths of the remaining operators as follows:
\beqa \label{eq:c5:cu1}
\tilde{c}_u [A,B,C,D] & = & \frac{1}{M_{\mathcal{S}}^2}\bigg[Y^{\cal S}_{AB} (Y^{\cal S}_{CD})^*\nl
\mi  \frac{6}{16\pi^2} \left[\left(Y_2^{\cal S}\right)_{AC}\left(Y_2^{\cal S}\right)_{BD} + \left(Y_2^{\cal S}\right)_{AD}\left(Y_2^{\cal S} \right)_{BC}\right]\bigg] \nonumber\\
& - & \frac{1}{M^2_{S_1}}\,\frac{6}{16\pi^2} \left[\left(Y_2^{S_1}\right)_{AC}\left(Y_2^{S_1}\right)_{BD} + \left(Y_2^{S_1}\right)_{AD}\left(Y_2^{S_1}\right)_{BC}\right],\,\nl\eeqa
\beqa \label{eq:c5:cd2}
c_d [A,B,C,D] &=& \frac{1}{M^2_{\mathbb{S}}}\,\bigg[Y^{\mathbb{S}}_{AB} (Y^{\mathbb{S}}_{CD})^*\nl
\mi \frac{30}{16\pi^2}\left[\left(Y_2^\mathbb{S}\right)_{AC}\left(Y_2^\mathbb{S}\right)_{BD} + \left(Y_2^\mathbb{S}\right)_{AD}\left(Y_2^\mathbb{S}\right)_{BC}\bigg]\right]\nonumber\\
& - & \frac{1}{M^2_{\overline{S}}}\frac{6}{16\pi^2}\,\left[\left(\bar{Y}_2^{\overline{S}}\right)_{AC}\left(\bar{Y}_2^{\overline{S}}\right)_{BD} + \left(\bar{Y}_2^{\overline{S}}\right)_{AD}\left(\bar{Y}_2^{\overline{S}}\right)_{BC}\right],\,\nl \eeqa 
and 
\beqa \label{eq:c5:cu2}
c_u [A,B,C,D]&=& \frac{1}{M^2_{\mathbb{S}}}\,\bigg[Y^{\mathbb{S}}_{AB} (Y^{\mathbb{S}}_{CD})^*\nl
\mi \frac{30}{16\pi^2}\left[\left(Y_2^\mathbb{S}\right)_{AC}\left(Y_2^\mathbb{S}\right)_{BD} + \left(Y_2^\mathbb{S}\right)_{AD}\left(Y_2^\mathbb{S}\right)_{BC}\right]\bigg]\nonumber\\
& - & \frac{1}{M^2_{\overline{S}}}\frac{6}{16\pi^2}\,\left[\left(Y_2^{\overline{S}}\right)_{AC}\left(Y_2^{\overline{S}}\right)_{BD} + \left(Y_2^{\overline{S}}\right)_{AD}\left(Y_2^{\overline{S}}\right)_{BC}\right].\,\nl\eeqa
From the given expressions for \(\tilde{c}_{u,d}\) and \(c_{u,d}\), it is evident that the sextet \(\Sigma\) contributes to flavour violations in the down-type quark sector, while \({\cal S}\) exclusively contributes to flavour violation in the up-quark sector. Additionally, \(\mathbb{S}\) contributes to flavour violations in both up and down quark sectors at the tree level. The contributions from \(S_1\) and \(\overline{S}\) to flavour violation occur only at the loop level. However, these contributions can be comparable to those at the tree level depending on the hierarchical structure of the Yukawa couplings.

For an estimation of the constraints on the masses and couplings of sextet scalars, it is essential to evolve the relevant strength \(c_{u,d}\) and \(\tilde{c}_{u,d}\) from the mass scale of the integrated-out scalar, denoted by \(\mu = M_\Phi\), down to the scale at which meson-antimeson mixing is experimentally observed. For \(K^0-\overline{K}^0\) oscillation, the pertinent scale is \(\mu=2\) GeV. The renormalisation group equation (RGE) evolved effective coefficient, for cases where \(M_\Phi > m_t\) (the mass of the top quark), is calculated as follows~\cite{Ciuchini:1998ix}:
\be \label{eq:c5:CK}
C^1_K = \left(0.82 - 0.016\, \frac{\alpha_s(M_\Phi)}{\alpha_s(m_t)}\right) \left(\frac{\alpha_s(M_\Phi)}{\alpha_s(m_t)} \right)^{0.29}\, c_d[1,1,2,2](M_\Phi).\,\ee
The analogous RGE equation for \(\tilde{C}^1_K\) can be derived by substituting \(c_d\) with \(\tilde{c}_d\) in the Eq.~\eqref{eq:c5:CK}. It is important to note that both \(C^1_K\) and \(\tilde{C}^1_K\) are coefficients for the effective operators, \(Q_1\) and \(\tilde{Q}_1\), respectively. These operators are listed  in~\cite{Ciuchini:1998ix} and correspond to our \({\cal O}_d\) and \({\cal \tilde{O}}_d\), with \(A=B=1\) and \(C=D=2\). These operators do not mix with other operators during RGE evolution, simplifying the running expressions, as shown in Eq.~(\ref{eq:c5:CK}).

Similarly, we can arrive at the expression for Wilson coefficients for \bds as follows;
\beqa \label{eq:c5:CB}
C^1_{B_d} &=& \left(0.865 - 0.017\, \frac{\alpha_s(M_\Phi)}{\alpha_s(m_t)}\right) \left(\frac{\alpha_s(M_\Phi)}{\alpha_s(m_t)} \right)^{0.29}\, c_d[1,1,3,3](M_\Phi)\,, \nonumber \\
C^1_{B_s} &=& \left(0.865 - 0.017\, \frac{\alpha_s(M_\Phi)}{\alpha_s(m_t)}\right) \left(\frac{\alpha_s(M_\Phi)}{\alpha_s(m_t)} \right)^{0.29}\, c_d[2,2,3,3](M_\Phi),\, \eeqa
at  $\mu=m_b$~\cite{Becirevic:2001jj}. In the case of the charm mixing governed by \dd oscillations, one finds~\cite{UTfit:2007eik},
\be \label{eq:c5:CD}
C^1_D = \left(0.837 - 0.016\, \frac{\alpha_s(M_\Phi)}{\alpha_s(m_t)}\right) \left(\frac{\alpha_s(M_\Phi)}{\alpha_s(m_t)} \right)^{0.29}\, c_u[1,1,2,2](M_\Phi),\,\ee
at $\mu=2.8$ GeV. Various coefficients then can be compared with the present limits obtained from a fit to the experimental data by UTFit collaboration~\cite{UTfit:2007eik}. The present upper limits on the effective strengths are as follows:
\beqa \label{eq:c5:const_C}
\begin{array}{lr}
    \left| C^1_K \right| < 9.6 \times 10^{-13}, & \left| C^1_{B_d} \right| < 2.3 \times 10^{-11} \\
    \left| C^1_{B_s} \right| < 1.1 \times 10^{-9},\hspace{1cm}\text{and} & \left| C^1_D \right| < 7.2 \times 10^{-13}. \\
\end{array}
\eeqa

The same upper bounds also apply to the corresponding $\tilde{C}^1$. One can constrain the masses of sextet scalars by knowing the bounds on different Wilson coefficients, as provided in Eq.~\eqref{eq:c5:const_C}, and Yukawa matrices.

\section{Neutron-Antineutron Oscillations}
\label{sec:c5:nnbar}
The colour sextet fields in renormalisable \(SO(10)\) models can also induce transitions between neutral baryons and their antiparticles. Unlike the flavour transitions described earlier, this process requires a source of baryon number violation. In \(SO(10)\), the source of baryon number violation is inherently present, and such oscillation typically emerges from gauge-invariant quartic couplings involving three sextet fields and an SM singlet from \(\overline{ 126}_{\hh}\), $i.e. $\(\sigma\), which has a \(B-L\) charge of \(-2\) (refer to Tab.~(\ref{tab:c2:scalars})). The $vev$ of \(\sigma\) not only breaks \(B-L\) and generates Majorana masses for the right-handed neutrinos, but also leads to \(B-L\) violating trilinear couplings among various sextet scalars. The \(n\)-\(\bar{n}\) oscillations via dimension-nine, electrically neutral and six-fermion operators~\cite{Rao:1982gt,Caswell:1982qs,Rao:1983sd} and violates baryon number by two units.

To derive the effective operators that mediate neutral baryon-antibaryon oscillations, we begin by constructing the most general quartic interaction terms that involve three sextet scalars and \(\sigma\), while respecting the SM gauge symmetry. Such combinations are listed below:
\beqa \label{eq:c5:qs}
& & \eta_{ij}\, \sigma\, \Sigma^{\alpha \beta}\, {S_i}^\gamma_{\alpha \gamma}\, {S_j}^\theta_{\beta \theta}\, \label{eq:c5:q1} \\
& & \eta_2\, \varepsilon_{\alpha \gamma \theta}\, \sigma\, \Sigma^{\alpha \beta}\, \Sigma^{\gamma \sigma}\, {\cal S}^\theta_{\beta \sigma} \label{eq:c5:q2}\\
& & \eta_3\, \sigma^*\, \Sigma^*_{\alpha \beta}\, \mathbb{S}^{\alpha \sigma a}_{\sigma b}\, \mathbb{S}^{\beta \rho b}_{\rho  a}\, \label{eq:c5:q3} \\
& & \eta_4\, \sigma^*\, \Sigma^*_{\alpha \beta}\, \overline{S}^{\alpha \sigma}_{\sigma}\, \overline{S}^{\beta \rho}_{\rho}\,. \label{eq:c5:q4} \eeqa
The interaction terms listed in the first two lines of the above Eq.~\eqref{eq:c5:qs} can originate from a quartic term involving four \(\overline{126}_{\hh}\) fields, $i.e.$ $\overline{126}_{\hh}$ and mixing between \(S\) and \(\tilde{S}\). The interaction terms on the third and fourth lines can result from gauge invariant combinations such as \((\overline{126}_{\hh}^\dagger \overline{126}_{\hh})^2\) and \((\overline{126}_{\hh}^\dagger {120}_{\hh})^2\), respectively. Further, we assume that in Eqs.~(\ref{eq:c5:q1}-\ref{eq:c5:q4}), all sextet fields are expressed in the mass basis. Eq.~\eqref{eq:c5:q1} identifies three distinct operators corresponding to \(i, j = 1, 2\) with \(\eta_{12} = \eta_{21}\). Given that \(M_{S_1} < M_{S_2}\), the most significant contribution to neutral baryon-antibaryon oscillations typically comes from the term corresponding to \(\eta_{11}\) in Eq.~\eqref{eq:c5:q1}. Consequently, our analysis will focus primarily on the operators induced by \(S_1\).

\begin{figure}[t]
    \centering
    \includegraphics[width=0.65\linewidth]{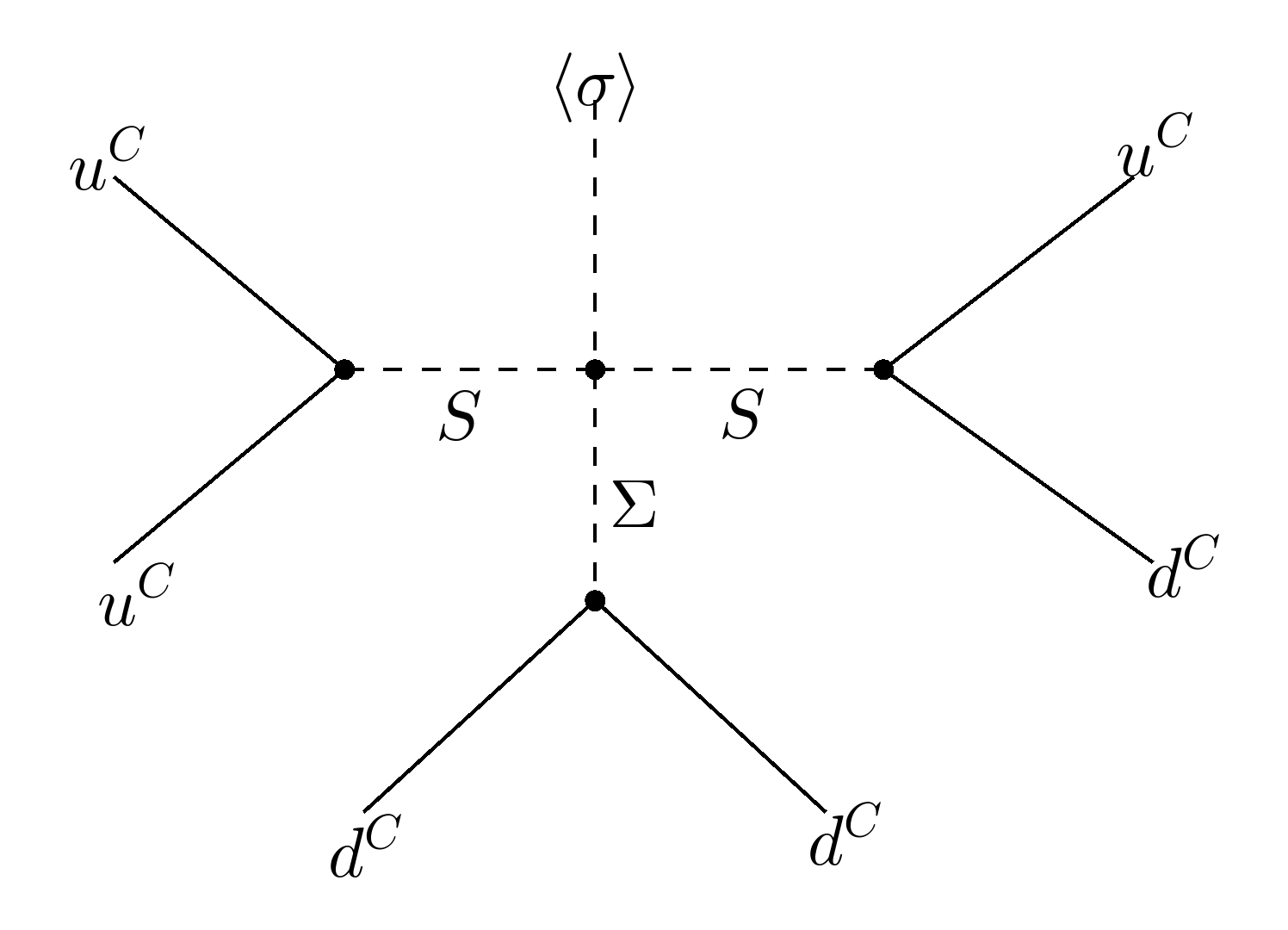}
    \caption{Feynman graph depicting \nn oscillation, drawn using~\cite{Harlander:2020cyh}}
    \label{fig:c5:nnbar}
\end{figure}


By integrating out various colour sextet fields from the vertices specified in Eqs.~(\ref{eq:c5:q1}-\ref{eq:c5:q4}) and their corresponding hermitian conjugates, and using the diquark couplings computed in Eqs.~\eqref{eq:c5:sextet_120}, we derive an effective Lagrangian inducing baryon-antibaryon oscillations at the leading order, as shown below;
\be \label{eq:c5:L_nn}
{\cal L}_{\rm eff}^{|\Delta B|=2} = \sum_{i=1}^3\,c_i\,{\cal O}_i\,\hc, \ee
with 
\beqa \label{eq:c5:O_i}
{\cal O}_1 & = & \frac{1}{2}\varepsilon_{\alpha \beta \gamma} \varepsilon_{\sigma \rho \eta}\,\left(u^{C \dagger}_{\alpha A}\,\cc\,d^{C *}_{\beta B}\right) \left(u^{C \dagger}_{\sigma C}\,\cc\,d^{C *}_{\rho D}\right)
\left(d^{C \dagger}_{\gamma E}\,\cc\,d^{C *}_{\eta F}\right)\,, \nonumber \\
{\cal O}_2 & = & \varepsilon_{\alpha \beta \gamma} \varepsilon_{\sigma \rho \eta}\,\left(d^{C \dagger}_{\alpha A}\,\cc\,d^{C *}_{\sigma B}\right) \left(d^{C \dagger}_{\beta C}\,\cc\,d^{C *}_{\rho D}\right)
\left(u^{C \dagger}_{\gamma E}\,\cc\,u^{C *}_{\eta F}\right)\,, \nonumber \\
{\cal O}_3 & = & \frac{1}{2}\varepsilon^{\alpha \beta \gamma} \varepsilon^{\sigma \rho \eta}\,\left(u^{\dagger}_{\alpha A}\,\cc\,d^{*}_{\beta B}\right) \left(u^{\dagger}_{\sigma C}\,\cc\,d^{*}_{\rho D}\right)
\left(d^{C T}_{\gamma E}\,\cc\,d^{C}_{\eta F}\right)\,.\eeqa
In the above Eq.~\eqref{eq:c5:O_i}, the quark fields are in their physical basis. The operators \({\cal O}_1\) and \({\cal O}_2\) are derived from the quartic terms found in Eq.~\eqref{eq:c5:q1} and Eq.~\eqref{eq:c5:q2}, respectively. The subsequent terms, outlined in Eqs.~(\ref{eq:c5:q3}, and \ref{eq:c5:q4}), both lead to a single operator, which is \({\cal O}_3\).

The effective strengths of the \nn operators listed in Eq.~\eqref{eq:c5:O_i} are integrating out the heavy sextet fields. We integrate out the sextet fields from Eq.~\eqref{eq:c5:sextet_all} and substitute the expression in the vertices mentioned in Eqs.~(\ref{eq:c5:q1}-\ref{eq:c5:q4}). This gives us the different six fermion operators mentioned in Eq.~\eqref{eq:c5:O_i} along with its strength as mentioned below;
\beqa \label{eq:c5:c_i}
c_1 & = & \frac{2 \eta_{11} v_\sigma}{M_\Sigma^2 M_{S_1}^4}\,\left(U_{d^C}^\dagger F^* U_{d^C}^* \right)_{EF}\,\Big[\frac{4 i c_\theta^2}{15 \sqrt{15}}\, \left(U_{u^C}^\dagger F^* U_{d^C}^* \right)_{AB} \left(U_{u^C}^\dagger F^* U_{d^C}^* \right)_{CD} \Big. \nonumber \\
& + &  \frac{8 i s_\theta^2}{3 \sqrt{15}}\, \left(U_{u^C}^\dagger G^* U_{d^C}^* \right)_{AB} \left(U_{u^C}^\dagger G^* U_{d^C}^* \right)_{CD} \nonumber \\
& - & \Big. \frac{8 \sqrt{2} i c_\theta s_\theta}{15 \sqrt{3}}\, \left(U_{u^C}^\dagger F^* U_{d^C}^* \right)_{AB} \left(U_{u^C}^\dagger G^* U_{d^C}^* \right)_{CD} \Big]\,, \nonumber \\
c_2 & = & \frac{4 i}{15 \sqrt{15}}\,\frac{\eta_2 v_\sigma}{M_\Sigma^4 M_{\cal S}^2}\,\left(U_{d^C}^\dagger F^* U_{d^C}^* \right)_{AB}\,\left(U_{d^C}^\dagger F^* U_{d^C}^* \right)_{CD}\,\left(U_{u^C}^\dagger F^* U_{u^C}^* \right)_{EF}\, \nonumber \\
c_3 & = & \frac{2 v_\sigma}{M_\Sigma^2}\,\left(U_{d^C}^T F U_{d^C} \right)_{EF}\,\Big[\frac{24 i \eta_3}{15 \sqrt{15} M_{\mathbb{S}}^4}\, \left(U_{u}^\dagger F^* U_{d}^* \right)_{AB} \left(U_{u}^\dagger F^* U_{d}^* \right)_{CD} \Big. \nonumber \\
& - &\Big. \frac{8 i \eta_4}{3 \sqrt{15} M_{\overline{S}}^4}\, \left(U_{u}^\dagger G^* U_{d}^* \right)_{AB} \left(U_{u}^\dagger G^* U_{d}^* \right)_{CD} \Big].\, \eeqa
When $\sigma$ acquires $vev$, $\langle \sigma \rangle\,\equiv\, v_{\sigma}$, $B-L$ is violated by two units, as also discussed in subsection (\ref{ss:c3:B-Lvd}). The unitary matrices \(U_f\) and \(U_{f^C}\), where \(f = u, d\), are the rotation matrices within the flavour space and can be calculated directly from the corresponding quark mass matrices, as elaborated in the section~(\ref{sec:c3:results_model}). Further, the coefficients \(c_1\) and \(c_3\) receive contributions from both \(120_{\hh}\) and \(\overline{126}_{\hh}\).

We intend to map the above-mentioned operators, given in Eq.~\eqref{eq:c5:O_i}, with those listed in~\cite{Grojean:2018fus,Buchoff:2015qwa} and thus transform quark fields into left- and right-chiral notations using the relations specified in the Eq.~\eqref{eq:c1:conjugation}. The resultant operators are shown below;
\beqa \label{eq:c5:c_i_2}
{\cal O}_1 & = & \frac{1}{2}\varepsilon_{\alpha \beta \gamma} \,\varepsilon_{\sigma \rho \eta}\,\left(\overline{(u^{\alpha}_{R A})^C}\, d^{\beta}_{R B}\right)\, \left(\overline{(u^{\sigma}_{R C})^C}\, d^{\rho}_{R D}\right)\, \left(\overline{(d^{\gamma}_{R E})^C}\, d^{\eta}_{R F}\right) \,, \nonumber \\
{\cal O}_2 & = & \varepsilon_{\alpha \beta \gamma}\, \varepsilon_{\sigma \rho \eta}\,\left(\overline{(d^{\alpha}_{R A})^C}\, d^{\sigma}_{R B}\right)\, \left(\overline{(d^{\beta}_{R C})^C}\, d^{\rho}_{R D}\right)\, \left(\overline{(u^{\gamma}_{R E})^C}\, u^{\eta}_{R  F}\right)\,, \nonumber \\
{\cal O}_3^{*} & = & \frac{1}{2}\varepsilon_{\alpha \beta \gamma}\, \epsilon_{\sigma \rho \eta}\,\left(\overline{(u^{\alpha}_{L A})^C}\, d^{\beta}_{L B}\right)\, \left(\overline{(u^{\sigma}_{L C})^C}\, d^{\rho}_{L  D}\right)\, \left(\overline{(d^{\gamma}_{R E })^C}\, d^{\eta}_{R F}\right).\,\eeqa
To compute \nn\, oscillation, we set \(A=B=C=D=E=F=1\) in the expression of  \({\cal O}_i\) and \(c_i\) given in Eqs.~(\ref{eq:c5:O_i} and \ref{eq:c5:c_i_2}) respectively. Further, the operators \({\cal O}_1\) and \({\cal O}^*_3\) correspond to \({\cal O}_1\) and \({\cal O}_3\) of~\cite{Grojean:2018fus,Buchoff:2015qwa}, respectively. Additionally, the operator \({\cal{O}}_2\) in our case, aligns with \({\cal O}^1_{RRR}\) from~\cite{Buchoff:2015qwa}, which is itself a linear combination of the operators \({\cal O}_1\) and \({\cal O}_4\), as shown below;
\be \label{eq:c5:O2_O1}
{\cal O}_2 = \frac{1}{4} {\cal O}^1_{RRR} = \frac{1}{5} \left({\cal O}_4 - 12\, {\cal \tilde{O}}_1 \right).\,\ee
Notably, \({\cal O}_4\) has a vanishing nuclear matrix element~\cite{Syritsyn:2016ijx}, whereas \({\cal \tilde{O}}_1\) has a nuclear matrix element that is identical to that of \({\cal O}_1\). Consequently, in our analysis, the operator \({\cal O}_2\) is directly linked to \({\cal O}_1\). This simplification results in only two linearly independent operators, \({\cal O}_1\) and \({\cal O}_3\),  written in Eq.~\eqref{eq:c5:c_i_2} in our scenario.

The operators \({\cal O}_{1,3}\) need to be evolved from the scale of the sextet masses, indicated by \(\mu = M_\Phi\), down to \(\mu_0 = 2\) GeV. $\mu_{0}$ typically represents the energy scale at which nuclear matrix elements are calculated using lattice techniques. However, the scale of $M_{\Phi}$ is still ad-hoc as more than one scalars induce \nn. Consequently, we define $M_{\Phi}$ as the mean of the masses of the different sextets scalars involved in the \nn\, oscillation. Also, a key characteristic of the basis in which \({\cal O}_{1,3}\) are expressed is that they are not going to mix with each other under RGE flow. The mean lifetime of the \nn\, oscillation can then be calculated using these effective operators as follows:
\beqa \label{eq:c5:tau_nn}
\tau_{n \bar{n}}^{-1} &=& \left|\sum_{i=1,3} \langle \bar{n}|{\cal O}_i(\mu_0)|n\rangle\, c_i(\mu_0) \right|\,,\nonumber\\
&=& \left|\sum_{i=1,3} \langle \bar{n}|{\cal O}_i(\mu_0)|n\rangle\left(\frac{\alpha_s^{(4)}(m_b)}{\alpha_s^{(4)}(\mu_0)}\right)^{\frac{3 \gamma^{(0)}_i}{50}}\left(\frac{\alpha_s^{(5)}(m_t)}{\alpha_s^{(5)}(m_b)}\right)^{\frac{3 \gamma^{(0)}_i}{46}}\right.\nl
& & \;\times\; \left.\left(\frac{\alpha_s^{(6)}(M_\Phi)}{\alpha_s^{(6)}(m_t)}\right)^{\frac{\gamma^{(0)}_i}{14}} c_i(M_\Phi) \right|,\, \eeqa
where \(\gamma^{(0)}_i\) represents the leading order anomalous dimension of the operator \({\cal O}_i\), and \(\alpha_s^{(n_f)}\) denotes the strong coupling constant for \(n_f\) flavours of light quarks to be computed at the top $(m_{t})$ and bottom $(m_{b})$ pole mass. We take \(\gamma^{(0)}_1 = 4\) and \(\gamma^{(0)}_3 = 0\) which are specified in~\cite{Buchoff:2015qwa}. Incorporating the results of the hadronic matrix elements of $O_{1}$ and $O_{3}$ computed from lattice computation, as  given in~\cite{Syritsyn:2016ijx} and the running effects as described in~\cite{Grojean:2018fus}, we arrive at the following expression for oscillation time of neutron to the antineutron in our case:
\be \label{eq:c5:tau_nn_2}
\tau_{n \bar{n}}^{-1} = \left|0.760\, \left(\frac{\alpha_s(M_\Phi)}{\alpha_s(10^5\,{\rm GeV})} \right)^{2/7}\, \left( c_1 - \frac{12}{5}\, c_2 \right)\,+\, 1.08\, c_3^* \right|\,\Lambda_{\rm QCD}^6\,. \ee
The $n$-$\bar{n}$ oscillation time in the given \so\, model can be explicitly calculated by substituting $c_{1,3}$ from Eq.~\eqref{eq:c5:c_i} in the aforementioned relation given in Eq.~\eqref{eq:c5:tau_nn_2}.

Evidently, \((U_{u}^\dagger G^* U_{d}^*)_{11} \neq 0\) despite \(G\) being antisymmetric in the flavour space, as \(U_u\) and \(U_d\) are in general different. As a result, the colour sextet scalars residing in \(120_{\hh}\) can also contribute to non-vanishing \nn\, oscillations together with the sextets residing in $\overline{126}_{\hh}$.

\section{Perturbativity of the Effective Quartic Couplings}
\label{sec:c5:quartic}
\begin{figure}[t]
\centering
\includegraphics[scale=0.29]{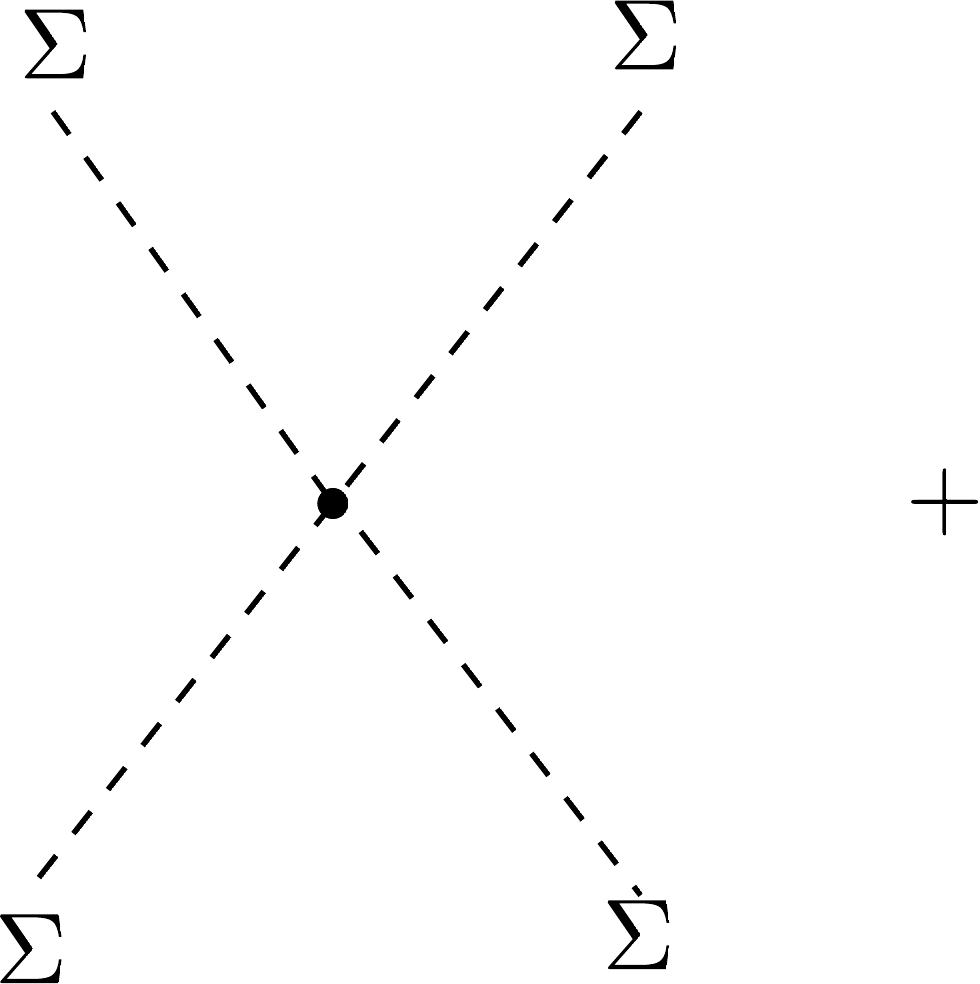}\hspace{0.5cm}\includegraphics[scale=1]{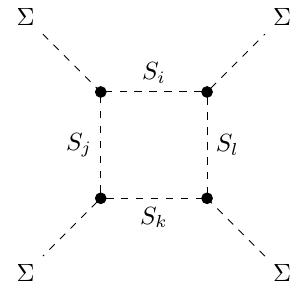}
\caption{Diagram showing quartic coupling of $\Sigma$ and a correction due to trilinear coupling among three colour sextet scalars. The vertex marked by a bullet point represents a trilinear coupling generated by the VEV of the \(B-L\) charged scalar \(\sigma\). }
\label{fig:c5:fig1}
\end{figure}

From the expressions for \(c_i\) in Eq.~\eqref{eq:c5:c_i}, it is evident that maximising the \nn\, transition rate would require a large \(v_\sigma\) and the presence of at least two of the three colour sextet fields at a lower scale. However, this scenario may lead to large negative effective quartic couplings for the light scalars~\cite{Babu:2002uu}. This issue stems from corrections induced by trilinear terms to the quartic coupling. These trilinear vertices are generated when \(\sigma\) acquires a $vev$, as shown in Eqs.~(\ref{eq:c5:q1}-\ref{eq:c5:q4}). We focus on the first term, Eq.~(\ref{eq:c5:q1}), to evaluate the constraints on the masses of these underlying scalars and calculate the correction to the quartic coupling of \(\Sigma\), $i.e.$ $\lambda$, as shown in the diagram depicted in Fig.~(\ref{fig:c5:fig1}).

The correction to the tree-level quartic coupling from trilinear vertices is as follows:
\be \label{eq:c5:lambda_eff}
-i\,\lambda_{\rm eff} = -i\,\lambda + i\, \delta \lambda,\,\ee
where the computation of $\delta \lambda$ for the quartic vertex appearing in Eq.~\eqref{eq:c5:q1} using the diagram shown in Fig.~(\ref{fig:c5:fig1})  in the limit of vanishing external momentum is given as follows:
\beqa \label{eq:c5:dl}
 i\, \delta\lambda \eq 4\sum_{i,j,k,l}\, \eta_{ij} \eta_{jk}^* \eta_{kl} \eta_{li}^*\, v_\sigma^4\, \nl
 & & \;\times\;\underbrace{\int \frac{d^4p}{(2 \pi)^4} \frac{1}{(p^2 - M_{S_i}^2) (p^2 - M_{S_j}^2) (p^2 - M_{S_k}^2) (p^2 - M_{S_l}^2)}}_{\equiv I_{ijkl}},\,\eeqa
where the factor 4 in the above Eq.~\eqref{eq:c5:dl} is a symmetry factor. Further, computation of the diagram by putting $i,j,k,l=1,2$ leads to the following;
\beqa \label{eq:c5:dl2}
i\, \frac{\delta \lambda}{4 v_\sigma^4} &=& |\eta_{11}|^4\, I_{1111} + |\eta_{22}|^4\, I_{2222} + 4|\eta_{12}|^2 \left( |\eta_{11}|^2\, I_{1112} + |\eta_{22}|^2\,I_{2221}\right) \, \nonumber \\
& + & \left(2 |\eta_{12}|^4 + 4\, {\rm Re}[\eta_{11} \eta_{22} \eta_{12}^* \eta_{12}^*]\right)\, I_{1122}.\,\eeqa
It is evident from the diagram, Fig.~(\ref {fig:c5:fig1}) that the amplitude of the diagram is finite. Additionally, the different loop integrals appearing in the above Eq.~(\ref{eq:c5:dl2}) are defined as follows~\cite{Buras:2002vd}:
\beqa \label{eq:c5:Int}
I_{iiii} &=& \frac{i}{4 \pi^2}\,\frac{1}{24 M_{S_i}^4}\,, \nonumber \\
I_{iiij} &=& \frac{i}{4 \pi^2}\,\frac{M_{S_i}^4 - M_{S_j}^4 + 2 M_{S_i}^2 M_{S_j}^2\, \log(M_{S_j}^2/M_{S_i}^2)}{8 M_{S_i}^2 (M_{S_i}^2 - M_{S_j}^2)^3}\,, \nonumber \\
I_{iijj} &=& \frac{i}{4 \pi^2}\,\frac{1}{4 (M_{S_i}^2 - M_{S_j}^2)^2}\,\left(\frac{M_{S_i}^2 + M_{S_j}^2}{M_{S_i}^2 - M_{S_j}^2}\, \log\left(\frac{M_{S_i}^2}{M_{S_j}^2}\right) - 2\right)\,.\eeqa

The integrals appearing in Eq.~\eqref{eq:c5:Int} modulo \(i\) yield positive results. The sole negative contribution to \(\delta \lambda\) could originate from the last term in Eq.~\eqref{eq:c5:dl2}. However, this term is insufficient to nullify the positive contributions from the other terms. As a result, \(\delta \lambda\) remains positive. For, \(M_{S_1} \ll M_{S_2}\), Eq.~\eqref{eq:c5:dl2} can be reduced to the following;
\be \label{eq:c5:dleff}
\lambda_{\rm eff} \simeq \lambda - \frac{|\eta_{11}|^4}{24 \pi^2}\,\frac{v_\sigma^4}{M_{S_1}^4}\,.\ee
If $\lambda_{\rm eff} < \lambda$, $SU(3)_C$ vacuum would not be stable. Consequently, both $\lambda_{\rm eff}$ and $\lambda$  should be positive and perturbative, which leads to a constraint on the mass of sextet propagating inside the loop, $i.e.$ $M_{S_1}$.
\be \label{eq:c5:MS}
M_{S_1} \geq \frac{|\eta_{11}| v_\sigma}{(24 \pi^2)^{1/4}}\,,\ee
provided $M_{\Sigma} < v_\sigma$. If $M_{\Sigma} > v_\sigma$, the effective theory below $v_{\sigma}$ is devoid of $\Sigma$ and hence its quartic term. Consequently, the constraint mentioned in Eq.~\eqref{eq:c5:MS} is not applicable in such a scenario.

The above discussion on the perturbativity of quartic coupling can be easily extended to include the other sextet scalars and their interactions, as mentioned in Eqs.~(\ref{eq:c5:q1}-\ref{eq:c5:q4}). Generally, when two or more sextet scalars are coupled through a \(B-L\) violating vertex, the $SU(3)_C$ potential becomes unstable if all are significantly below the \(B-L\) breaking scale. To avoid this issue, at least one of these sextets must have a mass greater than the \(B-L\) breaking scale. Consequently, this requirement would increase \nn \, oscillation time in models with a high \(B-L\) breaking scale.

\section{Baryogenesis}
\label{sec:c5:baryo}
So far, we have considered the contribution of sextet scalar fields in generating $\Delta F = \Delta B =2$ processes in the sections~(\ref{sec:c5:qfv} and \ref{sec:c5:nnbar}). Further, we now explore the potential of sextet scalars in generating the observed baryon asymmetry of the universe. The sextets scalar fields are also charged under $B-L$ and, consequently, can also play a role in generating baryon asymmetry of the universe. One of the fundamental prerequisites to generate the baryon asymmetry is the presence of $B-L$ violation, which is inherently in the \so\, GUT framework. Furthermore, once we have $B-L$ violation, it is immuned from its washout by electroweak sphalerons~\cite{Kuzmin:1987wn,Harvey:1990qw}.

The relevant sextet scalars for this include \(\Sigma\), \(S\), \(\tilde{S}\) (or \(S_{1,2}\) in the physical basis), as given in Eqs.~(\ref{eq:c5:sextet_all} and \ref{eq:c5:q1}), and the pertinent Lagrangian is shown as follows:
\beqa \label{eq:c5:Lbaryo1}
-{\cal L} &\supset &  Y^\Sigma_{AB}\,d^{C T}_{\alpha A}\,C^{-1}\,d^C_{\beta B}\, \Sigma^{\alpha \beta}\,+\, Y^{S_i}_{AB}\,\epsilon^{\alpha \beta \gamma}\, u^{C T}_{\gamma A}\,C^{-1}\,d^C_{\sigma B}\, {S_i}^\sigma_{\alpha \beta}\, \nonumber \\
& + & \eta_{ij}\, \sigma\, \Sigma^{\alpha \beta}\, {S_i}^\gamma_{\alpha \gamma}\, {S_j}^\theta_{\beta \theta}\,\hc\,.\eeqa
The Lagrangian described in the Eq.~\eqref{eq:c5:Lbaryo1} contains all the essential conditions to generate baryon asymmetry, called Sakharov conditions~\cite{Sakharov:1967dj}. It naturally includes parity (P) violation since \(\Sigma\) interacts exclusively with right-chiral fermions, while CP violation can emerge from the phases in \(\eta_{ij}\), which are elaborated upon below. Additionally, the $vev$ of \(\sigma\) leads to $B-L$ violation, as previously mentioned. The requirement for departure from thermal equilibrium is fulfilled through the out-of-equilibrium decays of \(\Sigma\) in an expanding universe, which will be further discussed below.

We assume the mass hierarchy $M_\Sigma \gg M_{S_2} > M_{S_1} \gg m_t$. The relevant scatterings and decays to compute baryogenesis can be read from Eq.~\eqref{eq:c5:Lbaryo1} and are given as follows:
\begin{itemize}
\item $B$ conserving: decay $\Sigma \to d^C_A\, d^C_B$ and scatterings $\Sigma\, S_i^* \to d^C_A\,\overline{u^C}_B$, $\Sigma\,u^C \to S_i\,d^C$ and $\Sigma\,\overline{d^C}_A \to S_i\,\overline{u^C}_B$,
\item $B$ violating: decay $\Sigma \to S_i^*\, S_j^*$ and scatterings $S_i^*\,S_j^* \to d^C_A\,d^C_B$, $\Sigma\, S_i \to \overline{d^C}_A\,\overline{u^C}_B$, $S_i^*\,\overline{d^C}_A \to S_j\,d^C_B$, $\Sigma\, u^C_A \to S_i^*\,\overline{d^C}_B$ and $\Sigma\,d^C_A \to S_i^*\,\overline{u^C}_B$,
\end{itemize}
together with their CP conjugate and inverse processes.

We begin by computing the CP asymmetry in the baryon number violating decay of $\Sigma \to S^{*}_i\,S^{*}_j$. As shown below, CP asymmetry can be generated from the interference between the tree and the one-loop process, which has an absorptive part;
\be \label{eq:c5:eps1}
\epsilon_{ij} = \frac{\Gamma[\Sigma \to S^*_i S^*_j] - \Gamma[\Sigma^* \to S_i S_j]}{\Gamma_{\rm tot}[\Sigma]}\,.\ee

\begin{figure}[t]
\centering
\includegraphics[width=\textwidth]{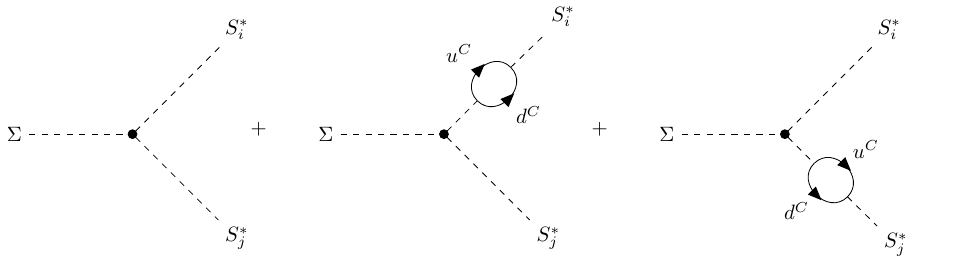}
\caption{The decays of the colour sextet scalar ($\Sigma$), both at tree level and through one-loop diagrams, contribute to the generation of CP asymmetry. The vertex marked by the bullet point represents a trilinear coupling, which is induced by the $vev$ of the \(B-L\) charged scalar, \(\sigma\), drawn using~\cite{Binosi:2008ig}}
\label{fig:c5:fig2}
\end{figure}

The relevant diagrams showing the tree and one loop level depicting the decay of $\Sigma$ are shown in Fig.~(\ref{fig:c5:fig2}).
The resulting leading order CP asymmetry is shown below; 
\be \label{eq:c5:eps2}
\epsilon_{ij} = -\frac{1}{\pi}\, \frac{{\rm BR}[\Sigma \to S^*_i S^*_j]}{|\eta_{ij}|^2}\,\left(\sum_{k\neq i} \frac{x_{i/k}}{1-x_{i/k}}\,{\rm Im}\left[\eta^*_{ij} \eta_{jk}\,{\rm Tr}[Y^{S_i \dagger} Y^{S_j}]\right]\,+\, i \leftrightarrow j \right)\,, \ee
with $x_{i/k} \equiv M_{S_i}^2/M_{S_k}^2$. Further, $\epsilon_{ij} = \epsilon_{ji}$ as $\eta_{ij} = \eta_{ji}$ in Eq.~\eqref{eq:c5:Lbaryo1}. The total CP asymmetry produced in decays of $\Sigma$ will be sum over all the decays of $\Sigma$ into different $S_{1,2}$:
\be \label{eq:c5:eps_tot}
\epsilon = \epsilon_{11} + \epsilon_{22} + 2 \epsilon_{12}\,,\ee

The asymmetry created between the number densities of \(S_i\) and \(S_i^*\), resulting from the out-of-equilibrium decay of \(\Sigma\), is subsequently transferred to the SM right-chiral quarks through the decays of \(S_i\). This transfer occurs at temperatures below the freeze-out point of the asymmetry, given that \(M_{S_i} \ll M_\Sigma\) and occurs when \(S_i\) exits thermal equilibrium.

The final baryon-to-entropy ratio is defined as;
\be \label{eq:c5:YB}
Y_B = \frac{4}{3}\,\epsilon\, \frac{\kappa(K)}{g_*}.\, \ee
where, the factor of \(4/3\) represents the total \(B-L\) charge generated after the decay of \(\Sigma\). The parameter \(g_*\) denotes the effective number of relativistic degrees of freedom present at the time of decay, which is approximately 125, including \(\Sigma\), \(S_{1,2}\), and the SM particles. The function \(\kappa(K)\) is an efficiency factor that accounts for the potential washout of asymmetries due to inverse decays and various scattering processes previously discussed. The parameter \(K\) is crucial as it measures the out-of-equilibrium condition and is defined as follows:
\be \label{eq:c5:K}
K = \frac{\Gamma[\Sigma \to S_i^*\,S_j^*]}{2 H(M_\Sigma)},\,\ee
together with the Hubble parameter as:
\be \label{eq:c5:H}
H(T) = 1.66\, g_*^{1/2}\,\frac{T^2}{M_P}\,.\ee

An exact value for \(\kappa\) is typically determined by numerically solving the full Boltzmann equations, as given in~\cite{Herrmann:2014fha,Fridell:2021gag}. However, an approximate analytical solution for \(\kappa\) is available and well-suited for the current setup. Assuming that \(\Sigma\) starts with an initial thermal abundance, the formula for \(\kappa\) is provided by~\cite{Buchmuller:2004nz}:
\be \label{eq:c5:kappa}
\kappa(x) = \frac{2}{x\, z_B(x)} \left(1-\exp\left[-\frac{1}{2}\, x\, z_B(x)\right]\right)\,,
\ee
such that,
\be \label{eq:c5:zB}
z_B(x) = 2 + 4\, x^{0.13}\, \exp\left[-\frac{2.5}{x} \right]\,.\ee
The solution for \(\kappa\) mentioned above in Eq.~\eqref{eq:c5:kappa} primarily accounts for washout effects from inverse decays and is applicable when \(K \leq 10^3\). For values of \(K\) exceeding \(10^3\), scattering processes become more significant, leading to an exponential decrease in \(\kappa\). It is important to note that when \(K \leq 1\), \(\kappa(K) \to 1\) indicates that there is no dilution in the baryon asymmetry due to the washout.

We aim to demonstrate the amount of CP violation, as described by Eq.~\eqref{eq:c5:eps_tot}, within our framework is sufficient to account for the observed baryon to entropy ratio of the universe, $i.e.$ \(Y_B^{\rm exp} = (6.10 \pm 0.04) \times 10^{-10}\)~\cite{Planck:2018vyg}. To achieve this, we first determine the maximum possible value of \(\epsilon\) and subsequently assess the amount of damping permitted through Eq.~\eqref{eq:c5:YB}, ensuring that \(Y_B \geq 6.0 \times 10^{-10}\). Assuming the complex nature of $\eta_{ij}$, parameterising $\eta_{ij}\,=\,|\eta|\,\exp\left(\phi_{ij}\right)$ and that the branching ratio \({\rm BR}[\Sigma \to S^*_i S^*_j] \approx 1/4\) for all \(i\) and \(j\) in Eqs.~(\ref{eq:c5:eps2} and \ref{eq:c5:eps_tot}), leads to the following maximum value of CP asymmetry; 
\be \label{eq:c5:eps_tot_2}
\epsilon = \frac{1}{2\pi} \left(\frac{1+x_{1/2}}{1-x_{1/2}}\right)\, \left|{\rm Tr}[Y^{S_1 \dagger} Y^{S_2}]\right|\,\left(\sin(\phi_{12}-\phi_{22} - \phi_{Y}) - \sin(\phi_{12}-\phi_{11}+\phi_{Y}) \right)\,, \ee
where $\phi_{Y} \equiv {\mathrm Arg}({\rm Tr}[Y^{S_1 \dagger} Y^{S_2}])$ is the phase that arise from the Yukawa couplings. For $x_{1/2} \ll 1$, implying a large hierarchy between in masses of the decay product species, the maximum possible value of $\epsilon$ becomes the following; 
\be \label{eq:c5:eps_max}
\epsilon_{\rm max} \simeq \frac{1}{\pi} \left|{\rm Tr}[Y^{S_1 \dagger} Y^{S_2}]\right|\,. \ee
Substituting \(Y^{S_{1,2}}\) from Eq.~\eqref{eq:c5:YPhi} and using \({\rm Tr}[F^\dagger G] = 0\), which  is due symmetricity and antisymmetricity of \(F\) and  \(G\) respectively. We find the following maximum possible value of CP asymmetry in this case as follows;
\be \label{eq:c5:eps_max2}
\epsilon_{\rm max} \simeq \frac{2}{3\pi} \left| \sin 2\theta\, {\rm Tr}\left[G^\dagger G - \frac{1}{10} F^\dagger F \right]\right|\,. \ee

\begin{figure}[t]
\centering
\includegraphics[scale=0.65]{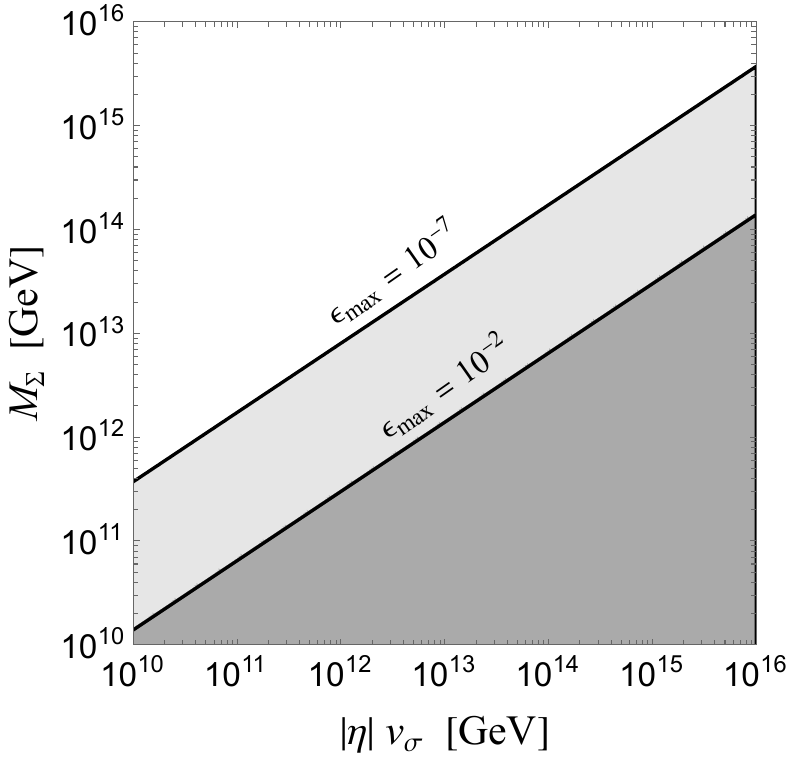}
\caption{Values of \(M_\Sigma\) and \(|\eta| v_\sigma\) that are ruled out by \(Y_B > 6.0 \times 10^{-10}\) are indicated for two scenarios: \(\epsilon_{\rm max} = 10^{-7}\) is represented by a lighter gray shade, and \(\epsilon_{\rm max} = 10^{-2}\) by a darker gray.}
\label{fig:c5:fig3}
\end{figure}

Further, for $M_{S_i} \ll M_\Sigma$, one finds $\Gamma[\Sigma \to S_i^*\,S_j^*] \simeq \frac{|\eta_{ij}|^2 v_\sigma^2}{16 \pi M_\Sigma}$. Substituting this in the expression of $K$ given in Eq.~\eqref{eq:c5:K} and setting $|\eta_{ij}| = |\eta|$, we get
\be \label{eq:c5:K2}
K = 1.07 \times \left(\frac{|\eta| v_\sigma}{10^{15}\,{\rm GeV}}\right)^2 \times \left(\frac{10^{15}\,{\rm GeV}}{M_\Sigma} \right)^3\,.\ee
Substituting $K$, shown in the above Eq.~\eqref{eq:c5:K2}, into \(\kappa(K)\) given in Eq.~\eqref{eq:c5:kappa}, the baryon to entropy ratio \(Y_B\) can be calculated based on the given value of \(\epsilon_{\rm max}\) in Eq.~\eqref{eq:c5:eps_max2}. Adhering to the constraint that \(Y_B \geq 6.0 \times 10^{-10}\) the allowed regions of $M_{\sigma}$ as a function quartic coupling times the $vev$ are shown in Fig.~(\ref{fig:c5:fig3})~\cite{Patel:2022nek}.

We infer from Fig.~(\ref{fig:c5:fig3}) that to have a viable observed baryon asymmetry of the universe from the decay of $M_{\sigma}$, its mass cannot be less than the $B-L$ scale.

The generation of baryon asymmetry discussed in this section relies on sextet fields from the Yukawa sectors of GUT scalars and requires both \(120_{\hh}\) and \(\overline{126}_{\hh}\). In cases where only one particular irrep, either \(120_{\hh}\) or \(\overline{126}_{\hh}\), is present in the Yukawa sector and gives rise to one sextet field, an additional sextet could emerge from \(54_{\hh}\). A similar mechanism to generate the asymmetry of the universe is still applicable~\cite{Babu:2012vc}. Further, scalars other than sextets can also be used to generate baryon assymetry~\cite{Hati:2018cqp}.

\section{Constraints on the Sextets}
\label{sec:c5:results}

We will now constraint the mass scales of various colour sextet scalars based on the several phenomenological implications discussed in previous sections, including quark flavour violation (section~(\ref{sec:c5:qfv})), \(n\)-\(\overline{n}\) oscillation (section~(\ref{sec:c5:nnbar})), constraints on the perturbativity of the quartic coupling (section~(\ref{sec:c5:quartic})), and the ability to generate cosmological baryon asymmetry (section~(\ref{sec:c5:baryo})). From the discussions of previous sections, it is apparent that the phenomena related to sextets depend on two critical parameters: (a) the Yukawa couplings with the quarks, $i.e.$ the matrices \(G\) and \(F\), and (b) the \(B-L\) breaking  \(v_\sigma\).
The Yukawa matrices of a realistic non-supersymmetric \so\, model based on $10_{\hh}$ and $\overline{126}_{\hh}$ and the matrix $F$ can be inferred from the Eq.~\eqref{eq:c3:HF_form} from the Chapter~(\ref{ch:3}), while their explicit form can be inferred from the Appendix~(\ref{app:2}). 

\be \label{eq:c5:FG_fit}
G \sim \frac{\lambda^4}{\alpha_3} \left(\ba{ccc} 0 & \lambda^4 & \lambda^3\\
- \lambda^4 & 0 & \lambda^2\\
- \lambda^3 & -\lambda^2 & 0 \ea\right)\,,~~F \sim \frac{\lambda^4}{\alpha_2} \left(\ba{ccc} \lambda^5 & \lambda^4 & \lambda^3\\
\lambda^4 & \lambda^3 & \lambda^2\\
\lambda^3 & \lambda^2 & \lambda \ea\right).\,\ee
We have omitted the \({\cal O}(1)\) coefficients for each element of the matrices \(F\) and \(G\). The matrix elements of \(G\) are assumed to be of similar magnitudes as of \(F\), as observed in an earlier fit~\cite{Joshipura:2011nn}. The parameters \(\alpha_{2,3}\), with \(|\alpha_{2,3}| \lesssim 1\), measure the extent of Higgs doublet mixing as discussed in the section~(\ref{sec:c3:results_model}) of the previous Chapter~(\ref{ch:3}).

Assuming that the light neutrino masses are generated via Type-I seesaw, consequently $v_{\sigma}$ is constrained as follows,
\be \label{eq:c5:vsgm_fit}
v_\sigma = \alpha_2\,v^\prime_S\, \simeq \alpha_2 \times 10^{15}\, {\rm GeV}\,, \ee
where \(v^\prime_S\) is explicitly defined in~\cite{Mummidi:2021anm} and is found to be in the range \(10^{14}\)-\(10^{15}\) GeV. In addition to \(v_\sigma\), \(F\), and \(G\), the analysis also requires unitary matrices, \(U_{f}\) and \(U_{f^C}\), which diagonalise the quark mass matrices \(M_f\) for \(f=u,d\). Their approximate forms can be referenced from the Eq.~(\ref{eq:c3:U_form}).

Further, we consider two example values for $\alpha_{2,3}$, which, in turn, constrains the $B-L$ scale and order of entries appearing in $G$ and $F$.
\begin{enumerate}[label=(\roman*)]
\item High-scale $B-L$ (HS): $\alpha_{2,3} = \lambda$. This implies
\be \label{eq:c5:FG_HS}
v_\sigma \simeq 10^{14}\,{\rm GeV}\,,~~G \simeq \left(\ba{ccc} 0 & \lambda^7 & \lambda^6\\
- \lambda^7 & 0 & \lambda^5\\
- \lambda^6 & -\lambda^5 & 0 \ea\right)\,,~~F \simeq \left(\ba{ccc} \lambda^8 & \lambda^7 & \lambda^6\\
\lambda^7 & \lambda^6 & \lambda^5\\
\lambda^6 & \lambda^5 & \lambda^4 \ea\right)\,.\ee

Given that \(\alpha_{2,3} = \lambda\), the lightest pair of electroweak doublet Higgs notably includes significant contributions from the doublets residing in \(120_{\hh}\) and \(\overline{126}_{\hh}\). Consequently, the entries of \(G\) and \(F\) must be relatively small in this context. Further, \(\alpha_{2,3}\) cannot be substantially greater than \(\lambda\) because, in such cases, the contribution of \(10_{\hh}\) to the fermion masses becomes negligible. And \(120_{\hh}\) and \(\overline{126}_{\hh}\) individaully cannot reproduce the observed fermion mass spectrum~\cite{Joshipura:2011nn}. {\label{pt:c5:aa}}
\item Intermediate-scale $B-L$ (IS): $\alpha_{2,3} = \lambda^5$. This leads to the following.

\be \label{eq:c5:FG_IS}
v_\sigma \simeq 10^{11}\,{\rm GeV}\,,~~G \simeq \left(\ba{ccc} 0 & \lambda^3 & \lambda^2\\
- \lambda^3 & 0 & \lambda\\
- \lambda^2 & -\lambda & 0 \ea\right)\,,~~F \simeq \left(\ba{ccc} \lambda^4 & \lambda^3 & \lambda^2\\
\lambda^3 & \lambda^2 & \lambda\\
\lambda^2 & \lambda & 1 \ea\right)\,.\ee

In this scenario, \(G\) and \(F\) exhibit relatively stronger couplings with the SM fermions, as the contributions from the doublets residing in \(120_{\hh}\) and \(\overline{126}_{\hh}\) to the lightest Higgs are suppressed. {\label{pt:c5:ab}}
\end{enumerate}

In a scenario where the \(B-L\) breaking scale, $i.e.$ (\(v_\sigma\)), is significantly lower than \(10^{11}\) GeV, \(\alpha_{2}\) would need to be much less than \(\lambda^5\). This condition might lead to non-perturbative values for some of the couplings in \(F\) within the class of realistic models. Additionally, a smaller \(\alpha_2\) with perturbative values for couplings in \(F\) suggests that the charged fermion masses predominantly originate from \( 10_{\hh}\) and \( 120_{\hh}\), a possibility which has been forbidden by the fits~\cite{Joshipura:2011nn}. Alternatively, the relationship between \(v_\sigma\) and the Yukawa couplings can be understood as follows: \(M\,=\, v_\sigma F\) is the right-handed neutrino Majorana mass matrix and the magnitude of the elements in \(M_R\) is dictated to reproduce the light neutrino masses via type I seesaw mechanism. Consequently, this necessitates smaller values for \(F\) when \(v_\sigma\) is close to the GUT scale and relatively larger values for \(F\) when \(v_\sigma\) assumes intermediate values.

Before estimating constraints on the sextet scalars for the HS and IS scenarios discussed in points~\ref{pt:c5:aa} and~\ref{pt:c5:ab}, it is crucial to consider a model-independent limit on their masses derived from direct search experiments. Colour sextets can be pair-produced at the Large Hadron Collider (LHC) via gluon fusion~\cite{Chen:2008hh,Han:2009ya,Richardson:2011df}. Notably, this production process is independent of the couplings with quarks, thus providing a stringent lower limit on the sextet's masses.

The non-observation of deviations from SM predictions so far implies~\cite{Richardson:2011df}:
\be \label{eq:c5:LHC}
M_{\Phi} \geq 1\,{\rm TeV}\,.\ee
Other direct search methods, such as resonant production and single top production, also put constraints on the sextet's couplings with quarks. These methods are known to provide relatively milder limits when the couplings are small~\cite{Pascual-Dias:2020hxo,Fridell:2021gag}. As a result, we will use the robust lower limit obtained from gluon fusion, given in Eq.~\eqref{eq:c5:LHC}, and proceed to constrain the sextets within the mass range of \(10^3\) to \(10^{16}\) GeV. The latter bound on the mass of sextets comes from the effective theory approach.  

\subsection{Light $\Sigma$, $S_1$}
{\label{ss:c5:l_sigma_S1}}

\begin{figure}[t]
\centering
\includegraphics[width=0.48\textwidth]{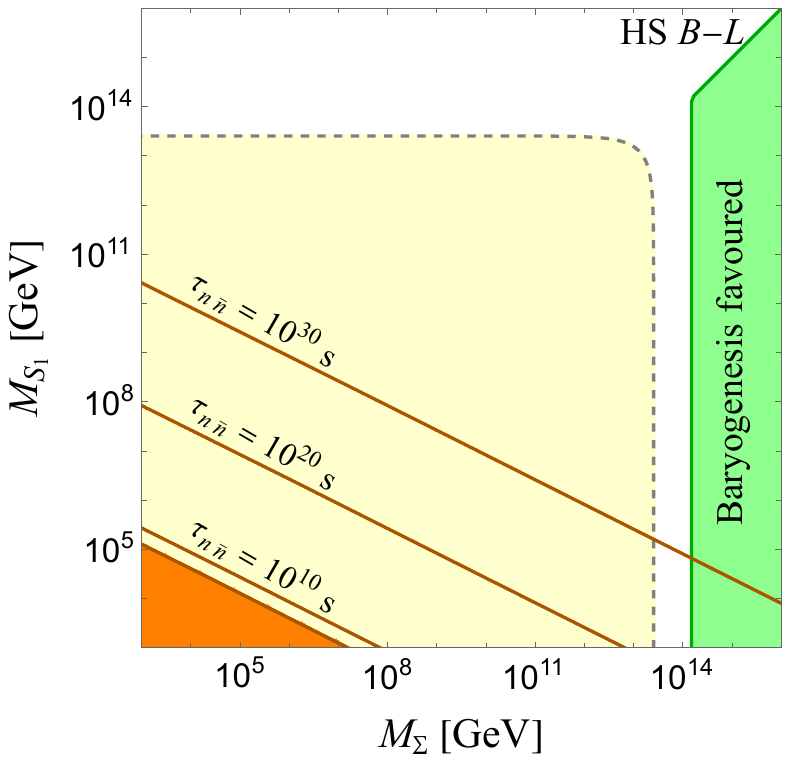}\hspace{0.2cm}
\includegraphics[width=0.48\textwidth]{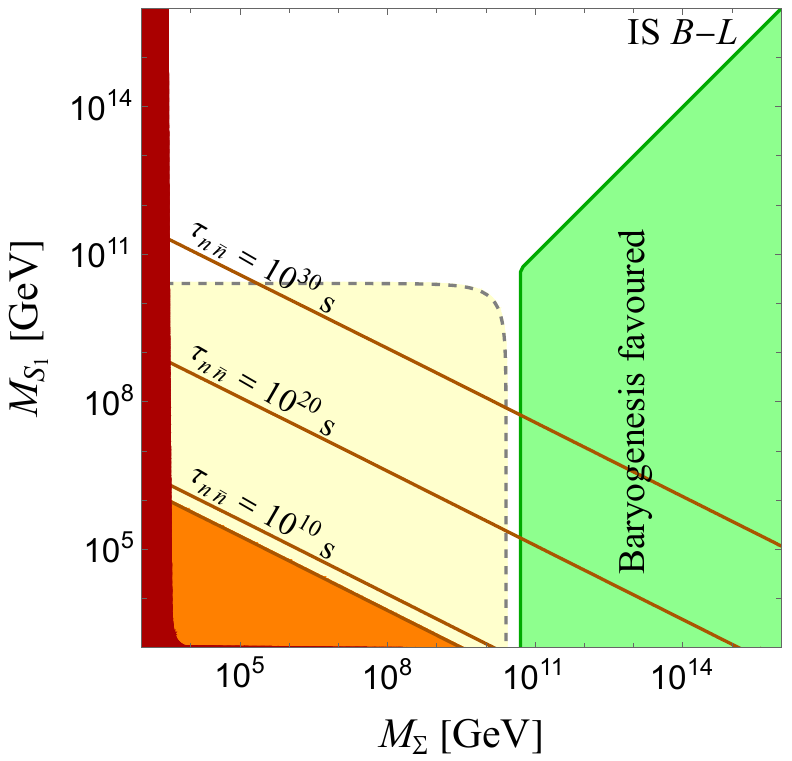}
\caption{The figure illustrates constraints on the masses of \(\Sigma\) and \(S_1\) for high (left panel) and intermediate (right panel) \(B-L\) breaking scales. The red-shaded area is ruled out based on neutral meson-antimeson oscillations and is the most stringent bound. The orange region is excluded due to the current limit on neutron-antineutron oscillation time, \(\tau_{n \bar{n}}>4.7\times 10^{8}\) seconds~\cite{Super-Kamiokande:2020bov}. The diagonal red lines, ascending from bottom to top, indicate \(\tau_{n\bar{n}} = 10^{10}\), \(10^{20}\), and \(10^{30}\) seconds, respectively. The yellow area outlined by the dashed contour is disfavoured due to the perturbativity of effective quartic couplings, while the green shaded area meets the criteria set by cosmological baryogenesis constraints.}
\label{fig:c5:fig4}
\end{figure}

To begin with, we assume $|\eta_{11}|=1$ and $\theta=\sfrac{\pi}{4}$ in Eq.~\eqref{eq:c5:c_i}. Further, together with the aforementioned assumption and using Yukawa couplings given in Eqs.~(\ref{eq:c5:FG_HS}, \ref{eq:c5:FG_IS}, and \ref{eq:c3:U_form}), substituting it Eq.~\eqref{eq:c5:tau_nn_2}), we compute the \nn\, transition time as shown below. Further, as discussed towards the end of the section~(\ref{sec:c5:results}), we restrict the $M_{\Sigma}$ as $1\,{\rm TeV} \le M_{\Sigma},M_{S_i} < M_{\rm GUT}$ while the remaining sextet fields stay close to the GUT scale~\cite{delAguila:1980qag}. 
\beqa \label{eq:c5:phb1}
\left|\left(U_{d^C}^\dagger F^* U_{d^C}^* \right)_{11}\right| \sim \left|\left(U_{u^C}^\dagger F^* U_{d^C}^* \right)_{11}\right| \sim \left|\left(U_{u^C}^\dagger G^* U_{d^C}^* \right)_{11}\right| \simeq \lambda^8\,, \nonumber\\
\left|\left(U_{d^C}^\dagger F^* U_{d^C}^* \right)_{11}\right| \sim \left|\left(U_{u^C}^\dagger F^* U_{d^C}^* \right)_{11}\right| \sim \left|\left(U_{u^C}^\dagger G^* U_{d^C}^* \right)_{11}\right| \simeq \lambda^4\,,
\eeqa
for both the HS and IS cases, respectively. Further,  we calculate the contributions of \( \Sigma \) and \( S_1 \) to meson-antimeson mixing as given in expressions in Eqs.~(\ref{eq:c5:CK}, \ref{eq:c5:CB}, and \ref{eq:c5:CD}) simultaneously adhering to the constraint given in Eq.~(\ref{eq:c5:const_C}). We set the matching scale at \( M_\Phi = (M_\Sigma + M_{S_1})/2 \), as also discussed in the previous section. The derived limits on \( M_\Sigma \) and \( M_{S_1 }\) from meson-antimeson and neutron-antineutron oscillation are shown in Fig.~(\ref{fig:c5:fig4}). 
Further, $S_{1}$ and $\Sigma$ cannot be simultaneously light as it will hamper the perturbativity of effectively generated quartic coupling, as discussed in the section (\ref{sec:c5:quartic}) and is also displayed in the Fig. (\ref{fig:c5:fig4})~\cite{Patel:2022nek}.

As shown in Fig.~(\ref{fig:c5:fig4}), in scenarios with a high-scale \(v_\sigma\), either \(\Sigma\) or \(S_1\) could be significantly lighter than the GUT scale. However, the perturbativity of the quartic coupling restricts both from being simultaneously lighter than \(10^{13}\) GeV. Their weak couplings with quarks mean that a TeV-scale \(\Sigma\) or \(S_1\) could remain largely unconstrained by \(|\Delta F|=2\) or \(|\Delta B|=2\) processes. When \(v_\sigma = 10^{11}\) GeV, \(M_{S_1}\) (or \(M_\Sigma\)) could be as low as 1 TeV (or 10 TeV) provided that the other sextet is heavier than \(v_\sigma\). In this scenario, a lighter \(M_{S_1}\) leads to a comparatively faster neutron-antineutron transition time, as observed in the right panel of Fig.~(\ref{fig:c5:fig4}). Overall, constraints from the perturbativity of quartic couplings largely forbid the possibility of observing \(n\)-\(\bar{n}\) oscillations in the near-future experiments for both the high and intermediate \(B-L\) breaking scales.

Sub-GUT scale \(\Sigma\) and \(S_i\) might also explain the universe's cosmological baryon asymmetry through thermal baryogenesis, as elaborated in the section~(\ref{sec:c5:baryo}). The maximum CP asymmetry calculated using Eq.~\eqref{eq:c5:eps_max2} for the High and Intermediate $B-L$ scale is reduced to be as follows:
\be \label{eq:c5:emax3}
\epsilon_{\rm max} =  \begin{cases}
      {\cal O} (10^{-7}), & \text{for HS}\\
      {\cal O} (10^{-2}), & \text{for IS}\\
    \end{cases}       \ee 
From the Fig. (\ref{fig:c5:fig3}), we infer following constraint on the mass of $\Sigma$ $(M_{\Sigma})$.
\be \label{eq:c5:msigma}
M_\Sigma \gtrsim  \begin{cases}
      2 \times 10^{14}\,{\rm GeV,} & \text{for HS}\\
      5 \times 10^{10}\,{\rm GeV,} & \text{for IS}\\
    \end{cases}       \ee 
The region where the baryon asymmetry, \(Y_B\), is at least \(6.0 \times 10^{-10}\), indicating it is favoured by baryogenesis generated by sextet fields, is also depicted in Fig.~(\ref{fig:c5:fig4}). For a very light \(S_1\) and \(v_\sigma \approx 10^{11}\) GeV, this region could potentially be investigated further through enhanced measurements of \(n\)-\(\bar{n}\) oscillations.

\subsection{Light $\Sigma$, ${\cal S}$}
{\label{ss:c5:sig_calS}}
Assuming \(|\eta_2| = 1\) in Eq.~\eqref{eq:c5:q2}, we evaluate constraints on a light \(\Sigma\) and \({\cal S}\) while keeping other sextets at the GUT scale. Unlike \(S_{1,2}\), \({\cal S}\) only interacts with up-type quarks and induces \dd oscillations at the tree level as well as the loop level, which significantly restricts strongly coupled TeV-scale \({\cal S}\).  
\begin{figure}[t]
\centering
\includegraphics[width=0.48\textwidth]{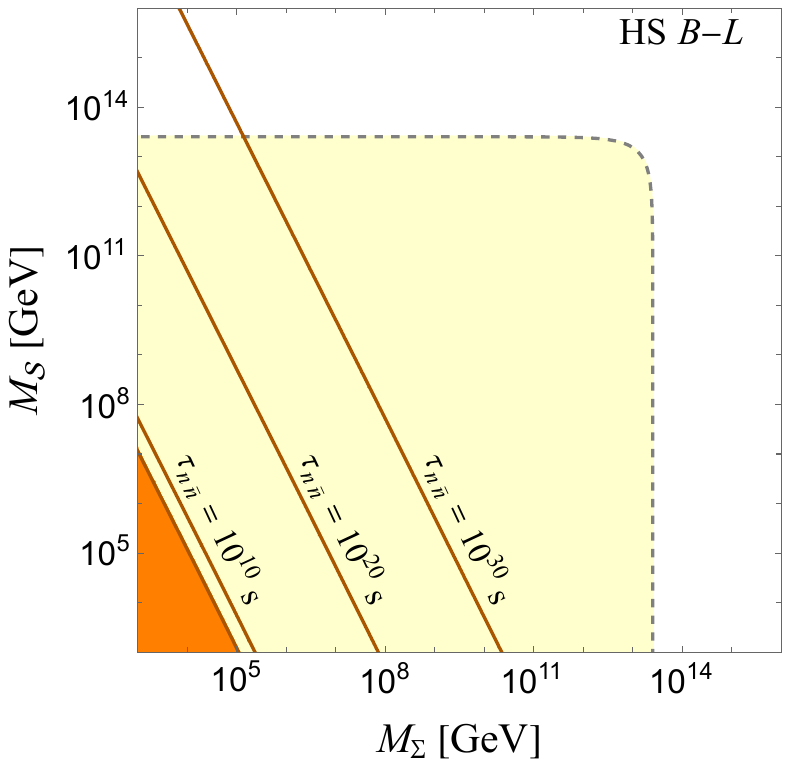}\hspace*{0.5cm}
\includegraphics[width=0.48\textwidth]{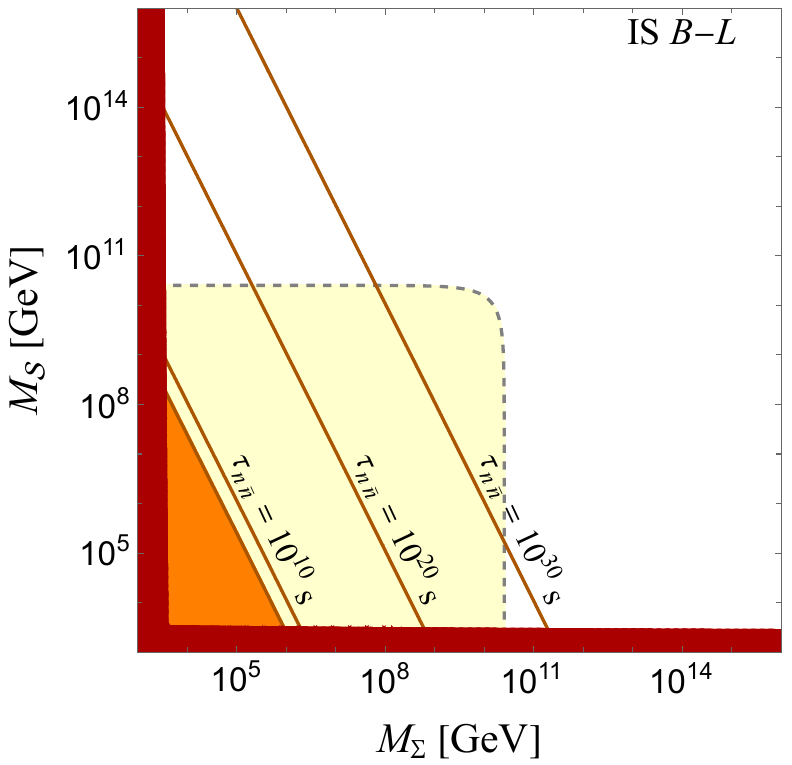}
\caption{Constraints on the masses of \(\Sigma\) and \({\cal S}\) for the high (left panel) and intermediate (right panel) \(B-L\) breaking scale. The additional details are the same as those in the caption of Fig. (\ref{fig:c5:fig4}).}
\label{fig:c5:fig5}
\end{figure}
It is evident from Fig.~(\ref{fig:c5:fig5}) that limits from meson-antimeson oscillations and the perturbativity of the effective quartic coupling suggest that an observable \(n\)-\(\bar{n}\) transition rate in near-future experiments is unlikely in realistic renormalisable \(SO(10)\) models.

\subsection{Light $\Sigma$, $\mathbb{S}$}
\begin{figure}[t]
\centering
\includegraphics[width=0.48\textwidth]{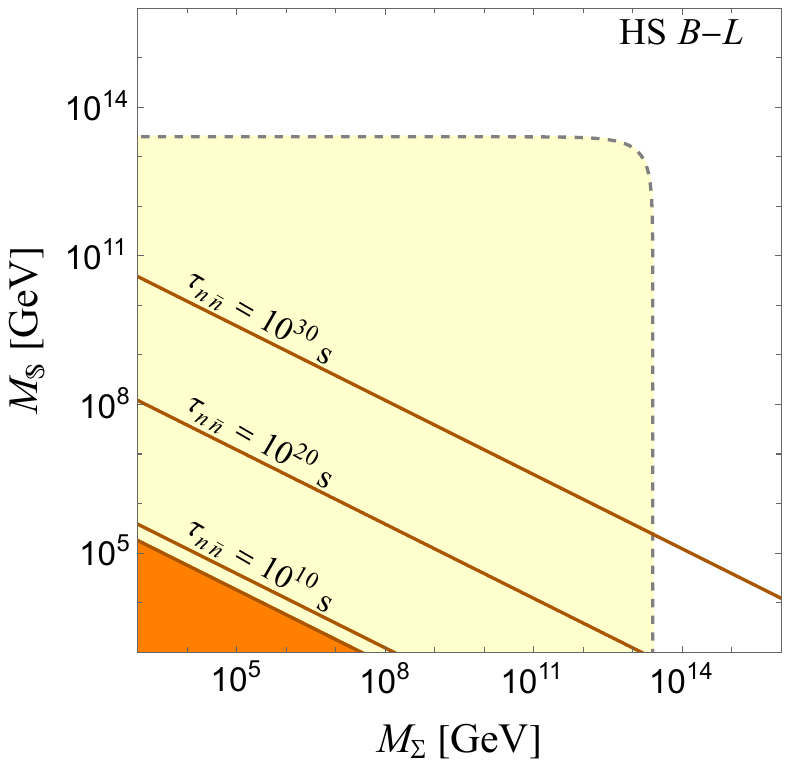}\hspace*{0.5cm}
\includegraphics[width=0.48\textwidth]{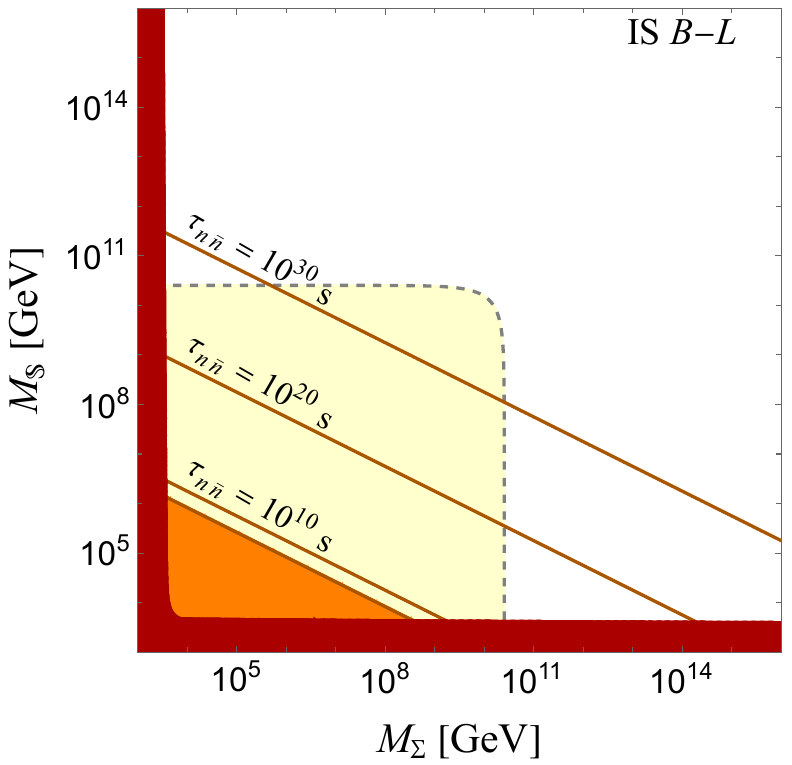}
\caption{Constraints on the masses of \(\Sigma\) and \(\mathbb{S}\) for the high (left panel) and intermediate (right panel) \(B-L\) breaking scale. The additional details are the same as those in the caption of Fig.~(\ref{fig:c5:fig4}).}
\label{fig:c5:fig6}
\end{figure}

{\label{ss:c5:sig_bbs}}
Along similar lines, we examine a scenario where light \(\Sigma\) and \(\mathbb{S}\) are considered while the remaining sextets have their masses at the GUT scale. The findings are presented in Fig.~(\ref{fig:c5:fig6}). For this analysis, we set \(|\eta_3| = 1\) as earlier cases and explore two scenarios regarding the values of \(F\) and \(v_\sigma\) as previously discussed. Unlike \(S_i\), the electroweak triplet \(\mathbb{S}\) mediates quark flavour violating interactions at tree level in both the up- and down-type quark sectors. This leads to a relatively high upper bound on \(M_{\mathbb{S}}\) when larger Yukawa couplings are involved. The constraints arising from \(n\)-\(\bar{n}\) oscillations and the perturbativity of the effective quartic coupling are comparable to those seen with light \(\Sigma\) and \(S_1\). Alone, sub-GUT scale \(\Sigma\) and \(\mathbb{S}\) are insufficient for generating baryon asymmetry, and an additional \(\Sigma\)-like scalar would be necessary for viable baryogenesis, as discussed in the section~(\ref{sec:c5:baryo}).

\subsection{Light $\Sigma$, $\overline{S}$}
{\label{ss:c5:msig_sbar}}
At last, 
\begin{figure}[t]
\centering
\includegraphics[width=0.48\textwidth]{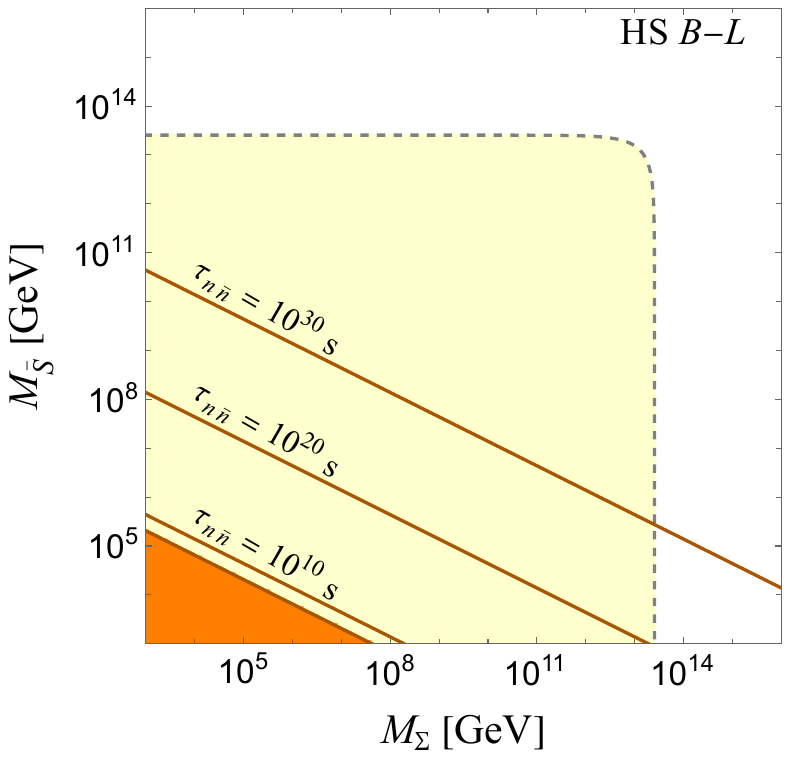}\hspace*{0.5cm}
\includegraphics[width=0.48\textwidth]{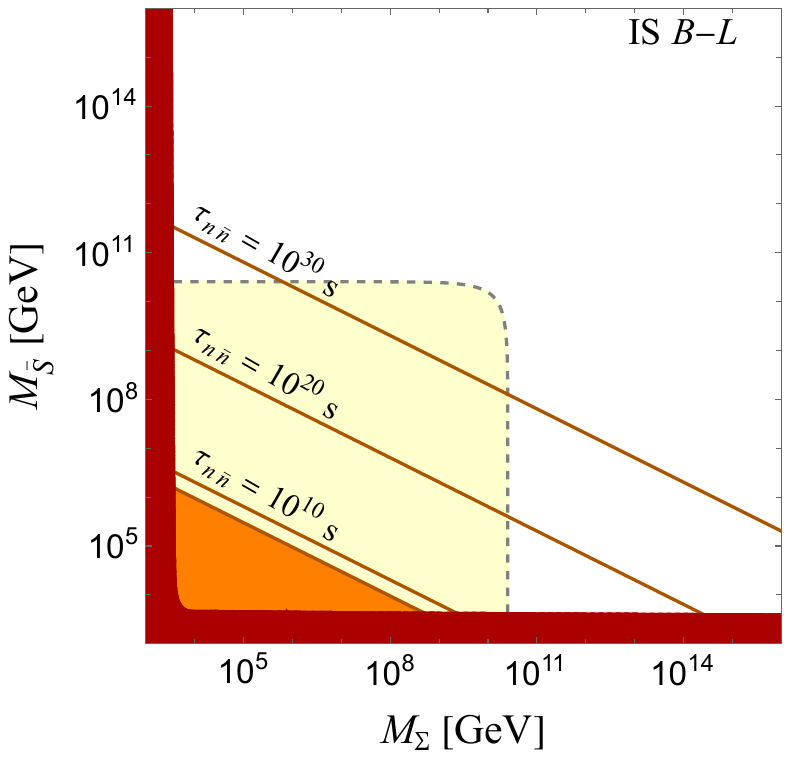}
\caption{Depiction of constraints on the masses of \(\Sigma\) and \(\overline{S}\) for the high (left panel) and intermediate (right panel) \(B-L\) breaking scale. The additional details remain the same as described in the caption of Fig.~\ref{fig:c5:fig4}.}
\label{fig:c5:fig7}
\end{figure}
 we explore the scenario where \(1\,{\rm TeV} \le M_{\Sigma}, M_{\overline{S}} < M_{\rm GUT}\), while remaining sextets are at the GUT scale. Setting \(|\eta_4| = 1\), the results after performing the similar as the previous cases are displayed in Fig.~(\ref{fig:c5:fig7}). Moreover, \(\overline{S}\) resides in \( 120_{\hh}\) and interacts with left-chiral up- and down-type quarks through flavour anti-symmetric couplings. As shown in \(Y^{\overline{S}}\) in Eq.~\eqref{eq:c5:YPhi}, its diagonal couplings disappear when \(U_u = U_d\); however, this condition does not appear in the usual cases. Consequently, effective diagonal couplings are generated by quark mixing, leading to a non-zero rate for \nn\, oscillations at the leading order. Similar to \(S_{1,2}\), \(\overline{S}\) also mediates the \(|\Delta F|=2\) processes at the 1-loop level. Nonetheless, these processes impose more stringent constraints on the light \(\overline{S}\) compared to \(S_{1,2}\), as illustrated in Figs.~(\ref{fig:c5:fig4}, and \ref{fig:c5:fig7}). This stricter limit is due to \(\overline{S}\)'s relatively large Yukawa coupling due to different Clebsch-Gordan factors. Overall, the constraints on the masses of sub-GUT scale \(\mathbb{S}\) and \(\overline{S}\) appear similar.

\section{Summary}
\label{sec:c5:summary}

\lettrine[lines=2, lhang=0.33, loversize=0.15, findent=0.15em]{S}CALARS TRANSFORMING UNDER the two-index symmetric representations of the \(SU(3)_{C}\) of the SM gauge group have been studied for their diverse phenomenological implications in the context of bottom-up approaches in this chapter. We assess the potential for these sextets to emerge from the realistic renormalisable \so\, which naturally accommodates five distinct colour sextet fields shown in Tab.~(\ref{tab:c5:sextetscalars}). We derive their interactions with quarks and calculate effective operators that contribute to electrically neutral meson-antimeson in the section~(\ref{sec:c5:qfv}) and baryon-antibaryon oscillations in the section~(\ref{sec:c5:nnbar}). The latter is influenced by \(B-L\) breaking, induced by the $vev$ of an SM singlet field \(\sigma\). Additionally, we analyse the effective quartic coupling of the sextet scalars, which is prone to the perturbativity constraints from \(B-L\) breaking effects, as discussed in section~(\ref{sec:c5:quartic}). Further, we propose a rather simple way of generating the baryon asymmetry with these sextets in the section~(\ref{sec:c5:baryo}). The key findings of this chapter are summarised below:

\begin{itemize}
\item Four pairs of sextets—\(\Sigma\)-\(S\), \(\Sigma\)-\(\mathcal{S}\), \(\Sigma\)-\(\mathbb{S}\), and \(\Sigma\)-\(\overline{S}\)— can induce neutron-antineutron observable oscillation in near-future experiments if both sextets in a pair are as light as \(10^5\) to \(10^8\) GeV. However, this scenario is largely ruled out by the need for perturbative quartic couplings in models with \(v_\sigma \geq 10^{11}\) GeV.
\item Observable \(n\)-\(\bar{n}\) oscillations along with perturbative effective quartic couplings could occur if \(v_\sigma < 10^8\) GeV and the sextets' couplings with quarks are around \({\cal O}(1)\). Yet, this typically results in too light right-handed neutrinos, making type I seesaw mechanism non-viable in realistic \so\, models.
\item If \(M_\Sigma > 10^{11}\) GeV, \(S\) could be as light as on the order of TeV, while \(\mathbb{S}\), \(\overline{S}\), and \({\cal S}\) might be \(\gtrsim 10\) TeV. Conversely, for \(S\), \(\mathbb{S}\), \(\overline{S}\), and \({\cal S}\) heavier than \(10^{11}\) GeV, \(M_\Sigma\) could be around \(10\) TeV or more. Such constraints for these lighter sextets arise mainly from meson-antimeson oscillations and/or direct searches.
\item \so\, models featuring both $120_{\hh}$ and $\overline{126}_{\hh}$ in the Yukawa sector exhibits a pair of sextets, \(S_{1,2}\), having the same quantum numbers as \(S\). With \(\Sigma\) heavier than \(v_\sigma\) and \(S_{1,2} \ll M_{\Sigma}\), this scenario offers a promising way to generate the cosmological universe's baryon asymmetry.
\end{itemize}
Many insights from this study are based on the strong correlation between the couplings of sextets with quarks and the \(B-L\) scale within renormalisable \(SO(10)\) GUTs. These parameters cannot vary freely as often presumed in bottom-up approaches.

Further, it is important to note that baryon asymmetry can also be produced through thermal leptogenesis, given that lepton number violation and right-handed neutrinos are integral components of \(SO(10)\) GUTs~\cite{Mummidi:2021anm}. In scenarios where thermal leptogenesis fails to account for the observed asymmetry—due to insufficient CP violation in the lepton sector or an inappropriate mass spectrum and couplings of the right-handed neutrinos—the sextet-driven cosmological baryogenesis outlined in this Chapter offers a feasible alternative.

In Chapters~(\ref{ch:3}) and (\ref{ch:4}), we constrained the scalar fields capable of destabilising a nucleon. Chapter~(\ref{ch:5}) explored the various phenomena a sextet field can contribute to. Some proton decay mediators could also contribute to flavour violation, \(n\)-\(\bar{n}\) oscillations, and potentially generate baryogenesis, but the proton decay constraints provide stringent mass limits. Broadly speaking, in Chapters (\ref{ch:3}) and (\ref{ch:4}), we investigated the phenomenology of all the scalar fields capable of contributing to the renormalisable and non-renormalisable Yukawa sector of \(SO(10)\) and constrained them in a top-down approach (see Tabs.~(\ref{tab:c2:scalars} and \ref{tab:c2:16scalars})). In the renormalisable sector, the scalar fields we have not yet studied include \(t (1,3,2)\) and \(O(8,2,\frac{1}{2})\) along with their conjugate partners (cf. Tab.~\ref{tab:c2:scalars}). The \(t\) field is known to contribute to the Type-II seesaw mechanism~\cite{Melfo:2010gf} and can contribute to lepton-flavour-violating processes~\cite{Calibbi:2009wk}. Additionally, the \(O\) field does not carry any \(B-L\) charge and can induce quark-flavour-violating processes~\cite{Manohar:2006ga}. We anticipate that the constraints on the \(SU(3)\) octets ($i.e.$, \(O\)) will be similar to those on the sextets coming from flavour violation, albeit \({\cal O} (1-10)\) suppressed or enhanced due to different Clebsch-Gordan coefficients.

The investigations performed in Chapters~(\ref{ch:3}, \ref{ch:4} and \ref{ch:5}) show how a well-defined model at higher energy levels can constrain new physics possibilities at lower energies, rendering the model testable and potentially falsifiable.

\clearpage

\chapter{Scalar Induced Corrections to the Yukawa Correlations}
\label{ch:6}
\graphicspath{{60_Chapter_6/fig_ch6/}}
\section{Overview}
\label{sec:intro}
\lettrine[lines=2, lhang=0.33, loversize=0.15, findent=0.15em]{G}RAND UNIFIED MODELS conventionally rely on multiple scalar irreps in the Yukawa sector to yield a realistic fermion mass spectrum. The presence of more than one scalar irrep in the Yukawa and Higgs sector of grand unification introduces numerous unknown and tunable parameters in the scalar potential. Additionally, as these scalar irreps consist of many scalar fields, as shown in Chapters~(\ref{ch:3}, \ref{ch:4}, and \ref{ch:5}), they contribute to hierarchy problems, which are among the least appealing features of grand unification. This chapter explores the role played by scalars in the construction of a minimal GUT model. Here, minimality means only a single irrep contributes to the Yukawa sector.

The smallest irrep contributing to the renormalisable Yukawa sector of \(SO(10)\) has four fields (cf. Tab.~(\ref{tab:c2:scalars})). The presence of only \(10_{\mathrm{H}}\) in the Yukawa sector results in mass degeneracy across all charged and neutral fermion sectors, \(Y_d=Y_u=Y_e=Y_{\nu^C}\), which can be corrected at tree level by introducing \(\overline{126}_{\mathrm{H}}\)~\cite{Joshipura:2011nn}. If the Yukawa sector consists of a single five-dimensional Higgs, as in the case of \su, the mass degeneracy occurs only in the down quark and charged lepton sectors, $i.e.$ $Y_{d}\,=\,Y_{e}^T$. In this chapter, a minimal \(SU(5)\) model is used to study the indirect implication of scalar field(s) in its ability to render the minimal model phenomenologically viable.

Starting with identifying the problem of generational mass degeneracy in the tree-level mass relation in the minimal \su\, GUT in section~(\ref{sec:c6:yuk-t}), we set the Feynman rules to compute quantum corrections in a minimally extended \su\, model with singlet(s) fermions in section~(\ref{sec:c6:Yuk-1}). We derive the general expression for one loop matching in section~(\ref{sec:c6:1loop}). In section~(\ref{sec:c6:loopd}), we provide the essential elements (vertex corrections and wave function renormalisation factors) of the matching condition for different SM particles and elaborate on their different contribution to $Y_{d}$ and $Y_{e}$. We use the computed corrections along with the matching relation to asses its viability in reproducing viable Yukawa ratio in section~(\ref{sec:c6:dep_ydye}). Further, we perform a comprehensive three-generations $\chi^2$ analysis in section~(\ref{sec:c6:res}) to check the ability of the approach to generate a fermionic spectrum including the leptonic spectrum (in case of more than singlets) and conclude the chapter in section~(\ref{sec:c6:concl}).

\section{Tree Level Yukawa Relations}
{\label{sec:c6:yuk-t}}
We begin this section by identifying the problem in the minimal $SU(5)$ model. Consider the following Lagrangian with only $5_{\hh}$ dimensional irrep participating in the Yukawa sector:
\beqa{\label{eq:c6:ltree}}
-{\cal L}^{\rm minimal}_{\rm Y} \eq \frac{1}{8}\,Y_{1\,AB}\,\varepsilon_{pqrst}\,\ft^{pq\,T}_A\,\cc\,\ft^{rs}_B\,5^t_{\hh} \;+\; Y_{2\,AB}\,\ft^{T\,pq}_A\,\cc\,\ff_{p\,B}\,5^{\dagger}_{q\,\hh} \nl\hcn,
\eeqa
where, we follow the same conventions for $SU(3)_{\mathrm{C}}$, $SU(2)_{\mathrm{L}}$, and generation indices as mentioned in the section~(\ref{ssec:c2:OESO}). We also begin to denote the spinor indices by the uppercase dotted and undotted Latin alphabets $i.e.\,1\,\leq\,A,\,\Dot{A},B\,\Dot{B}...\,\leq\, 2$. The factor $\sfrac{1}{8}$ adjacent to $Y_1$ vertex is chosen to overcome the decomposition and combinatorics factors. Additionally, $Y_1$ is symmetric in the generation indices.

We decompose the Lagrangian mentioned in Eq.~\eqref{eq:c6:ltree} using the Eqs.~(\ref{eq:c2:fn}, \ref{eq:c2:ff}, \ref{eq:c2:ft}, and \ref{eq:c2:5dec}) and obtain the interaction of the SM-Higgs with the fermions residing in the $\ft$ and $\ff$ irreps of \su, which is shown below: 
\beqa{\label{eq:c6:ltree2}}
-{\cal L}^{\rm (h)}_{\rm Y} \eq \big(-Y_{1\,AB}\big)\,\varepsilon_{ab}\,q^{a\alpha\,T}_A\,\cc\,u^C_{\alpha\,B}\,D^b\;\nl
\ad \big(-Y_{2\,AB}\big)\,\left(q^{a\alpha\,T}_A\,\cc\,d^C_{\alpha\,B}\;+\; e^{C\,T}_A\,\cc\,l_B^a\right)\,D^{*}_a. 
\eeqa
The Eq.~(\ref{eq:c6:ltree2}) yields the following Yukawa relations valid at the GUT scale:
\beqa {\label{eq:c6:yuktree}}
Y_u \eq Y_1, \nl
Y_d \eq Y_2,\nl
Y_e \eq Y_2^T. \eeqa
Choosing a basis such that $Y_{2}$ is diagonal implies $\frac{y_d}{y_e}\,=\,\frac{y_s}{y_\mu}\,=\, \frac{y_b}{y_{\tau}}\,=1$ at the GUT scale, where $y_{u,d,e}$ and $y_{e,\mu,\tau}$ are the respective eigenvalues of matrix $Y_{d}$ and $Y_{e}$.

\begin{figure}[t]
    \centering
    \includegraphics[width=0.8\linewidth]{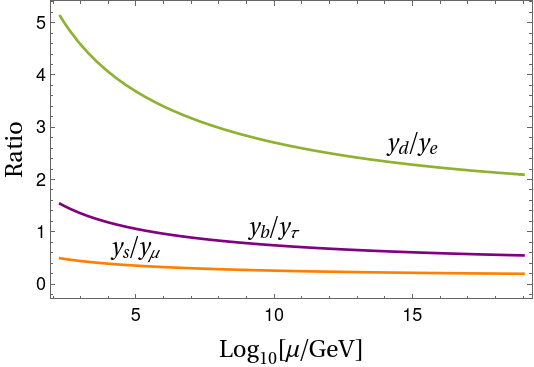}
    \caption{Running of the ratios of different generations of Yukawas of down-quarks and charged leptons at two-loop from top-pole mass to Plank scale, assuming no intermediate scales during the running. }
    \label{fig:c6:yukrun}
\end{figure}

The graph shown in Fig.~(\ref{fig:c6:yukrun}) shows the renormalisation group running of ratios of Yukawas of different generations of down quark and lepton. We observe for two loops at $M_{\rm{GUT}}$, we require the ratios to be $\frac{y_d}{y_e}\,\sim\,2.0$, $\frac{y_s}{y_\mu}\,\sim\,0.2$ and $\frac{y_b}{y_{\tau}}\,\sim\,0.6$. So, the required ratios are off by $50\%$, $400\%$ and $67\%$ for the first, second and third generations, respectively\footnote{To compute the deviation in the ratios, we have used the relation $\delta \equiv \frac{y_{\rm{e}}-y_{o}}{y_{e}}\times 100$, where $y_{e}$ and $y_{o}$ are the expected and observed Yukawa ratios respectively.}. A correction of more than $100\%$ implies the necessity of heavier leptons than the quarks at the GUT scale. We infer that a single five-dimensional Higgs contributing to the Yukawa sector yields mass degeneracy in the down-quark and lepton sectors, which is inconsistent with the observations.

To remedy such a situation illustrated by Eq.~\eqref{eq:c6:yuktree}, one adheres to the following solutions: (a) Expanding the scalar sector~\cite{Georgi:1979df,Barbieri:1981yw,Giveon:1991zm,Dorsner:2006dj,FileviezPerez:2007bcw,Goto:2023qch}, such as by adding a $45_{\hh}$-dimensional Higgs field; (b) Incorporating higher-dimensional\; non-renormalisable operators~\cite{Ellis:1979fg,Berezinsky:1983va,Altarelli:2000fu,Emmanuel-Costa:2003szk,Dorsner:2006hw,Antusch:2021yqe,Antusch:2022afk}; or (c) Introducing vector-like fermions that mix with the charged leptons and/or down-type quarks from the chiral multiplets of $SU(5)$~\cite{Hempfling:1993kv,Shafi:1999rm,Barr:2003zx,Oshimo:2009ia,Babu:2012pb,Dorsner:2014wva,FileviezPerez:2018dyf,Antusch:2023mxx,Antusch:2023kli,Antusch:2023mqe}. Each approach modifies the tree-level matching condition, Eq.~\eqref{eq:c6:yuktree}, and introduces new couplings that can rectify the effective Yukawa couplings and yield those observed in the SM. As we know, GUTs have other scalar fields that are part of common multiplets and can modify tree-level matching conditions.

In further sections, we present a simple method to address the mass degeneracy between charged leptons and down-type quarks. Our approach involves applying higher-order corrections to the tree-level matching conditions of the Yukawa couplings.

\section{Modified Yukawa Relations}
\label{sec:c6:Yuk-1}
We start with the Yukawa sector of the model as described in the Eq.~\eqref{eq:c6:ltree} together with $N$ generations of the gauge singlet, i.e. ${\bf 1}$, Weyl fermion(s). Including singlet(s) does not increase additional parameters in the Higgs sector and can be used to generate neutrino mass via the Type-I seesaw mechanism. The most general renormalisable interactions between these fields can be parametrised as follows;
\beqa \label{eq:c6:LY}
{\cal L}_{\rm Y} &=& \frac{1}{8} (Y_1)_{AB}\, {\bf 10}_{A}^T\, \cc\, {\bf 10}_{B}\, 5_H +  (Y_2)_{AB}\, {\bf 10}_{A}^T\,\cc\, \overline{\bf 5}_B\, 5^*_{\hh} \nonumber \\
& + & (Y_3)_{A\alpha}\, {\bf{\overline{5}}}_A^T\, \cc\, {\bf 1}_\alpha\, 5_H  -\frac{1}{2}\, (M_N)_{\alpha \beta}\,{\bf 1}^T_\alpha\,\cc\, {\bf 1}_\beta \hc,\, \eeqa
where \(\alpha=1,...,N\) denotes the generations of singlet fermions, a notation specifically reserved for the same. Additionally, \({\cal C}\) denotes the usual charge-conjugation matrix. For clarity, we have omitted the gauge and Lorentz indices, although they follow the same conventions discussed previously. Further, \(M_N\) refers to the gauge-invariant Majorana mass of the singlet fermions, alias the right-handed neutrinos.

Decompositions of Eq.~\eqref{eq:c6:LY} yields the following interactions of the Higgs Triplet $(T)$ (cf. Tab.~(\ref{tab:c2:16scalars}) for its SM charge) with the SM fermions:
\beqa \label{eq:c6:L_T}
-{\cal L}_{\rm Y}^{(T)} &=& -(Y_1)_{AB}\, \left(u^{C\,T}_{\gamma\,A}\,\cc\, e^C_B + \frac{1}{2}\,\varepsilon_{\alpha\beta\gamma}\, q^{a\alpha\,T}_A\,\cc\, q^{b\beta}_B\right) T^{\gamma} \nl 
&+& (Y_2)_{AB}\, \left(\varepsilon_{\alpha\beta\gamma}\,u^{CT}_{\alpha}\,\cc\, d^C_{\beta\,B} +\,\varepsilon_{ab}\,q^{a\,\gamma\,T}_{A}\,\cc\, l^b_B\right)\, T_{\gamma}^* \nl
\ad (Y_3)_{A\beta}\, d^{CT}_{\alpha\,A}\,\cc\, \nu^C_\beta\, T^{\alpha} \hc,\eeqa
and the new vertex of SM-Higgs, in addition to the vertices specified in Eq.~\eqref{eq:c6:ltree2}, is as follows: 
\beqa \label{eq:c6:L_h}
-{\cal L}_{\rm Y}^{(D)} &=& (Y_{3})_{A\alpha}\,\varepsilon_{ab}\,l^b_A\,\cc\,\nu^C_{\alpha}\,D^a.\eeqa
Matching the Lagrangian \({\cal L}_{\rm Y}^{(D)}\), given in Eqs.~(\ref{eq:c6:ltree2} and \ref{eq:c6:L_h}), with the SM Yukawa Lagrangian at tree level results in \(Y_u = Y_1\) and \(Y_d = Y_e^T = Y_2\) at the renormalisation scale \(\mu = M_{\rm GUT}\), as discussed in the previous section. 

In principle, corrections from all the degrees of freedom (heavy and light) should be considered while computing quantum corrections. However, after matching, it is the contribution of these heavy degrees of freedom that makes the difference. In the current scenario, light degrees of freedom include all the SM particles and heavy degrees of freedom consist of scalar triplets, singlet fermion(s) and heavy gauge bosons. Consequently, we also consider the correction induced by the gauge bosons. The following Lagrangian depicts the gauge bosons' interaction with SM fermions:
\beqa{\label{eq:c6:GB}}
{\cal{L}}_G \eq \ft^{\dagger}\,i\,\overline{\sigma}^{\Dot{\mu}}\,D_{\Dot{\mu}}\ft + \ff^{\dagger}\,i\,\overline{\sigma}^{\Dot{\mu}}\,D_{\Dot{\mu}}\,\ff.
\eeqa
The action of the covariant derivative on the fermion fields are defined as follows~\cite{Langacker:1980js}:
\beqa{\label{eq:c6:covd}}
D_{\Dot{\mu}}\ff_p \eq \left(\partial_{\Dot{\mu}}\,\delta_p^q - i \frac{g}{\sqrt{2}}\,X_{\Dot{\mu}\,p}^q\right)\,\ff_q,\nl
D_{\Dot{\mu}}\ft^{pq} \eq \left(\partial_{\Dot{\mu}}\,\delta^p_k + i \frac{g}{\sqrt{2}}\,X^{\,p}_{\Dot{\mu}k}\right)\,\ft^{kq}.
\eeqa
Using the action of covariant derivative given in Eq.~\eqref{eq:c6:covd}, we compute the interaction of $\overline{X}_\alpha^a$ with the SM fermions as it has the appropriate leptoquark and diquark coupling, which is shown below~\cite{Buras:1977yy,Langacker:1980js};
\beqa \label{eq:c6:L_X}
-{\cal L}_{\rm G}^{(X)} &=& \frac{g}{\sqrt{2}} \overline{X}^a_{\Dot{\mu}\,\alpha} \left( \big(d^C_{\alpha\,A}\big)^{\dagger}\, \overline{\sigma}^{\Dot{\mu}}\, l_A\,-\, \varepsilon^{\alpha\beta\gamma}\,\big(q^{a\,\beta}_A\big)^{\dagger}\, \overline{\sigma}^{\Dot{\mu}}\, u^C_{\gamma\,A} -\,\varepsilon_{ab}\, \big(e^C_A\big)^{\dagger}\, \overline{\sigma}^{\Dot{\mu}}\, q^{b\alpha}_A \right) \nl\hcn\,,
\eeqa
where $X$ transforms as $(3,2,-5/6)$ under the SM gauge symmetry, and $\overline{X}$ is conjugate partner of $X$. The computation of coupling of $\overline{X}$ with SM fermions can also be inferred from the Eq.~\eqref{eq:c2:XandY}.

\begin{figure}[t!]
    \centering
    \includegraphics[width=0.3\linewidth]{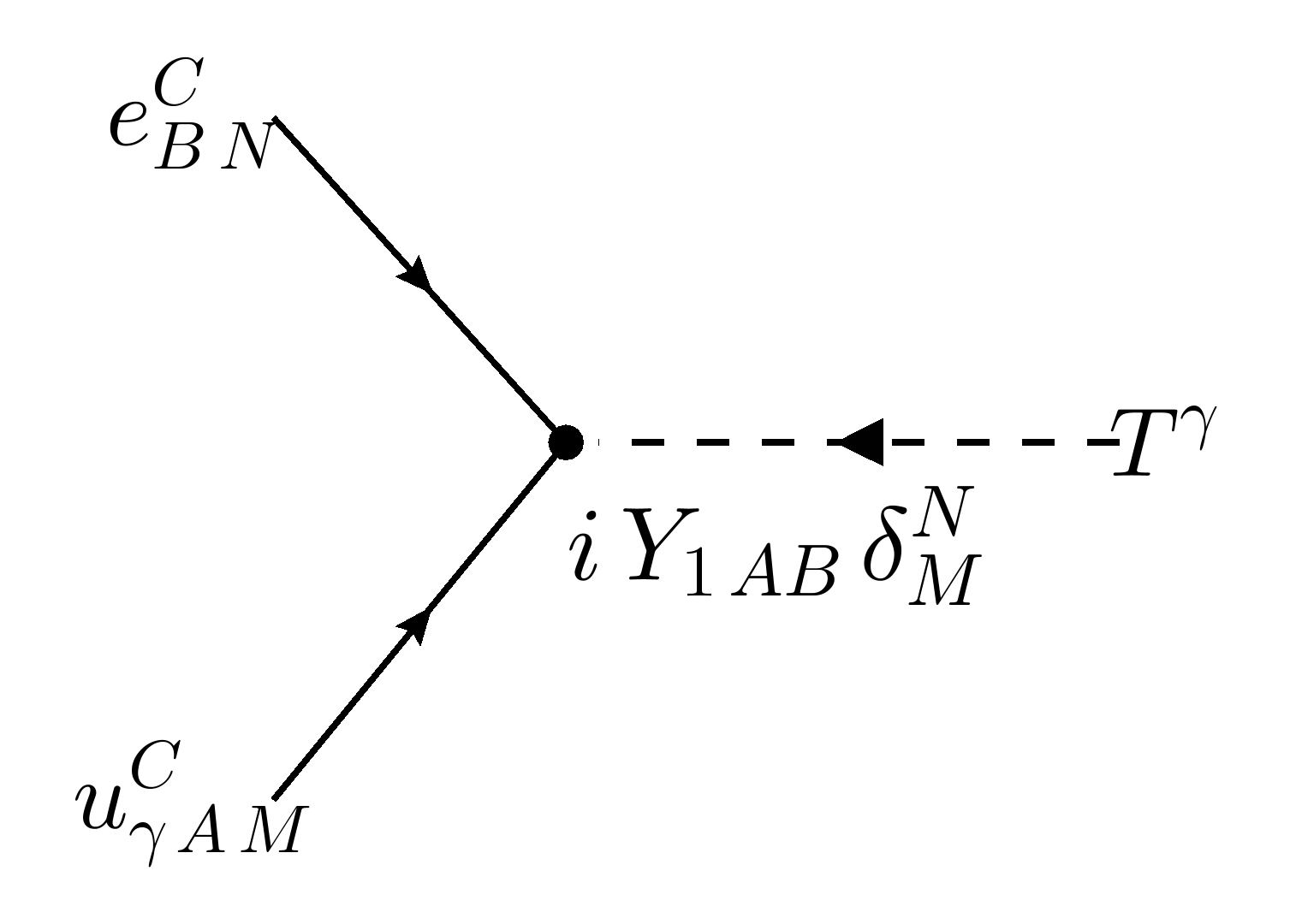} \hspace{0.5cm}
    \includegraphics[width=0.3\linewidth]{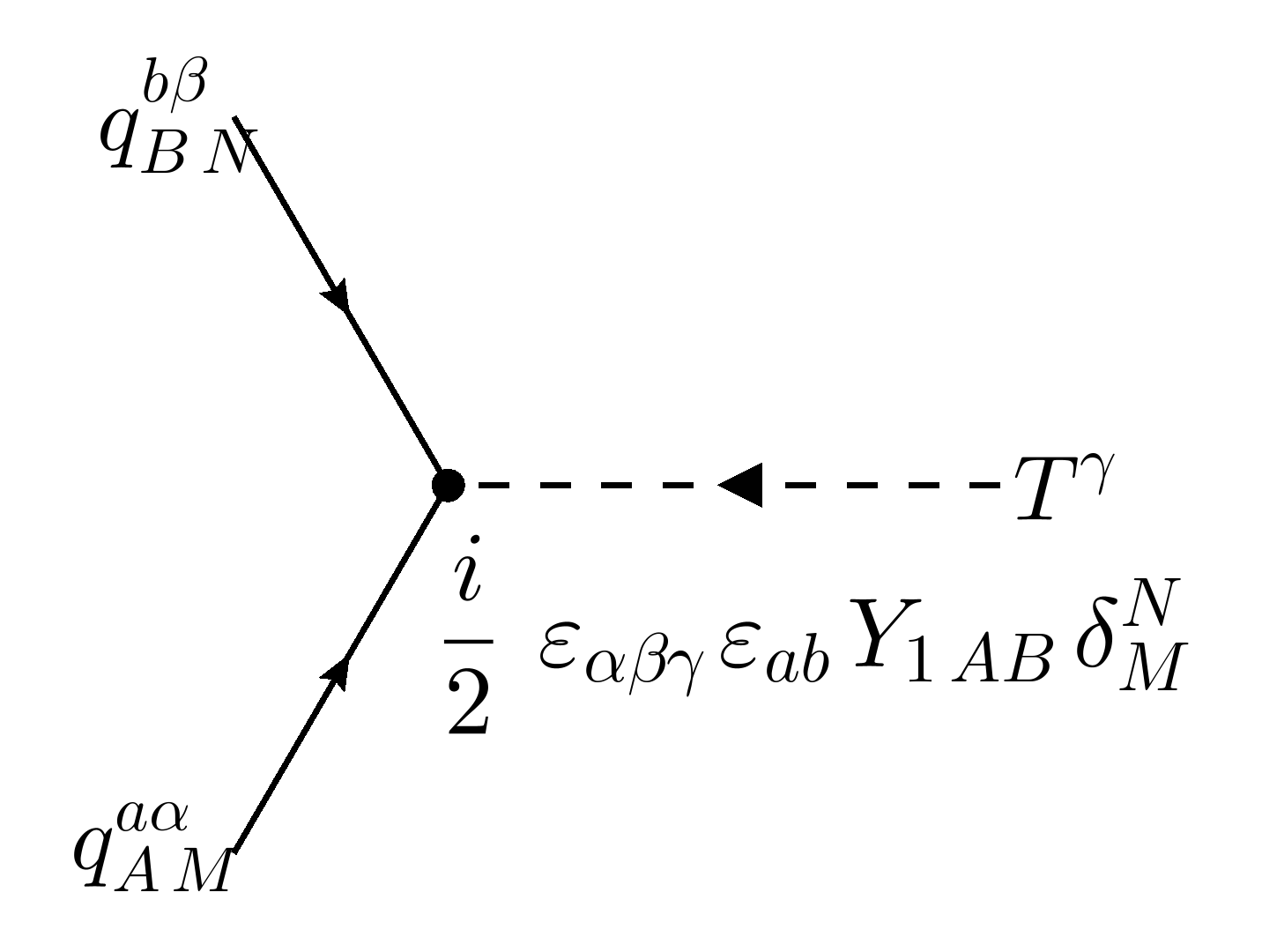} \hspace{0.5cm}
    \includegraphics[width=0.3\linewidth]{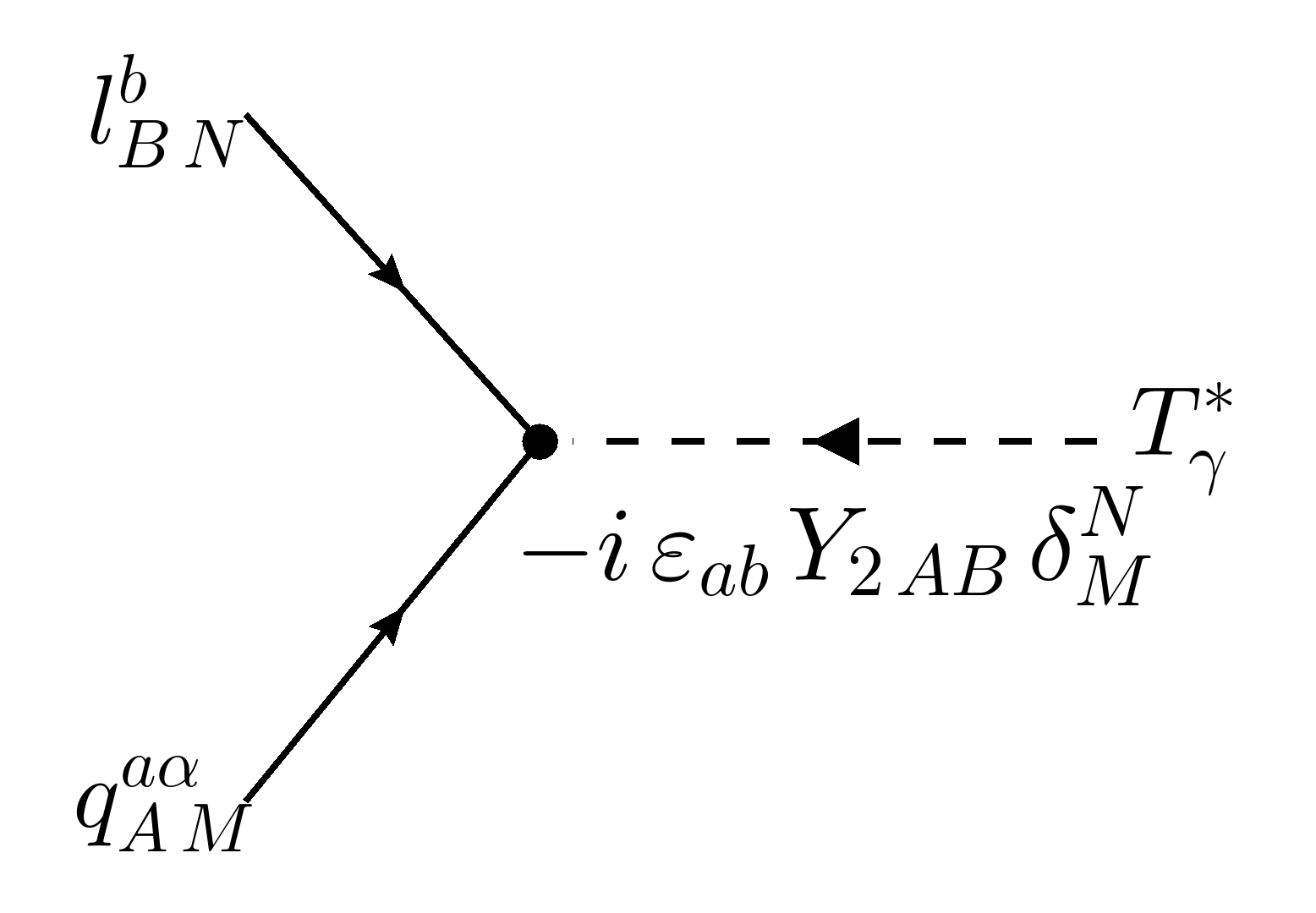}\vspace{1cm}
    \includegraphics[width=0.3\linewidth]{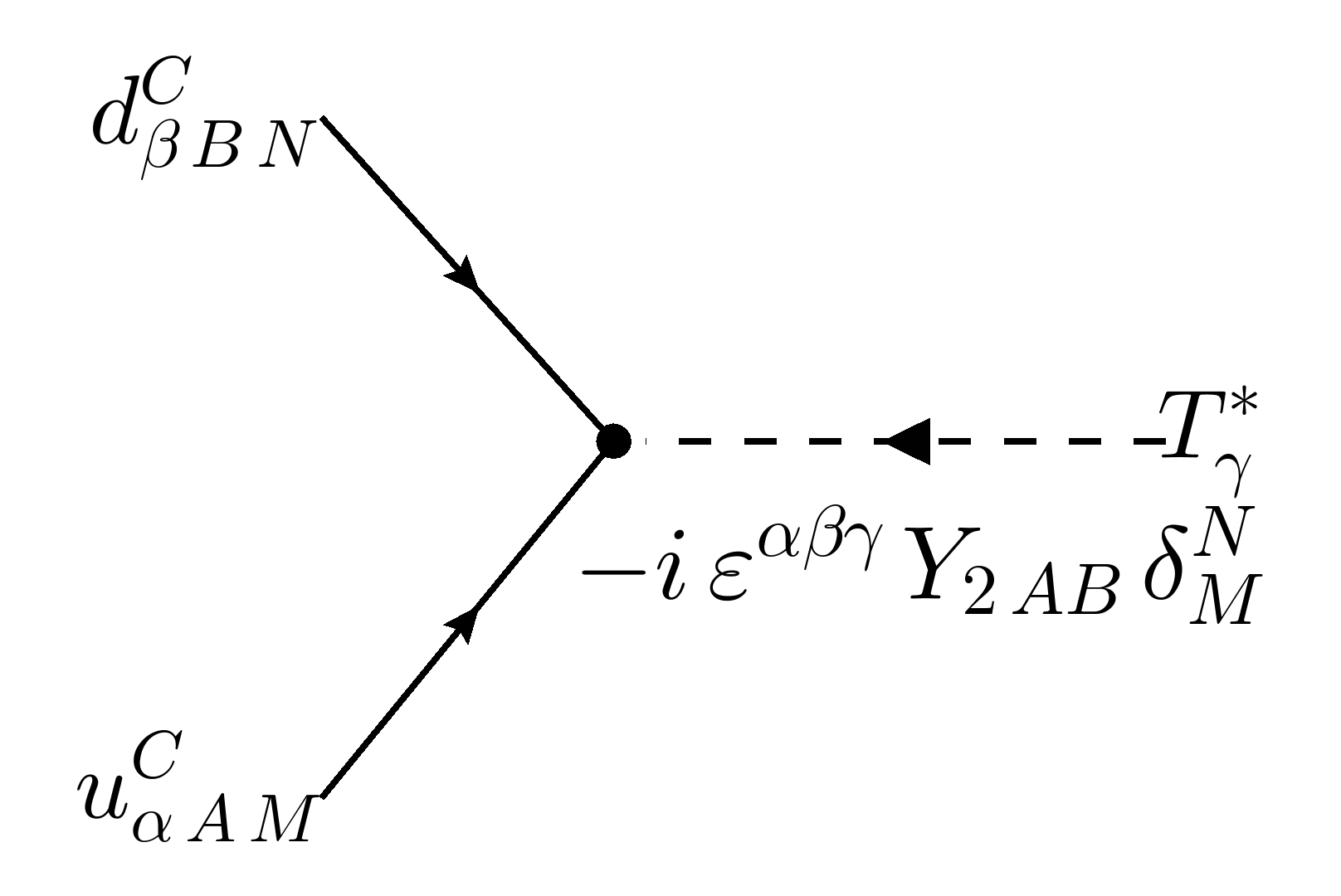} \hspace{0.5cm}
    \includegraphics[width=0.3\linewidth]{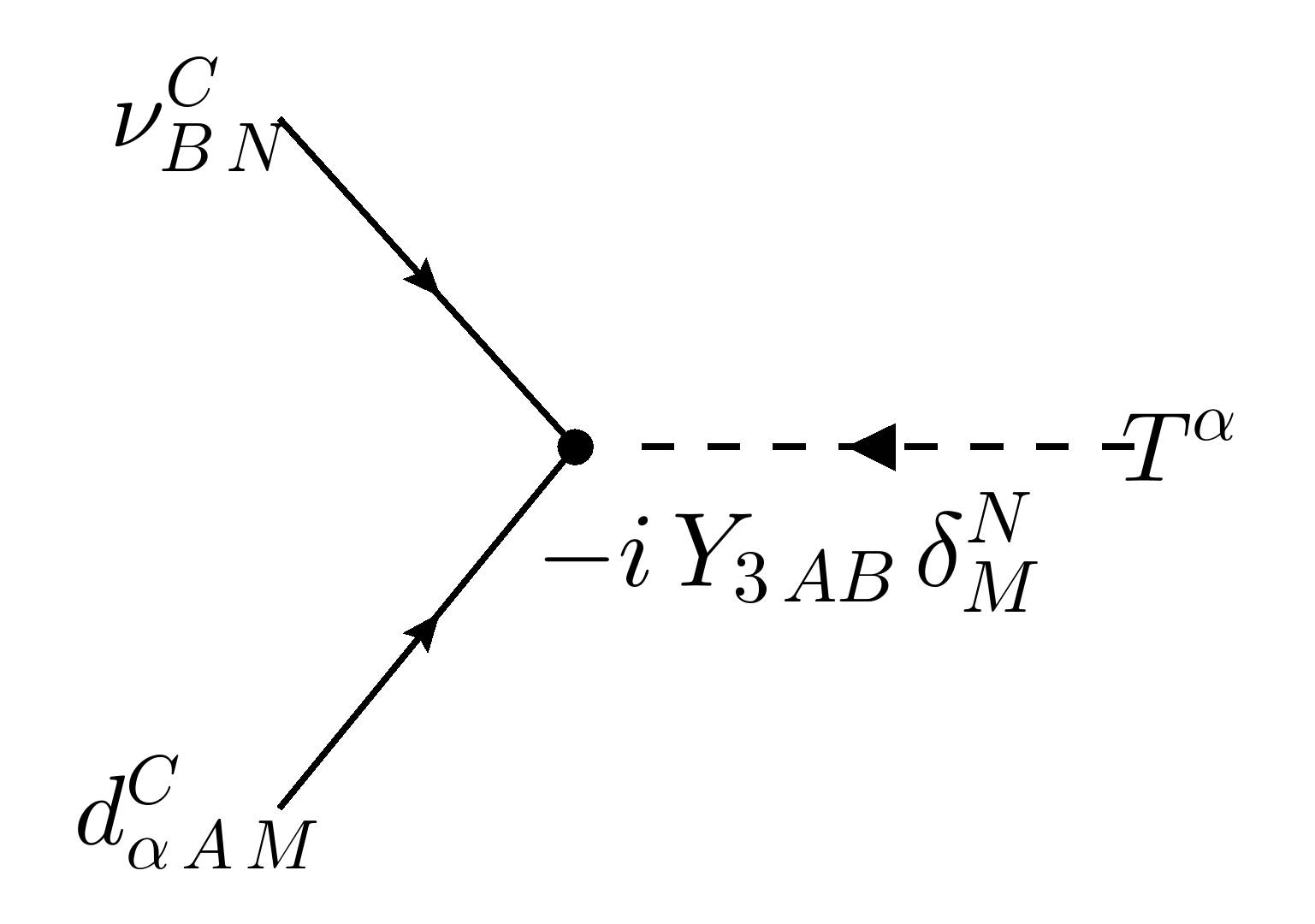} \hspace{0.5cm}
    \includegraphics[width=0.3\linewidth]{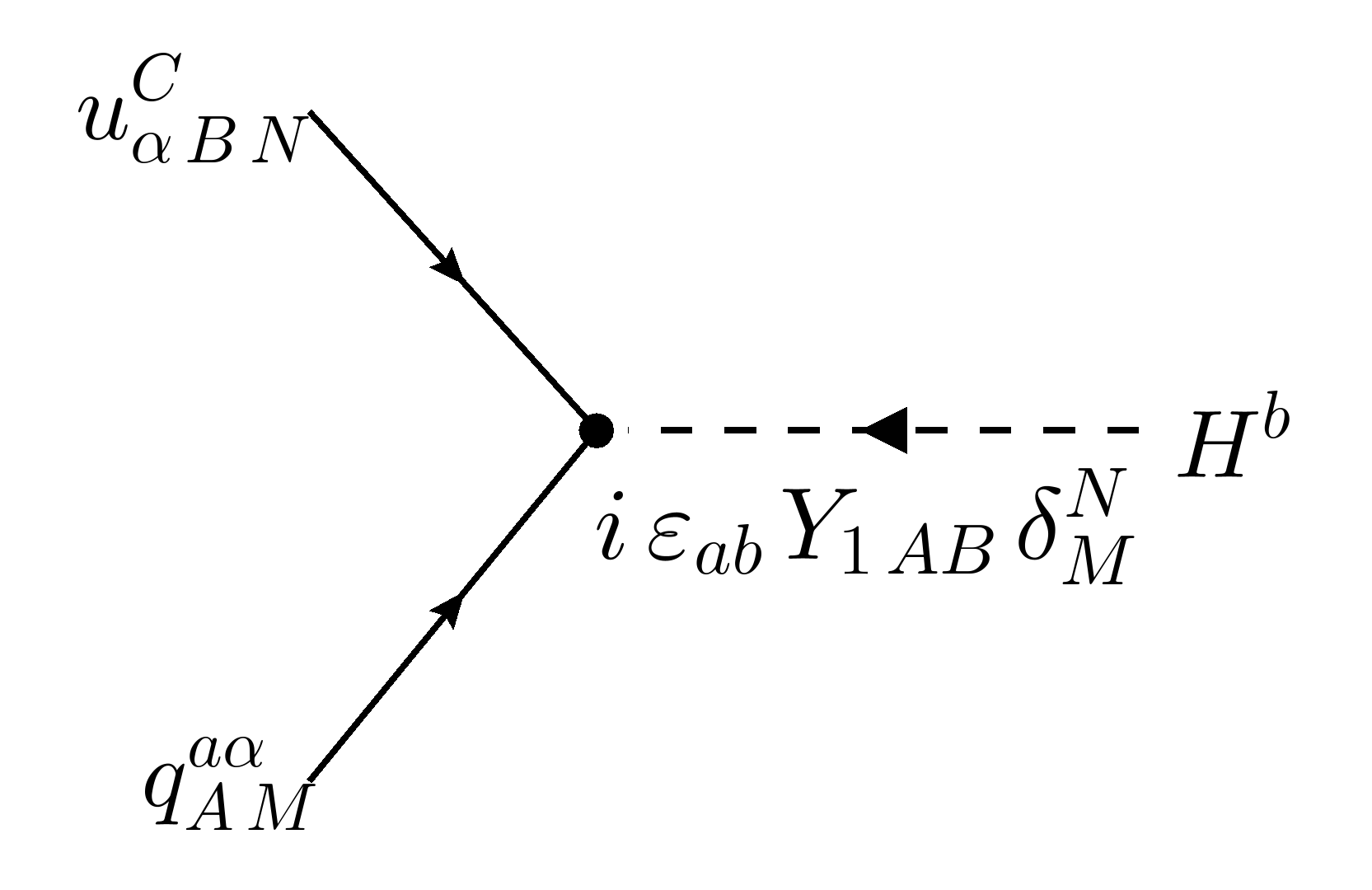}\vspace{1cm}
    \includegraphics[width=0.3\linewidth]{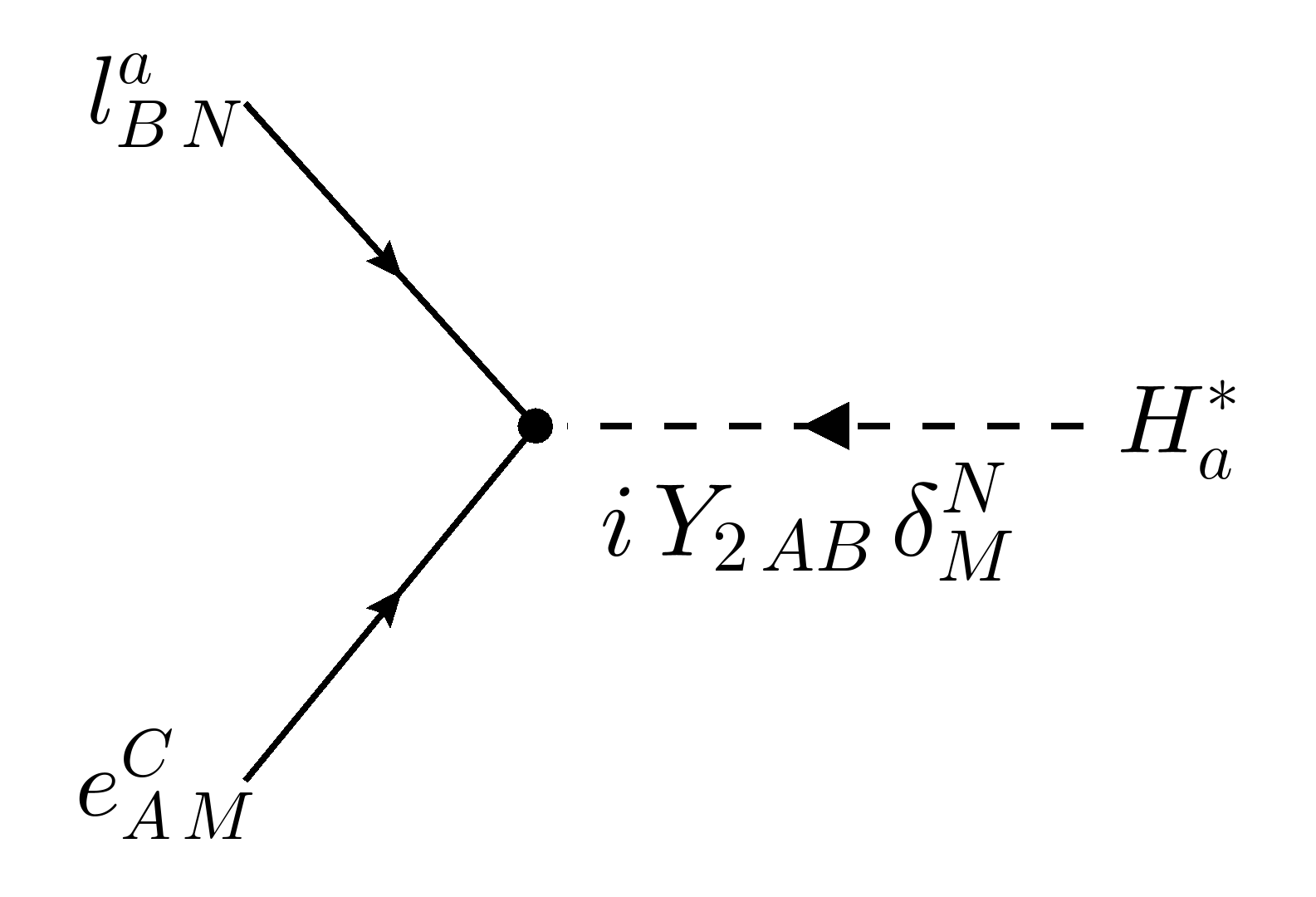} \hspace{0.5cm}
    \includegraphics[width=0.3\linewidth]{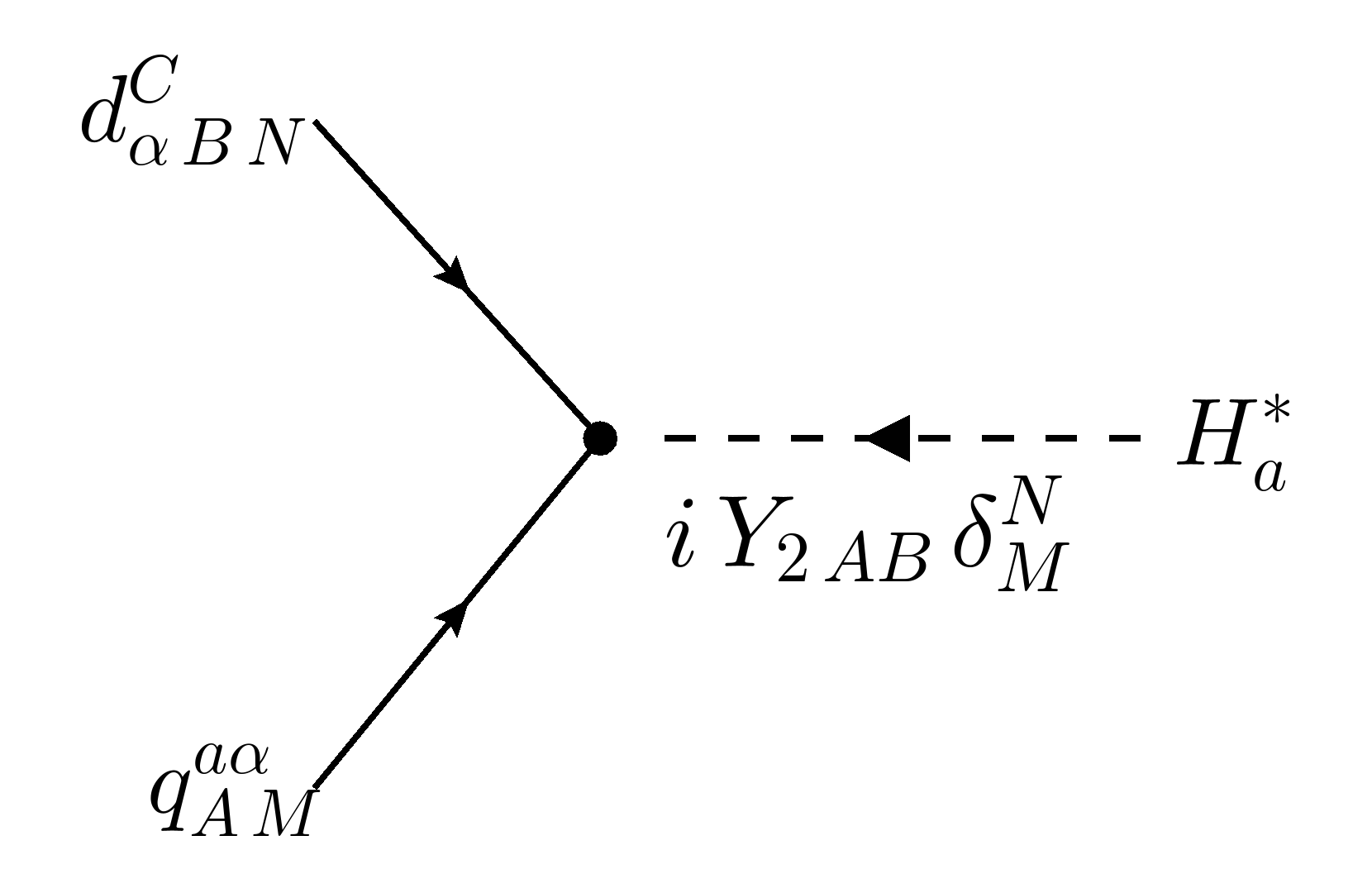} \hspace{0.5cm}
    \includegraphics[width=0.3\linewidth]{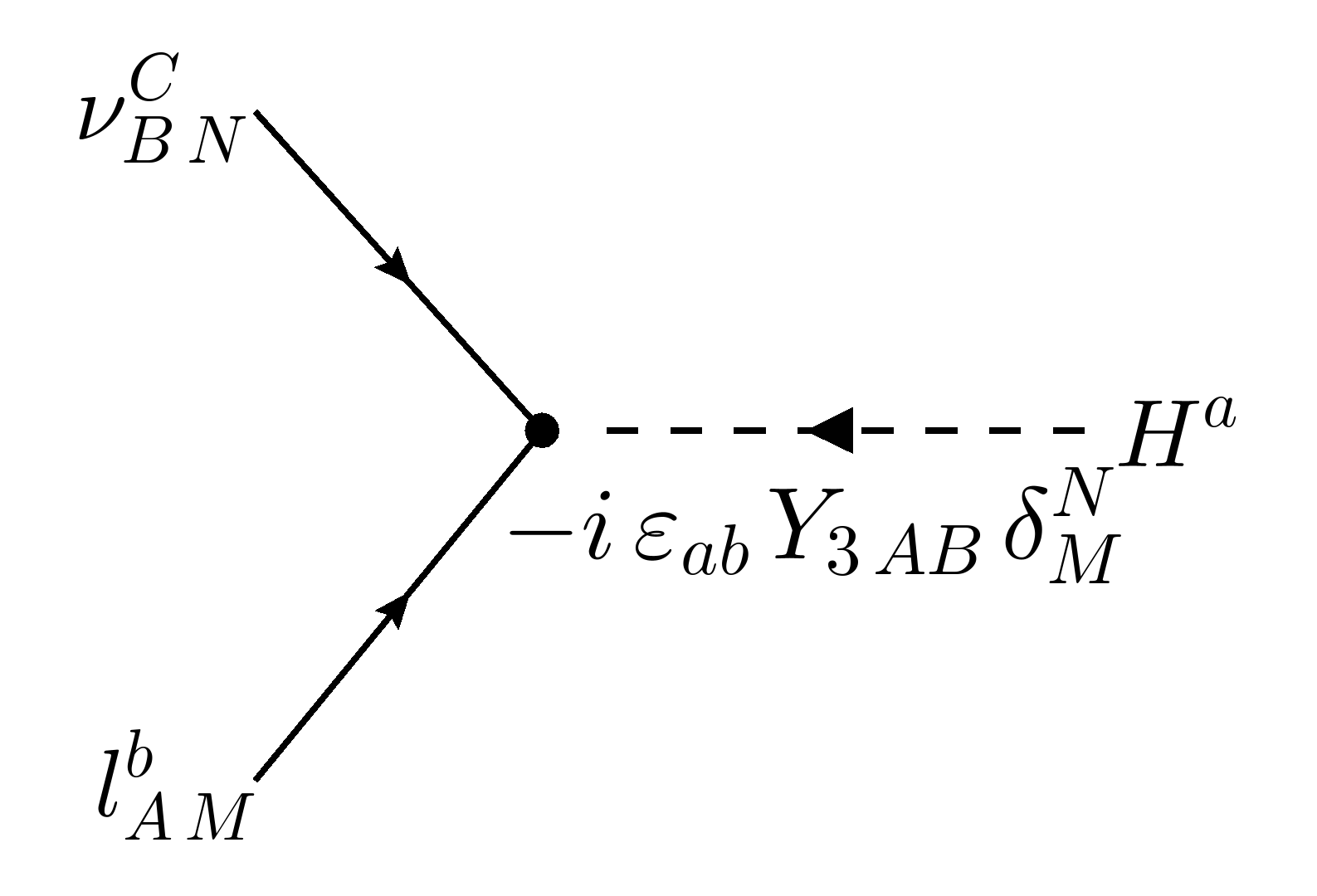}\vspace{1cm}    \includegraphics[width=0.3\linewidth]{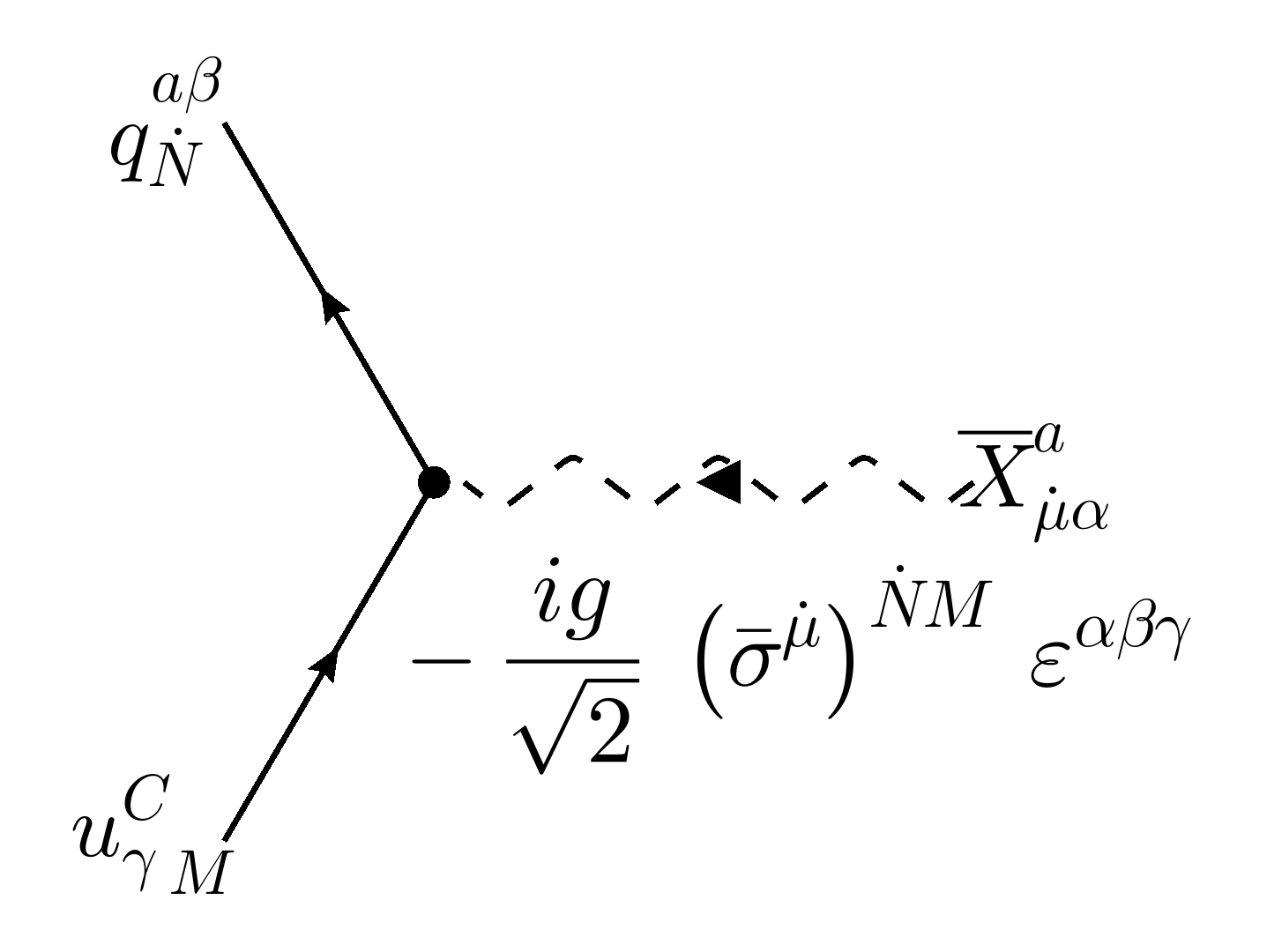} \hspace{0.5cm}
    \includegraphics[width=0.3\linewidth]{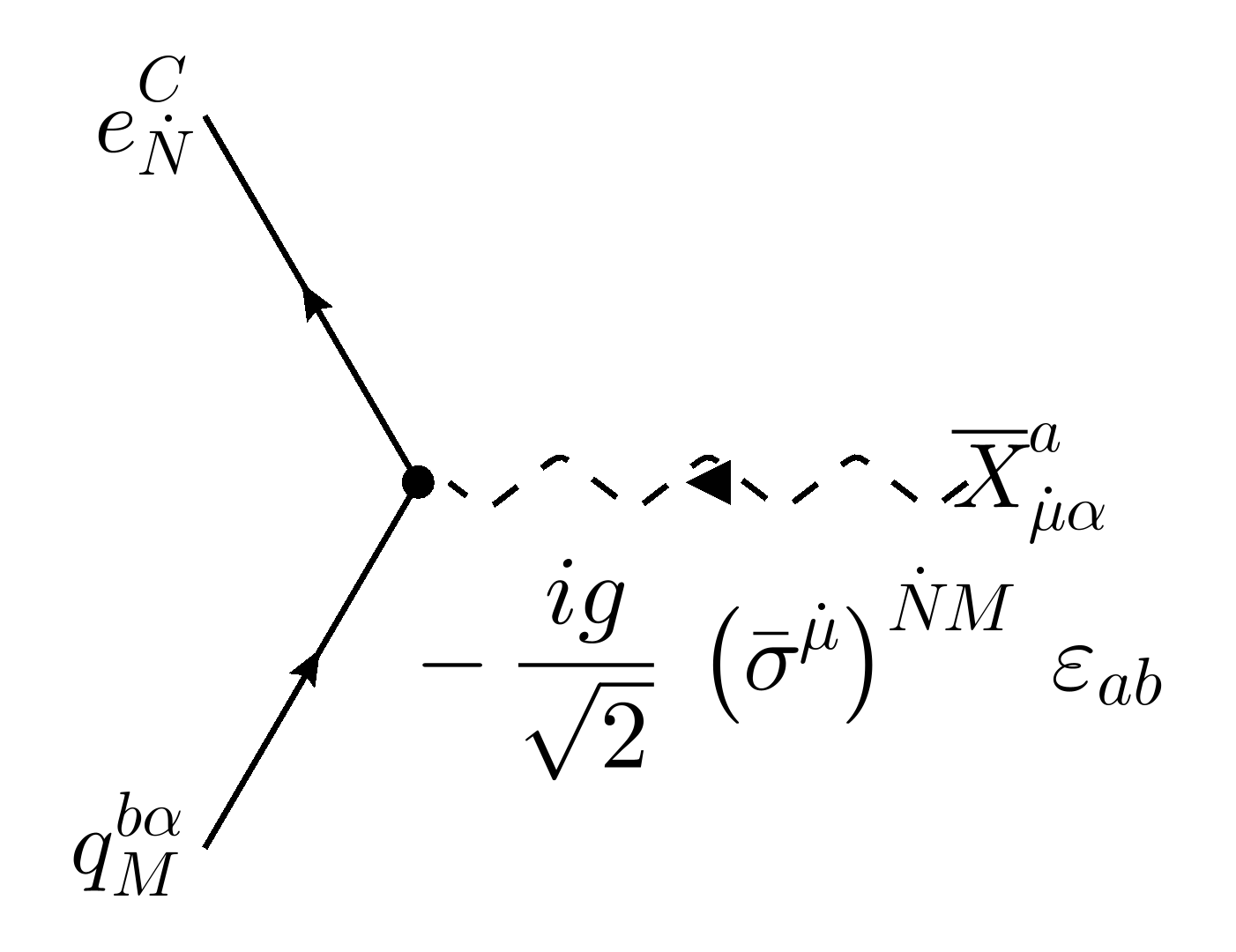} \hspace{0.5cm}
    \includegraphics[width=0.3\linewidth]{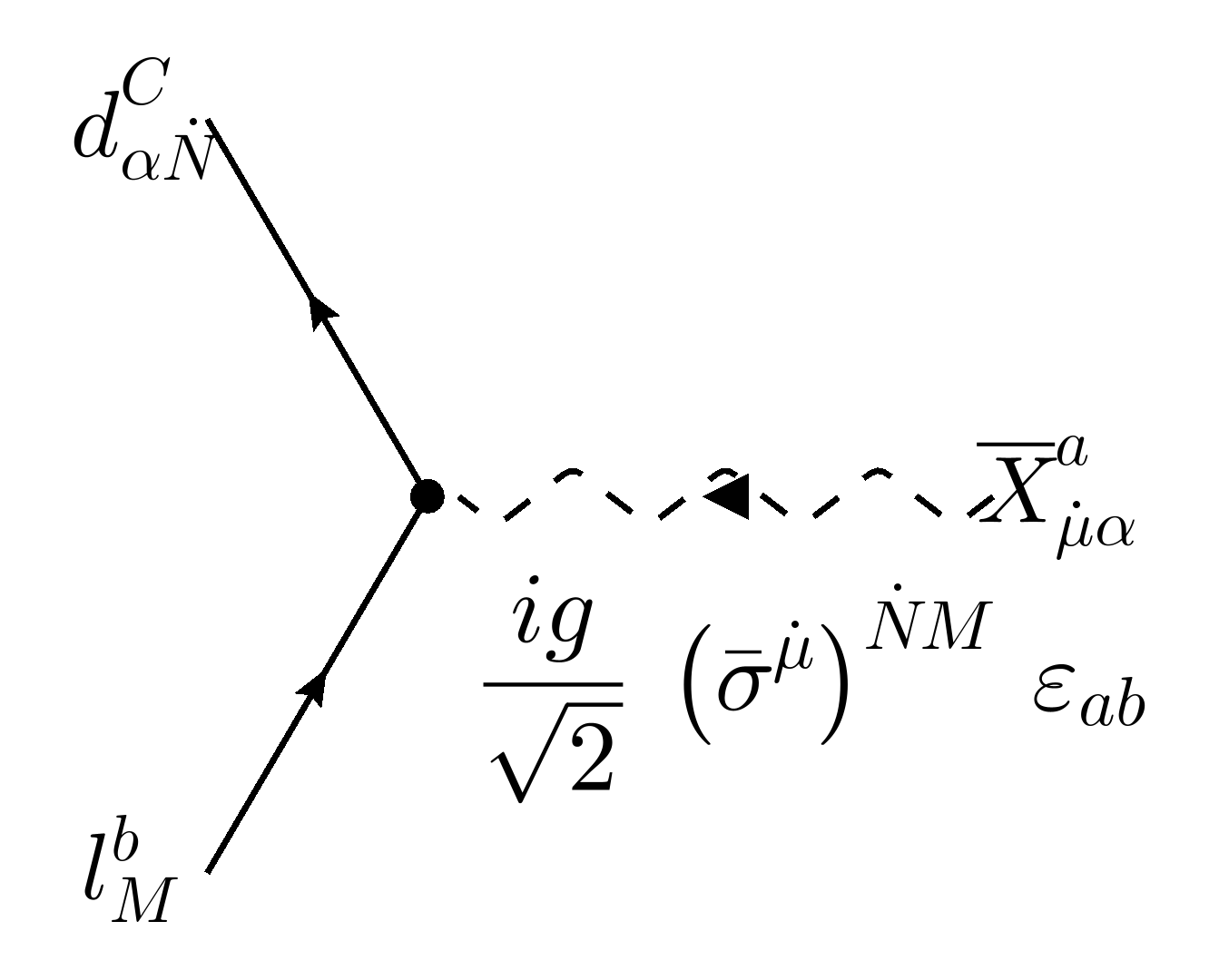}\vspace{1cm}    
\caption{Feynman graphs of the various vertices appearing in Eqs. (\ref{eq:c6:ltree2}, \ref{eq:c6:L_T}, \ref{eq:c6:L_h}, and \ref{eq:c6:L_X}). Here $M,\,N,\,\Dot{M},\,\Dot{N}\,=\,1,2$ and are the spinor indices. These diagrams are drawn using~\cite{Harlander:2020cyh}}
\label{fig:c6:feynmanrules} 
\end{figure}

In Fig.~(\ref{fig:c6:feynmanrules}), we draw the Feynman graphs and set the various Feynman rules for the various vertices appearing in Eqs.~(\ref{eq:c6:ltree2}, \ref{eq:c6:L_T}, \ref{eq:c6:L_h}, and \ref{eq:c6:L_X}). In further sections, we shall use these rules to compute the loop corrections to the Yukawa relations.

\section{Formulae for one-loop Matching}
{\label{sec:c6:1loop}}

This section presents a detailed derivation of the one-loop matching expression for the Yukawa relations. Yukawa threshold corrections were initially computed in~\cite{PhysRevD.28.194, Kane:1993td, Hempfling:1993kv, Wright:1994qb}, utilising a methodology developed for gauge couplings as done in~\cite{Weinberg:1980wa, Hall:1980kf}. For a comprehensive evaluation, we closely adhere to the general formalism outlined in~\cite{Wright:1994qb} and thoroughly explain the treatment applied to the Yukawa couplings.  

Consider chiral fermions $\psi_i$, $\chi_i$, and a scalar $\phi$, having the following gauge and Yukawa interactions in the complete theory;
\be \label{eq:c6:full_l}
{\cal L} =  \psi^{\dagger}_i\, i\slashed{D}\, \psi_i +  \chi^{\dagger}_i\,i\slashed{D}\, \chi_i + D_\mu \phi^\dagger D^\mu \phi - \left\{Y_{ij} \psi_i^T\, \cc\, \chi_j \phi \hc \right\}\,,\ee
where, $\slashed{D}\equiv D_{\Dot{\mu}}\,\overline{\sigma}^{\Dot{\mu}}$. Let $\psi_i$, $\chi_i$, and $\phi$ decompose into light fields, denoted $\psi_{li}$, $\chi_{li}$, and $\phi_l$, and heavy fields, denoted $\psi_{hi}$, $\chi_{hi}$, and $\phi_h$. After integrating out the heavy fields, the effective Lagrangian for the light fields is given by the following expression:
\beqa \label{eq:c6:light_L}
{\cal L}_{\rm eff} \eq \psi^{\dagger}_{l i}\,i\slashed{D}\, (Z_\psi)_{ij} \psi_{l j} +  \chi^{\dagger}_{li}\,i \slashed{D} (Z_\chi)_{ij} \chi_{lj} + D_{\Dot{\mu}} \phi_l^\dagger Z_\phi D^{\Dot{\mu}} \phi_l \hc \nl \mi \left\{\tilde{Y}_{ij} \psi_{li}^T\, \cc\, \chi_{lj} \phi_l \hc \right\} + \cdot\cdot\cdot\,,\eeqa
where $"..."$ represents the non-renormalisable operators induced by the fields that have been integrated out. The wave-function renormalisation factor $Z$ can be parameterised as follows:
\be \label{eq:c6:Z}
(Z_{\psi,\chi})_{ij} = \delta_{ij} + (K_{\psi,\chi})_{ij}\,,~~Z_\phi = 1 + K_\phi,\,\ee
where $K_{\psi,\chi,\phi}$ is determined by the wavefunction renormalisation of the corresponding light field at one loop, incorporating at least one heavy field in the loop. Likewise, $\tilde{Y}$ in Eq.~\eqref{eq:c6:light_L} can also receive quantum corrections, which can be written as follows:
\be \label{eq:c6:Y_tilde}
\tilde{Y} = Y + \delta Y\,,\ee
where $\delta Y$ is the 1-loop Yukawa vertex correction with heavy fields in the loop.

We canonically normalise the kinetic terms of the fields appearing in the Eq.~\eqref{eq:c6:light_L} as follows:
\beqa \label{eq:c6:filed_red}
\psi_{li} = \left(U_\psi \tilde{Z}_\psi^{-1/2} U_\psi^\dagger \right)_{ij}\, \tilde{\psi}_{l j}\,,~~\chi_{li} = \left(U_\chi \tilde{Z}_\chi^{-1/2} U_\chi^\dagger \right)_{ij}\, \tilde{\chi}_{l j}\,,~~\phi_l = Z_\phi^{-1/2}\,\tilde{\phi}_l.\,\eeqa
Here, $\tilde{Z}_{\psi,\chi} = U_{\psi,\chi}^\dagger Z_{\psi,\chi} U_{\psi,\chi}$ are diagonal matrices. By substituting the above into Eq.~\eqref{eq:c6:light_L}, we achieve canonically normalised kinetic terms for the light fermions $\tilde{\psi}_{li}$, $\tilde{\chi}_{li}$, and the scalar $\tilde{\phi}_l$. Additionally, by applying Eq.~\eqref{eq:c6:filed_red} to the last term of Eq.~\eqref{eq:c6:light_L}, the effective Yukawa couplings in the new basis are derived as follows:
\be \label{eq:c6:eff_LY}
{\cal L}_{\rm eff} \supset (Y_{\rm eff})_{ij}\, \tilde{\psi}_{li}^T\,\cc\, \tilde{\chi}_{lj} \phi_l \hc\,, \ee
with,
\beqa \label{eq:c6:Y_eff_1}
Y_{\rm eff} &=& U_\psi^* \tilde{Z}_\psi^{-1/2} U_\psi^T\,  \tilde{Y}\, U_\chi \tilde{Z}_\chi^{-1/2} U_\chi^\dagger\,Z_\phi^{-1/2}.\, \eeqa

One can perturbatively expand the $Z$ using the definitions in Eq.~\eqref{eq:c6:Z} as follows;
\be \label{eq:c6:tilz}
\tilde{Z}_{\psi,\chi}^{-1/2} = ({\bf 1} + \tilde{K}_{\psi,\chi})^{-1/2} \simeq {\bf 1} - \frac{1}{2} \tilde{K}_{\psi,\chi},\, \ee
where $\tilde{K}_{\psi,\chi} = U_{\psi,\chi}^\dagger K_{\psi,\chi} U_{\psi,\chi}$ are diagonal and real matrices with $(\tilde{K}_{\psi,\chi})_{ii} < 1$. Similarly, $Z_\phi^{-1/2} \approx 1-\frac{1}{2} K_\phi$. By substituting these into Eq.~\eqref{eq:c6:Y_eff_1} and retaining only the leading order terms in $\delta Y$ and $K$, we find~\cite{Patel:2023gwt};
\beqa \label{eq:c6:Y_eff_2}
Y_{\rm eff}\big(\mu\big) & =  & Y \left(1-\frac{1}{2}K_\phi\big(\mu\big) \right) + \delta Y -\frac{1}{2} K^T_\psi\big(\mu\big) Y - \frac{1}{2} Y K_\chi\big(\mu\big).\, \eeqa
The above expression in Eq.~\eqref{eq:c6:Y_eff_2} can be employed as a one-loop corrected matching condition at a renormalisation scale $\mu$, through the substitution of $\delta Y$ and $K_{\psi,\chi,\phi}$ with their renormalised part defined in the $\overline{\rm MS}$ scheme. Then, the one-loop corrected formulation for the effective Yukawa couplings is yielded, incorporating the original Yukawa couplings from the full theory and the predominant corrections introduced by the heavy particles.

In the present scenario case, the one-loop matching condition for the Yukawa couplings at a renormalisation scale $\mu$, given in Eq.~\eqref{eq:c6:Y_eff_2}, can be reparameterised as follows;
\beqa \label{eq:c6:dY_gen}
Y_f = Y_f^{0}\left(1-\frac{K_h}{2}\right) + \delta Y_f - \frac{1}{2} \left(K^T_f Y_f^{0} + Y_f^{0} K_{f^C}\right),\eeqa
where $f=u,d,e,\nu$. Together with,
\be \label{eq:c6:Y0}
Y_u^0 = Y_1\,,~~Y_d^0 = Y_2\,,~~Y_e^0 = Y_2^T\,,~~Y_\nu^0 = Y_3,\, \ee 
at $\mu = M_{\rm GUT}$.

For the matching at the 1-loop level, the Yukawa couplings receive two types of contributions. The first type comes from vertex corrections that involve the colour triplet or the heavy gauge boson in the loop. The second type of contribution to the Yukawa threshold correction arises from the wavefunction renormalisation of fermions and scalars. Both contributions require at least one heavy field propagating inside the loop.

\section{Computation of One-Loop Corrected Matching Conditions}
{\label{sec:c6:loopd}}
\begin{figure}[t!]
        \centering
        \includegraphics[width=0.9\linewidth]{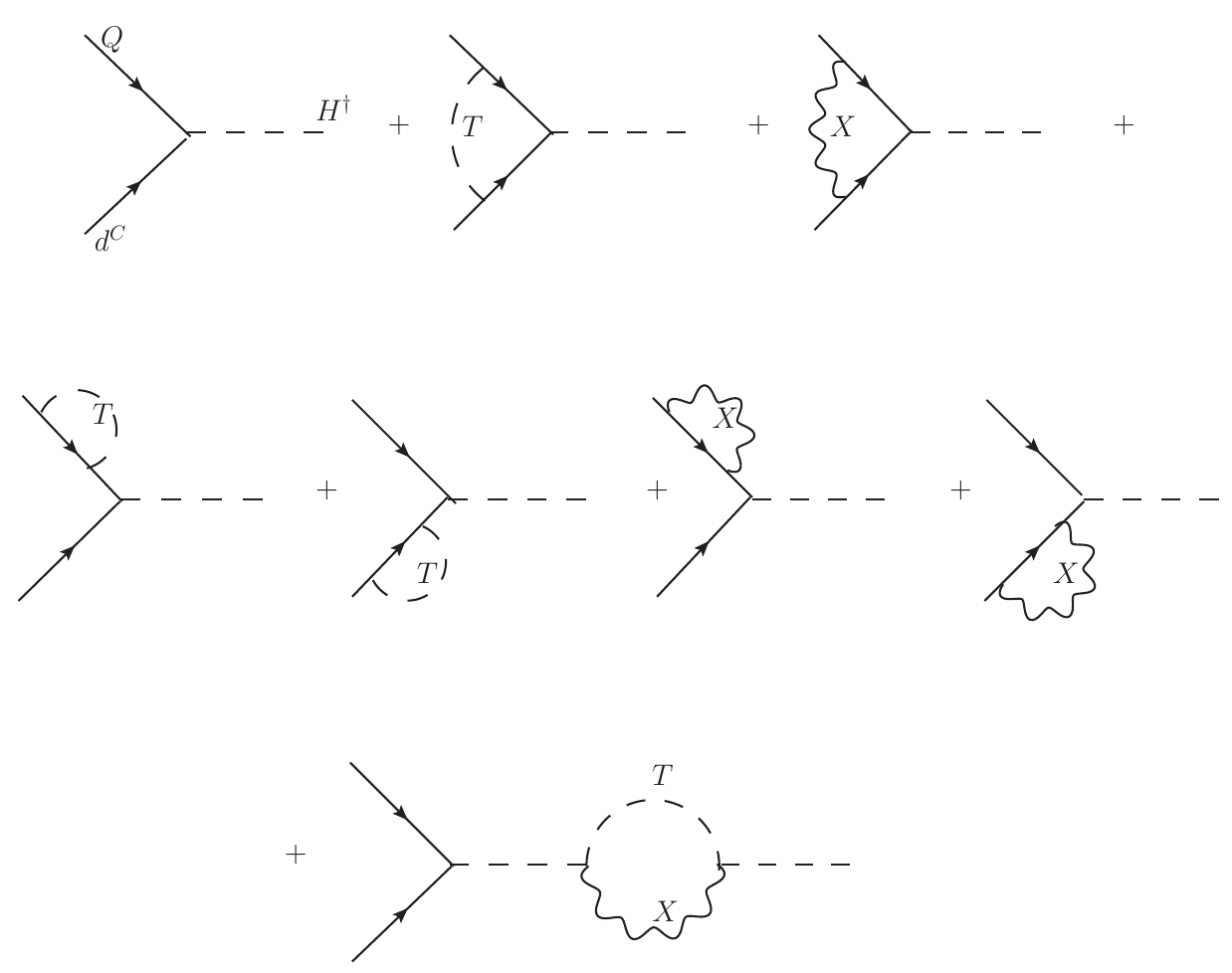}
        \caption{One-loop correction diagrams for \(Y_d\), with similar diagrams applicable to \(Y_e\), \(Y_u\), and \(Y_{\nu^C}\). However, \(\nu^C\) contributes exclusively to the corrections in \(Y_d\).}
        \label{fig:c6:loopcorrtoYd}
    \end{figure}
Fig.~(\ref{fig:c6:loopcorrtoYd}) shows the different diagrams contributing to the corrections in $Y_d$ at one-loop. The diagrams shown in the first line, in Fig.~(\ref{fig:c6:loopcorrtoYd}), are the vertex corrections, while the set of diagrams shown in the second line shows correction to the external leg. In contrast, the diagram shown in the last line is the correction to the Higgs line and is common for all the fermions. Further, we consider only those degrees of freedom to be massive, which are not included in the spectrum of the SM, and all the SM particles propagating inside the loop are massless.

\subsection{Vertex Corrections}
\begin{figure}[t!]
    \centering
    \includegraphics[width=1.05\linewidth]{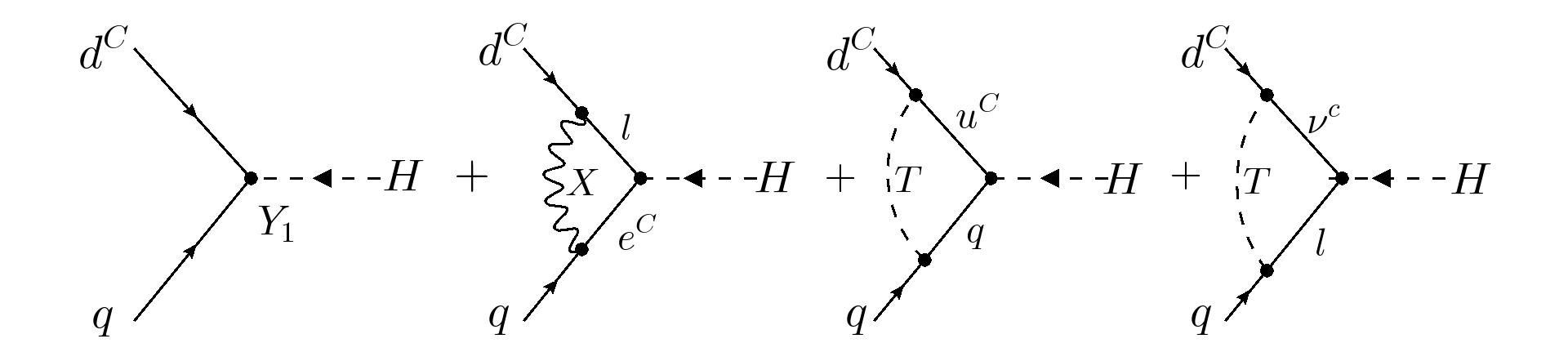}
    \caption{Feynman diagrams depicting vertex Corrections to $Y_{d}$ induced by heavy particles $(X,\,T,\,\nu^C)$ present in the framework.}
    \label{fig:c6:delyd}
\end{figure}

{\label{sec:c6:vercorr}}
The diagrams in Fig.~(\ref{fig:c6:delyd}) contribute to the vertex corrections to $Y_{d}$ at one loop. Their computation yields the following result;
\beqa{\label{eq:c6:delYd}}
(\delta Y_d)_{AB} &=& 2 g^2 (Y_2)_{AB} f[M_X^2,0]+\left(Y_1 Y_1^* Y_2 \right)_{AB} f[M_T^2,0],\nl
& + & \sum_\alpha \left(Y_2 Y_3^* \right)_{i \alpha} \left(Y_3^T\right)_{\alpha j} f[M_T^2,M_{N_\alpha}^2],
\eeqa
with,
\beqa \label{eq:c6:LF_f}
f[m_1^2,m_2^2] &\equiv& -\frac{1}{16 \pi^2} \left(\frac{m_1^2 \log \frac{m_1^2}{\mu^2}-m_2^2\log \frac{m_2^2}{\mu^2}}{m_1^2-m_2^2} - 1 \right).\, \eeqa
The first diagram in Fig.~(\ref{fig:c6:delyd}) is the tree-level diagram. The second diagram in the same figure is the one-loop contribution induced by the heavy gauge bosons, whose computation is outlined in Eq.~\eqref{eq:c6:app:gbc} of the Appendix~(\ref{app:6}). The resultant of such computation is the first term of the Eq.~\eqref{eq:c6:delYd}. The second term of the Eq.~(\ref{eq:c6:delYd}) is the outcome of computing the third diagram of Fig.~(\ref{fig:c6:delyd}) when the heavy triplet is propagating inside the loop, and loop computation of the same is done in Eqs.~(\ref{eq:c6:app:compyk1} and \ref{eq:c6:app:comp2}). The last diagram of Fig.~(\ref{eq:c6:delYd}) results in the expression given in the last line of Eq.~\eqref{eq:c6:delYd}, with $\nu$ being the number of singlets. In this diagram, $\nu^C$ and $T$ propagate inside the loop, and its computation is relegated to the Eq.~\eqref{eq:c6:app:nuc}.

\begin{figure}[t!]
    \centering
    \includegraphics[width=1.03\linewidth]{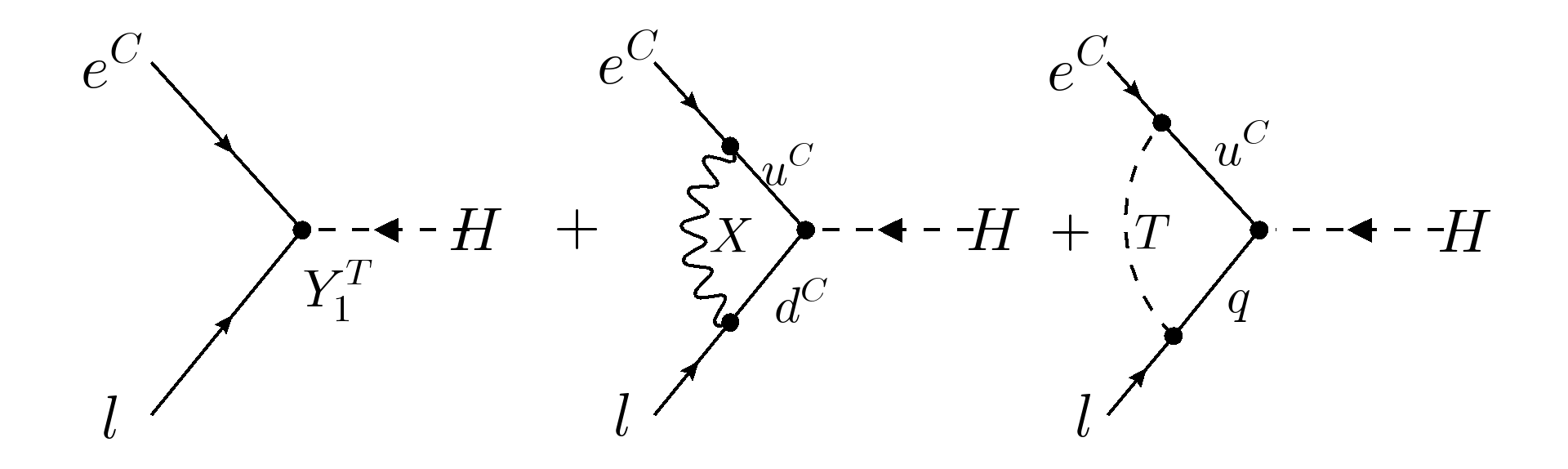}
    \caption{Feynman diagrams depicting vertex corrections to $Y_{e}$ induced by $X$ and $T$.}
    \label{fig:c6:delye}
\end{figure}

The Feynman diagrams given in Fig.~(\ref{fig:c6:delye}) depict the vertex corrections to $Y_{e}$, and their computation yields the following result;
\beqa{\label{eq:c6:delye}}
(\delta Y_e)_{AB} &=& 6 g^2 (Y_2^T)_{AB} f[M_X^2,0]+ 3 \left(Y_2^T Y_1^* Y_1 \right)_{AB} f[M_T^2,0],\eeqa
where the loop function $f[M_T^2,0]$ is already defined in Eq.~(\ref{eq:c6:LF_f}). The tree-level contribution to $Y_{e}$ is the first diagram in the Fig.~(\ref{fig:c6:delye}). The second diagram is the loop contribution induced by heavy gauge bosons $X$, and its evaluation is the first term of the Eq.~\eqref{eq:c6:delye}. The computation of this diagram can be performed analogously by following the evaluation outlined in Eq.~\eqref{eq:c6:app:gbc}. The third diagram is induced by Triplets, and its computed result is the last term of the Eq.~\eqref{eq:c6:delye}. This computation can also be done using the evaluation outlined in Eqs.~(\ref{eq:c6:app:compyk1} and \ref{eq:c6:app:comp2}).

The Feynman diagrams in Figs. (\ref{fig:c6:delyd}) and (\ref{fig:c6:delye}), depicting corrections to \(Y_{d}\) and \(Y_{e}\), receive similar contributions from \(X\) and \(T\). The evident difference between these two figures is the contribution of the \(\nu^C-T\) diagram, which is present only in \(Y_{d}\). This is reflected in the last line of Eq.~(\ref{eq:c6:delye}) and is the sole source of alleviating the degeneracy between \(Y_{d}\) and \(Y_{e}\). If we switch off this diagram, the condition \(Y_{d}\,=\,Y_{e}^T\) will hold again. This is the role that the quantum corrections induced by a scalar can play in modifying the tree-level Yukawa relations. 

The heavy triplets and gauge bosons can also induce loop correction to $Y_{u}$ and $Y_{\nu}$. However, $\nu^C$ cannot contribute to any of these corrections to these vertices. The vertex corrections to $Y_{u}$ and $Y_{\nu}$ is shown below~\cite{Patel:2023gwt};
\beqa \label{eq:c6:dY}
(\delta Y_u)_{AB} &=& 4 g^2 (Y_1)_{AB} f[M_X^2,0] \,+\, \left(Y_1 Y_2^* Y_2^T + Y_2 Y_2^\dagger Y_1^T \right)_{AB} f[M_T^2,0], \nonumber\\ 
(\delta Y_\nu)_{A\alpha} &=& 3 \left(Y_2^T Y_2^* Y_3 \right)_{A\alpha} f[M_T^2,0],\eeqa
at the scale \(\mu\), \(M_{N_\alpha}\) represents the mass of \(\nu^C_\alpha\).

\subsection{Wave Function Renormalisation Corrections}
{\label{sec:c6:WFR}}
Further, we compute the correction to the external legs due to the heavy particles in the theory. This particular correction modifies the wave function of the external particles and can be related to the self-energy correction as shown in Fig.~(\ref{fig:c6:WFR}):
\begin{figure}[t!]
    \centering
    \includegraphics[width=0.9\linewidth]{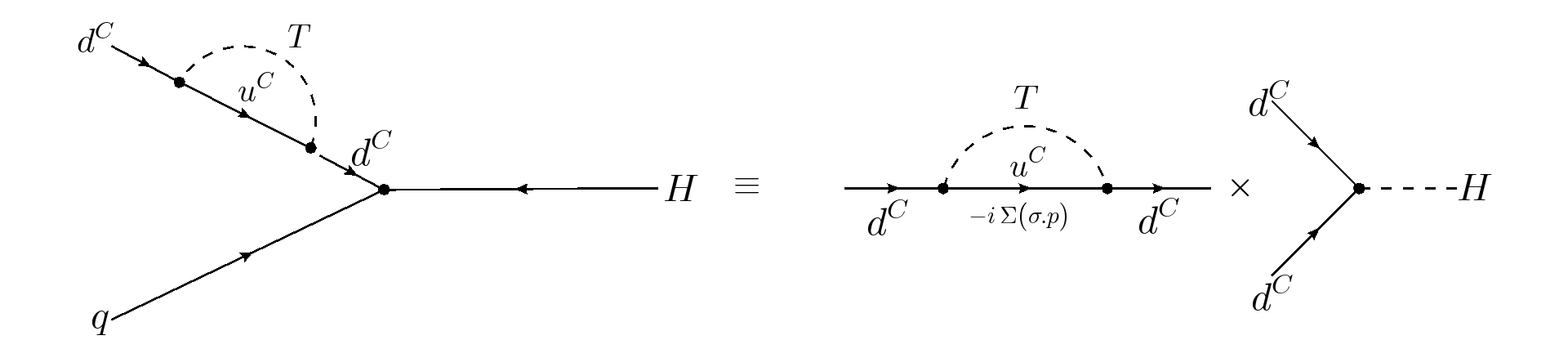}
    \caption{Wave function renormalisation and its connection to self-energy correction}
    \label{fig:c6:WFR}
\end{figure}
The wave function renormalisation factor \(K\), a matrix in generation space, is the derivative of the self-energy correction with respect to the square of the outgoing four-momentum, evaluated at zero four-momentum. Its mathematical formulation is shown below,
\beqa{\label{eq:c6:defWFR}}
\big(K\big)_{IJ} &\equiv & \frac{\partial\Sigma}{\partial\big(p.\sigma_{IJ}\big)}\Big|_{p^2=0}.
\eeqa

\begin{figure}
    \centering
    \includegraphics[width=1.0\linewidth]{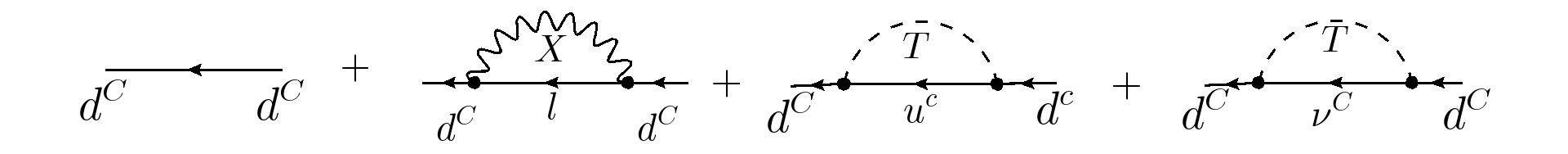}
    \caption{Feynman diagrams depicting correction to the external leg $d^C$.}
    \label{fig:c6:kdc}
\end{figure}

The Feynman diagrams depicting corrections to the external fermion leg of $d^C$ are shown in Fig.~(\ref{fig:c6:kdc}).
The wave function renormalisation computation result for the diagrams shown in Fig.~(\ref{fig:c6:kdc}) is shown below;
\beqa{\label{eq:c6:kdc}}
(K_{d^C})_{AB} &=& 2g^2 \delta_{ij} h[M_X^2,0] - 2 \left(Y_2^\dagger Y_2\right)_{AB} h[M_T^2,0]\nl
\mi \sum_\alpha \left(Y_3^*\right)_{A \alpha} \left(Y_3^T\right)_{\alpha B} h[M_T^2,M_{N_\alpha}^2],\eeqa
where, the loop function $h[m_1^2,m_2^2]$ has the following definition.
\beqa \label{eq:c6:LF_h}
h[m_1^2,m_2^2] &=& \frac{1}{16 \pi^2} \left(\frac{1}{2} \log\frac{m_1^2}{\mu^2} +  \frac{\frac{1}{2} r^2 \log r -\frac{3}{4}r^2 + r - \frac{1}{4}}{(1-r)^2}\right).\, \eeqa
The first diagram shown in the Fig.~(\ref{fig:c6:kdc}) is just the tree-level propagator or external fermion line. The second diagram is correction induced by heavy gauge boson $(X)$, and its result is the first term of the Eq.~(\ref{eq:c6:kdc}). The third self-energy correction diagram is induced by $T$, and the outcome is the second term of the Eq.~(\ref{eq:c6:kdc}). The computation of this diagram is shown in the Appendix~(\ref{app:6}) in Eq.~\eqref{eq:c6:app:wfr}. The last self-energy correction diagram is induced by $\nu^C$, and its computed result is shown in the last line of the Eq.~(\ref{eq:c6:kdc}).

The finite parts of wavefunction renormalisation computations for the light fermions and scalars at 1-loop, involving at least one heavy field in the loop, yield the following wavefunction renormalisation factors of different SM particles and are shown below;
\beqa \label{eq:c6:K}
(K_{q})_{AB} &=& 3g^2 \delta_{AB} h[M_X^2,0] - \frac{1}{2}\left( Y_1^{*} Y_1^T +2  Y_2^* Y_2^{T}\right)_{AB} h[M_T^2,0], \nonumber\\
(K_{u^C})_{AB} &=& 4g^2 \delta_{AB} h[M_X^2,0] - \left(Y_1^* Y_1^T +2 Y_2^* Y_2^T\right)_{AB} h[M_T^2,0], \nonumber\\
(K_l)_{AB} &=& 3g^2 \delta_{AB} h[M_X^2,0] - 3 \left(Y_2^\dagger Y_2\right)_{AB} h[M_T^2,0], \nonumber\\
(K_{e^C})_{AB} &=& 6g^2 \delta_{AB} h[M_X^2,0] - 3 \left(Y_1^\dagger Y_1\right)_{AB} h[M_T^2,0], \nonumber\\
(K_{\nu^C})_{\alpha\beta } &=& - 3 \left(Y_3^\dagger Y_3\right)_{\alpha\beta} h[M_T^2,0], \nonumber\\ \eeqa 
at the scale $\mu$. The loop integration factors are defined in Eqs.~(\ref{eq:c6:LF_f} and \ref{eq:c6:LF_h}). Again, only $K_{d^C}$ receives a contribution from the singlet fermions. The next sections show that these contributions from singlet fermions are crucial for uplifting degeneracy between the charged lepton and down-type quarks.

The external Higgs leg can also be corrected due to contributions from \(X\) and \(T\). This correction can be inferred from the last diagram shown in Fig.~(\ref{fig:c6:loopcorrtoYd}), with the result shown below;
\beqa{\label{eq:c6:kh}}
K_{h} &=& \frac{g^2}{2}\,\left(f[M_X^2,M_T^2] + g[M_X^2,M_T^2]\right),\eeqa
where, $g$ is the gauge coupling. The loop function $g[m_1^2,m_2^2]$ has the following definition,
\beqa \label{eq:c6:LF_g}
g[m_1^2,m_2^2] & = & \frac{1}{16 \pi^2} \frac{\frac{r^3}{6}-r^2+\frac{r}{2}+ r \log r + \frac{1}{3}}{(1 -r)^3}\,,\eeqa
provided, $r=m_2^2/m_1^2$, as defined earlier. The wave-function renormalisation for the Higgs, provided in Eq.~(\ref{eq:c6:kh}), is independent of flavour indices and thus contributes uniformly in all the loop corrections. 

The vertex corrections given in Eqs.~ (\ref{eq:c6:delYd}, \ref{eq:c6:delye}, and \ref{eq:c6:dY}), along with the wave-function renormalisation factors provided in Eqs. (\ref{eq:c6:kdc}), (\ref{eq:c6:K} and \ref{eq:c6:kh}), and the various loop functions defined in Eqs. (\ref{eq:c6:LF_f}, \ref{eq:c6:LF_h}, and (\ref{eq:c6:LF_g}), are essential for computing the Yukawa relations at one loop.

\section{Deviation from $Y_d = Y_e^T$}
\label{sec:c6:dep_ydye}
Substituting Eqs.~(\ref{eq:c6:delYd}, \ref{eq:c6:delye}, \ref{eq:c6:dY}, \ref{eq:c6:kdc}, \ref{eq:c6:K}, and \ref{eq:c6:kh}) in Eq.~\eqref{eq:c6:dY_gen}, it becomes evident that the 1-loop corrections break the degeneracy between $Y_e$ and $Y_d$. We can arrive at the following expression that is valid at the GUT scale:
\beqa  \label{eq:c6:ydye_ana}
\left(Y_d - Y_e^T\right)_{ij} &=& -2g^2 (Y_2)_{ij}\, \left(2f[M_X^2,0] - h[M_X^2,0] \right) \nonumber \\
&-& \left(Y_1 Y_1^* Y_2\right)_{ij}\, \left(f[M_T^2,0] +\frac{5}{8} h[M_T^2,0] \right) \nonumber \\
&+& \sum_\alpha \left(Y_2 Y_3^*\right)_{i \alpha} \left(Y_3\right)_{j \alpha} \Big(f[M_T^2,M_{N_\alpha}^2] + \Big. \frac{1}{2} h[M_T^2,M_{N_\alpha}^2] \Big) \,.\eeqa
 Eq.~\eqref{eq:c6:ydye_ana} is the crux of this chapter and demonstrates that $Y_d \neq Y_e^T$ due to the advent of loop corrections. This also indicates that the disparity between the two matrices, $i.e.$ $Y_{d}$ and $Y_{e}$, can be calculated based on the masses of the heavy scalar, gauge boson, right-handed neutrinos, and their couplings. As we will explore in the subsequent section, these parameters also dictate the masses of other fermions and thus can be significantly constrained.

\begin{figure}[t]
\centering
\includegraphics[width=0.65\textwidth]{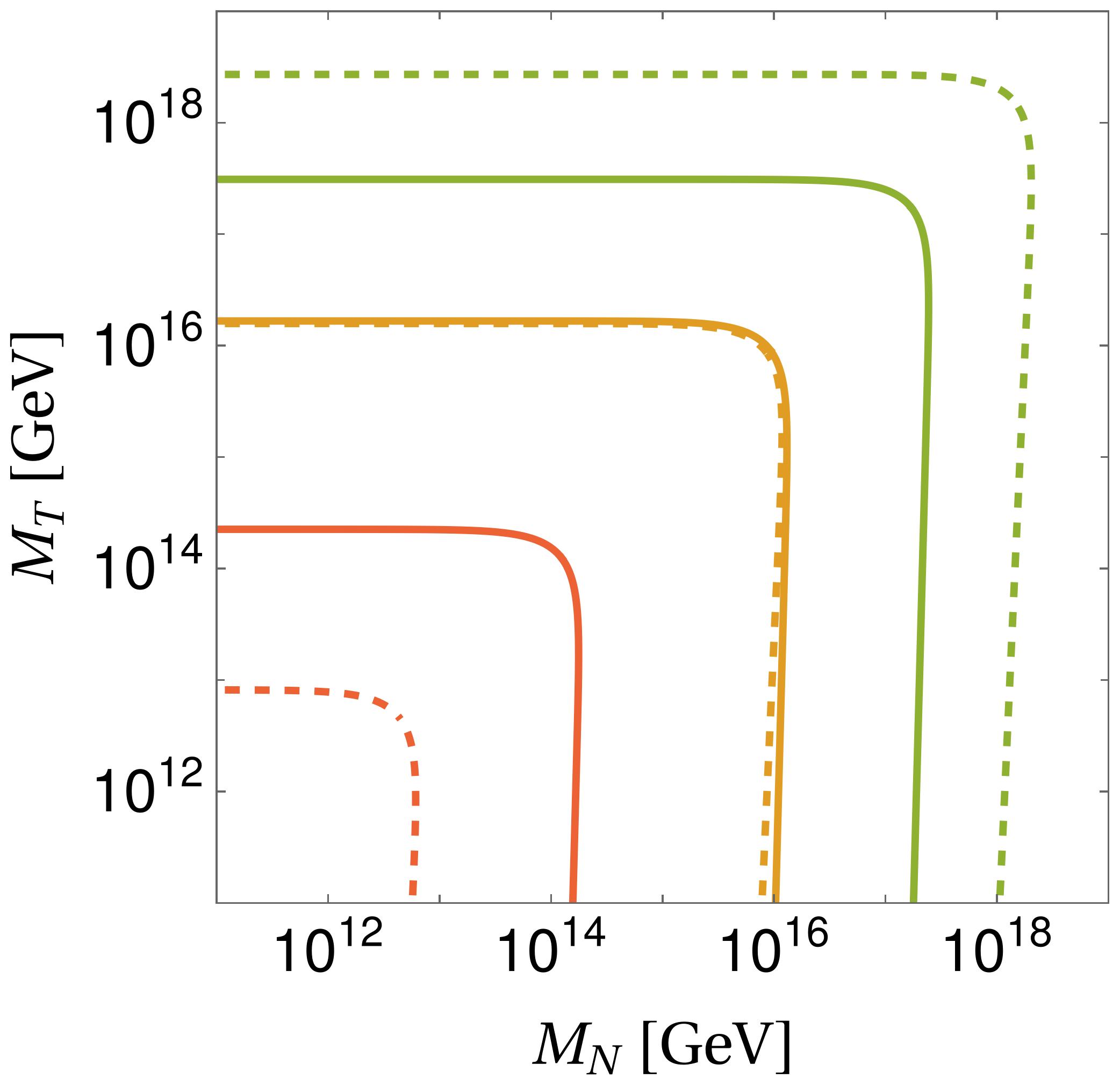}
\caption{Contours of $y_b/y_\tau = 3/2$ (red), $y_b/y_\tau = 1$ (orange), and $y_b/y_\tau = 2/3$ (green) are drawn using Eq.~\eqref{eq:c6:btau_ana}. These contours are based on parameters $y_t=0.427$, $g=0.53$, and $\mu=M_X=10^{16}$ GeV, and for $y_\nu = \sqrt{4 \pi}$ (solid lines) and $y_\nu = 2.7$ (dashed lines).}
\label{fig:c6:fig1}
\end{figure}

Before evaluating the ability of Eq.~\eqref{eq:c6:ydye_ana} in reproducing the full and realistic fermion mass spectrum, we first examine its ability to yield viable Yukawa ratio for third-generation Yukawa couplings, i.e. $y_b$ and $y_\tau$, at two loops. Here, we fix the number of RH neutrinos in the theory as $one$, $i.e.\, \alpha=1$, in Eq.~\eqref{eq:c6:ydye_ana}.
\beqa  \label{eq:c6:btau_ana}
\frac{y_b}{y_\tau} &\simeq& 1-2g^2 \left(2f[M_X^2,0] - h[M_X^2,0] \right) \nonumber \\
&-& 2y_t^2 \left(f[M_T^2,0] +\frac{5}{8} h[M_T^2,0] \right) \nonumber \\
&+& y_\nu^2 \left(f[M_T^2,M_N^2] +\frac{1}{2} h[M_T^2,M_N^2] \right)\,,\eeqa
where $y_t$ represents the top-quark Yukawa coupling and $y_\nu = (Y_3)_{31}$ at the GUT scale. We have considered the dominant contribution while deriving the above expression. For some specific values of $y_t$, $y_\nu$, and $\mu=M_X=10^{16}$ GeV, the contours representing different values of the ratio $y_b/y_\tau$ are displayed on the $M_T$-$M_N$ plane in Fig.~(\ref{fig:c6:fig1}).

Extrapolating the GUT scale of the observed fermion mass data necessitates $y_b/y_\tau \approx 2/3$ at two-loops, as can be inferred from Fig.~(\ref{fig:c6:yukrun}). As depicted in Fig.~(\ref{fig:c6:fig1}), achieving this ratio is only possible if either $M_T$ or $M_N$ exceeds $\mu=M_X$ by at least one to two orders of magnitude and large $y_\nu$ is also required. With $g, y_t < 1$, it is predominantly the third term in Eq.~(\ref{eq:c6:btau_ana}) that is required to significantly alter the degeneracy between $y_b$ and $y_\tau$, thus requiring the value of $y_\nu$ to be large. $M_T \gg M_{\rm GUT}$ or $M_N \gg M_{\rm GUT}$, in conjunction with a large $y_\nu$, are necessary to counteract the loop suppression factor of $1/(16 \pi)^2$. This clear, simple and qualitative understanding of the favourable mass scales for the colour triplet scalar and RH neutrino is also largely applicable when considering the full three-generation fermion spectrum, as will be demonstrated in the following section~(\ref{sec:c6:res}).

Moreover, RH neutrino couples to lepton doublet via SM-Higgs and contributes to the light neutrino mass via the typical type I seesaw mechanism~\cite{Minkowski:1977sc,Yanagida:1979as,Mohapatra:1979ia,Schechter:1980gr}, which turns out be  $m_\nu = v^2 y_\nu^2/M_N$. For the requirement to reproduce the atmospheric neutrino mass scale, then one finds:
\be \label{eq:c6:MN_1gen}
M_N \leq 7.6 \times 10^{16}\,{\rm GeV}\,\left(\frac{y_\nu}{\sqrt{4 \pi}}\right)^2\,\left(\frac{0.05\,{\rm eV}}{m_\nu} \right)\,. \ee
As evident, $M_N$ cannot significantly exceed $M_{\rm GUT}$, a phenomenologically viable ratio of $y_b/y_\tau$ can only be achieved if $M_T > M_{\rm GUT}$. Conversely, when considering perturbative values of $y_\nu$ and a scenario where $M_N$ greatly surpasses $M_{\rm GUT}$, the contribution of the RH neutrino to the light neutrino mass becomes relatively negligible. This insufficiency in reproducing a viable atmospheric neutrino mass scale necessitates the inclusion of an additional source for neutrino masses, which will be highlighted later.

\section{Full Three Generations Analysis}
\label{sec:c6:res}
We intend to determine whether the computed vertex corrections, different wave function renormalisation factors and the matching condition given in the Eqs.~(\ref{eq:c6:delYd}, \ref{eq:c6:delye}, \ref{eq:c6:dY}, \ref{eq:c6:kdc}, \ref{eq:c6:K}, \ref{eq:c6:kh} and \ref{eq:c6:dY_gen}) can reproduce the realistic values of the SM Yukawa couplings and the quark mixing (CKM) matrix, for which we implement a $\chi^2$ optimisation process. Initially, we consider just one RH neutrino with mass $M_{N_1} \equiv M_N$, as discussed earlier. Confer to the Appendix~(\ref{app:6}) for the definition of $\chi^2$ and optimisation process. 

\begin{figure}[t]
\centering
\includegraphics[width=0.65\textwidth]{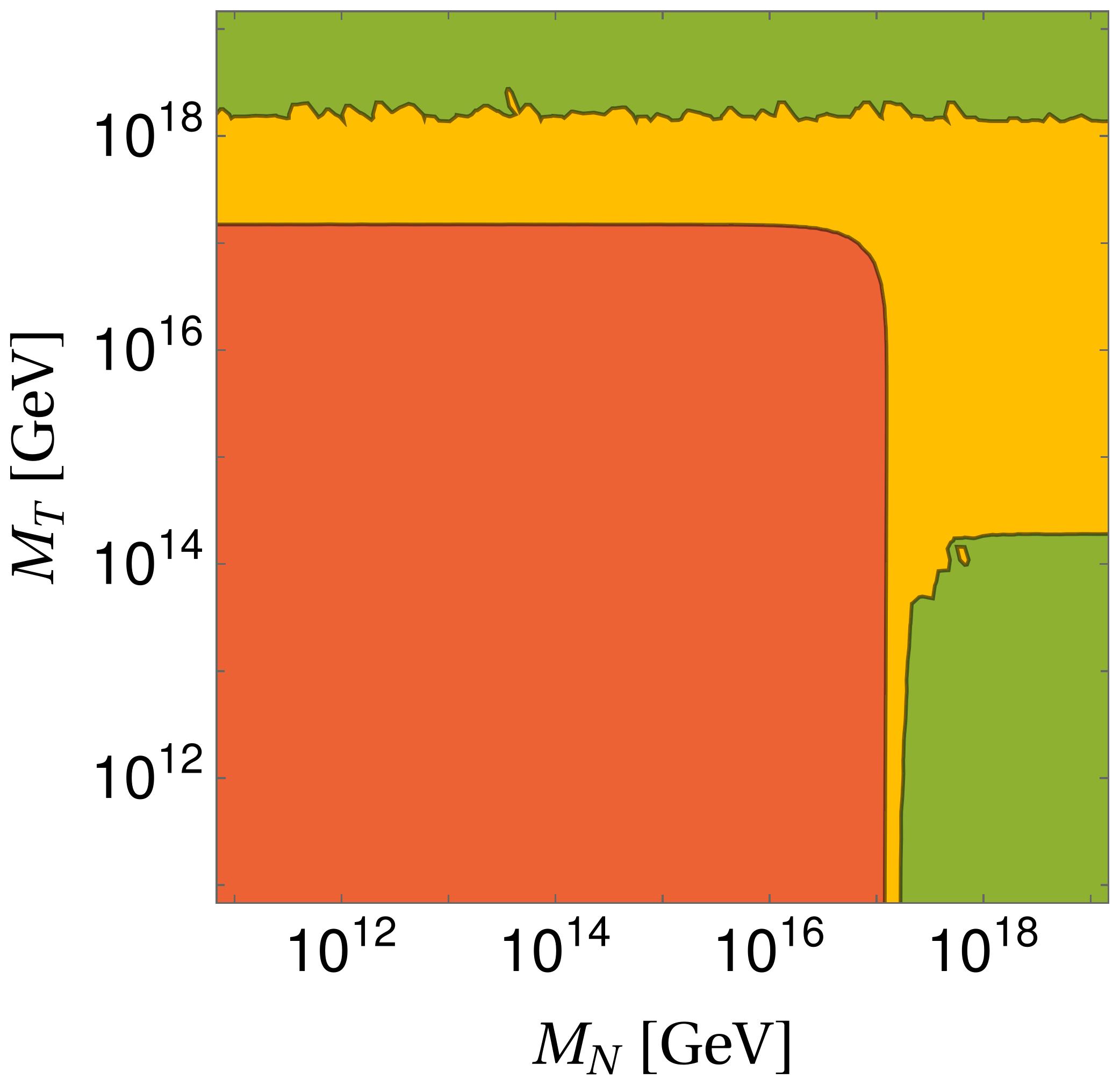}
\caption{The distribution of minimized $\chi^2$ values shown for various$M_T$ and $M_N$. The regions are color-coded as follows: green for $\chi^2_{\rm min} \leq 3$, yellow for $3 < \chi^2_{\rm min} \leq 9$, and red for $\chi^2_{\rm min} > 9$. We have set $\mu=M_X=10^{16}$ GeV and $g=0.53$, with the constraint that $|(Y_{1,2,3})_{ij}|<\sqrt{4 \pi}$, for these fits.} 
\label{fig:c6:fig2}
\end{figure}

Incidentally, the absence of 1-loop corrections, that is, when $Y_d = Y_e^T$, the minimum value of $\chi^2$ obtained is $53$. Therefore, values of $\chi^2_{\rm min}$ less than $53$ indicate improvements due to quantum-corrected matching conditions in the framework. Specifically, if $\chi^2_{\rm min}$ is less than $9$, it guarantees that no observable deviates from its central value by more than $\pm 3\sigma$, and thus, the model can be considered to provide a viable charged fermion mass spectrum and quark mixing.

As illustrated in Fig.~(\ref{fig:c6:fig2}), a very good fit for the entire charged fermion mass spectrum and the quark mixing parameters can be achieved if $M_T$ or $M_N \geq 10^{17.2}$ GeV. These results align well with the constraints on $M_T$ and $M_N$ obtained for $y_b/y_\tau \lesssim 2/3$ in the simplified scenario previously discussed in the section~(\ref{sec:c6:dep_ydye}) and as depicted in Fig.~(\ref{fig:c6:fig1}).

The three-generation $\chi^2$ analysis presented in Fig.~(\ref{fig:c6:fig2}) demonstrates that all the $13$ observables (detailed in the Appendix~\ref{app:6}) can be appropriately fitted within their respective $\pm 1\sigma$ ranges (corresponding to $\chi_{\rm min}^2 \leq 3$) under two conditions: (i) $M_T \leq 10^{14.5}$ GeV and $M_N \geq 10^{17.2}$ GeV, or (ii) $M_T \geq 10^{18.2}$ GeV. The latter scenario places $M_T$ near the Planck scale, intensifying the doublet-triplet splitting problem~\cite{Masiero:1982fe,Grinstein:1982um,Babu:2006nf}. Conversely, the former scenario is comparatively feasible and avoids these technical difficulties as $M_N$ is unrelated to $M_{\rm GUT}$, allowing a significant hierarchy between them. Furthermore, $M_T$ can remain substantially below $M_{\rm GUT}$ while adhering to the proton decay lifetime constraint, $M_T \gtrsim 10^{11}$ GeV~\cite{Dorsner:2012uz,Patel:2022wya}.  The case (ii), which sets \(M_{\rm{T}} \sim 10^{18}\) GeV for a viable fit, also requires \(M_{5_{\rm{H}}}\) be close to the Planck scale~($M_{\rm{pl}}\sim 10^{18.5}$ GeV). The presence of a Planck-scale parameter in the Lagrangian renders this case unfeasible.

Further, we list the benchmark solution of the scenario discussed above and label it as solution (I) in Tab.~(\ref{tab:c6:tab1_part1}). We also provide the fitted values of the corresponding input parameters. In this case, the leptonic spectrum is unavailable as we have considered only one generation of the singlet.

\begin{table*}[t]
    \begin{center} 
        \begin{math} 
            \begin{tabular}{cccr}
                \hline
                \hline 
                \multirow{2}{*}{~~Observable~~} & \multirow{2}{*}{ \hspace*{0.5cm}  $O_{\rm exp}$ \hspace*{0.5cm} } &
                \multicolumn{2}{c}{\bf \hspace*{0.5cm} Solution (I) \hspace*{0.5cm} } \\
                 &   & $O_{\rm th}$ & pull \\
                \hline
                $y_u$  & $2.81 \times 10^{-6}$ & $2.92\times 10^{-6}$ & $ 0$ \\
                $y_c$   & $1.42\times10^{-3}$& $1.42\times 10^{-3}$& $0$\\
                $y_t$   & $4.27 \times 10^{-1} $ & $4.35 \times 10^{-1}$& $0.2$\\
                $y_d$  & $ 6.14 \times 10^{-6} $ & $3.60\times10^{-6}$  & $-1.2$\\
                $y_s$   & $1.25 \times10^{-4}$ & $1.26\times10^{-4}$ & $\sim 0$ \\
                $y_b$   & $5.80 \times10^{-3}$ & $5.77 \times10^{-3}$& $\sim 0$\\
                $y_e$   & $2.75 \times 10^{-6}$ & $2.76 \times 10^{-6}$ & $0.2$ \\
                $y_{\mu}$  &$5.72 \times10^{-4}$ & $5.71\times10^{-4}$ & $\sim0$\\
                $y_{\tau}$   & $9.68 \times10^{-3}$ & $9.83 \times10^{-3}$& $ 0.2$\\
                $|V_{us}|$ & $0.2286$ & $0.2303$ & $0.1$ \\
                $|V_{cb}|$ & $0.0457$ & $0.0461$ & $0.1$\\
                $|V_{ub}|$ & $0.004$2 & $0.0043$ & $\sim 0$ \\
                $\sin\delta_{\rm CKM}$ & $0.78$ & $0.78$ & $ 0$  \\
                \hline
                $\chi^2_{\rm min}$    & & & $2.0$ \\
                \hline
                $M_T$ [GeV]& &$10^{12}$ & \\
                $M_{N_1}$ [GeV]& &$4.7\times 10^{18}$ & \\
                \hline
                \hline 
            \end{tabular}
        \end{math}
    \end{center}
    \caption{The benchmark best-fit values of different observables for solution (I) for the charged fermion mass spectrum and quark mixing parameters. The extrapolated values at the scale of $\mu=10^{16}$ GeV are provided along with the values reproduced through $\chi^2$ minimization and the corresponding pulls. The masses of the leptoquark scalar and RH neutrino for solution (I) are also specified at the bottom of the table.} 
    \label{tab:c6:tab1_part1} 
\end{table*}

Below, we present the fitted values for the Yukawa coupling matrices $Y_{1,2,3}$ associated with benchmark solution (I), as outlined in Tab.~(\ref{tab:c6:tab1_part1}). These values are obtained at a scale of $\mu=10^{16}$ GeV;
\beqa{\label{eq:c6:sol1}}
Y_1 &=&
\,\left(
\begin{array}{ccc}
 2.79\times10^{-6}  & 0 & 0 \\
 0 & 0.141\times10^{-3}  & 0 \\
 0 & 0 & 0.436 \\
\end{array}
\right)\,,~~~
Y_3 = \left(
\begin{array}{c}
 6.63 \times 10^{-2}  \\
 -3.46  \\
 2.93  \\
\end{array}
\right)\,, \nonumber\\
Y_2 &=& 10^{-4}\,\times\, \left(
\begin{array}{ccc}
-0.0424 - i\,0.107 & -1.84 + i\,0.481 & 1.24 - i\,0.40 \\
0.898 + i\,0.771 & 2.76 - i\,0.798 & -3.57 + i\,3.84 \\
-7.58 - i\,7.53 & 86.0 - i\,14.5 & -11.5 - i\,25.5 \\
\end{array}
\right).\nl\eeqa

Further, we extend the RH neutrino generation to include the leptonic spectrum. At least two RH neutrinos are required to explain the solar and atmospheric neutrino mass scales. The light neutrino masses are subsequently generated via the standard Type I seesaw mechanism, as shown below:
\be \label{eq:c6:seesaw}
M_\nu = - v^2\, Y_\nu M_N^{-1} Y_\nu^T\,. \ee
Here, $M_\nu$ is $3 \times 3$ light neutrino mass matrix while $M_N$ is $2\times2$ heavy neutrino mass matrix. $Y_\nu$ is $3 \times 2$ matrix which can be computed using Eqs.~ (\ref{eq:c6:delYd}, \ref{eq:c6:delye}, \ref{eq:c6:dY}, \ref{eq:c6:kdc}, \ref{eq:c6:K}, \ref{eq:c6:kh}, and \ref{eq:c6:dY_gen}). The above expression, in Eq.~(\ref{eq:c6:seesaw}), leads to one massless light neutrino as $M_{\nu}$ is a rank two matrix. Additionally, if we have three RH neutrinos in the framework, $M_{\nu}$, $M_{N}$ and $Y_{\nu}$ are all $3\times 3$ matrix.

In contrast to solution (I), we have expanded the definition of \(\chi^2\) for solution (II and III) to incorporate the solar and atmospheric squared mass differences, three mixing angles, and a Dirac CP phase. This approach evaluates whether the framework described by Eqs.~(\ref{eq:c6:seesaw}, \ref{eq:c6:dY}, \ref{eq:c6:K}, and (\ref{eq:c6:dY_gen}) can accurately reproduce the charged and neutral fermion mass-spectrum and mixing. For the input values of neutrino observables, we have taken the latest results from~\cite{Esteban:2020cvm}, applying a \(\pm 10\%\) uncertainty (cf. Appendix~(\ref{app:6}) for details). Further, we disregard the RGE effects in the running of neutrino parameters as they are known to yield insignificant deviations~\cite{Chankowski:1993tx,Babu:1993qv,Antusch:2005gp,Mei:2005qp}. We have considered the normal hierarchy of neutrino masses to fit the neutral leptonic spectrum, in which mass eigenvalues follow the conventional pattern $M_1<M_2<M_3$ and label this particular scenario as the solution (II).

The results of the \(\chi^2\) minimisation for the above scenario are presented in Tab.~(\ref{tab:c6:tab1}) as a solution (II). This analysis reveals a good agreement with all fermion masses and mixing parameters, resulting in a \(\chi^2_{\rm min} = 4\). The determined values for \(M_{N_1}\) and \(M_{N_2}\) are found to be smaller than \(M_{\rm GUT}\), necessitating \(M_T > 10^{17.2}\) GeV as predicted by Fig.~(\ref{fig:c6:fig2}). 

Building upon the discussions above, a straightforward extension can be considered where more than two RH neutrinos are involved. In such a scenario, at least one RH neutrino, strongly coupled with the SM fermions, possesses a mass exceeding \(M_{\rm GUT}\). This configuration is crucial for providing the necessary threshold corrections to achieve a viable charged fermion spectrum, although its impact on neutrino mass generation remains secondary. Conversely, other RH neutrinos with masses below the GUT scale contribute significantly to the realistic light neutrino spectrum while minimally affecting the aforementioned threshold corrections and is labelled by solution (III) in Tab.~(\ref{tab:c6:tab1}), where \(N_3\), with \(M_{N_2} > M_{\rm{GUT}}\). This interaction provides essential threshold corrections, particularly to the down-type quark sector. Notably, this scenario also highlights that the colour triplet scalar need not have high-scale mass, offering a potentially more realistic approach to model building within this framework.

\begin{table*}[t!]
	\begin{center} 
		\begin{math} 
			\begin{tabular}{cccccc}
				\hline
				\hline 
      \multirow{2}{*}{~~Observable~~} & \multirow{2}{*}{ \hspace*{0.5cm}  $O_{\rm exp}$ \hspace*{0.5cm} } & 
      \multicolumn{2}{c}{\bf \hspace*{0.5cm}  Solution II \hspace*{0.5cm} } &
      \multicolumn{2}{c}{\bf \hspace*{0.5cm}  Solution III \hspace*{0.5cm} }\\
				 &   & $O_{\rm th}$ & pull & $O_{\rm th}$ & pull \\
				\hline
				$y_u$  & $2.81 \times 10^{-6}$ & $2.81 \times10^{-6}$& $ 0$ & $2.81\times 10^{-6}$ & $0$ \\
				$y_c$   & $1.42\times10^{-3}$ & $1.42\times10^{-3}$& $0$ & $1.42 \times 10^{-3}$ & $0$\\
				$y_t$   & $4.27 \times 10^{-1} $ & $4.30\times10^{-1}$& $0.1$ & $4.28 \times 10^{-1}$& $\sim 0$\\
				$y_d$  & $ 6.14 \times 10^{-6} $ & $ 2.85\times 10^{-6}$ & $-1.8$ & $2.91 \times 10^{-6}$ & $-1.8$\\
				$y_s$   & $1.25 \times10^{-4}$ & $1.24\times 10^{-4}$ & $\sim 0$ & $1.25\times 10 ^{-4}$ & $\sim 0$\\
				$y_b$   & $5.80 \times10^{-3}$ & $6.09\times 10^{-3}$& $0.5$ & $5.79\times 10^{-3}  $ & $\sim 0$\\
				$y_e$   & $2.75 \times 10^{-6}$ & $2.81 \times 10^{-6}$& $0.2$ & $2.82 \times 10^{-6}$ & $0.3$\\
				$y_{\mu}$  &$5.72 \times10^{-4}$ & $5.65 \times 10^{-4}$ & $-0.1$ & $5.71 \times 10^{-4}$& $\sim 0 $\\
				$y_{\tau}$   & $9.68 \times10^{-3}$ & $9.06 \times 10^{-3}$ & $-0.6$ & $9.70\times 10^{-3}$ & $\sim 0$\\
				$|V_{us}|$ & $0.2286$ & $0.2292$& $\sim 0$ & $0.2291$ & $\sim 0$\\
				$|V_{cb}|$ & $0.0457$ & $0.0458$& $\sim 0$ & $0.0458$ & $\sim 0$\\
				$|V_{ub}|$ & $0.004$2 & $0.0042$ & $\sim 0$ & $0.0042$ & $\sim 0$ \\
				$\sin\delta_{\rm CKM}$ & $0.78$ & $0.78$ & $ 0$ & $0.78$ & $\sim 0$  \\
				$\Delta m^2_{\text{sol}}~ [{\rm eV}^2]$ & $7.41\times10^{-5}$& $7.53 \times 10^{-5}$ & $\sim 0$ & $7.51\times 10^{-5}$ & $\sim 0 $\\
				$\Delta m^2_{\text{atm}}~ [{\rm eV}^2]$ & $2.511\times10^{-3}$& $2.586 \times 10^{-3}$ & $\sim 0$ & $2.572 \times 10^{-3}$ & $\sim 0$ \\
				$\sin^2 \theta _{12}$ & $0.303$ & $0.303$ & $\sim 0$ & $0.303$ & $\sim 0$\\
				$\sin^2 \theta _{23}$ & $0.572$ & $0.558$ & $-0.2 $ & $0.571$ & $\sim 0$  \\
				$\sin^2 \theta _{13}$ & $0.02203$ & $0.02194$& $\sim 0$ & $0.02201$ & $\sim 0$ \\
                $\delta_{\rm{MNS}}\,[^\circ]$ & $197$ & $192$ & $-0.2$ & $197$ & $\sim 0$\\
    \hline
                $\chi^2_{\rm min}$    & & $4.0$ & & $3.1$ & \\
    \hline
                $M_T$ [GeV]& & $7.7 \times 10^{17}$& & $10^{13} $ &\\
                $M_{N_1}$ [GeV]& & $ 4.1 \times 10^{16}$ & & $5.6 \times 10^{12}$&\\
                $M_{N_2}$ [GeV] & & $2.3 \times 10^{12}$& & $6.9 \times 10^{17}$ &\\
                $M_{N_3}$ [GeV] & & $-$ & &$1.2\times 10^{13}$ & \\ 
				\hline
				\hline 
			\end{tabular}
		\end{math}
	\end{center}
	\caption{The benchmark best-fit solutions obtained for solutions (II) and (III). In addition to the charged fermion sector observable, $\chi^2$ also includes observables of neutral leptons. Other details are same as Tab.~(\ref{tab:c6:tab1_part1})} 
	\label{tab:c6:tab1} 
\end{table*}


For the case of Solution (II), best-fit values of $Y_{1,2,3}$ are as follows;
\beqa{\label{eq:c6:solii}}
Y_1&=&\left(
\begin{array}{ccc}
 2.78 \times 10^{-6} & 0 & 0 \\
 0 & 1.40\times 10^{-3} & 0 \\
 0 & 0 & -4.24\times 10^{-1} \\
\end{array}
\right)\,,~~~\nl  
Y_2 \eq  \begin{pmatrix}
-0.0154 + i\, 0.0492 & -0.265 - i\, 1.05 & 0.139 - i\, 1.20 \\
-0.245 - i\, 0.334 & 9.05 + i\, 0.273 & 5.95 - i\, 1.33 \\
3.31 + i\, 4.24 & -61.9 + i\, 47.4 & 21.5 + i\, 39.8
\end{pmatrix}\times10^{-4},\nl
Y_3 \eq \left(
\begin{array}{cc}
 2.96\times 10^{-1} & -2.43\times 10^{-2} \\
 -3.44 & 8.92\times 10^{-3} \\
 -3.50 & -4.25\times 10^{-2} \\
\end{array}
\right).
\eeqa

Similarly, for Solution III, we find the following best-fit $Y_{1,2,3}$;
\beqa{\label{eq:c6:sol3}}
Y_1&=&\left(
\begin{array}{ccc}
 2.79 \times 10^{-6} & 0 & 0 \\
 0 & 1.41 \times 10^{-3} & 0 \\
 0 & 0 & -4.28\times 10^{-1} \\
\end{array}
\right)\,,\nl
Y_2 \eq \begin{pmatrix}
0.0610 - i\, 0.00622 & 0.433 - i\, 1.62 & -0.393 + i\, 1.27 \\
0.126 + i\, 0.440 & 8.54 - i\, 4.19 & -5.30 + i\, 5.20 \\
2.26 - i\, 4.77 & -70.3 + i\, 9.19 & 8.49 - i\, 57.4
\end{pmatrix}\times10^{-4},\nl
Y_3 \eq \left(
\begin{array}{ccc}
 3.40 \times 10^{-2}  & 9.83 \times 10^{-2} & 1.98 \times 10^{-2} \\
 -4.23\times 10^{-3} & 3.50 & -2.64 \times 10^{-1} \\
 -5.33\times 10^{-2} & -3.44 &9.50\times 10^{-2} \\
\end{array}
\right).
\eeqa

\section{Summary}
\label{sec:c6:concl}

\lettrine[lines=2, lhang=0.33, loversize=0.15, findent=0.15em]{P}URSUIT OF FINDING a minimal model is only possible if we have an outcome which acts as a common solution to multiple problems. This chapter is along a similar line and shows that the observationally inconsistent relation in the minimal \su\, GUT, i.e. \(Y_d = Y_e^T\), can be rectified without expanding the scalar sector. The incorporation of one-loop corrections to the initial tree-level matching conditions in the presence of one or more fermion singlets, the colour triplet scalar, and the leptoquark vector can impart substantial threshold corrections that can render viable the aforementioned inconsistent relation. These added singlets can also account for generating a viable leptonic spectrum; hence, we solve two problems simultaneously. This approach is crucial to making a minimal/optimal GUT model.

Our quantitative analysis indicates that to obtain a realistic spectrum for the charged fermion Yukawa couplings and quark mixing, one must either significantly increase the mass of the colour triplet scalar (\(M_T \gg M_X\)) or substantially elevate the masses of the RH neutrinos (\(M_{N_\nu} \gg M_X\)), assuming the mass of the leptoquark gauge boson (\(M_X\)) sets the unification scale. The latter option is less favoured if the same fermion singlets are expected to contribute to a viable light neutrino spectrum through the traditional type I seesaw mechanism. However, the scenario with \(M_{N_\alpha} \gg M_X\) remains viable if neutrinos receive their masses through alternative methods. This includes the possibility of a type I seesaw mechanism employing additional copies of RH neutrinos, which possess sub-GUT scale masses and relatively smaller couplings with the SM leptons.

Invoking quantum corrections induced by scalars to modify the tree-level Yukawa relations can find its immediate application in \so\, GUTs. An \so\ model consisting of only a single irrep in the Yukawa sector is also known to yield mass degeneracy in the charged fermion sector. For instance, the inclusion of $10_{\hh}$-dimensional scalar irrep in the Yukawa sector yields $Y_{u}\,=\,Y_{d}\,=\,Y_{e}\,=\,Y_{\nu^C}$. To rectify such a situation, one may include more than one copy of $10_{\hh}$ and compute the one-loop corrections to the tree-level matching condition due to the scalars and gauge bosons. Such aforementioned degeneracy may get alleviated, which may result in a minimal \so\ model. Further, an \so\, model based on $10_{\hh}\,+\,120_{\hh}$ is also known to yield inconsistent tree-level mass relations~\cite{Joshipura:2011nn}. Loop correction induced by scalars within such a framework may remove inconsistent generational mass degeneracy.

The quantum corrections to the Yukawa relations can alter the conventional understanding of the minimal Yukawa sector of the conventional GUTs and offer deeper insights into the same. Consequently, it provides enough motivation and evidence to dig and explore other variants of GUTs, be it SUSY \su\, or \so. Radiative corrections to $Y_d = Y_e^T$ stemming from the superpartners of the SM fields propagating inside the loop have been considered~in~\cite{Diaz-Cruz:2000nvf,Enkhbat:2009jt}. As we have seen in the previous chapters, \so\ exhibits a vast spectrum of scalar particles, thus extending an interesting opportunity to revisit all the conventional GUT models to minimise the Yukawa sector.

\clearpage
\thispagestyle{empty}\newpage
\vspace*{\fill}
\begin{Huge}
\begin{center}
    \textbf{This page is intentionally left blank.}
\end{center}
\end{Huge}\vspace*{\fill}
\newpage
\thispagestyle{empty}
\chapter{Conclusions and Outlook}\label{chap:summary}
\label{ch:7}

\lettrine[lines=2, lhang=0.33, loversize=0.15, findent=0.15em]{I}N THIS THESIS, we have explored various facets of scalar fields and their direct and indirect implications in the framework of GUTs, demonstrating new particles, in particular the scalar fields, can guide and dictate the future direction of hidden physics, alias the New Physics. 

GUTs, which aim to unify the known fundamental interactions, also place quarks and leptons within the same multiplet. This unification necessitates the presence of scalar irreps, whose dimensionality is sometimes larger than that of the fermion and (occasionally) gauge boson irreps. These scalar irreps, comprised of many scalars including SM Higgs, are essential for elementary requirements to break the bigger gauge symmetry into SM gauge symmetry, able to reproduce observed fermion masses and mixing angles. Each scalar field is associated with a physical energy scale and imparts significant implications. These implications can be direct and indirect, manifesting at lower energies.

In Chapter~(\ref{ch:2}), Tabs.~(\ref{tab:c2:scalars} and \ref{tab:c2:16scalars}) list the scalars that have been focused on in this thesis. Reviewing these tables will also summarise the Chapters~(\ref{ch:3}, \ref{ch:4}, and, \ref{ch:5}). We begin with Tab.~(\ref{tab:c2:scalars}), which lists the scalars residing in the $10_{\hh}$, $120_{\hh}$, and $\overline{126}_{\hh}$ dimensional irreps. Many of these scalars are charged under $B-L$, meaning they will induce processes that affect the Baryon and Lepton numbers. There are only four distinct $B-L$ charges that different scalars can have, and the relationship between $B-L$, $U(1)_{X}$, and $U(1)_{Y}$ is dictated by Eq.~(\ref{eq:c2:B-LRel}). These four distinct $B-L$ charges are : $0$, $\pm \sfrac{2}{3}$, $\pm \sfrac{4}{3}$ and $\pm 2$. 
Below, we consider each $B-L$ charge individually and explain the associated phenomenology of the scalars with that specific $B-L$ charge, which has also been summarised in Tabs.~(\ref{tab:c7:scalars} and \ref{tab:c7:scalarsum16H}).

The scalars \( D \) and \( O \), together with their conjugate partners, exhibit \( B-L = 0 \). The scalar \( D \) is present in all the irreps contributing to the renormalisable Yukawa sector and can be responsible for Electroweak symmetry breaking and the Dirac mass of charged and neutral fermions. The scalars \( O \) and \(\overline{O} \) couple solely to quarks and can be constrained by quark flavour violation.

The scalars $T$, $\Theta$, ${\cal T}$, and ${\mathbb{T}}$ along with their conjugates have $B-L\,=\, \big| \sfrac{2}{3}\big|$. The scalars $T$, $\Theta$, ${\cal T}$, and ${\mathbb{T}}$ exhibits diquark and leptoquark couplings, which can be obtained through the methodology outlined in Chapter~(\ref{ch:2}). These scalars can affect the stability of the proton and induce proton decay through mass dimension six $(D=6)$ operators. Our analysis reveals that only $T$ and $\mathbb{T}$ can induce proton decays at tree level, and ${\cal T}$ can induce proton decays at loop level. The induced proton decays by these scalars are $B-L$ conserving. Further, $\Theta$ only exhibits diquark coupling whose Yukawa couplings are antisymmetric and can induce $B-L$ violating proton decays. In a realistic \so\, model based on $10_{\hh}$ and $\overline{126}_{\hh}$, only $T$ and $\overline{T}$ can induce $B-L$ conserving proton decays and proton favours to decay into the second generation mesons as elaborated in Chapter~(\ref{ch:3}). As these scalars induce proton decay, they cannot remain light in a realistic \so\, scenario and the minimum mass $T$ can acquire adhering to proton decay constraint is $10^{11}$ GeV.

The non-renormalisable interactions in GUTs, which offer an alternate solution to the inconsistencies of renormalisable GUTs in UV, can also affect proton stability. In non-renormalisable \(SO(10)\) models, the smallest irrep \(16_{\rm{H}}\) couples with the \(\fs\)-plet, contributing to the fermion masses at the non-renormalisable level. $16_{\hh}$ exhibit pairs of scalars that also have $B-L\,=\,\big|\sfrac{2}{3}\big|$ and exhibits diquark and leptoquark vertices, which are shown in Tab.~(\ref{tab:c7:scalarsum16H}). In Chapter~(\ref{ch:4}), we studied the implications of such pairs of scalars in inducing $D=6$ proton decays.  It was argued that \(16_{\hh}\) can mimic the role of \(126_{\hh}\), and the proton decay spectrum was computed in a realistic model based on the same. The pattern of branching of the proton is similar to the findings of Chapter~(\ref{ch:3}). Unlike renormalisable \(SO(10)\) GUTs, where a single scalar mediates decay at leading order, non-renormalisable models involve pairs of scalars, which can potentially remain closer to the electroweak scale than the GUT scale.

The scalars $S$, $\Sigma$, ${\cal S}$, and ${\mathbb{S}}$, which also have $B-L\,=\,\big|\sfrac{2}{3}\big|$, are termed as sextet scalars. These sextets solely couple to diquarks and hence cannot induce leading order proton decay and have been extensively studied in Chapter~(\ref{ch:5}). They induce flavour violations and baryon number-violating phenomena like neutral baryon-antibaryon oscillation. The \nn\, oscillation is induced by three sextets and requires a source of baryon number violation in the theory. Further, $S$ is the only the only sextet which is degenerate and is present both in $120_{\hh}$ and $\overline{126}_{\hh}$. This speciality of $S$ is harnessed in Chapter~(\ref{ch:5}) to study the cosmological implications of sextet scalars in generating the observed baryon-asymmetry of the universe.

Scalars with \(B-L = \big|\sfrac{4}{3}\big|\) exhibit only leptoquark couplings and can be a source of loop-induced lepton and quark flavour violations. These scalars are $\Delta$ and $\Omega$, along with their conjugate partners. However, they can interact with scalars having \(B-L = \big|\sfrac{2}{3}\big|\), which exhibit diquark couplings and induce nucleon decays. These nucleon decays differ from the \(D=6\), \(B-L\) conserving nucleon decays, occurring instead at \(D=7\) and violating \(B-L\) by two units. Such nucleon decays in the framework of realistic GUTs have been thoroughly examined in Chapter~(\ref{ch:3}) and is summarised in Tab.~(\ref{tab:c7:scalars}).

\begin{landscape}
\begin{longtable}{>{\raggedright\arraybackslash}p{1.75cm} >{\centering\arraybackslash}p{1.5cm} >{\raggedleft\arraybackslash}p{1.3cm} >{\centering\arraybackslash}p{0.8cm} >{\centering\arraybackslash}p{0.8cm} >{\centering\arraybackslash}p{1cm} >{\raggedright\arraybackslash}p{11.5cm} >{\raggedright\arraybackslash}p{2cm}} 
\hline
\hline
SM charges & Symbol & $B-L$ & $10_{\hh}$ & $120_{\hh}$ & $\overline{126}_{\hh}$ & Phenomenology & Mass Bounds [GeV] \\
\hline
\endfirsthead

\multicolumn{8}{c}%
{\tablename\ \thetable\ -- \textit{Continued from previous page}} \\
\hline
SM charges & Symbol & $B-L$ & $10_{\hh}$ & $120_{\hh}$ & $\overline{126}_{\hh}$ & Phenomenology & Mass Bounds [GeV] \\
\hline
\endhead

\hline \multicolumn{8}{r}{\textit{Continued on next page}} \\
\endfoot

\hline
\endlastfoot

$\left(1,1,0\right)$ & $\sigma$ & $-2$ & 0 & 0 & 1 & Breaks $B-L$ & $\langle \sigma \rangle \geq 10^{11}$ \\
$\left(1,1,1\right)$ & $s$ & 2 & 0 & 1 & 0 & --- & --- \\
$\left(1,1,-1\right)$ & $\overline{s}$ & $-2$ & 0 & 1 & 1 & --- & --- \\
$\left(1,1,-2\right)$ & $X$ & $-2$ & 0 & 0 & 1 & --- & --- \\
$\left(1,2,-\frac{1}{2}\right)$ & $\overline{D}_a$ & 0 & 1 & 2 & 1 & EW symmetry breaking & 125 \\
$\left(1,2,\frac{1}{2}\right)$ & $D^a$ & 0 & 1 & 2 & 1 & EW symmetry breaking & 125 \\
$\left(1,3,1\right)$ & $t^{ab}$ & 2 & 0 & 0 & 1 & --- & --- \\
$\left(3,1,-\frac{1}{3}\right)$ & $T^{\alpha}$ & $-\frac{2}{3}$ & 1 & 2 & 2 & $B-L$ conserving and violating nucleon decays at tree level & $\geq 10^{11}$ \\
$\left(\overline{3},1,\frac{1}{3}\right)$ & $\overline{T}_\alpha$ & $\frac{2}{3}$ & 1 & 2 & 1 & $B-L$ conserving and violating nucleon decays at tree level & $\geq 10^{11}$ \\
$\left(3,1,\frac{2}{3}\right)$ & $\Theta_{\alpha\beta}$ & $-\frac{2}{3}$ & 0 & 1 & 1 & $B-L$ violating nucleon decays & --- \\
$\left(\overline{3},1,-\frac{2}{3}\right)$ & $\overline{\Theta}^{\alpha\beta}$ & $\frac{2}{3}$ & 0 & 1 & 0 & $B-L$ violating nucleon decay & --- \\
$\left(3,1,-\frac{4}{3}\right)$ & $\mathcal{T}^\alpha$ & $-\frac{2}{3}$ & 0 & 1 & 1 & $B-L$ conserving nucleon decay at loop level & --- \\
$\left(\overline{3},1,\frac{4}{3}\right)$ & $\overline{\mathcal{T}}_\alpha$ & $\frac{2}{3}$ & 0 & 1 & 0 & $B-L$ conserving nucleon decay at loop level & --- \\
$\left(3,2,\frac{1}{6}\right)$ & $\Delta^{\alpha a}$ & $\frac{4}{3}$ & 0 & 1 & 1 & $B-L$ violating nucleon decay at tree level & $\geq 10^{6}$ \\
$\left(\overline{3},2,-\frac{1}{6}\right)$ & $\overline{\Delta}_{\alpha a}$ & $-\frac{4}{3}$ & 0 & 1 & 1 & $B-L$ violating nucleon decay at tree level & $\geq 10^{6}$ \\
$\left(3,2,\frac{7}{6}\right)$ & $\Omega^a_{\alpha \beta}$ & $\frac{4}{3}$ & 0 & 1 & 1 & $B-L$ violating nucleon decay at tree level & --- \\
$\left(\overline{3},2,-\frac{7}{6}\right)$ & $\overline{\Omega}_a^{\alpha \beta}$ & $-\frac{4}{3}$ & 0 & 1 & 1 & $B-L$ violating nucleon decay at tree level & --- \\
$\left(3,3,-\frac{1}{3}\right)$ & $\mathbb{T}{^{a\alpha}_{b}}$ & $-\frac{2}{3}$ & 0 & 1 & 0 & $B-L$ conserving and violating nucleon decay at tree level & --- \\
$\left(\overline{3},3,\frac{1}{3}\right)$ & $\overline{\mathbb{T}}{^a_{b\alpha}}$ & $\frac{2}{3}$ & 0 & 1 & 1 & $B-L$ conserving and violating nucleon decay at tree level & --- \\
$\left(6,1,-\frac{2}{3}\right)$ & $\Sigma^{\alpha\beta}$ & $\frac{2}{3}$ & 0 & 0 & 1 & $n-\bar{n}$ at tree and QFV at tree level & $\geq 10^{12}$ \\
$\left(6,1,\frac{1}{3}\right)$ & $S{^{\alpha}_{\beta\gamma}}$ & $\frac{2}{3}$ & 0 & 1 & 1 & $n-\bar{n}$ at tree and QFV at loop level & $\geq 10^{5}$ If $M_{\Sigma}\geq 10^{12}$ \\
$\left(\overline{6},1,-\frac{1}{3}\right)$ & $\overline{S}{^{\beta\gamma}_{\alpha}}$ & $-\frac{2}{3}$ & 0 & 1 & 0 & $n-\bar{n}$ at tree and QFV at loop level & \ditto \\
$\left(6,1,\frac{4}{3}\right)$ & ${\cal S}^{\alpha}_{\beta\gamma}$ & $\frac{2}{3}$ & 0 & 0 & 1 & $n-\bar{n}$ at tree and QFV at tree level &\ditto\\
$\left(\overline{6},3,-\frac{1}{3}\right)$ & $\mathbb{S}^{\alpha\beta a}_{\gamma b}$ & $-\frac{2}{3}$ & 0 & 0 & 1 & $n-\bar{n}$ at tree and QFV at tree level & \ditto \\
$\left(8,2,\frac{1}{2}\right)$ & $O^{\alpha a}_{\beta}$ & 0 & 0 & 1 & 1 & --- & --- \\
$\left(8,2,-\frac{1}{2}\right)$ & $\overline{O}_{\alpha a}^{\beta}$ & 0 & 0 & 1 & 1 & --- & --- \\
\hline\hline
\caption{Classification of different scalar fields with their charges under the SM gauge group ($SU(3)_C$, $SU(2)_L$, $U(1)_Y$), $B-L$ charge, multiplicities in $10_{\hh}$, $120_{\hh}$, $\overline{126}_{\hh}$ representations, associated phenomenology, and mass bounds. Further, $\langle v\rangle$ denotes $vev$. }
\label{tab:c7:scalars}
\end{longtable}
\end{landscape}

Scalars with \(B-L = \big|2\big|\), such as $\sigma$, $s$, $\overline{s}$, $X$, and $t$, exhibit only dilepton couplings. The $vev$ of $\sigma$ violates \(B-L\) by two units, generates the Majorana mass of RH neutrinos, and cannot attain values less than \(10^{11}\) GeV in a minimal realistic \so\, models as the Yukawa couplings would become non-perturbative, discussed in Chapter~(\ref{ch:5}). The $vev$ of $\sigma$ also contributes to \(B-L\) violating processes such as proton decay and neutron-antineutron oscillation, as demonstrated in Chapters~(\ref{ch:3} and \ref{ch:5}). Further, $t$ is known to yield light neutrino masses via the Type-II seesaw mechanism.
In Chapters~(\ref{ch:3}, \ref{ch:4}, and \ref{ch:5}), studied direct implications induced by the scalars have been summarised in Tabs.~(\ref{tab:c7:scalars} and \ref{tab:c7:scalarsum16H}). 

 One of the immediate consequences of the mass bounds summarised in Tabs.~(\ref{tab:c7:scalars} and \ref{tab:c7:scalarsum16H}) can be studied in the context of gauge coupling unification. As can be inferred from Fig.~(\ref{fig:c1:SMgaugerunning}), the SM gauge couplings do not precisely unify in this case. The masses of the scalars affect the running of the gauge couplings, and the exact expressions can be found in~\cite{Cheng:1973nv}. For simplicity, we consider a one-loop analysis for gauge coupling unification. The running depends on the one-loop coefficients \(b_i\), which vary for different scalars. To account for the effects of scalar masses, \(b_i\) is replaced by \(B_i = b_i + \sum_I b_{iI}\,\frac{\ln\left(\sfrac{M_{\rm{GUT}}}{M_I}\right)}{\ln\left(\sfrac{M_{\rm{GUT}}}{M_Z}\right)}\), where \(M_{\rm{GUT}}\), \(M_I\), and \(M_Z\) represent the GUT scale, the mass of the scalar, and the Z-boson mass scale, respectively~\cite{Giveon:1991zm}. For successful unification at one loop, the following conditions are to be achieved~\cite{Dorsner:2005fq}:
\beqa{\label{eq:c6:gcu}}
\frac{B_{23}}{B_{12}} &\approx& 0.719, \nl
\ln\left(\frac{M_{\rm{GUT}}}{M_{\rm{Z}}}\right) &\approx& \frac{184.9}{B_{12}},
\eeqa
where \(B_{23} = B_2 - B_3\) and \(B_{12} = B_1 - B_2\). The values of \(B_{12}\) and \(B_{23}\) for various scalars appearing in Tab.~(\ref{tab:c7:scalars}) can be inferred from~\cite{Patel:2011eh}. The SM value of the ratio \(\frac{B_{23}}{B_{12}} = 0.53\) implies that for successful unification, \(B_{23}\) must increase and \(B_{12}\) should decrease. Scalars such as \(\Delta\), \(\overline{\Delta}\), \(\mathbb{S}\), \(\overline{\mathbb{S}}\), \(t\), \(\mathbb{T}\), and \(\overline{\mathbb{T}}\) has this property and contribute to gauge coupling unification if they remain light. 
These scalars can also impart threshold corrections at an intermediate scale. The study of precision gauge coupling unification requires two-loop analysis and one-loop threshold correction which is beyond the scope of this thesis.

In Chapter~(\ref{ch:6}), we explored the indirect implication of scalars in affecting low-energy observables like Yukawa couplings. The observationally inconsistent relationship \(Y_d = Y_e^T\) in non-supersymmetric \(SU(5)\) GUTs can be corrected at the one-loop level by switching on quantum corrections induced by heavy scalars in an otherwise minimally extended \su\, framework by singlet fermion(s). A mass hierarchy of at least two orders is needed between triplet and singlet(s) to yield ample threshold corrections. These quantum corrections can significantly alter conclusions about the minimal Yukawa sector, motivating further investigation in supersymmetric and non-supersymmetric variants of \(SU(5)\) and \(SO(10)\) GUTs.

\begin{table}[t]
    \centering
    \begin{tabular}{cccc}
    \hline
        ~~Pair of Scalar~~ & ~~$\big|B-L\big|$~~ & ~~Proton Decay:~~& ~~Bounds~~ \\
        & & ~~Tree/Loop~~ & ~~ in GeV~~ \\
    \hline\hline
     $\hat{\sigma} - \hat{T}$ & $\sfrac{2}{3}$ & Both & $10^{11}$  \\
     $\hat{T} - \hat{H}^*$& $\sfrac{2}{3}$ & Both & --- \\
     $\hat{T} - \hat{\Theta}$& $\sfrac{2}{3}$  & Loop & ---\\
     $\hat{H} - \hat{\Delta}$& $\sfrac{2}{3}$  & Both & $10^3$\\
     $\hat{\Delta} - \hat{\Delta}$& $\sfrac{2}{3}$ & Loop & --- \\
     $\hat{\Delta} - \hat{\Theta}^*$ & $\sfrac{2}{3}$ & Loop & --- \\
     $\hat{\Delta} - \hat{t}^*$& $\sfrac{2}{3}$ & Loop & ---   \\
     $\hat{\Theta} - \hat{t}$& $\sfrac{2}{3}$  & Loop & ---\\
     \hline
    \end{tabular}
    \caption{Pair of scalars residing in $16_{\hh}$ capable of mediating tree and loop level proton decays. The last column shows the bound on the scalar, which does not acquire $vev$ from the tree-level proton decays, where $\langle \hat{H} \rangle\,\sim\,246$ GeV has been used. "---" represents the respective pair of scalars is unconstrained from proton decay.}
    \label{tab:c7:scalarsum16H}
\end{table}


This thesis presents a systematic, careful, and comprehensive analysis of the scalar sector of grand unified models in its ability to directly influence the new physics within the context of concrete minimal and realistic $SO(10)$ models and their variations. The implications of scalars, together with the stringent bounds, are computed in Chapters~(\ref{ch:3}, \ref{ch:4}, and \ref{ch:5}).
 These stringent bounds are indirect and are due to various phenomenological implications. Another way to compute the bounds on the scalars is to evaluate one loop structure of the \so\,scalar potential~\cite{Bertolini:2009es,Bertolini:2012im,Bertolini:2013vta,Held:2022hnw,Jarkovska:2023zwv}, and obtain the bounds in terms of several parameters entering in the scalar potential. These studies have also revealed the impacts of quantum corrections in enabling breaking chains that are not allowed at the tree level. Through our computed bounds, one can constrain the parameters of the scalar potential, leading to a more precise understanding of the scalar potential of realistic $SO(10)$ models. 

The obtained bounds on the scalars are also important as scalars are associated with a physical energy scale that can affect gauge coupling unification~\cite{Dorsner:2012pp} and influence the running of baryon number and lepton number violating operators, including their flavour dependence. Such precision computations are necessary for constraining and detecting new physics.

As exemplified in Chapter~(\ref{ch:6}), the scalars play a significant role in constructing a minimal $SU(5)$ model, and the discussed $SU(5)$ model employs the Type-I seesaw mechanism to yield light neutrino masses. Another variant of this framework would be to implement the Type-II seesaw mechanism and to investigate whether loop-corrected Yukawa relations in a minimal $SU(5)$ model with $5_\hh$ and $15_\hh$ contributing to the Yukawa sector can yield a realistic fermion mass spectrum and whether the seesaw scale can be lowered compared to the Type-I scenario. Furthermore, a yet another variant would be to expand the fermion sector of the minimal $SU(5)$ model with a $\mathbf{24}$-fermion multiplet and to examine whether loop-corrected Yukawa relations, together with neutrino masses due to the Type-III seesaw mechanism, can yield the observed fermion mass spectrum. It would be interesting to check the possibility of achieving gauge coupling unification in both of the above-mentioned scenarios of the minimal $SU(5)$ model with either the Type-II or III seesaw mechanism or both.

The other extension of the loop-corrected Yukawa relations would be considering the variants of $SO(10)$ model. As mentioned, a single $10_{\hh}$-dimensional Higgs in the Yukawa sector yields generational mass degeneracy across all charged and neutral fermions. An interesting question arises about whether this issue could be rectified by introducing two or more copies of the $10_{\hh}$-dimensional irrep. An $SO(10)$ model with $10_{\hh}+120_{\hh}$ is also known to yield inconsistent tree-level mass relations. It would be a significant improvement if quantum corrections can render this model viable. Further, evaluating one-loop level Yukawa relations in an $SO(10)$ model based on $10_{\hh}+\overline{126}_{\hh}$-dimensional Higgs fields would be highly interesting and valuable for precision calculations. 
Furthermore, the supersymmetric extensions of the GUT model will be another case to consider in which there are known solutions to the doublet-triplet splitting problem.

\lettrine[lines=2, lhang=0.33, loversize=0.15, findent=0.15em]{S}CALARS SHALL PLAY a pivotal role in constructing a calculable, optimal/minimal, realistic, and renormalisable GUT model, testable through gravitational wave observations, baryon number-violating processes like nucleon decay and baryon-antibaryon oscillation, and can also explain observational cosmological evidence. The scalar sector of any GUT model is vast, and it offers a Pandora's box of many interesting and challenging outcomes.

\clearpage
\thispagestyle{empty}\newpage
\vspace*{\fill}
\begin{Huge}
\begin{center}
    \textbf{This page is intentionally left blank.}
\end{center}
\end{Huge}\vspace*{\fill}
\newpage

\appendix

\renewcommand{\chaptername}{Appendix}
\addtocontents{toc}{\protect\renewcommand{\protect\cftchappresnum}{\chaptername~}}

\pagestyle{fancy}
\fancyhf{} 
\fancyfoot[LO,RE]{\thepage} 

\fancyhead[RO]{%
  \begin{tikzpicture}[remember picture, overlay]
    \node[fill=gray!30, text width=3cm, align=center, font=\bfseries\Large, rotate=90, anchor=north east, minimum height=1.5cm, text=black] 
      at ([xshift=-1.5cm]current page.east) {\strut Appendix-\thechapter};  
  \end{tikzpicture}
}
\fancyhead[RE]{\nouppercase{\rightmark}} 
\fancyhead[LO]{\nouppercase{\rightmark}} 

\thispagestyle{empty}
\renewcommand{\chaptermark}[1]{ \markboth{Appendix \thechapter: #1}{}}
\renewcommand{\sectionmark}[1]{\markright{Section  \thesection: #1}}

\chapter{Algebra of $SO(N)$ and $SU(N)$}
\label{app:2}
\graphicspath{{100_Appendices/}}
The formal definition of a mathematical object called a \textit{group} is as follows. Let $G$ be a set (finite or infinite) and $o$ be a binary operation. $\left(G,o\right)$ is called a group if it satisfies the following axioms;
\begin{enumerate}
    \item \textbf{Closure.}\;\;
    $\forall$ $g_{1,2}\,\epsilon\,G$, $g_1\,o\,g_2\, \epsilon\, G$.
    \item \textbf{Existence of Identity}.\;\;
    $\exists$ $e\,\epsilon\, G$ such that $e\,o\,g\,=\,g\,o\,e\,=\,e$, $\forall\, g\, \epsilon\, G $.
    \item \textbf{Existence of Inverse}.\;\; $\forall\,g\,\epsilon\,G$\, $\exists\,g^{-1}\,\epsilon\,G$, such that $g\,o\,g^{-1}\,=\,g^{-1}\,o\,g\,=\,e$.
    \item \textbf{Associative}.\;\; $\forall\, g_{1,2,3}\,\epsilon\,G,$ $\left(g_1\,o\,g_2\right)\,o\,g_3\,=\,g_1\,\left(g_2\,o\,g_3\right)$. 
    \end{enumerate}

Below, we review some properties of $SO(N)$ and $SU(N)$ groups.
\section{$SO(N)$ in nutshell}
\label{sec:c2:SON}

This section overviews the algebra associated with the $O(N)$ group~\cite{Langacker:1980js,Slansky:1981yr,Wilczek:1981iz,Georgi:2000vve,Zee:2016fuk,Pal_2019}). $O(N)$ denotes a group of orthogonal transformations in an $N$-dimensional space. These transformations preserve the norm (length)  of the vector during rotations within an $N$-dimensional space. Let $R$ be the $N\times N$ matrix defined in an $N$-dimensional space with real elements. Then, the definition of orthogonal transformation is as follows;
\beqa{\label{eq:c2:SONdef}}
R^{T}\,R\eq \mathds{1} ,
\eeqa
where, $R^T_{ij}\,\equiv\,R_{ji}$ and ${\mathds{1}}$ is the $N\times N$ identity matrix. We call the orthogonal transformation a special-orthogonal transformation when the determinant of $R$ is $1$. Set of all $N\times N$ matrices satisfying Eq.~(\ref{eq:c2:SONdef}) with the additional constraint of $\left| R\right|\,=\, 1$ forms a group under the binary operation of matrix multiplication and is called $SO(N)$. Of course, a set of matrices with only $\left | R \right |\,=\,-1$ do not constitute a group as it lacks an identity element. Further, it is straightforward to compute the number of independent components in an orthogonal matrix, as shown below;
\beqa{\label{eq:c2:SONdog}}
\#_{\text{Independent elements}} \eq  \frac{N\big(N-1\big)}{2}. 
\eeqa
The number of independent components equals the total number of planes in an \(N\)-dimensional space where a rotation could be carried out. A $N$-dimensional vector transforms as follows;
\beqa{\label{eq:c2:trans}}
X_{i}\to X'_{i} \eq R_{ij}\,X_j,
\eeqa
where, $i$ is the i$^{\rm th}$-component of the N-dimensional vector in \so.  $R_{ij}$ is an element of $SO(N)$, which is a Lie group and thus can be parameterised as a perturbation near an identity element, as shown below:
\beqa{\label{eq:c2:Rij}}
R_{ij}\left(\omega\right)\,\eq \delta_{ij} + \omega_{ij}\,M_{ij},
\eeqa
where, $M_{ij}$ are the linearly independent generators of the group with $\omega_{ij}$s being the infinitesimal transformation parameters.  In a sense, \(M_{ij}\) represents the direction normal to the plane formed by the axis \(i-j\) through which a rotation occurs, and the magnitude of rotation is determined by \(\omega\).  Substituting Eq.~(\ref{eq:c2:Rij}) into Eq.~(\ref{eq:c2:SONdef}) constricts the $M_{ij}$s to obey the following commutation relation,  sometimes also referred to as the \textit{algebra} of the group:
\beqa{\label{eq:c2:algebra}}
\left[M_{ij},M_{kl}\right]\; \equiv\; M_{ij}\,M_{kl} - M_{kl}\,M_{ij} \eq -i \left( \delta_{ik}\,M_{jl} - \delta_{il}\,M_{jk} - \delta_{jk}\,M_{il} + \delta_{jl}\,M_{ik}\right)\nl
\eeqa
It is evident that when the indices \(i \neq j\) and \(k \neq l\), the commutator in Eq.~(\ref{eq:c2:algebra}) vanishes identically, meaning $\left[M_{12},M_{13}\right]\,=\,0$ and so forth. The number of generators that mutually commute is said to be the \textit{rank} of the algebra. For both $SO(2N)$ and $\,SO(2N+1)$, this rank remains the same and equals $N$. The matrix representation of $M_{ij}$ satisfying the algebra mentioned in Eq.~(\ref{eq:c2:algebra}) is as follows;
\beqa{\label{eq:c2:Rrep}}
\left(M_{ij}\right)_{mn} \eq i\,\left(\,\delta_{im}\,\delta_{jn} - \delta_{in}\,\delta_{jm}\,\right).
\eeqa
The above relation represents the $mn$-th element of the generator $M_{ij}$, with $i\,\neq\,j$. The general element $R$ of the group $SO(N)$ can be parameterised in terms of $\omega$ and $M$ as follows; 
\beqa{\label{eq:c2:Rdef}}
R\left(\omega\right) \eq \exp\left( -\frac{i}{2}\,\sum_{i,j=1}^{i,j=N}\omega_{ij}\,M_{ij}\right),
\eeqa
where, $\omega$ and $M$ are antisymmetric in $ij$ and the factor $\frac{1}{2}$ is due to the over counting. The expression above Eq.~(\ref{eq:c2:Rdef}) shows the transformation of a vector representation. One can also build similar transformation rules for tensors built from vector representations.

Further, let $r_{i}$ be the elements of a vector defined in an $2N$-dimensional space, where $1\leq\,i\,\leq\,2N$. Under orthogonal transformation the quadratic form $\sum_{i=1}^{i=2N}\,r_i^2$ remains invariant. Moreover, we can demand the following invariance:
\beqa{\label{eq:c2:spinor}}
\sum_{i=1}^{i=2N}\,r_i^2\eq \sum_{i=1}^{i=2N}\,\left(\gamma_{i}\,r_i\right)^2,
\eeqa
where $\gamma_i$ are the $2N$ Hermitian matrices. The above condition given in Eq.~(\ref{eq:c2:spinor}) is only satisfied if the $\gamma$ matrices fulfil the following constraint:
\beqa{\label{eq:c2:clifford}}
\left\{\gamma_i,\gamma_j\right\}\;\equiv\;\gamma_i\,\gamma_j + \gamma_j\,\gamma_i\eq 2\delta_{ij},
\eeqa
The algebra given in Eq.~(\ref{eq:c2:clifford}) is called \textit{Clifford Algebra} and says that different $\gamma$ matrices anticommute with each other. An example of matrices satisfying the Clifford algebra is the Pauli matrices. With the introduction of the Pauli matrices ($\sigma_{1,2,3}$), the correspondence between \(SU(2)\) and \(SO(3)\) becomes evident as follows;
\beqa{\label{eq:c2:SU2-SO3}}
x^2 + y^2 + z^2 \eq \left(\sigma_1\,x \,+\,\sigma_2\,y\,+\,\sigma_3\,z\right)^2.
\eeqa

The immediate consequence of the relation, given in Eq.~(\ref{eq:c2:clifford}), on $\gamma_i$s are: i) $Tr\big(\gamma_i\big)\,=\,1\big)$, ii) $\gamma_i^2\,=\,1$. Since the \(\gamma\) matrices are traceless and their square is the identity, it implies that their eigenvalues must be \(\pm 1\). Consequently, \(\gamma\) matrices are only defined in even dimensions.  

The transformation given in Eq.~(\ref{eq:c2:trans}) induces a transformation in $\gamma$ matrices as follows;
\beqa{\label{eq:c2:gammatrans}}
\gamma_{i}\;\to\; \gamma'_{i} \eq R_{ij}\,\gamma_j\nl
\left\{\gamma'_i,\gamma'_j\right\}\eq R_{ik}\,\left\{\gamma_k,\gamma_l\right\}\,R_{lj} \;=\;2\,\delta_{ij}.
\eeqa
Since the transformed \(\gamma'\) matrices also satisfy the same Clifford algebra as the untransformed ones, they must be related to the original \(\gamma\) matrices via a similarity transformation as follows;
\beqa{\label{eq:c2:similarity}}
\gamma'\eq {\mathcal{T}}\big(R\big)^{-1}\,\gamma\,{\mathcal{T}}\big(R\big).
\eeqa
Thus for every transformation $R$ in $2N-$dimensional space, we can associate ${\mathcal{T}}\big(R\big)$ in $2^{N}$-dimensional space which acts on an $2^N$ dimensional object and is referred to as a \textit{spinor representation}. The transformed object in that $2^{N}$ dimensional space is called a \textit{spinor}. The transformation rule of a spinor is as follows:
\beqa{\label{eq:c2:spinortrans}}
\varphi_i\;\to\;\varphi'_i\eq {\mathcal{T}}\big(R\big)_{ij}\,\varphi_j\eeqa

Analogous to Eq.~(\ref{eq:c2:Rij}), the ${\cal {T}}\big(R\big)$ can be expressed as follows.
\beqa{\label{eq:c2:Tij}}
{\cal{T}}_{ij}\eq \delta_{ij}\, + \frac{i}{2}\,\Omega_{ij}\,N_{ij},\,  
\eeqa
Again, $N_{ij}$ generates the rotation in $i-j$ plane and the amount of rotation is $\Omega_{ij}$. A solution of $N_{ij}$ satisfying the Eq.~(\ref{eq:c2:similarity}) is as follows;
\beqa{\label{eq:c2:solNij}}
N_{ij}\eq \frac{i}{4}\,\left[\gamma_i,\gamma_j\right]\nl
\Rightarrow {\cal{T}} \eq \exp\left(-\frac{i}{4}\,\Omega_{ij}\,N_{ij}\right).
\eeqa
${\cal{T}}$ is the $2^N\times 2^N$-dimensional matrix through which a $2^{N}$-dimensional spinor transforms.  As it is evident from the transformation rules of vector representation given in Eq.~(\ref{eq:c2:Rdef}) and spinor transformation rule given in Eq.~(\ref{eq:c2:solNij}), under a \(2\pi\) rotation the vector representation remains invariant, while the spinor representation picks up a negative sign. Moreover, a spinor representation is invariant under a \(4\pi\) rotation.

Along the similar lines, we can define $\gamma_F\,\equiv\,\big(-i\big)^{N}\,\prod_{i=1}^{i=2N}\,\gamma_i$, and $P_{\pm}\equiv \frac{1}{2}\,\left(1\pm \gamma_F\right)$. $P_{\pm}$ are called the projection operators with the properties; i) $P_{\pm}^2\,=\,P_{\pm}$, ii) $P_{+}\,P_{-}\,=\,P_{-}\,P_{+}\,=\,0$, and iii) $P_{+}\,+\,P_{-}\,=\,1$. Additionally, $\gamma_F$ anticommutes with $\gamma_{i}$ and commutes with $N_{ij}$. The action of \(P_{\pm}\) on \(\varphi'\) yields \(P_{\pm} \varphi' = P_{\pm} \exp\left(-\frac{i}{4} \Omega_{ij} N_{ij}\right) \varphi\). Since \(P_{\pm}\) commutes with \(N_{ij}\), \(P_{\pm} \varphi' = \exp\left(-\frac{i}{4} \Omega_{ij} N_{ij}\right) P_{\pm} \varphi\), thus, the projected spinor transforms similarly to the original spinor. Consequently, one spinor representation of \(SO(2N)\) decomposes into two irreducible spinor representations of dimension \(2^{N-1}\). The connection between $P_{+}\varphi\equiv \varphi_{+}$ and $P_{-}\varphi\,\equiv\,\varphi_{-}$ is established via charge conjugation operator (${\cal{C}}$). The exact nature of ${\cal{C}}$ depends upon the number of dimensions ($N$) and its relation with the spinorial generators is as follows~\cite{Zee:2016fuk}:
\beqa{\label{eq:c2:conjugation}}
{\cal{ C}}^{-1}\,\left(N_{ij}\,\left(1\pm \gamma_{\rm{F}}\right)\,\right)^*\,{\cal{C}} \eq -N_{ij}\,\left(1\pm \big(-1\big)^N\,\gamma_{F}\right).
\eeqa
The expression in Eq.~(\ref{eq:c2:conjugation}) specifies that for \(SO(8m)\), where \(m \in \mathbb{Z}^+\), \(\varphi_{\pm}^* = \varphi_{\pm}\); for \(SO(8m + 4)\), where \(m \in \mathbb{Z}^+\), \(\varphi_{\pm}^* = -\varphi_{\pm}\); and for \(SO(4m+2)\), where \(m \in \mathbb{Z}^+\), \(\varphi_{\pm}^* = \varphi_{\mp}\). In the case of \(SO(8m)\), the irreps are termed real; for \(SO(8m+4)\), they are termed pseudo-real; and for \(SO(4m+2)\), they are complex conjugates of each other.

The direct product of two spinorial irreps of \(SO(4m+2)\) can be expressed in terms of the antisymmetric product of \(\gamma\)'s mentioned in Eq.~(\ref{eq:c2:clifford}). For \(SO(2N)\), there are \(2N\) \(\gamma\)~matrices~\cite{Zee:2016fuk};
\beqa{\label{eq:c2:phidec}}
\varphi_{\pm} \times \varphi_{\pm} \eq [1] \oplus [3] \oplus [5] ... \oplus [2N-1],\nl
\varphi_{\pm}\times \varphi_{\mp} \eq [0] \oplus [2] \oplus [4] ... \oplus [2N],
\eeqa
where, $[i]$ represent antisymmetric product of $i$ number of $\gamma$ matrices. $SO(2N)$ comprises various subgroups derived from orthogonal groups and unitary groups. The decomposition of irreps of $SO(2M+2N) \to SO(2M) \times SO(2N)$ is as follows~\cite{Zee:2016fuk};
\beqa{\label{eq:c2:subSON}}
2_{+}^{M+N-1}&\to& \left(2_{+}^{M-1},2_{+}^{N-1}\right) \oplus \left(2_{-}^{M-1},2_{-}^{N-1}\right),\nl
2_{-}^{M+N-1}&\to& \left(2_{+}^{M-1},2_{-}^{N-1}\right) \oplus \left(2_{-}^{M-1},2_{+}^{N-1}\right),
\eeqa
where, $2_{\pm}^{M-1}$ and $2_{\pm}^{N-1}$ are the irreps of $SO(2M)$ and $SO(2N)$ respectively. Similarly, $U(N)$ can be naturally embedded in $SO(2N)$ and the decomposition of the fundamental irrep of $SO(2N)$ into the fundamental irrep of $U(N)$ is as follows:
\beqa{\label{eq:c2:SONtoSU5}}
2N&\to& N \oplus \bar{N}.\eeqa

So far, we have discussed the characteristics of the $SO(2N)$ group. Now, we will focus on a specific case, the $SO(10)$ group, and outline its properties, which are enumerated below:
\begin{enumerate}[label=(\roman*)]
    \item \so\, is a rank five group and exhibits complex spinorial irreps.
    \item The fundamental irrep in \so\, is of $10$- dimensional. It also exhibits a $32$-dimensional spinorial representation, which can be decomposed into two sixteen-dimensional irreps $16_{\pm}$.
    \item The tensor multiplication of two-$16$-plet of \so\, is as follows:
    \beqa{\label{eq:c2:16t16}}
    16_{\pm}\times 16_{\pm} \eq 10 \oplus 120 \oplus \overline{126},\nl
    16_{\pm}\times 16_{\mp}\eq 1 \oplus 45 \oplus 210,
    \eeqa
    where \(10\), \(120\), and \(126\) denote the number of independent degrees of freedom in one, three, and five-indexed antisymmetric tensors, respectively. Similarly, \(45\), \(210\), and \(1\) represent the number of independent degrees of freedom in two, four, and entirely antisymmetric tensors, respectively.
\end{enumerate}

We now conclude this section and discuss the properties of $U(N)$ group in the next section.

\section{$U(N)$ in nutshell }
\label{sec:c2:UN}
Unitary transformations are the sets of transformations that preserve the norm of a vector in a complex vector space. Let \(U\) be an \(N \times N\) matrix with complex elements; the definition of a unitary transformation is as follows:
\beqa{\label{eq:c2:UN}}
U^{\dagger}\,U\eq {\mathds{1}},\eeqa
where, $U^{\dagger}_{ij}\equiv U^{*}_{ji} $ and ${\mathds{1}}$ in $N\times N$ identity matrix. Likewise, similar to orthogonal matrices, unitary matrices also have a determinant of \(\pm 1\), that is, \(\left| U\right| = \pm 1\). All sets of \(N\times N\) matrices satisfying Eq.~(\ref{eq:c2:UN}) form a group under the binary operation of matrix multiplication and are termed as the unitary group $U(N)$. Furthermore, the set of all matrices with determinant 1, i.e., \(\left| U\right| =  1\), is known to form the special unitary group $(SU(N))$ under matrix multiplication. The number of independent components in a unitary matrix in $N^2$ and the number of independent components in $SU(N)$ matrix is, thus, $N^2-1$.

A general $SU(N)$ transformation can be parameterised as follows:
\beqa{\label{eq:c2:SUNlaw}}
U \eq \exp\left(-i\,\sum_{i=1}^{i=N^2-1} \theta_i\,T_i\right),
\eeqa
where, $\theta$ are real parameters determining the amount of transformation and $T_a$ are Hermitian and traceless generators. The fundamental representation $\varrho$ of $SU(N)$ transforms accordingly with the following relation;
\beqa{\label{eq:c2:SUNfunlaw}}
\varrho^i\,\to\,\varrho^{i'}\,\eq\,U^i_j\,\varrho^j.
\eeqa

Similarly, the conjugate representation, denoted as \(\bar{N}\), is defined as follows:
\beqa{\label{eq:c2:SUNconlaw}}
\varrho_i\,\equiv\,\big(\varrho^i\big)^*\,\to\,\varrho_i'\,\equiv\,U_i^j\,\varrho_j, \hspace{1cm}U_i^j\,=\,\big(U^i_j\big)^*. 
\eeqa

The tensor product of two $N$-dimensional fundamental irrep of $SU(N)$ transforms into two indexed antisymmetric and symmetric tensors as shown below;
\beqa{\label{eq:c2:NtN}}
\varrho^i\times \varrho^j \eq \Phi^{\left[ ij \right]}
 \oplus \Phi^{\{ij\}},\nl
 N\times_{A,S} N \eq \frac{N(N-1)}{2} \oplus \frac{N(N+1)}{2}.
 \eeqa
The tensor $\Phi^{[ij]}$ is antisymmetric in \(ij\) with \(\frac{N(N-1)}{2}\) independent degrees of freedom, while $\Phi^{\{ij\}}$ is symmetric in \(ij\) with \(\frac{N(N+1)}{2}\) independent degrees of freedom. Moreover, the symbol $\times_{A}$ $(\times_{S})$ denotes the antisymmetric (symmetric) tensor product.

The tensor product of fundamental and conjugate representation is as follows;
\beqa{\label{eq:c2:NtNb}}
\varrho^i\times \varrho_j \eq \Phi^{i}_j
 \oplus \delta^{i}_j\,\phi,\nl
 N\times \overline{N} \eq N^2-1 \oplus 1.
 \eeqa
The tensor \(\Phi^i_j\) is traceless with \(N^2-1\) independent degrees of freedom, while \(\phi\) is a singlet and has only one degree of freedom.

The decomposition of $SU(N)$ into its subgroups is as follows~\cite{Zee:2016fuk}:
\beqa{\label{eq:c2:SU(N)dec}}
SU(M+N)&\to& SU(M)\times SU(N) \times U(1)\nl
M+N &\to& \left(M,1, \frac{1}{M}\right) \oplus \left(1,N,-\frac{1}{N}\right),\nl
(M+N)\times \overline{\big(M+N\big)} \eq \left(\big(M+N\big)^2 - 1\right) \oplus 1 \nl &\to& \big(M^2-1,1,0\big) \oplus \big(1,N^2-1,0\big), \nl 
 &\oplus &\,  \big(M,\bar{N},\frac{1}{M}-\frac{1}{N}\big) \oplus \big(\bar{M},N,\frac{1}{N}-\frac{1}{M}\big) \oplus \big(1,1,0\big),\nl
 \big(M+N\big)\times_{A,S}\big(M+N\big) \eq \big(M_{A,S},1,\frac{1}{M}\big) \oplus \big(1,N_{A,S},\frac{1}{N}\big) \oplus \big(M,N,-\frac{1}{M}-\frac{1}{N}\big),\nl
\eeqa
where, $M_S$ $\big(M_A\big)$ are the symmetric (antisymmetric) components of $M$.

\newpage
\thispagestyle{empty}

\clearpage
\thispagestyle{empty}\newpage
\vspace*{\fill}
\begin{Huge}
\begin{center}
    \textbf{This page is intentionally left blank.}
\end{center}
\end{Huge}\vspace*{\fill}
\newpage


\chapter{Decay Width Expressions \& $\chi^2$ Definition}
\label{app:3}
\graphicspath{{100_Appendices/}}
\section{Proton Decay Width Relations}
The proton decay width expressions computed from chiral perturbation theory are as follows~\cite{Nath:2006ut,Beneito:2023xbk};{\label{app:c3:decaywidths}}
\beqa \label{eq:app:c3:decay_width}
\Gamma[p \to e_i^+\pi^0] &=& \frac{(m_p^2 - m_{\pi^0}^2)^2}{32\, \pi\, m_p^3 f_\pi^2} A^2 \times \left( \frac{1+\tilde{D}+\tilde{F}}{\sqrt{2}}\right)^2 \nl & & \Big(\left| \alpha\, y[u_1,d_1,e^C_i,u^C_1] + \beta\, y^{\prime *}[u^C_1,d^C_1,e^C_i,u^C_1] \right|^2 \Big. \nonumber \\
& + & \Big. \left| \alpha\, y^*[u^C_1,d^C_1,e_i,u_1] + \beta\, y^\prime[u_1,d_1,e_i,u_1]\right|^2 \Big), \nonumber\\
\Gamma[p \to \overline{\nu}\pi^+] &=& \frac{(m_p^2 - m_{\pi^\pm}^2)^2}{32\, \pi\, m_p^3 f_\pi^2} A^2 \left(1+\tilde{D}+\tilde{F} \right)^2 \nl & & \times \sum_{i=1}^3 \left| \alpha\, y^*[u^C_1,d^C_1,\nu_i, d_1] + \beta\, y^\prime[u_1,d_1,\nu_i, d_1] \right|^2\,, \nonumber\\
\Gamma[p \to e_i^+K^0] &=& \frac{(m_p^2 - m_{K^0}^2)^2}{32\, \pi\, m_p^3 f_\pi^2} A^2\, \frac{1}{2} \Big[\Big|C_{Li}^- - C_{Ri}^- +\frac{m_p}{m_B}(\tilde{D}-\tilde{F})\left(C_{Li}^+ - C_{Ri}^+\right)\Big|^2 \Big. \nonumber \\
& + & \Big. \Big|C_{Li}^- + C_{Ri}^- +\frac{m_p}{m_B}(\tilde{D}-\tilde{F})\left(C_{Li}^+ + C_{Ri}^+\right)\Big|^2 \Big]\,, \nonumber\\
\Gamma[p \to \overline{\nu}K^+] &=& \frac{(m_p^2 - m_{K^\pm}^2)^2}{32\, \pi\, m_p^3 f_\pi^2}\, A^2\, \sum_{i=1}^3 \Big| \frac{2\tilde{D}}{3} \frac{m_p}{m_B} C^\nu_{L1i} + \left(1+\frac{\tilde{D}+3\tilde{F}}{3} \frac{m_p}{m_B}\right) C^\nu_{L2i} \Big|^2, \nonumber\\
\Gamma[p \to e_i^+\eta] &=& \frac{(m_p^2 - m_{\eta}^2)^2}{32\, \pi\, m_p^3 f_\pi^2}\, A^2\, \frac{1}{6}\Big[ \Big|C_{L i}^+ (1-\tilde{D}+3\tilde{F}) - 2 C_{L i}^-\Big|^2 \, \nonumber\\
& + & \Big. \Big| C_{R i}^+ (1-\tilde{D}+3\tilde{F}) - 2 C_{R i}^-\Big|^2\Big]\, \Big.,\eeqa
where,
\beqa \label{C_dw}
C^\pm_{L i} &=& \alpha\, y^*[u^C_1,d^C_2,e_i,u_1] \pm \beta\, y^\prime[u_1,d_2,e_i,u_1]\,,
\nonumber\\
C^\pm_{R i} &=& \alpha\, y[u_1,d_2,e^C_i,u^C_1] \pm \beta\, y^{\prime *}[u^C_1,d^C_2,e^C_i,u^C_1]\,,
\nonumber\\
C^\nu_{L1i} &=& \alpha\, y^*[u_1^C,d_2^C,\nu_i,d_1] + \beta\, y^\prime[u_1,d_2,\nu_i,d_1]\,,\nonumber \\
C^\nu_{L2i} &=& \alpha\, y^*[u_1^C,d_1^C,\nu_i,d_2] + \beta\, y^\prime[u_1,d_1,\nu_i,d_2]\,. \eeqa
Here, $m_H$ denotes the mass of hadron $H$ ($H=p,\pi^0,\pi^\pm,K^0,K^\pm,\eta$), $m_B$ is average baryon mass and $f_\pi$ is pion decay constant. $\alpha$, $\beta$, $\tilde{D}$ and $\tilde{F}$ are the parameters of the chiral Lagrangian. The factor $A$ accounts for the renormalisation effects in hadronic matrix elements from the weak scale to the $m_p$.

\section{Coefficients for Computing Proton Decays}
The definition of $C$ given in Eq.~(\ref{C_dw}) is slightly modified to compute the branching pattern for proton decays in non-renormalisable \so\, model and is shown as follows;
\beqa \label{eq:app4:C_dw}
C^\pm_{L i} &=& \alpha\, c_3^*[u^C_1,d^C_2,u_1,e_i] \pm \beta\, c_1[u_1,d_2,u_1,e_i]\,,
\nonumber\\
C^\pm_{R i} &=& \alpha\, c_2^*[u_1,d_2,u^C_1,e^C_i] \pm \beta\, c_4[u^C_1,d^C_2,u^C_1,e^C_i]\,,
\nonumber\\
C^\nu_{L1i} &=& -\alpha\, c_3^*[u_1^C,d_2^C,d_1,\nu_i] - \beta\, c_1[u_1,d_2,d_1,\nu_i]\,,\nonumber \\
C^\nu_{L2i} &=& -\alpha\, c_3^*[u_1^C,d_1^C,d_2] - \beta\, c_1[u_1,d_1,d_2,\nu_i]\,. \eeqa

\section{Solution for ${ 10}_{\hh}+ \overline{ 126}_{\hh}$ model}
\label{app:c3:fit}
 This section explains elementary aspects of fermion mass fittings in a class of renormalisable \so\,model with $10_{\hh}$ and $\overline{126}_{\hh}$ participating in the Yukawa sector. As pointed out in chapter~(\ref{ch:3}), the model requires complex $10_{\hh}$ instead of real $10_{\hh}$. The complexification of $10_{\hh}$ introduces extra couplings in the Yukawa sector~($H$ and $\tilde{H}$). The extra Yukawa coupling $(\tilde{H})$ can be forbidden by imposing a Peccei-Quinn symmetry under which $\fs\,\to\,\exp(i\alpha)\,\fs$ and $\big(10_{\hh}, \overline{126}_{\hh}\big)\,\to\, \exp(-i2\alpha)\,\big(10_{\hh}, \overline{126}_{\hh}\big)$. 

The Yukawa coupling of SM-like Higgs doublet with SM fermions can be inferred from Eqs.~(\ref{eq:c2:10/5}, \ref{eq:c3:126/5}, \ref{eq:c3:126/45:2}). As the doublets, $D$ and $\overline{D}$, resides both in $10_{\hh}$ and $\overline{126}_{\hh}$, one can define $h_{u}$ and $h_{d}$ as their lightest linear combination as follows:
\beqa{\label{eq:app3:huhd}}
D_{10_{\hh}} \eq \beta_{1}\, h_{d}\;\;\text{and}\;\; D_{\overline{126}_{\hh}}\;=\; \beta_{2}\,h_{u}\nl
\overline{D}_{10_{\hh}} \eq \alpha_{1}\, h_{d}\;\;\text{and}\;\; \overline{D}_{\overline{126}_{\hh}}\;=\; \alpha_{2}\,h_{u}\eeqa
where, $\alpha_{i}$ and $\beta_{i}$ is subjected to follow $|\alpha_{1}^2| + |\alpha_{2}^2|\,\leq\,1\,\geq\,|\beta_{1}^2| + |\beta_{2}^2|$

The effective Yukawa relations can be written using the Eqs.~(\ref{eq:c2:10/5}, \ref{eq:c3:126/5}, \ref{eq:c3:126/45:2}) as follows:
\beqa \label{Eff_Yuk}
Y_d  &=& H^\prime + F^\prime \,,~~Y_u  =  r\,(H^\prime + s F^\prime)\,, \nonumber \\
Y_e  &=& H^\prime - 3 F^\prime \,, ~~Y_ \nu = r\,(H^\prime -3 s F^\prime)\,, ~~M_R  = v_S^\prime\, F^\prime \,, \eeqa
where,
\beqa
H^\prime \eq i\, \alpha_{1}\,2\sqrt{2}\,H,\;\;F^\prime \,=\, i\,\alpha_{2}\,2\sqrt{\frac{2}{3}}\,F,\;\;r\,= \,\frac{\beta_1}{\alpha_1},\;\;s\,=\,\frac{\alpha_1\beta_2}{\alpha_2\beta_1},\;\;v^\prime_{s}\,=\,\frac{v_s}{\alpha_2}\nl\eeqa

$v_{s}$ is the $vev$ of the singlet field $\sigma$~(cf. Tab.~(\ref{tab:c2:scalars})). The light neutrino mass matrix is obtained as \(M_\nu = - v_u^2\, Y_\nu\,M_R^{-1}\, Y_\nu^T\), where \(v_{u,d} = \langle h_{u,d} \rangle\), \(v_u^2 + v_d^2 \equiv v^2 = (174\, {\rm GeV})^2\) and \(v_u/v_d \equiv \tan\beta = 1.5\). The mass matrices for the charged fermions are given by \(M_{d,e} \equiv v_d\, Y_{d,e}\) and \(M_u \equiv v_u Y_u\).





The viability of the model in its  capacity to yield observed charged and neutral fermion masses and mixing angles is done using the $\chi^2$ optimisation method whose definition is as  follows:
\beqa{\label{eq:app2:chisqdef}}
\chi^2 \eq \sum_{i}\Bigg(\frac{O^{\rm{th}}_{i}\,-\,O^{\rm{exp}}_{i}}{\sigma_{i}}\Bigg)^2\eeqa
The index \( i \) runs over a total of 19 observables, including nine charged lepton masses, four quark mixing parameters, two neutral fermion mass-squared differences, and three mixing angles and one Dirac phase~(or two Majorana phases) in the neutrino sector. The theoretically computed value of the \( i \)-th observable is denoted by \( O_{i}^{\text{th}} \), while \( O_{i}^{\text{exp}} \) represents the corresponding experimental value at the GUT scale. The experimental values of these 19 observables at the GUT scale can be found in~\cite{Mummidi:2021anm}. The uncertainty associated with the \( i \)-th observable, \( \sigma_i \), is taken as $30\%$ for the light quarks (\( y_u \), \( y_d \), and \( y_s \)) and $10\%$ for the other observables. The fits performed in~\cite{Mummidi:2021anm} also account for the baryon-to-photon ratio but only consider three mixing angles in the neutrino sector.

Below, the necessary parameters of the best-fit solution obtained in the recent work~\cite{Mummidi:2021anm} are given. One finds the best-fit parameters for the Yukawa couplings as;
\beqa \label{input_prm}
H^\prime &=& \left(
\begin{array}{ccc}
 0.00023 & 0 & 0 \\
 0 & -0.04811 & 0 \\
 0 & 0 & -5.79504 \\
\end{array}
\right) \times 10^{-3}\,, \nonumber \\
F^\prime &=&  \left(
\begin{array}{ccc}
 -0.0088+0.0178 i & 0.0475\, -0.0889 i & 0.4635\, +0.6797 i \\
 0.0475\, -0.0889 i & 1.1279\, +0.5108 i & -1.2218-2.5921 i \\
 0.4635\, +0.6797 i & -1.2218-2.5921 i & 5.4683\, -5.9856 i \\
\end{array}
\right)\times 10^{-4}\,. \nl\eeqa
Also, \(r =  77.4189\), \(s = 0.3140 - 0.0282\, i\) and \(v_S^\prime = 9.84 \times 10^{14}\,{\rm GeV}\) \\

The unitary matrices \(U_f\) and \(U_{f^C}\) diagonalise \(M_f\) (\(f=d,u,e,\nu\)) such that \(U_f^\dagger M_f M_f^\dagger U_f = U_{f^C}^\dagger M_f^\dagger M_f U_{f^C} \equiv D^2_f\). Since \(M_f\) are symmetric, it follows that \(U_f = U_{f^C}^*\). The numerically obtained \(U_f\) are as follows.\beqa \label{input_Uf}
U_d &=& U_{d^C}^* =  \left(
\begin{array}{ccc}
 -0.9497-0.2658 i & -0.0895+0.1388 i & -0.0101-0.0116 i \\
 -0.1053+0.1272 i & -0.5434-0.8213 i & 0.0276\, +0.0463 i \\
 -0.0157 & 0.0538 & 0.9984 \\
\end{array}
\right)\,, \nonumber\\ 
U_u &=& U_{u^C}^* =  \left(
\begin{array}{ccc}
 -0.4028-0.8952 i & 0.1456\, +0.123 i & -0.003-0.0034 i \\
 -0.0213-0.1895 i & -0.5203-0.8323 i & 0.0086\, +0.0135 i \\
 -0.0016 & 0.0165 & 0.9999 \\
\end{array}
\right)\,, \nonumber\\ 
U_e &=& U_{e^C}^* = \left(
\begin{array}{ccc}
 -0.3995-0.916 i & -0.0093+0.0126 i & 0.0115\, +0.03 i \\
 -0.0161-0.0074 i & -0.2511-0.9618 i & -0.0268-0.1043 i \\
 0.031 & -0.108 & 0.9937 \\
\end{array}
\right)\,, \nonumber\\ 
U_\nu & = & \left(
\begin{array}{ccc}
 -0.4683+0.689 i & 0.3113\, -0.4343 i & 0.0721\, -0.1233 i \\
 0.151\, -0.2689 i & 0.3553\, -0.5592 i & -0.357+0.5818 i \\
 0.4591 & 0.5249 & 0.7168 \\
\end{array}
\right)\,. \eeqa
The matrices $H^\prime$, $F^\prime$ and $U_f$ are used to compute the proton decay spectrum as explained in the sections~(\ref{sec:c3:results_model}) and spectrum of sextets in section~(\ref{sec:c5:results}). 
\newpage
\thispagestyle{empty}
\clearpage
\thispagestyle{empty}\newpage
\vspace*{\fill}
\begin{Huge}
\begin{center}
    \textbf{This page is intentionally left blank.}
\end{center}
\end{Huge}\vspace*{\fill}
\newpage

\thispagestyle{empty}
\chapter{Loop Computation}
{\label{app:6}}
\graphicspath{{100_Appendices/}}

\section{Loop Computations}
\begin{figure}[t]
    \centering
    \includegraphics[width=0.8\linewidth]{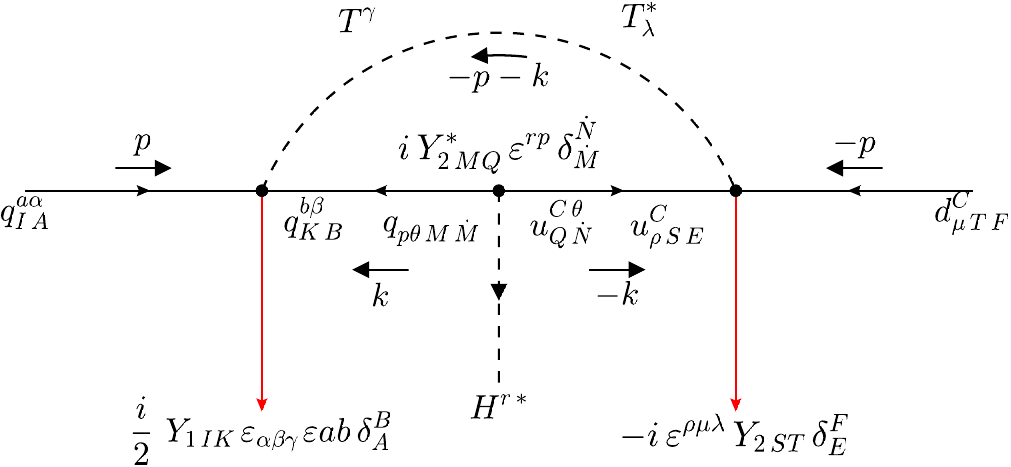}
    \caption{Vertex Correction to $Y_d$. The uppercase Greek alphabet represents Generation indices and spinor indices.}
    \label{fig:c6:ykc1}
\end{figure}

This section shows the computation of loop diagrams used in Chapter~(\ref{ch:6}) using the various conventions used in the main text. The diagram given in Fig.~(\ref{fig:c6:ykc1}) shows the triplet-induced correction to $Y_{d}$ and its computation is as follows;
\beqa{\label{eq:c6:app:compyk1}}
-i\Sigma_{IT} \eq \left(\frac{i}{2}\,Y_{1\,IK}\,\varepsilon_{\alpha\beta\gamma}\,\varepsilon_{ab}\,\delta^B_A\right)\,\delta_{KM}\,\delta^{\beta}_{\theta}\,\delta^{b}_p\,\left(i\,Y^*_{2\,MQ}\,\varepsilon^{rp}\,\delta^{\Dot{N}}_{\Dot{M}}\right)\nl
& & \,\times\;\delta^{\theta}_{\rho}\,\delta_{QS}\,\left(-i\,\varepsilon^{\rho\mu\lambda}\,Y_{2\,ST}\,\delta^F_E\right)\,\delta^{\gamma}_{\lambda}\nl
& & \,\times\; \int \frac{d^4\,k}{\big(2\pi\big)^4}\,\frac{i\,k.\sigma_{B\Dot{M}}}{k^2}\,\frac{i\,k.\overline{\sigma}^{\Dot{N}E}}{k^2}\,\frac{i}{\big(p+k\big)^2-M^2_T}
\eeqa
where $-i\Sigma$ is the amplitude of the diagram.
Using $k.\overline{\sigma}\,=\,k_{\Dot{\mu}}\,\overline{\sigma}^{\Dot{\mu}}$, $k_{\Dot{\mu}}\,k_{\Dot{\nu}}\to \frac{g_{\Dot{\mu}\Dot{\nu}}}{D}\,k^2$ and $\sigma_{\Dot{\mu}\,B\Dot{M}}\,\overline{\sigma}^{\Dot{\mu}\,\Dot{N}E}\,=\,2\,\delta^E_B\,\delta^{\Dot{N}}_{\Dot{M}}$ and along with different contractions, the expression written in the Eq.~(\ref{eq:c6:app:compyk1}) gets reduced to the following:
\beqa{\label{eq:c6:app:comp2}}
\Sigma_{IT} \eq -\left(Y_1\,Y_2^*\,Y_2\right)_{IT}\,\delta^F_A\,\delta_{\alpha}^{\mu}\,\delta_a^r\,f[M^2_T,0]
\eeqa
where the loop function has already been defined in Eq.~(\ref{eq:c6:LF_f}).

\begin{figure}[t]
    \centering
    \includegraphics[width=0.8\linewidth]{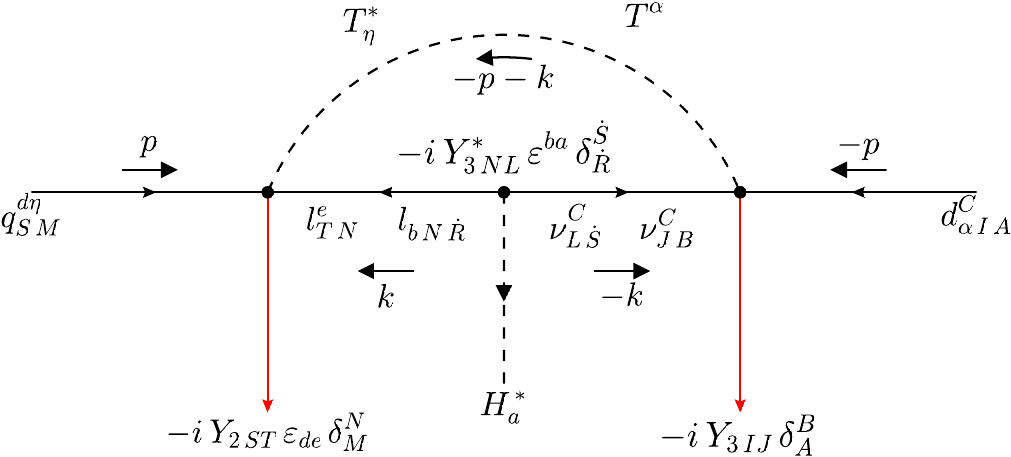}
    \caption{Diagram depicting correction to $Y_d$ with the contributions from $\nu^C$.}
    \label{fig:c6:nuc}
\end{figure}

Further, the contribution of $\nu^C$ and $T$ to the $Y_d$ can be inferred and computed from the diagram shown in Fig.~(\ref{fig:c6:nuc}), where we still have not fixed the generation of singlets $(\nu^C)$;
In the following computation, we assume that $\nu^C$(s) is (are) in the physical basis:
\beqa{\label{eq:c6:app:nuc}}
-i\Sigma_{SI} \eq \sum_J\,\sum_L\,\left(-i\,Y_{2\,ST}\,\varepsilon_{de}\,\delta_M^N\right)\,\delta^e_b\,\delta_{TM}\,\left(-i\,Y^*_{3\,NL}\,\varepsilon^{ba}\,\delta_{\Dot{T}}^{\Dot{S}}\right)\nl
& & \,\times\;\delta_{LJ}\,\left(-i\,Y_{3\,IJ}\,\delta^B_A\right)\,\delta^{\alpha}_{\eta}\nl
& & \,\times\; \int \frac{d^4\,k}{\big(2\pi\big)^4}\,\frac{i\,k.\sigma_{N\Dot{R}}}{k^2}\,\frac{i\,k.\overline{\sigma}^{\Dot{S}B}}{\big(k^2-M^2_{N\,J}\big)}\,\frac{i}{\big(p+k\big)^2-M^2_T}\nl
\Sigma_{SI}\eq \,\sum_{L}\left(\big(Y_2\,Y_3^*\big)_{SL}\,Y^T_{3\,L\,I}\right)\,\delta^{\alpha}_{\eta}\,\delta^a_d \,\delta^N_A\,f[M^2_T,M^2_{N_{L}}]
\eeqa

We also outline the correction computation to the $Y_d$ induced by gauge bosons, as shown below:
\begin{figure}[t]
    \centering
    \includegraphics[width=0.8\linewidth]{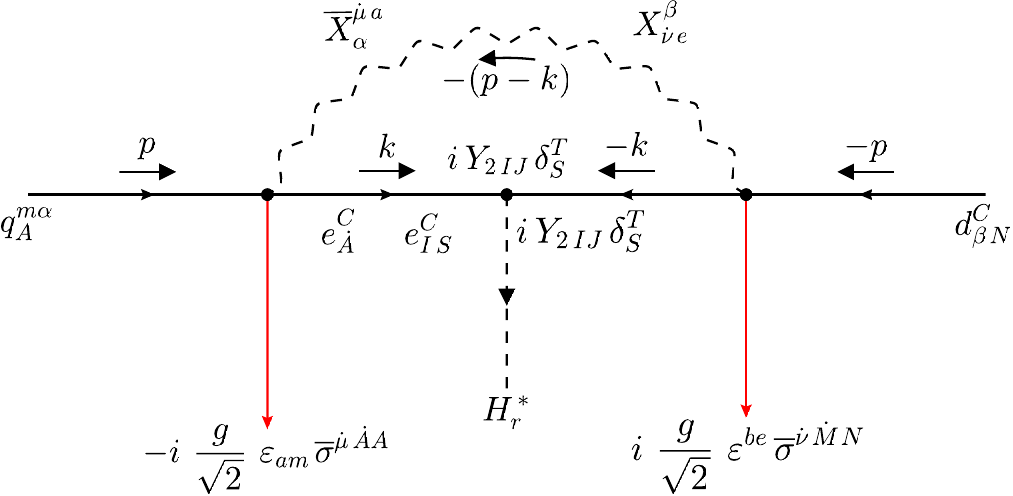}
    \caption{Diagram depicting correction to $Y_d$ induced by the heavy gauge bosons $X$ and $\overline{X}$.}
    \label{fig:c6:gbc}
\end{figure}
\beqa{\label{eq:c6:app:gbc}}
-i\Sigma_{IJ}\eq \left(-i\frac{g}{\sqrt{2}}\right)\,\left(i\,Y_{2\,IJ}\right)\,\left(i\frac{g}{\sqrt{2}}\right)\,\varepsilon_{am}\,\delta^r_b\,\varepsilon^{be}\,\delta^a_e\,\delta^{\beta}_{\alpha}\,\delta^T_S\nl
& & \times \int \frac{d^4\,k}{\big(2\pi\big)^4}\,\overline{\sigma}^{\Dot{\mu}\,\Dot{A}\,A}\,\left(\frac{i\,k.\sigma^{S\Dot{A}}}{k^2}\right)\,\left(\frac{ik.\overline{\sigma}^{\Dot{M}T}}{k^2}\right)\,\left(-\sigma_{\Dot{\nu}}^{N\Dot{M}}\right)\,\frac{-i}{\big(p-k\big)^2-M^2_X}\nl
\Sigma_{IJ} \eq 2\,g^2\,Y_{2\,IJ}\,\delta_m^r\,\delta_A^N\,f[M^2_X,0],
\eeqa
where the definition of the loop function, $f$, has already been defined in the Eq.~(\ref{eq:c6:LF_f}). 

\begin{figure}[t!]
    \centering
    \includegraphics[width=0.6\linewidth]{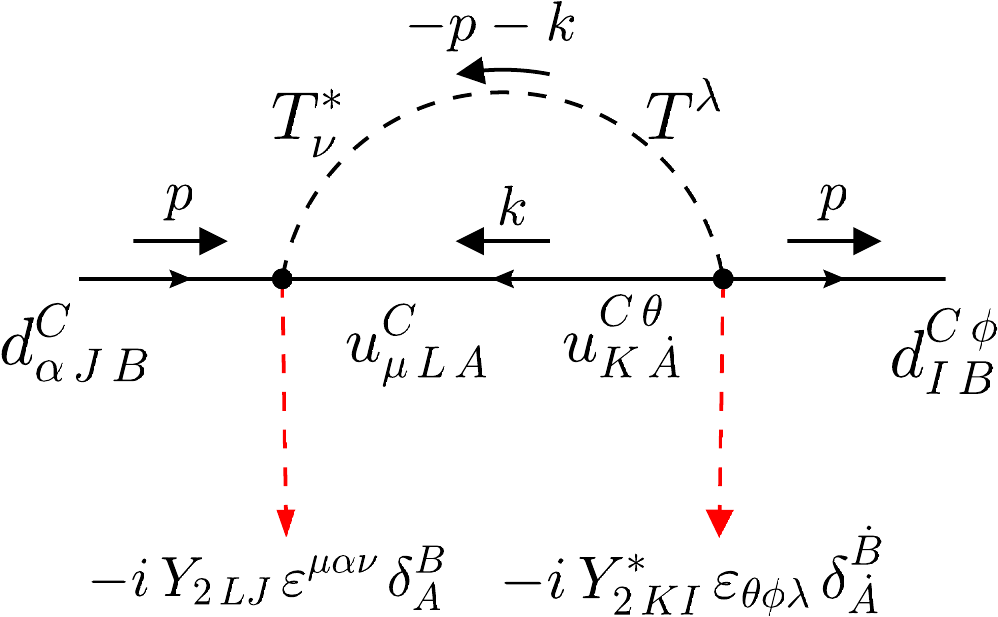}
    \caption{Self energy correction to the $d^C$. }
    \label{fig:c6:app:wfr2}
\end{figure}

We show the steps for a template exercise to compute the wave function renormalisation for the Feynman graph, shown in Fig.~(\ref{fig:c6:app:wfr2}) contributing to the correction in $Y_d$.
\beqa{\label{eq:c6:app:wfr}}
-i\Sigma_{IJ}\big(p.\sigma) \eq \left(-i\,Y_{2\,KI}^*\,\varepsilon_{\theta\phi\lambda}\,\delta_{\Dot{A}}^{\Dot{B}}\right)\,\delta_{KL}\,\delta^{\theta}_{\mu}\,\nl
& & \times\, \left[\int \frac{d^4k}{\big(2\pi\big)^4}\, \frac{ik.\sigma_{A\Dot{A}}}{k^2}\,\frac{i}{\big(p+k\big)^2-M^2_T}\,\right] \times\, \left(-i\,Y_{2\,LJ}\,\varepsilon^{\mu\alpha\nu}\,\delta_A^B\right)\,\delta_\nu^{\lambda}\nl
\eq -2\big(Y_2^{\dagger}\,Y_2\big)_{IJ}\,\delta^{\alpha}_{\phi}\,\delta^{\Dot{B}}_{\Dot{A}}\,\delta^B_A  \times\, \big(-i\big)\,\int_0^1\, dx\, x\,p.\sigma_{A\Dot{A}}\,\log\left(\frac{x\,M^2_T}{\mu^2}\right),\nl
\eeqa
Wave function renormalisation factor $k_{d^C}$ is defined as follows;
\beqa{\label{eq:c6:def}}
\big(k_{d^C}\big)_{IJ} &\equiv & \frac{\partial\Sigma}{\partial\big(p.\sigma_{IJ}\big)}\Big|_{p^2=0}\;=\; - 2 \left(Y_2^\dagger Y_2\right)_{IJ} h[M_T^2,0],
\eeqa
where $h$ is a loop function already defined in the Eq.~\eqref{eq:c6:LF_h}.

\section{$\chi^2$ Optimisation}

The \(\chi^2\) function defined in Eq.~(\ref{eq:app2:chisqdef}) for solution (I) includes 9 diagonal charged fermion Yukawa couplings and 4 CKM parameters. To get the input values of these parameters at the GUT scale, the SM Yukawa couplings are evolved from \(\mu = M_t\) (where \(M_t\) is the top pole mass) to \(\mu = M_{\rm GUT} = 10^{16}\) GeV using 2-loop RGE equations in the \(\overline{\rm MS}\) scheme, following the method outlined in~\cite{Mummidi:2021anm}. The 2-loop SM RG equations are computed using the {\sc PyR$@$TE}-3 package~\cite{Sartore:2020gou}. The values of the SM Yukawa and gauge couplings at \(\mu = M_t\) are taken from~\cite{Surya:2020ydm}. The RGE extrapolated values at the GUT scale are listed as \(O_{\rm exp}\) in Tab.~(\ref{tab:c6:tab1}). For standard deviations, \(\pm 30\%\) is used for light quark Yukawa couplings \((y_{u,d,s})\) and \(\pm 10\%\) for the other observables.

By choosing a basis in Eq.~(\ref{eq:c6:LY}), \(Y_1\) is set to be diagonal and real. The right-handed neutrino mass matrix, \(M_N\), in general, \(N\)-flavour case, can also be chosen to be real and diagonal simultaneously. \(Y_{2,3}\) are complex in this basis, and the matrices \(Y_{u,d,e}\) are computed and then diagonalised to obtain the nine diagonal Yukawa couplings and quark mixing parameters. These quantities are fitted to the extrapolated data at \(\mu = M_{\rm GUT}\) by minimizing the \(\chi^2\) function. \(M_X\) is set to \(M_{\rm GUT}\) and \(g=0.53\),  at \(\mu = 10^{16}\) GeV. By fixing \(M_T\) and \(M_N\) to specific values, the \(\chi^2\) is minimised with a constraint \( |(Y_{1,2,3})_{ij}| < \sqrt{4 \pi} \) on all input Yukawa couplings to ensure they remain within perturbative limits \cite{Allwicher:2021rtd}. This procedure is repeated for several values of \(M_T\) and \(M_N\). The distribution of the minimized \(\chi^2\) (\(\equiv \chi^2_{\rm min}\)) is displayed in Fig.~(\ref{fig:c6:fig2}).

\newpage
\pagestyle{fancy}
\fancyhf{} 
\fancyfoot[LE,RO]{\thepage} 

\fancyhead[RO]{%
  \begin{tikzpicture}[remember picture, overlay]
    \node[fill=gray!30, text width=3cm, align=center, font=\bfseries\Large, rotate=90, anchor=north east, minimum height=1.5cm, text=black] 
      at ([xshift=-1.5cm]current page.east) {\strut Bibliography};  
  \end{tikzpicture}
}
\clearpage
\thispagestyle{empty}\newpage
\vspace*{\fill}
\begin{Huge}
\begin{center}
    \textbf{This page is intentionally left blank.}
\end{center}
\end{Huge}\vspace*{\fill}
\newpage


\setstretch{1}

\thispagestyle{empty}
\renewcommand{\bibname}{References}
\printbibliography[heading=bibintoc]

\pagestyle{fancy}
\fancyhf{}
\fancyhead[RE]{\bfseries \bibname}
\fancyhead[LO]{\bfseries \bibname}
\fancyhead[LE,RO]{\thepage}

\end{document}